\documentclass[acmsmall,screen,nonacm]{acmart}

\usepackage{thmtools}

\usepackage{stmaryrd,mathtools} %
\SetSymbolFont{stmry}{bold}{U}{stmry}{m}{n}

\usepackage[inline]{enumitem}
\usepackage{algorithm}
\usepackage{algorithmic}

\newlist{proofrules}{itemize}{1}
\setlist[proofrules]{label=\textbullet,leftmargin=1.5em,
  topsep=5pt,itemsep=5pt,parsep=0pt}
\newlist{proofbranches}{description}{1}
\setlist[proofbranches]{style=unboxed,font=\normalfont,
  leftmargin=1.8em,labelindent=0pt,labelsep=0.5em,
  topsep=3pt,itemsep=4pt,parsep=0pt}
\newlist{proofsubcases}{enumerate}{3}
\setlist[proofsubcases]{leftmargin=2em,
  topsep=2pt,itemsep=2pt,parsep=0pt}
\setlist[proofsubcases,1]{label=(\alph*),ref=(\alph*)}
\setlist[proofsubcases,2]{label=(\roman*),ref=(\roman*)}
\setlist[proofsubcases,3]{label=(\Alph*),ref=(\Alph*)}

\usepackage{subcaption}

\usepackage{multirow}

\usepackage{cleveref}

\AddToHook{env/conjecture/begin}{\crefalias{theorem}{conjecture}}
\AddToHook{env/proposition/begin}{\crefalias{theorem}{proposition}}
\AddToHook{env/lemma/begin}{\crefalias{theorem}{lemma}}
\AddToHook{env/corollary/begin}{\crefalias{theorem}{corollary}}
\AddToHook{env/example/begin}{\crefalias{theorem}{example}}
\AddToHook{env/definition/begin}{\crefalias{theorem}{definition}}

\makeatletter
\newcommand*{\STLDeferTheoremAnchor}{%
  \let\stl@saved@hyperstep\hyper@refstepcounter
  \def\hyper@refstepcounter##1{%
    \let\hyper@refstepcounter\stl@saved@hyperstep
    \hyper@makecurrent{##1}%
    \global\let\Hy@dth@currentHref\@currentHref
    \Hy@theorem@makelinktarget{\Hy@dth@currentHref}%
  }%
}
\@ifundefined{Hy@theorem@makelinktarget}{}{%
  \AddToHook{cmd/@thm/before}{\STLDeferTheoremAnchor}%
}
\makeatother

\graphicspath{{fig/}}

\newif\ifenableproof
\enableprooftrue   %

\newcommand{\updownarrowDisplayT}[5]{%
  \mathrel{%
    \begingroup
    \setlength{\arraycolsep}{0.2pt}%
    \renewcommand{\arraystretch}{0.5}%
    \begin{array}{r l c}
      \smash{\scriptscriptstyle #1} & \smash{\scriptscriptstyle #2} & \multirow{3}{*}{$\mkern-1mu \textstyle #5$} \\
      \scriptstyle\uparrow & \scriptstyle\downarrow & \\
      \smash{\scriptscriptstyle #3} & \smash{\scriptscriptstyle #4} &
    \end{array}%
    \endgroup
  }%
}

\newcommand{\updownarrowInlineT}[5]{%
  \mathrel{%
    \begingroup
      \setlength{\arraycolsep}{0.2pt}%
      \renewcommand{\arraystretch}{0.5}%
      \vcenter{%
        \hbox{$%
          \begin{array}{r c l l}
            \smash{\scriptscriptstyle #1} &
            \multirow{2}{*}{$\uparrow\mkern-2mu\downarrow$} &
            \smash{\scriptscriptstyle #2} & \multirow{2}{*}{$\mkern-1mu #5$} \\
            \smash{\scriptscriptstyle #3} & &
            \smash{\scriptscriptstyle #4} &
          \end{array}%
        $}%
      }%
    \endgroup
  }%
}

\newcommand{\updownarrowInlineScriptT}[5]{%
  \mathrel{%
    \begingroup
      \setlength{\arraycolsep}{0.2pt}%
      \renewcommand{\arraystretch}{0.5}%
      \vcenter{%
        \hbox{$%
          \begin{array}{r c l l}
            \smash{\scriptscriptstyle #1} &
            \multirow{2}{*}{$\scriptstyle\uparrow\mkern-2mu\downarrow$} &
            \smash{\scriptscriptstyle #2} & \multirow{2}{*}{$\mkern-1mu \scriptstyle #5$} \\
            \smash{\scriptscriptstyle #3} & &
            \smash{\scriptscriptstyle #4} &
          \end{array}%
        $}%
      }%
    \endgroup
  }%
}

\newcommand{\updownarrowSymbolT}[5]{%
  \mathchoice
    {\updownarrowDisplayT{#1}{#2}{#3}{#4}{#5}}%
    {\updownarrowInlineT{#1}{#2}{#3}{#4}{#5}}%
    {\updownarrowInlineScriptT{#1}{#2}{#3}{#4}{#5}}%
    {\updownarrowInlineScriptT{#1}{#2}{#3}{#4}{#5}}%
}

\newcommand{\uparrowDisplayT}[3]{%
  \mathrel{%
    \begingroup
    \setlength{\arraycolsep}{0.2pt}%
    \renewcommand{\arraystretch}{0.5}%
    \begin{array}{c c}
      \smash{\scriptscriptstyle #1} & \multirow{3}{*}{$\mkern-1mu \textstyle #3$} \\
      \scriptstyle\uparrow & \\
      \smash{\scriptscriptstyle #2} &
    \end{array}%
    \endgroup
  }%
}

\newcommand{\uparrowInlineT}[3]{%
  \mathrel{%
    \begingroup
      \setlength{\arraycolsep}{0.2pt}%
      \renewcommand{\arraystretch}{0.5}%
      \vcenter{%
        \hbox{$%
          \begin{array}{c c l}
            \smash{\scriptscriptstyle #1} & 
            \multirow{2}{*}{$\uparrow$} &
            \multirow{2}{*}{$\mkern1mu #3$} \\
            \smash{\scriptscriptstyle #2} & &
          \end{array}%
        $}%
      }%
    \endgroup
  }%
}

\newcommand{\uparrowInlineScriptT}[3]{%
  \mathrel{%
    \begingroup
      \setlength{\arraycolsep}{0.2pt}%
      \renewcommand{\arraystretch}{0.5}%
      \vcenter{%
        \hbox{$%
          \begin{array}{c c l}
            \smash{\scriptscriptstyle #1} & 
            \multirow{2}{*}{$\scriptstyle \uparrow$} &
            \multirow{2}{*}{$\mkern-1mu \scriptstyle #3$} \\
            \smash{\scriptscriptstyle #2} & &
          \end{array}%
        $}%
      }%
    \endgroup
  }%
}

\newcommand{\uparrowSymbolT}[3]{%
  \mathchoice
    {\uparrowDisplayT{#1}{#2}{#3}}%
    {\uparrowInlineT{#1}{#2}{#3}}%
    {\uparrowInlineScriptT{#1}{#2}{#3}}%
    {\uparrowInlineScriptT{#1}{#2}{#3}}%
}

\newcommand{\dashInline}[2]{%
  \mathrel{%
    \begingroup
    \setlength{\arraycolsep}{0.1pt}%
    \renewcommand{\arraystretch}{0.15}%
    \vcenter{%
      \hbox{$
      \begin{array}{@{}c@{}c@{}}
        & \\
        & \\
        \smash{\scriptstyle #1} & \\[0.25ex]
        \scriptstyle \cdots & \smash{\scriptstyle #2}
      \end{array}%
      $}%
    }%
    \endgroup
  }%
}

\newcommand{\dashDisplay}[2]{%
  \mathrel{%
    \begingroup
    \setlength{\arraycolsep}{0.1pt}%
    \renewcommand{\arraystretch}{0.35}%
    \vcenter{%
      \hbox{$
      \begin{array}{@{}c@{}}
        \\
        \\
        \smash{\scriptstyle #1} \\
        \scriptstyle \cdots \\
        \smash{\scriptstyle #2} 
      \end{array}%
      $}%
    }%
    \endgroup
  }%
}

\newcommand{\ltDash}[1]{%
  \mathchoice
    {\dashDisplay{<}{#1}}%
    {\dashInline{<}{#1}}%
    {\dashInline{<}{#1}}%
    {\dashInline{<}{#1}}%
}

\newcommand{\gtDash}[1]{%
  \mathchoice
    {\dashDisplay{>}{#1}}%
    {\dashInline{>}{#1}}%
    {\dashInline{>}{#1}}%
    {\dashInline{>}{#1}}%
}

\newcommand*\true{\mathit{true}}

\newcommand{\ttrue}{\top}
\newcommand{\ffalse}{\bot}

\newcommand{\strue}{\mathop{\vcenter{\hbox{$\scriptstyle\bullet$}}}}
\newcommand{\sfalse}{\mathop{\vcenter{\hbox{$\scriptstyle\circ$}}}}

\newcommand{\realset}{\mathbb{R}}
\newcommand{\boolset}{\mathbb{B}}
\newcommand{\natset}{\mathbb{N}}
\newcommand{\oset}{\mathbb{O}}
\newcommand{\nonnegrealset}{\realset_{\geq 0}}
\newcommand{\power}[1]{2^{#1}}

\newcommand{\var}[1]{\mathop{\ensuremath{\mathit{vars}(#1)}}}
\newcommand{\fvar}[1]{\mathop{\ensuremath{\mathit{fv}(#1)}}}

\newcommand{\xto}[1]{\ensuremath{\mathbin{\raisebox{-1pt}{$\xrightarrow{#1}$}}}}
\newcommand{\xtosup}[2]{\ensuremath{\mathbin{\raisebox{-1pt}{$\xrightarrow{\smash{#1}}^{#2}$}}}}

\newcommand{\until}[1]{\mathop{\mathbf{U}_{#1}}}
\newcommand{\release}[1]{\mathop{\mathbf{R}_{#1}}}
\newcommand{\always}[1]{\mathop{\square_{#1}}}
\newcommand{\eventually}[1]{\mathop{\Diamond_{#1}}}

\DeclareFontEncoding{LS1}{}{}
\DeclareFontSubstitution{LS1}{stix}{m}{n}
\DeclareSymbolFont{stixsymbols}{LS1}{stixscr}{m}{n}
\DeclareMathSymbol{\dottedcircle}{\mathord}{stixsymbols}{"E3}

\newcommand{\bothtemp}{%
  \mathop{%
    \vcenter{\hbox{\scalebox{0.75}{$\dottedcircle$}}}%
  }\nolimits
}

\newcommand{\robustness}[4]{\rho_{#4}(#1, #2, #3)}

\newcommand{\partition}[1]{\mathcal{#1}}
\newcommand{\intvOf}[2]{\partition{#1}(#2)}

\makeatletter
\newcommand{\ul}[2]{\mathpalette\ul@{{#1}{#2}}} %
\newcommand{\ul@}[2]{\ul@@{#1}#2}               %
\newcommand{\ul@@}[3]{%
  \underbracket[0.25pt]{\m@th#1 #2}_{\hbox{$\m@th\ul@demote{#1}{#3}$}}%
}

\newcommand{\ul@demote}[2]{%
  \ifx#1\displaystyle     \scriptstyle      #2\else
  \ifx#1\textstyle        \scriptstyle      #2\else
  \ifx#1\scriptstyle      \scriptscriptstyle#2\else
  \scriptscriptstyle      #2%
  \fi\fi\fi
}
\makeatother

\newcommand{\linf}[1]{\mathrm{inf}\mkern2mu#1}
\newcommand{\rsup}[1]{\mathrm{sup}\mkern2mu#1}

\newcommand{\lend}[1]{\ensuremath{\mathrm{inf}(#1)}}

\newcommand{\intv}[4]{%
  {\textstyle\intervalSymbolT{#2,#3}{#1}{#4}}}

\newcommand{\intvro}[3]{%
  \intervalRoSymbol{#2,#3}{#1}}

\newcommand{\pair}[4]{
  \langle #1,#2,#3,#4 \rangle}

\newcommand{\domain}[2]{
  \mathinner{[#1,#2)}}

\newcommand{\ipred}{
  \mathop{\natural}}

\newcommand{\scope}[3]{%
    \mathop{\iota_{#1}^{#2}}(#3)}

\newcommand{\scopetuple}[4]{%
\scopeSymbolT{#2,#3}{#1}{#4}
}

\newcommand{\scopeSymbolT}[3]{%
  \mathchoice
    {\intervalDisplayT{\langle}{\rangle}{#1}{#2}{#3}}%
    {\intervalInlineT{\langle}{\rangle}{#1}{#2}{#3}}%
    {\intervalInlineScriptT{\langle}{\rangle}{#1}{#2}{#3}}%
    {\intervalInlineScriptT{\langle}{\rangle}{#1}{#2}{#3}}%
}

\newcommand{\ipstl}[1]{[#1]}

\newcommand{\intervalRoSymbol}[2]{%
  \mathchoice
    {\intervalDisplayT{\ivert}{)}{#1}{#2}{}}%
    {\intervalInlineT{\ivertb\mkern3mu}{)}{#1}{#2}{}}%
    {\intervalInlineScriptT{\ivertsmall}{)}{#1}{#2}{}}%
    {\intervalInlineScriptT{\ivertsmall}{)}{#1}{#2}{}}%
}

\newcommand{\ivert}{\mathbin{\vcenter{\hbox{$\rule{0.4pt}{0.9em}$}}}}
\newcommand{\ivertsmall}{\mathbin{\vcenter{\hbox{$\rule{0.2pt}{0.6em}$}}}}
\newcommand{\ivertb}{\mathbin{\vcenter{\hbox{$\rule{0.65pt}{1.17em}$}}}}

\newcommand{\intervalDisplayT}[5]{%
  \mathrel{%
    \begingroup
    \setlength{\arraycolsep}{0.2pt}%
    \renewcommand{\arraystretch}{0.3}%
    \begin{array}{c c c}
      \smash{#1} & \smash{\textstyle #3} & \smash{#2} \\
      & & \\
      \smash{\scriptstyle #4} & & \smash{\scriptstyle #5}
    \end{array}%
    \endgroup
  }%
}

\newcommand{\intervalInlineT}[5]{%
  \mathrel{%
    \begingroup
      \setlength{\arraycolsep}{0.2pt}%
      \renewcommand{\arraystretch}{0.5}%
      \vcenter{%
        \hbox{$%
          \begin{array}{r c c c l}
          & \multirow{2}{*}{$#1$} & \multirow{2}{*}{$\textstyle #3$} & \multirow{2}{*}{$#2$} & \\
            \smash{\scriptstyle #4} & & & & \smash{\scriptstyle #5}
          \end{array}%
        $}%
      }%
    \endgroup
  }%
}

\newcommand{\intervalInlineScriptT}[5]{%
  \mathrel{%
    \lower 0.20ex\hbox{$\scriptscriptstyle #4$}\,
    \scriptstyle #1\,
    \scriptstyle #3\,
    \scriptstyle #2\,
    \lower 0.20ex\hbox{$\scriptscriptstyle #5$}%
  }%
}

\newcommand{\intervalSymbolT}[3]{%
  \mathchoice
    {\intervalDisplayT{\ivert}{\ivert}{#1}{#2}{#3}}%
    {\intervalInlineT{\vert}{\vert}{#1}{#2}{#3}}%
    {\intervalInlineScriptT{\ivertsmall}{\ivertsmall}{#1}{#2}{#3}}%
    {\intervalInlineScriptT{\ivertsmall}{\ivertsmall}{#1}{#2}{#3}}%
}

\newcommand{\comp}[1]{\lVert#1\rVert}

\newcommand{\creset}[2]{#1  \mathop{\vcentcolon=} #2}
\newcommand{\lty}[1]{\mathop{\mathsf{lty}}(#1)}

\newcommand{\nextop}{\mathop{{\textbf{X}}}}
\newcommand{\atomset}{\mathit{AJ}}

\newcommand{\intvpredT}[5]{\updownarrowSymbolT{#1}{#2}{#3}{#4}{#5}}
\newcommand{\intvpredupT}[3]{\uparrowSymbolT{#1}{#2}{#3}}

\newcommand{\cset}[1]{\{#1\}}

\newcommand{\tableau}[2]{{\mathcal{T}}_{#1}^{#2}}
\newcommand{\tabG}[1]{\mathrm{tableau}(#1)}
\newcommand{\simptabG}[1]{{\mathrm{tableau}^*}(#1)}

\newcommand{\labelcur}[1]{\ensuremath{\mathop{{\mathcal{L}}(#1)}}}

\newcommand{\trace}[1]{\mathit{trace}(#1)}
\newcommand{\ftrace}[1]{\mathit{ftrace}(#1)}

\newcommand{\hatableau}[2]{\mathop{\mathcal{A}}_{#1}^{#2}}

\newcommand{\composeha}[2]{\mathop{H}_{#1}^{#2}}

\newcommand{\goal}{\ensuremath{\mathit{Goal}}}

\newcommand{\predInv}{\mathrm{pInv}}
\DeclareRobustCommand{\invmark}[1]{%
  \begingroup\scriptsize
  \raisebox{0pt}[\dimexpr 2.05ex+0.2pt\relax][\dimexpr 0.55ex+0.2pt\relax]{%
    \makebox[\dimexpr 2.6ex+0.4pt\relax]{\small\textcircled{\scriptsize #1}}}%
  \endgroup}

\newcommand{\curlabels}{\mathrm{NL}}

\newcommand{\ranklvl}{\mathsf{ExpRank}}

\newcommand{\expansion}{\mathop{Exp}}

\newcommand{\lexpansion}{\mathrm{E}}

\newcommand{\init}{\mathit{init}}

\newcommand{\inv}{\mathit{inv}}

\newcommand{\flow}{\mathit{flow}}

\newcommand{\jump}{\mathit{jump}}

\newcommand{\vertc}[1]{%
  \mathopen{}\bigl(%
  \smash{\renewcommand{\arraystretch}{0.85}%
   \begin{array}{@{\,}c@{\,}}#1\end{array}}%
  \bigr)\mathclose{}%
}

\newcommand{\bindot}{%
  \mathbin{\mkern-3mu.\mkern-3mu}%
}

\newcommand{\nextrl}[1]{\nextop\bigl(\smash{#1}\bigr)}

\newcommand{\Alt}{\mathrel{\;\mid\;}}

\newcommand{\newLc}{\mathit{m_{t}}}

\newcommand{\ltyVar}{\mathit{m}}

\newcommand{\upLc}{\mathit{m}}
\newcommand{\dnLc}{\mathit{m'}}

\newcommand{\ltLc}{\mathit{l}}
\newcommand{\rtLc}{\mathit{r}}
\newcommand{\intvLty}[1]{\mathit{m_{#1}}}
\newcommand{\intvRty}[1]{\mathit{m'_{#1}}}
\newcommand{\intvLe}[1]{\mathit{u_{#1}}}
\newcommand{\intvRe}[1]{\mathit{u'_{#1}}}

\newcommand{\TemporalOpen}{%
  $\ipstl{\always{I} \varphi}
  \rightarrow \cset{\creset{c}{0}} \, . \, (\intvpredupT{c}{\upLc}{\always{I} \varphi}, \lty{\upLc})
  \qquad 
    \ipstl{\eventually{I} \varphi}
  \rightarrow \cset{\creset{c}{0}} \, . \, (\intvpredupT{c}{\upLc}{\eventually{I} \varphi}, \lty{\upLc})
  $
}

\newcommand{\RuleAlwaysUpBranchOne}{%
  \nextrl{c \leq \inf I,\ast} 
}

\newcommand{\RuleAlwaysUpBranchTwo}{%
\ipstl{\varphi},
\nextrl{c > \inf I, \ast}
}

\newcommand{\RuleAlwaysUpBranchThree}{%
\nextrl{%
  c \ltDash{\neg \upLc \vee \dnLc} \inf I,
  \cset{\creset{d}{0}} \bindot \vertc{\intvpredT{c}{d}{\upLc}{\neg{\dnLc}}{\always{I} \varphi}, \lty{\dnLc}}%
}}

\newcommand{\RuleAlwaysUpBranchFour}{%
\ipstl{\varphi},
\nextrl{
  c \gtDash{\upLc \wedge \neg \dnLc} \inf I,
  \cset{\creset{c}{\inf I}, \creset{d}{0}} \bindot
    \vertc{\intvpredT{c}{d}{\dnLc}{\neg{\dnLc}}{\always{I} \varphi},\lty{\dnLc}}}
}

\newcommand{\alwaysUpPred}{\intvpredupT{c}{\upLc}{\always{I} \varphi}}

\newcommand{\RuleAlwaysUp}{%
\(
\setlength{\arraycolsep}{1.2pt}%
\renewcommand{\arraystretch}{0.45}%
\begin{array}{r c l}
  \alwaysUpPred
  & \rightarrow
  & \RuleAlwaysUpBranchThree\Alt\RuleAlwaysUpBranchOne\\
  & \Alt &\RuleAlwaysUpBranchTwo\Alt\RuleAlwaysUpBranchFour
\end{array}
\)
}

\newcommand{\RuleAlwaysUpDownBranchOne}{%
  \nextrl{
    \lty{\newLc},
    c \ltDash{\neg \upLc \vee \newLc} \inf I,
    \intvpredT{c}{d}{\upLc}{\dnLc}{\always{I} \varphi}}
}

\newcommand{\RuleAlwaysUpDownBranchTwo}{%
 \nextrl{c \leq \inf I, \ast}
}

\newcommand{\RuleAlwaysUpDownBranchThree}{%
\ipstl{\varphi}, \nextrl{c > \inf I, \ast} 
}

\newcommand{\RuleAlwaysUpDownBranchFour}{%
\ipstl{\varphi}, 
\nextrl{
(c \gtDash{\upLc \wedge \neg \newLc} \inf I,  %
  d \ltDash{\dnLc \wedge \newLc} \sup I), %
  \cset{\creset{c}{\inf I}} \bindot (\intvpredT{c}{d}{\newLc}{\dnLc}{\always{I} \varphi}, \lty{\newLc})}
}

\newcommand{\RuleAlwaysUpDownBranchFive}{%
\ipstl{\varphi}, 
\nextrl{\lty{\newLc}, 
c \gtDash{\upLc \wedge \neg \newLc} \inf I, d \gtDash{\neg \dnLc \vee \neg \newLc} \sup I}
}

\newcommand{\alwaysUpDownPred}{\intvpredT{c}{d}{\upLc}{\dnLc}{\always{I} \varphi}}

\newcommand{\RuleAlwaysUpDown}{%
\(
\setlength{\arraycolsep}{1.2pt}%
\renewcommand{\arraystretch}{0.2}%
\begin{array}{r c l}
  \alwaysUpDownPred
  & \rightarrow
  & \RuleAlwaysUpDownBranchOne\Alt\RuleAlwaysUpDownBranchTwo\Alt\RuleAlwaysUpDownBranchThree \\
  & \multicolumn{2}{l}{\Alt\RuleAlwaysUpDownBranchFour} \\
  & \multicolumn{2}{l}{\Alt\RuleAlwaysUpDownBranchFive}
\end{array}
\)
}

\newcommand{\RuleFinallyUpBranchOne}{%
\nextrl{
  c \ltDash{\neg \upLc \vee \dnLc} \sup I,
  \cset{\creset{d}{0}} \bindot \vertc{\intvpredT{c}{d}{\upLc}{\neg{\dnLc}}{\eventually{I} \varphi}, \lty{\dnLc}}}
}

\newcommand{\RuleFinallyUpBranchTwo}{%
\ipstl{\varphi}, \inf I > 0,
\nextrl{%
      c \gtDash{\upLc \wedge \neg \dnLc} \inf I,
        \cset{\creset{c}{\inf I}, \creset{d}{0}} \bindot \vertc{\intvpredT{c}{d}{\dnLc}{\neg{\dnLc}}{\eventually{I} \varphi},\lty{\dnLc}}}
}

\newcommand{\RuleFinallyUpBranchThree}{%
\ipstl{\varphi},\inf I = 0
}

\newcommand{\finallyUpPred}{\intvpredupT{c}{\upLc}{\eventually{I} \varphi}}

\newcommand{\RuleFinallyUp}{%
\(
\setlength{\arraycolsep}{1.2pt}%
\renewcommand{\arraystretch}{0.5}%
\begin{array}{r c l}
  \finallyUpPred
  & \rightarrow
  & \RuleFinallyUpBranchOne\Alt\RuleFinallyUpBranchThree \\\
  & \multicolumn{2}{l}{\Alt\RuleFinallyUpBranchTwo}
\end{array}
\)
}

\newcommand{\RuleFinallyUpDownBranchOne}{%
  \nextrl{
    \lty{\newLc},
    c \ltDash{\neg \upLc \vee \newLc} \sup I,
    \intvpredT{c}{d}{\upLc}{\dnLc}{\eventually{I} \varphi}} 
}

\newcommand{\RuleFinallyUpDownBranchTwo}{%
\ipstl{\varphi},
\nextrl{
  (
  c \gtDash{\upLc \wedge \neg \newLc} \inf I,
  d \ltDash{\dnLc \wedge \newLc} \inf I),
  \cset{\creset{c}{\inf I}} \bindot \vertc{\intvpredT{c}{d}{\newLc}{\dnLc}{\eventually{I} \varphi}, \lty{\newLc}}}
}

\newcommand{\RuleFinallyUpDownBranchThree}{%
\ipstl{\varphi},
\nextrl{
  d \gtDash{\neg \dnLc \vee \neg \newLc} \inf I, \lty{\newLc}
  }
}

\newcommand{\finallyUpDownPred}{\intvpredT{c}{d}{\upLc}{\dnLc}{\eventually{I} \varphi}}

\newcommand{\RuleFinallyUpDown}{%
\(
\setlength{\arraycolsep}{1.2pt}%
\renewcommand{\arraystretch}{0.5}%
\begin{array}{r c l}
  \finallyUpDownPred
  & \rightarrow
  & \RuleFinallyUpDownBranchOne\Alt \RuleFinallyUpDownBranchThree \\
  & \multicolumn{2}{l}{\Alt \RuleFinallyUpDownBranchTwo}
\end{array}
\)
}

\newcommand{\vars}{\mathop{\ensuremath{\mathit{vars}}}}

\newcommand{\eqrl}[1]{\texttt{#1}}

\newcommand{\comm}{\eqrl{com}}
\newcommand{\xmerge}{\eqrl{mrg-n}}
\newcommand{\cmerge}{\eqrl{mrg-r}}
\newcommand{\cnestedmerge}{\eqrl{mrg-nr}}

\usepackage{tikz}
\usetikzlibrary{positioning, calc, arrows.meta, bending}

\AtBeginDocument{%

}

\makeatletter
\ExplSyntaxOn
\cs_new_protected:Npn \stl_capture_theorem:n #1
  {
    \cs_set_eq:Nc \stl_original_theorem: {#1}
    \cs_set_eq:cN {#1} \stlrecordedtheorem
  }

\NewDocumentEnvironment{stlrecordedtheorem}{O{} +b}
  {
    \cs_set_eq:cN {\@currenvir} \stl_original_theorem:
    \regex_extract_once:nnNTF
      { \A \c{label} \s* (?: \[ [^\]]* \] \s* )? \cB\{ ([^\{\}]+) \cE\} }
      {#2} \l_tmpa_seq
      {
        \exp_args:Nee \stl_record_theorem:nnnn
          {\@currenvir} {\seq_item:Nn \l_tmpa_seq {2}} {#1} {#2}
      }
      {
        \stl_original_theorem:[#1]
        #2
        \use:c {end\@currenvir}
      }
    \cs_set_eq:cN {end\@currenvir} \prg_do_nothing:
  }{}

\cs_new_protected:Npn \stl_record_theorem:nnnn #1#2#3#4
  {
    \begin{restatable}[#3]{#1}{stl-restatement-#2}
    #4
    \end{restatable}
  }

\AddToHook{env/theorem/begin}[stl-restatements]{\stl_capture_theorem:n{theorem}}
\AddToHook{env/lemma/begin}[stl-restatements]{\stl_capture_theorem:n{lemma}}
\AddToHook{env/proposition/begin}[stl-restatements]{\stl_capture_theorem:n{proposition}}
\AddToHook{cmd/appendix/before}
  {
    \RemoveFromHook{env/theorem/begin}[stl-restatements]
    \RemoveFromHook{env/lemma/begin}[stl-restatements]
    \RemoveFromHook{env/proposition/begin}[stl-restatements]
  }

\NewDocumentCommand{\restatetheorem}{m}
  {
    \cs_if_exist:cTF {stl-restatement-#1}
      { \use:c {stl-restatement-#1} * }
      {
        \PackageError{techrep}
          {No~statement~recorded~for~label~'#1'}
          {Put~its~label~at~the~start~of~the~statement.}
      }
  }
\ExplSyntaxOff
\makeatother

\newcommand{\annotatedInterval}[2]{%
  \underset{\text{\small\color{black!45}$\textstyle
    \vphantom{\intvpredT{c}{d}{\upLc}{\neg\dnLc}{\always{I}\varphi}}%
    \smash{#2}$}}%
    {\tikz[baseline=(interval.base)]{%
      \node[inner sep=0pt,outer sep=0pt,text=black] (interval)
        {\mathsurround=0pt$\textstyle\smash[b]{#1}\vphantom{X_{g'}}$};
      \useasboundingbox (interval.north west)
        rectangle ([yshift=-2.8pt]interval.south east);
      \draw[black!30,line width=0.25pt,densely dotted,line cap=round]
        ([yshift=-1.8pt]interval.south west) --
        ([yshift=-1.8pt]interval.south east);
    }}}

\newcommand{\supplementref}[2][]{\Cref{#2}}
\newcommand{\arxivonly}[1]{#1}

\begin{document}
\title{Unfolding Continuous Time: Interval-Based Tableaux for Signal Temporal Logic over Finite Signals}

\author{Geunyeol Yu}
\orcid{0000-0002-6171-9911}
\affiliation{%
  \institution{Pohang University of Science and Technology}
  \city{Pohang}
  \country{South Korea}}
\email{rgyen@postech.ac.kr}

\author{Jia Lee}
\orcid{0000-0002-3263-2550}
\affiliation{%
  \institution{Pohang University of Science and Technology}
  \city{Pohang}
  \country{South Korea}
}
\email{cee5539@postech.ac.kr}

\author{Kyungmin Bae}
\orcid{0000-0002-6430-5175}
\affiliation{%
  \institution{Pohang University of Science and Technology}
  \city{Pohang}
  \country{South Korea}}
\authornote{Corresponding author}
\email{kmbae@postech.ac.kr}

\begin{abstract}
Temporal reasoning in discrete time commonly decomposes a temporal formula into what must hold now and what must hold next, 
forming the basis of tableau constructions and automata-theoretic methods for logics such as LTL.
However, this decomposition does not directly extend to temporal logics interpreted over continuous-time signals, 
such as Signal Temporal Logic (STL), which is widely used to specify properties of continuous and hybrid dynamical systems.
To address this challenge, we introduce \emph{interval predicates}, which express the truth of STL formulas over time intervals rather than at individual instants.
They enable interval-based reasoning: temporal operators unfold across a finite partition of a signal, one interval at a time.
We formalize this unfolding as expansion rules, establish their
correctness, and construct finite tableaux for STL formulas over finite
signals.
We translate such a tableau into a hybrid automaton, enabling robust STL model checking of hybrid systems through reachability analysis.
The translation is independent of any particular reachability backend; experiments with four off-the-shelf analyzers on benchmarks 
with linear and nonlinear dynamics show substantial gains over state-of-the-art SMT-based robust STL model checking.

\end{abstract}


\maketitle              %

\section{Introduction}
\label{sec:intro}

Temporal reasoning in discrete time rests on a simple principle:
a temporal formula can be \emph{unfolded} into what must hold now and what
must hold at the next instant. For example,
$\varphi \until{} \psi \,\equiv\,
  \psi \vee \bigl(\varphi \wedge \bigcirc(\varphi \until{} \psi)\bigr)$,
where $\varphi \until{} \psi$ (``until'') requires $\psi$ at some
future instant with $\varphi$ holding until then, and $\bigcirc$ (``next'')
defers a formula to the following instant.
This equivalence splits an until-obligation into what must hold now and
what remains for the next instant.
Iterating such decompositions underlies tableau constructions and
automata-theoretic methods: automaton states represent sets of pending
obligations, and transitions follow their
unfolding~\cite{gerth1995simple,schwendimann1998,ltlf-ijcai13}.
 
However, this principle does not directly extend to temporal logics %
over continuous-time signals.
Real-time logics such as metric temporal logic
(MTL)~\cite{ouaknine2008some,bouyer2018model} admit a continuous
interpretation over signals, and signal temporal logic
(STL)~\cite{maler2004monitoring} extends this setting with real-valued
predicates.
STL is %
widely used to specify properties of continuous and
hybrid dynamical systems.
Over a continuous-time signal, there is no distinguished next instant
through which temporal obligations can advance, so the stepwise unfolding
above---and the tableau constructions built on it---have no direct
counterpart.
The central question is therefore what should play the role of ``next''
in continuous time: how can temporal
obligations advance over a continuous-time signal?
 
Existing approaches suggest two possible ways to advance temporal
obligations.
The first is based on the notion of \emph{next events}: under pointwise
semantics, MTL is interpreted over timed words, which provide a discrete
sequence of event positions, so temporal obligations can be deferred to
the next event~\cite{ouaknine2008some,ho2026mightyppl}.
Continuous signal semantics also admits a next-like modality, definable
from strict until as
$\bigcirc\varphi \equiv \varphi \until{} \varphi$~\cite{ferrere2019real};
unlike discrete next, however, it expresses truth over a right-neighborhood
rather than providing a successor position through which until can
unfold.
In sum, pointwise constructions rely on explicitly given event positions,
whereas continuous signal semantics provides no distinguished successor
at an arbitrary time.

The second approach uses clocks to track the timing bounds of temporal obligations.
MITL, the fragment of MTL with non-singular timing
intervals~\cite{alur1996benefits}, has been translated into automata under
both pointwise and continuous signal
semantics~\cite{brihaye2017m,mitlautomata,ferrere2019real,brihaye2017signal,geilen2003cav}.
In particular, under continuous signal semantics, clock-based
constructions may require discrete transitions at timing bounds 
\emph{even when no relevant subformula changes truth value}~\cite{mitlautomata,ferrere2019real}.
A natural route to STL model checking is to adapt these
continuous-semantics constructions to hybrid automata.
Once composed with a hybrid system, however, the additional discrete
transitions may increase the cost of reachability analysis, particularly
under nonlinear dynamics~\cite{chen2012taylor}.

This paper proposes an alternative approach: temporal obligations advance
to the \emph{next interval of stable truth}.
As is common in real-time and hybrid-systems analysis, we assume that
the relevant truth values change only finitely many times over a bounded
horizon~\cite{maler2004monitoring,ouaknine2008some}.
This induces a finite partition of the signal into intervals on which
those truth values remain stable.
Temporal obligations can then be unfolded across the partition, one
interval at a time, recovering the stepwise decomposition of discrete-time
tableaux over continuous time.
Realizing this unfolding, however, is not immediate: an obligation's
checking window need not align with the partition---it may cross interval
boundaries and, under nested operators, shrink or invert.

To address this challenge,  
we introduce \emph{interval predicates}\arxivonly{\footnote{The term also
appears in the real-time refinement literature~\cite{dongol2012algebra},
where interval predicates describe the behavior of real-time programs
over time intervals; ours are logical formulas for reasoning about STL
over intervals (see \Cref{sec:related}).}}
to express the truth of STL formulas over time intervals rather than at
individual instants.
An interval-predicate judgement records which subformula obligations are
active on the current interval, the windows over which they must be
checked, and the position of the current interval relative to those
windows.
To represent this information, interval predicates use clocks measuring
distances to window endpoints, boundary markers distinguishing open from
closed endpoints, and scopes identifying the intervals over which
obligations apply.
Over these judgements, we define expansion rules and
establish their correctness, obtaining sound and complete tableaux for signals of finite variability
(\Cref{sec:rule}). %

Interval predicates enable a \emph{tableau construction} analogous to that in discrete time:
states collect pending obligations, and transitions advance from one
partition interval to the next.
Each expansion step, however, is local to a single interval, whereas a
tableau state must capture all pending obligations at once.
We therefore build states by repeatedly applying the expansion rules
and merging interval predicates that track overlapping or adjacent intervals,
which keeps the construction finite.
For STL with non-singular timing intervals, as in standard MITL
constructions~\cite{alur1996benefits}, the tableau size is singly 
exponential in the formula size and the resolution of its timing
intervals.
The tableau advances with the signal's logical behavior rather than
a fixed clock grid, unlike the clock-driven constructions above (\Cref{sec:tableau}).

We translate such a tableau into a hybrid automaton.
The translation must reconcile a subtle semantic mismatch: the tableau
and hybrid automata interpret the boundaries of time intervals
differently.
We identify a mild \emph{well-behavedness} condition,
satisfied by typical hybrid systems in practice, that resolves this mismatch.
Composing the automaton for $\neg\varphi$ with a system automaton $H$
then enables robust STL model checking through reachability analysis, up
to a given robustness threshold~\cite{scp2025-stl}.
The translation is independent of any particular reachability analyzer,
allowing us to use off-the-shelf tools, including analyzers designed for
nonlinear dynamics (\Cref{sec:transform}).

We have implemented our approach and evaluated it on benchmarks with
linear and nonlinear dynamics.
Four reachability analyzers verify robust STL properties with substantial
performance gains over state-of-the-art SMT-based model
checking~\cite{cav2022-stl,scp2025-stl}, indicating that the resulting
property automata are effective for STL model checking
(\Cref{sec:expr}).

In summary, this paper makes the following contributions:
\begin{enumerate*}[label=(\arabic*)]
\item \emph{Interval-based decomposition}: interval predicates
  for continuous-time STL over finite signals, with sound and complete
  expansion rules.
 
 \item \emph{Finite tableau construction}: a tableau of size singly exponential 
  in the formula size and timing interval resolution,
  whose transitions follow changes in subformula truth.
  To our knowledge, this is the first tableau for STL under continuous
  signal semantics: prior dense-time tableaux target fragments of
  MITL~\cite{geilen2000onthefly,geilen2003cav}, and prior STL tableaux
  assume discrete time~\cite{stl-tecs25}.
 
\item \emph{Robust STL model checking}: a faithful translation to hybrid
  automata, enabling robust STL model checking through reachability 
  analysis without relying on a particular backend.
\end{enumerate*}

Due to space limitations, we refer the reader to
\supplementref[our longer report~]{sec:app-rule,sec:app-tableau,sec:app-automata}
for complete proofs.
The main theorems and their proofs are also formalized in
Lean~\cite{stl-tableaux-lean}.

\section{Preliminaries: STL and Hybrid Automata}
\label{sec:prelim}

\paragraph{Intervals.}

A \emph{real interval} is a subset of $\realset$ containing all values between two \emph{endpoints}.
For two intervals $I_1, I_2 \subseteq \realset$ and a binary operator $\star \in \{+, -\}$, let
$I_1 \mathbin{\star} I_2 = \{i_1 \mathbin{\star} i_2 \mid i_1 \in I_1, i_2 \in I_2\}$.
For a singleton $\{t\}$, we write $\{t\}\star I$ as $t \star I$.
For example, 
$[a, b] + [c, d] = [a + c, b + d]$.

We use \emph{boundary markers} $\oset=\{\strue,\sfalse\}$ to specify interval endpoints and inequality bounds:
\begin{enumerate*}[label=(\roman*)]
\item $\intv{\upLc}{l}{r}{\dnLc}$ for an interval with endpoints $l$ and $r$,
where $\upLc,\dnLc\in\oset$ indicate whether $l$ and $ r$ are included
($\strue$ closed, $\sfalse$ open); and
\item $a \ltDash{\upLc} a'$ for an inequality, meaning $a\le a'$ if $\upLc=\strue$ and $a<a'$ if $\upLc=\sfalse$.
\end{enumerate*}
For example, $[1,2)=\intv{\strue}{1}{2}{\sfalse}$ and $c<1$ is $c \ltDash{\sfalse} 1$.
Markers combine by Boolean operations in the usual way
(e.g., 
$\neg\strue = \sfalse$,
and
$\strue \lor \sfalse = \strue$).

\paragraph{Signal Temporal Logic.}

A signal represents an execution of a hybrid system over continuous time. 
Let $X$ be a finite set of real-valued variables representing the physical parameters of hybrid systems.
A \emph{signal} $\sigma : \domain{0}{\tau} \to \realset^X$ is a function
that provides the value of each variable in $X$ over time,
where $\realset^X$ is the set of  real-valued assignments over $X$,
and $\tau > 0$ is the \emph{time horizon} of $\sigma$.

The syntax of signal temporal logic (STL)
 is defined as a variation of MITL as follows \cite{maler2004monitoring}:%
\quad
$
\varphi \coloneqq
    \mathit{true}
    \,\mid\, f(x_1,\ldots, x_n) \geq 0
    \,\mid\, \neg \varphi
    \,\mid\, \varphi \wedge \varphi
    \,\mid\, \varphi \until{I}\varphi
$,%
\quad
where 
$f : {\realset}^n \to {\realset}$ is a real-valued function
over %
$x_1, \ldots, x_n \in X$,
and
$I \subseteq \nonnegrealset$ is a closed, non-singular interval  of time. %
We use untimed notations, such as $\until{}$,
as  shorthands for  %
operators with $[0, \infty)$,
such as $\until{[0, \infty)}$.
Other common %
operators can be defined as syntactic sugar,
e.g.:
$\mathit{false} \equiv \neg \mathit{true}$,
\;
$\varphi \vee \varphi' \equiv \neg (\neg \varphi \wedge \neg \varphi')$,
\;
$\eventually{I} \varphi \equiv \mathit{true} \until{I} \varphi$,
\;
$\always{I} \varphi \equiv \neg \eventually{I} \neg \varphi$,
\;
$\varphi \release{I} \varphi' \equiv \neg((\neg \varphi) \until{I} (\neg \varphi'))$,
and so on.

For a real-valued function $f$, let $\vars(f)$ denote the set of variables
on which $f$ depends. We write $\sigma_f$ for the restriction of
$\sigma$ to $\vars(f)$.
For %
$v \in {\realset}^X$,
let 
$f(v) = f(v(x_1), \ldots, v(x_n))$,
and write
$(x_1,\ldots,x_n)$ 
as 
$\vec{x}$.
Let $\boolset = \{\ttrue,\ffalse\}$ %
and $\bar{\realset} = \realset \cup \boolset$, 
ordered so that $\ffalse < r < \ttrue$ for all $r \in \realset$.
The semantics of STL is given by the following satisfaction relation.

\begin{definition}
Given a signal $\sigma: \domain{0}{\tau} \to {\realset}^X$,
the satisfaction of an STL formula 
$\varphi$ at a time $t \in [0, \tau)$,
denoted by $\sigma, t \models_{\tau} \varphi$,
is defined inductively as follows, where ${\bowtie} \in \{\geq,>,=,<,\leq\}$:%
\begin{itemize}
    \item $\sigma, t \models_{\tau} \mathit{true}$;

    \item $\sigma, t \models_{\tau} f(\vec{x}) \bowtie 0$  iff $f(\sigma(t)) \bowtie 0$;

    \item $\sigma, t \models_{\tau} \neg \varphi$
    iff $\sigma, t \not\models_{\tau}  \varphi$;

    \item $\sigma, t \models_{\tau} \varphi \wedge \varphi'$
    iff $\sigma, t \models_{\tau} \varphi \;\;\mbox{and}\;\; \sigma, t \models_{\tau} \varphi'$; and
    
    \item $\sigma, t \models_{\tau} \varphi \until{I} \varphi'$ iff
    $(\exists t' \in (t+I)\cap[t,\tau))\,
    (\sigma, t' \models_{\tau} \varphi' \text{ and }
    (\forall t'' \in [t,t'])\,\sigma, t'' \models_{\tau} \varphi)$.\footnote{%
We follow the standard STL semantics of $\varphi \until{I} \varphi'$
\cite{maler2004monitoring},
which requires both $\varphi$ and $\varphi'$ to hold at the
witness time $t' \geq t$.
}

\end{itemize}
\end{definition}

It is well known that 
any STL formula can be transformed 
into an equivalent formula
containing only the operators
$\wedge$, $\vee$, $\until{}$, $\release{}$, $\eventually{I}$, and $\always{I}$
\cite{donze2013efficient}. %
In particular, 
$\until{I}$ can be expressed using $\until{}$, $\eventually{I}$,
and $\always{[0,\lend{I}]}$, 
based on the equivalences
$\varphi \until{[a, b]} \varphi' \equiv 
    \eventually{[a, b]} \varphi'  \wedge \varphi \until{[a, \infty)} \varphi'$
and 
$\varphi \until{[a, \infty)} \varphi' \equiv
    \always{[0, a]}(\varphi \until{} \varphi')$.
From now on,
we assume that STL formulas contain 
only 
$\wedge$, $\vee$, $\until{}$, $\release{}$,
$\eventually{I}$, and $\always{I}$.

\paragraph{Partitions and Stability.}

A partition of an interval $D$ is a family of nonempty disjoint intervals
$\partition{J}=\{J_0,\dots,J_{N-1}\}$
such that
$D=\bigcup_{i=0}^{N-1} J_i$,
where 
$\sup{J_i} \leq \inf{J_k}$
for any $0\le i<k<N$.
We write $|\partition{J}|=N$ for the size of the partition.
For any $0\le i<|\partition{J}|$,
let $\intvOf{J}{i}=J_i$.

An interval $J$ is said to be \emph{stable} for an STL formula $\varphi$
with respect to a signal $\sigma$
if the truth value of $\varphi$ remains constant throughout $J$.
That is, for any $u,u'\in J$,
$\sigma,u\models_\tau\varphi$
iff
$\sigma,u'\models_\tau\varphi$.
An interval $J$ is said to be \emph{fully stable} for an STL formula
$\varphi$ if it is stable for every subformula of $\varphi$.
A partition $\partition{J}$ is \emph{fully stable} for
$\varphi$ if each of its intervals is fully stable for $\varphi$.

A signal $\sigma$ is said to have \emph{finite variability} for
an STL formula $\varphi$ iff there exists a finite fully stable partition
of the time domain of $\sigma$ with respect to $\varphi$.

\paragraph{Hybrid Automata.}

A \emph{hybrid automaton} %
is 
a finite state machine with %
continuous variables 
\cite{henzinger2000theory},
defined as a tuple  
$H = (Q, G, E, X, \mathit{init}, \mathit{inv}, \mathit{flow}, \mathit{jump})$,
where:
\begin{enumerate*}[label=(\roman*)]
    \item 
    $Q$ is a %
    set of modes;
    \item 
    $G$ is a set of labels;
    \item 
    $E \subseteq Q \times G \times Q$ is a set of edges;
    \item 
    $X$ is a %
    set of %
    variables;
    \item 
    $\mathit{init} \subseteq Q \times \realset^X$ 
        is a set of initial states;
    \item 
    $\mathit{inv} \subseteq Q \times \realset^X$
    is a set of valid states;
    \item 
    $\mathit{flow} \subseteq (Q \times \realset^X) \times \nonnegrealset \times \realset^X$
    is a flow condition,
    defining how the values of $X$ evolve continuously over time;
    and    
    \item 
    $\mathit{jump} : E \to 2^{(Q \times \realset^X) \times (Q \times \realset^X)}$
    specifies discrete transitions.
\end{enumerate*}

Hybrid automata induce infinite-state transition systems \cite{doyen2018verification}.
A state  is a pair $\langle q, v \rangle$ %
satisfying $\mathit{inv}(q, v)$,
where $q$ is a mode and $v$ is a valuation  for $X$.
A continuous transition 
$\langle q, v \rangle
\xto{r; \gamma} 
\langle q, v' \rangle$
of duration $r$
can occur
iff
there exists
a %
function $\gamma : [0,r] \to \realset^X$
such that 
$\gamma(0) = v$, 
$\gamma(r) = v'$, 
and
for each %
$t \in [0, r]$, %
$\mathit{flow}(\langle q, v \rangle, t, \gamma(t))$
and
$\mathit{inv}(q, \gamma(t))$ hold.
A discrete 
transition 
$\langle q, v \rangle \xto{\boldsymbol{\cdot}} \langle q', v' \rangle$
can occur
iff 
there exists an edge
$e=(q,g,q')\in E$ such that
$(\langle q,v \rangle, \langle q',v' \rangle)\in \mathit{jump}(e)$.
Let $\xtosup{\boldsymbol{\cdot}}{+}$ and
$\xtosup{\boldsymbol{\cdot}}{*}$
denote the transitive closure and the reflexive transitive closure of
$\xto{\boldsymbol{\cdot}}$, respectively.

A \emph{trajectory} is a signal
that describes a valid behavior of a hybrid automaton,
which alternates discrete transitions defined by $\mathit{jump}$ 
and continuous evolutions by $\mathit{flow}$ while satisfying $\mathit{inv}$. 
\begin{definition}
A signal $\sigma$ is a trajectory of $H$,
denoted by $\sigma \in H$,
iff there exists a sequence of discrete and continuous transitions
\[
\langle q_0, v_0 \rangle
\xtosup{\boldsymbol{\cdot}}{*}
\langle q_1, v_1 \rangle
\xto{r_1; \gamma_1}
\langle q_1, v_1' \rangle
\xtosup{\boldsymbol{\cdot}}{+}
\langle q_2, v_2 \rangle
\xto{r_2; \gamma_2}
\cdots
\xtosup{\boldsymbol{\cdot}}{+}
\langle q_n, v_n \rangle
\xto{r_n; \gamma_n}
\langle q_n, v_n' \rangle
\]
and times
$0=t_1<t_2<\cdots<t_n<t_{n+1}=\tau$
such that, for $1\le i\le n$,
$\mathit{init}(q_0,v_0)$ holds,
$r_i=t_{i+1}-t_i$,
and
$\sigma(t_i+u_i)=\gamma_i(u_i)$
for each $u_i\in[0,r_i)$.\arxivonly{\footnote{%
The execution has total duration $\tau$ and a final valuation at that time.
The signal $\sigma$ is defined on $\domain{0}{\tau}$.}}
\end{definition}

\begin{figure}[t]
\setlength{\fboxsep}{0pt}
\begin{minipage}[t]{0.4\textwidth}
\centering
\includegraphics{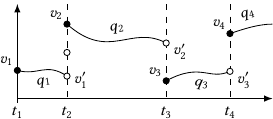}
\caption{A trajectory $\sigma$ of $H$}
\label{fig:signal}
\end{minipage}
\hfill
\begin{minipage}[t]{0.55\textwidth}
\centering
\includegraphics{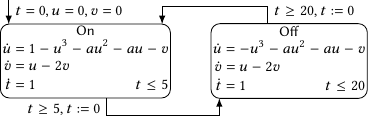}
\caption{A hybrid automaton}
\label{fig:cardiac-automaton} 
\end{minipage}
\Description{The left figure shows signal values across mode changes:
filled points are observed; hollow points are skipped. The right figure
shows On and Off modes with continuous dynamics and transitions
that reset the clock.}
\end{figure}

\Cref{fig:signal}
illustrates a trajectory $\sigma \in H$ of a hybrid automaton $H$.
During each interval $[t_i, t_{i+1}]$, 
a continuous transition 
$\langle q_i, v_i \rangle
\xto{r_i; \gamma_i}
\langle q_i, v_i' \rangle$ occurs.
\footnote{%
Flow conditions are typically specified by ordinary differential
equations, in which case $\gamma$ is continuous in time.
We call a flow condition \emph{continuous} if every
function $\gamma$ satisfying the flow condition is continuous.
}
Between these intervals, (successive) 
discrete transitions  
$\langle q_{i}, v_{i}' \rangle \xtosup{\boldsymbol{\cdot}}{+} \langle q_{i+1}, v_{i+1} \rangle$
take place, %
where the value of $\sigma(t_{i+1})$ is assigned the last value $v_{i+1}$ (black circles),
while the intermediate values (hollow circles) are ``skipped'' in the signal $\sigma$.

For a valuation $v \in \mathbb{R}^X$ and $Y \subseteq X$, 
let $v \restriction Y \in \realset^Y$ be a valuation
restricted to $Y$
such that $(v \restriction Y)(y) = v(y)$ for any variable $y \in Y$.
The parallel composition $H_1 \parallel H_2$ of hybrid automata $H_1$ and $H_2$ is 
defined as a single hybrid automaton in the standard way~\cite{henzinger2000theory,lynch2003hybrid}
(see also \supplementref{sec:prelim-app-ha} for its formal definition).
Intuitively, a jump in $H_1 \parallel H_2$ corresponds to a jump in at least one component, 
while the other component may either jump simultaneously or stutter.

\begin{example}
\Cref{fig:cardiac-automaton} shows 
an example of a hybrid automaton
for a cardiac cell with a pacemaker \cite{cardiac2015}.
The pacemaker operates in two modes:
\textsf{On}, where the cell receives electrical stimulation, 
and
\textsf{Off}, where no stimulus is present. 
The  variables $u$ and $v$
model %
electrical characteristics of the cell,
with their continuous dynamics defined by the nonlinear ordinary differential equations (ODEs):
\[
(\mathsf{On})
\left[
\begin{aligned}
\dot{u} &= 1 -u^3 - a u^2 - a u - v  
\\
\dot{v} &= u - 2 v
\end{aligned}
\right]
\qquad    
(\mathsf{Off})
\left[
\begin{aligned}
\dot{u} &= -u^3 - a u^2 - a u - v %
\\
\dot{v} &= u - 2 v
\end{aligned}
\right]
\]

The pacemaker changes its mode at regular time intervals,
tracked by the clock $t$ (with $\dot{t} = 1$).
The following are examples of STL properties:
\begin{enumerate*}[label=(\roman*)]
    \item $\always{[1, 5]} (u \geq 4 \to \eventually{[8, 10]} u \leq 2)$:
    between 1 and 5 seconds, whenever $u \geq 4$, 
    $u \leq 2$ holds at some time 8 to 10 seconds later; and
        
    \item $(\eventually{[6,12]} u \leq 4) \until{} t \geq 5$:
    the condition $t \geq 5$ eventually holds,
    and until it does,
    $u \leq 4$ holds at some time 6 to 12 seconds later.
    
\end{enumerate*}
\end{example}

\section{Interval Predicates and Expansion Rules}
\label{sec:rule}

\subsection{Interval Predicates}

Separation has long been a useful tool in reasoning about STL formulas.
At a high level, separation decomposes the satisfaction of an STL formula into fragments
with distinct temporal dependencies, such as past, present, and future.
Thanks to its ability to resolve these fragments independently,
separation has become a core technique for STL analysis~\cite{hodkinson2005separation,geilen2003cav,bae2019bounded}.

Our goal is to extend this separation to an interval-based setting.
To this end, we introduce \emph{interval predicates}, each defined over a partition interval,
to characterize the stable truth of STL formulas.
A key requirement of these predicates is to provide a \emph{divide-and-conquer} characterization:
they must decompose a stable truth into the parts associated with the current and subsequent intervals (\emph{divide}),
with each part handled independently in the same way (\emph{conquer}).

\begin{figure}[t]
\centering
\includegraphics{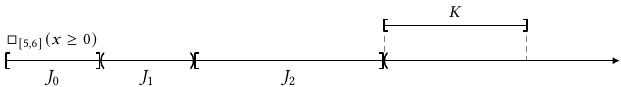}%
\caption{
$\always{[5,6]}(x\ge0)$ is stably true on $J_0 = [0,1]$ of $\partition{J}=\{[0,1],(1,2),[2,5],\ldots\}$.
$x\ge0$ holds on the checking interval $K=[5,7]$.
The dots denote further partition intervals omitted from the figure.
}
\label{fig:intv_pred_motivation}
\Description{The always formula on the closed interval from 0 to 1 checks
the closed interval from 5 to 7. Dashed guides place the checking
interval among later partition intervals.}
\end{figure}

A natural starting point is the interval predicate $\ipstl{\varphi}$, which represents
that $\varphi$ is stably true on $J$.
This is sufficient for formulas whose stable truth is determined solely on $J$, e.g., atomic propositions.
However, the situation becomes more subtle in the presence of temporal operators.
For example, consider
$\always{[5,6]}(x\ge0)$, which is stably true on $J_0=[0,1]$, as shown in \Cref{fig:intv_pred_motivation}.
To represent this stable truth on later intervals $J_i$ using a predicate of the form $\ipstl{\always{I}(x\ge0)}$,
the interval $I$ must be chosen so that $J_i+I = K$.
This obviously fails on $J_1=(1,2)$, since no $I$ can shift the open
interval $J_1$ to the closed interval $K$, and on $J_2=[2,5]$, which
would require $I=[3,2]$, invalid since $3>2$.

This example illustrates that representing such stable truth requires the endpoints of $J_0$, together with their openness or closedness, to be tracked from $J_i$.
Based on this observation, we use clocks and boundary markers to represent $J_0$:
two clocks $c$ and $d$ encode the distances from $J_i$ to the endpoints of $J_0$, and boundary markers $\upLc$ and $\dnLc$ indicate whether the endpoints are open or closed.

Accordingly, we define \emph{clocked} interval predicates:
$\intvpredupT{c}{\upLc}{\varphi}$
and $\intvpredT{c}{d}{\upLc}{\dnLc}{\varphi}$.
Intuitively, given a clock valuation $\alpha$\footnote{
A \emph{clock valuation} is a function
$\alpha : X_c \to \realset_{\ge 0}$
assigning a nonnegative real value to each clock variable in $X_c$.},
$\intvpredupT{c}{\upLc}{\varphi}$ over an interval $J$ represents that
$\varphi$ is stably true on an interval
starting $\alpha(c)$ time units before the left endpoint of $J$
and ending at the right endpoint of $J$.
Likewise,
$\intvpredT{c}{d}{\upLc}{\dnLc}{\varphi}$ over $J$ represents that
$\varphi$ is stably true on an interval
whose left and right endpoints are
$\alpha(c)$ and $\alpha(d)$ time units before the left endpoint of $J$, respectively.
Their syntax is given below.

\begin{definition}
An \emph{interval predicate} of an STL formula $\varphi$ is one of the following forms:
\begin{enumerate*}[label=(\roman*)]
  \item $\ipstl{\varphi}$,
  \item $\intvpredupT{c}{\upLc}{\bothtemp_I \varphi}$,
  \item $\intvpredT{c}{d}{\upLc}{\dnLc}{\bothtemp_I \varphi}$,
\end{enumerate*}
where $\bothtemp \in \{\always{}, \eventually{}\}$,
$\upLc, \dnLc \in \oset$, $I$ is an interval, and $c,d$ are clock variables.
\end{definition}

\noindent
An interval predicate is \emph{atomic} if its formula is an atomic proposition.
That is, among the interval predicates above, only $\ipstl{\varphi}$ can be atomic when $\varphi$ is atomic.

\begin{figure}[t]
\centering
\begin{subfigure}[b]{0.48\textwidth}
\centering
\includegraphics{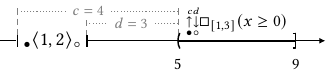}
\caption{
Scope %
when $c = 4$ and $d = 3$}
\end{subfigure}
\begin{subfigure}[b]{0.48\textwidth}
\centering
\includegraphics{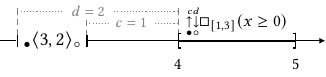}
\caption{
Scope %
when $c = 1$ and $d = 2$}
\end{subfigure}
\caption{
Scopes of
$\protect\intvpredT{c}{d}{\strue}{\sfalse}{\always{[1,3]}(x \ge 0)}$
}
\label{fig:intv_scope}
\Description{Clock values give scope endpoints 1 and 2 in the left
figure, but inverted endpoints 3 and 2 in the right figure. In scope
quadruples, filled markers indicate inclusion and hollow markers exclusion.}
\end{figure}

The interval tracked by an interval predicate is specified by its \emph{scope}.
If the interval is
$\intv{\upLc}{u}{v}{\dnLc}$\arxivonly{\footnote{Recall that this notation
indicates the endpoints of the interval together with whether they are open or closed.}},
its scope is the quadruple
$\scopetuple{\upLc}{u}{v}{\dnLc}$.
For example, consider \Cref{fig:intv_scope}~(a).
The scope of $\intvpredT{c}{d}{\strue}{\sfalse}{\always{[1,3]}(x \ge 0)}$ on $(5,9]$ with $c=4$ and $d=3$ is
$\scopetuple{\strue}{1}{2}{\sfalse}$,
since moving $4$ and $3$ time units backward from the left endpoint $5$
yields the endpoints $1$ and $2$, respectively.
Similarly, in \Cref{fig:intv_scope}~(b), the scope on $[4,5]$ with $c=1$ and $d=2$ becomes
$\scopetuple{\strue}{3}{2}{\sfalse}$.
In (b), the endpoints are inverted.
This illustrates why a scope is represented as a quadruple rather than an interval.
The definition is given as follows.

\begin{definition}
For an interval $J=\intv{\ltLc}{u}{v}{\rtLc}$,
the \emph{scope} of an interval predicate $p$
with respect to $J$ and a clock valuation $\alpha$,
denoted by $\scope{J}{\alpha}{p}$,
is defined as follows, where $\bothtemp \in \{\always{}, \eventually{}\}$:
\begin{align*}
\scope{J}{\alpha}{\ipstl{\varphi}}
= \pair{\ltLc}{u}{v}{\rtLc},
\quad
\scope{J}{\alpha}{\intvpredupT{c}{\upLc}{\bothtemp_I \varphi}}
= \pair{\upLc}{u-\alpha(c)}{v}{\rtLc},
\quad
\scope{J}{\alpha}{\intvpredT{c}{d}{\upLc}{\dnLc}{\bothtemp_I \varphi}}
= \pair{\upLc}{u-\alpha(c)}{u-\alpha(d)}{\dnLc}.
\end{align*}
\end{definition}

\begin{figure}[t]
\centering
\begin{subfigure}[b]{0.46\textwidth}
\centering
\includegraphics{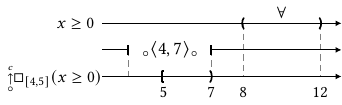}
\caption{Satisfaction of $\protect\intvpredupT{c}{\sfalse}{\always{[4,5]}(x \ge 0)}$ when $c=1$}
\end{subfigure}
\begin{subfigure}[b]{0.53\textwidth}
\centering
\includegraphics{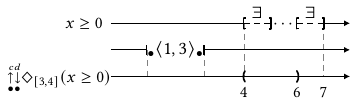}
\caption{Satisfaction of $\protect\intvpredT{c}{d}{\strue}{\strue}{\eventually{[3,4]}(x \ge 0)}$ when $c=3$ and $d=1$}
\end{subfigure}
\caption{Examples of interval predicate satisfaction}
\label{fig:intv_semantics}
\Description{The left figure illustrates always: x is nonnegative
throughout the checking interval. The right figure illustrates
eventually: each shifted window contains a point where x is nonnegative;
intermediate windows are omitted.}
\end{figure}

The satisfaction of an interval predicate on an interval $J$ 
is defined by the stable truth of its formula over the interval specified by its scope.

\begin{definition}
Given a signal $\sigma$ with time horizon $\tau$, an interval $J \subseteq \domain{0}{\tau}$, a clock valuation $\alpha$, and an interval
predicate $p$, let $\langle \ltLc,u,v,\rtLc\rangle=\scope{J}{\alpha}{p}$. 
Then, the satisfaction of $p$, denoted $\sigma, J, \alpha \models_\tau p$, is defined as follows,
where $\ipred \in \{\intvpredupT{c}{\upLc}{}, \intvpredT{c}{d}{\upLc}{\dnLc}{}\}$.\footnote{%
When the scopes are inverted, i.e., $u > v$,
$\ipred\eventually{I}\varphi$ is vacuously satisfied,
since $\intv{\ltLc}{u}{v}{\rtLc}$ is empty.
In contrast,
$\ipred\always{I}\varphi$ is not necessarily vacuously satisfied,
since satisfaction is evaluated over
$\intv{\ltLc}{u + \inf I}{v + \sup I}{\rtLc}$,
which may not be empty.
}
\[
\begin{alignedat}{3}
\sigma, J, \alpha
  &\models_\tau \ipstl{\varphi}
  &\quad \text{iff} \quad&
  (\forall t \in \textstyle\intv{\ltLc}{u}{v}{\rtLc})\,
  \sigma, t \models_\tau \varphi \\
\sigma, J, \alpha
  &\models_\tau \ipred\,\always{I}\varphi
  &\quad \text{iff} \quad&
  (\forall t \in \textstyle\intv{\ltLc}{u+\linf{I}}{v+\rsup{I}}{\rtLc} \cap \domain{0}{\tau})\,
  \sigma, t \models_\tau \varphi \\
\sigma, J, \alpha
  &\models_\tau \ipred\,\eventually{I}\varphi
  &\quad \text{iff} \quad&
  (\forall t \in \textstyle\intv{\ltLc}{u}{v}{\rtLc}) 
  (\exists t' \in (t + I) \cap \domain{0}{\tau})\,
  \sigma, t' \models_\tau \varphi
\end{alignedat}
\]
\end{definition}

\begin{example}
Consider \Cref{fig:intv_semantics}~(a).
Suppose that
$\intvpredupT{c}{\sfalse}{\always{[4,5]}(x \ge 0)}$
is satisfied on the interval $[5,7)$ with $c=1$.
Its scope is
$\scopetuple{\sfalse}{4}{7}{\sfalse}$, so
$x \ge 0$ holds at every point in
$(4+4,7+5)=(8,12)$.
Similarly, consider \Cref{fig:intv_semantics}~(b).
Suppose that
$\intvpredT{c}{d}{\strue}{\strue}{\eventually{[3,4]}(x \ge 0)}$
is satisfied on $(4,6)$ with $c=3$ and $d=1$.
Its scope is
$\scopetuple{\strue}{1}{3}{\strue}$.
By definition, for every
$t \in [1,3]$,
there exists
$t' \in t+[3,4]$
such that $x \ge 0$ holds.
\end{example}

\subsection{Interval Sequence Judgements}

\begin{figure}[t]
\centering
\begin{minipage}[b]{0.14\textwidth}
\centering
\includegraphics[trim=8pt 8pt 8pt 8pt]{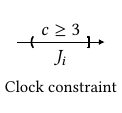}
\end{minipage}
\hfill
\begin{minipage}[b]{0.14\textwidth}
\centering
\includegraphics[trim=8pt 8pt 8pt 8pt]{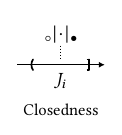}
\end{minipage}
\hfill
\begin{minipage}[b]{0.2\textwidth}
\centering
\includegraphics[trim=8pt 8pt 8pt 8pt]{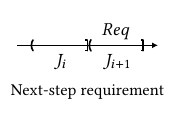}
\end{minipage}
\hfill
\begin{minipage}[b]{0.14\textwidth}
\centering
\includegraphics[trim=8pt 8pt 8pt 8pt]{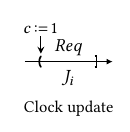}
\end{minipage}
\hfill
\begin{minipage}[b]{0.14\textwidth}
\centering
\includegraphics[trim=8pt 8pt 8pt 8pt]{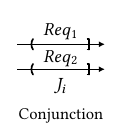}
\end{minipage}
\hfill
\begin{minipage}[b]{0.14\textwidth}
\centering
\includegraphics[trim=8pt 8pt 8pt 8pt]{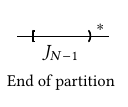}
\end{minipage}
\caption{
Components for decomposing stable truth over partition intervals.
}
\label{fig:intv_judgement_motivation}
\Description{Six timelines illustrate clock constraints, boundary
markers, requirements for the next interval, clock resets, conjunction,
and termination. Labels represent each requirement on its current or next
partition interval.}
\end{figure}

So far, interval predicates allow us to represent stable truth on a partition interval.
However, for decomposition, we need to represent
clock constraints, boundary
information, clock updates, next-interval requirements, conjunction, and
the end of the partition, as summarized in \Cref{fig:intv_judgement_motivation}.
To represent those components,
we introduce \emph{interval sequence judgements}.
Concretely,
an interval predicate itself is a judgement,
and so is a clock constraint.
The left-closedness of the interval is represented
by $\lty{m}$\footnote{%
The right-closedness of an interval can be determined
from the left-closedness of the next interval.
},
where $m$ specifies whether it is open or closed.

Clock updates using a clock reset set $\Lambda$
are written as $\Lambda.\Psi$, where $\Psi$ is a judgement.
A judgement $\Psi$ on the next interval
is represented by $\nextop\Psi$.
The conjunction of two judgements $\Psi_1$ and $\Psi_2$
is written as $\Psi_1,\Psi_2$.
Finally, the absence of a following partition interval is represented by $\ast$;
thus, $\nextop\ast$ indicates that the current interval is the last one.
The syntax is given as follows.

\begin{definition}
Given a set of clock variables $X_c$ and an STL formula $\varphi$,
the syntax of interval sequence judgements is given by:
\[
    \Psi \coloneqq
    \ast
    \,\mid\, c \bowtie a
    \,\mid\, \lty{\ltyVar}
    \,\mid\, p
    \,\mid\, \nextop \Psi
    \,\mid\, \Lambda.\Psi
    \,\mid\, \Psi, \Psi
\]
where
$p$ is an interval predicate for a subformula of $\varphi$,
$c \in X_c$,
$a \in \nonnegrealset$,
$\ltyVar \in \oset$,
${\bowtie} \in \{\geq,>,=,<,\leq\}$,
and $\Lambda$ is a set of clock resets of the form $\creset{c}{a}$.
\end{definition}

An interval sequence judgement is \emph{atomic} if it is one of
$\ast$, $c \bowtie a$, $\lty{\ltyVar}$, or $p$.
A clock variable occurring in a judgement $\Psi$ is \emph{free}
if it is not bound by any clock reset in $\Psi$.
We write $\fvar{\Psi}$ for the set of all free clock variables in $\Psi$.
We also write $\var{\Lambda}$ and $\var{\Psi}$ for the sets of clock variables occurring
in $\Lambda$ and $\Psi$, respectively.
Clock-reset sets $\Lambda_1$ and $\Lambda_2$
are \emph{consistent} if, for every $c\in\var{\Lambda_1}\cap\var{\Lambda_2}$,
$\creset{c}{a}\in\Lambda_1$ and
$\creset{c}{b}\in\Lambda_2$ imply $a=b$.

\begin{example}
The judgement
$\cset{\creset{c}{0}}.(\intvpredupT{c}{\strue}{\always{[2,4]}(x \ge 0)}),\lty{\strue}$
represents the interval predicate
$\intvpredupT{c}{\strue}{\always{[2,4]}(x \ge 0)}$
with clock value $c=0$ on a left-closed interval.
As another example, the judgement
$\cset{\creset{c}{7},\creset{d}{3}}.(\intvpredT{c}{d}{\strue}{\sfalse}{\always{[2,4]}(x \ge 0)}),
\ipstl{x \ge 0},
\nextop\ast$
represents the interval predicate
$\intvpredT{c}{d}{\strue}{\sfalse}{\always{[2,4]}(x \ge 0)}$
with clock values $c=7$ and $d=3$,
and the stable truth of $x \ge 0$ on the last partition interval.
\end{example}

As in interval predicates, interval sequence judgements are interpreted over partition intervals.
Thus, their satisfaction is naturally defined with respect to
a partition and a position indicating the current interval.
Before giving the formal definition, we introduce some notation on clock valuations.
For a clock valuation $\alpha$, let $\alpha \models c \bowtie a$ iff $\alpha(c)\bowtie a$.
For a constant $a \in \nonnegrealset$, let $\alpha + a$ denote the valuation defined by
$(\alpha + a)(c)=\alpha(c)+a$
for all $c\in X_c$.
For a set of clock resets $\Lambda$, let $\alpha[\Lambda]$ denote the valuation such that
$\alpha[\Lambda](c)=a$ if $\creset{c}{a}\in\Lambda$, and
$\alpha[\Lambda](c)=\alpha(c)$ otherwise.
The satisfaction relation is defined as follows.

\begin{definition}
For a signal $\sigma$ with time horizon $\tau$, a partition $\partition{J}$ of $\domain{0}{\tau}$,
a clock valuation $\alpha$, and a position
$k \in \{0,\ldots,|\partition{J}|\}$,
the satisfaction relation
$\sigma, \partition{J}, \alpha, k \models_\tau \Psi$
is defined as follows,
where $i \in \{0,\ldots,|\partition{J}|-1\}$ denotes an interval index.\footnote{\label{footnote:intv_index}%
Although $\lty{\ltyVar}$ is only defined for interval indices
$i$, for convenience we also allow it to be evaluated at
position $|\partition{J}|$. In this case, we define
$\lty{\strue}$ to hold and $\lty{\sfalse}$ not to hold,
since every partition of $\domain{0}{\tau}$ ends with a
right-open interval.
}
\begin{align*}
\sigma, \partition{J}, \alpha, k \models_\tau \ast \quad
& \text{iff} \quad k=|\partition{J}| \qquad 
\sigma, \partition{J}, \alpha, k \models_\tau c \bowtie a \;\; \text{iff} \;\; \alpha \models c \bowtie a \\
\sigma, \partition{J}, \alpha, i \models_\tau \lty{\ltyVar}\quad
& \text{iff} \quad \intvOf{J}{i} =\textstyle\intv{\ltyVar}{u}{u'}{\dnLc} \\
\sigma, \partition{J}, \alpha, i \models_\tau p \quad
&\text{iff} \quad \sigma, \intvOf{J}{i}, \alpha \models_\tau p \\
\sigma, \partition{J}, \alpha, i \models_\tau \nextop\Psi \quad
&\text{iff} \quad \sigma, \partition{J}, \alpha+|\intvOf{J}{i}|, i + 1 \models_\tau \Psi \\
\sigma, \partition{J}, \alpha, k \models_\tau \Lambda.\Psi \quad
&\text{iff} \quad \sigma, \partition{J}, \alpha[\Lambda], k \models_\tau \Psi\\ 
\sigma, \partition{J}, \alpha, k \models_\tau \Psi_1,\Psi_2 \quad
&\text{iff} \quad \sigma, \partition{J}, \alpha, k \models_\tau \Psi_1 \;\; \text{and} \;\; \sigma, \partition{J}, \alpha, k \models_\tau \Psi_2 
\end{align*}
\end{definition}

\begin{example}
Let $\partition{J}=\{[0,1],(1,3],(3,4)\}$.
Consider %
$\Psi_1 = \cset{\creset{c}{0}}.(\intvpredupT{c}{\strue}{\always{[2,4]}(x \ge 0)}),\lty{\strue}$ on $J_0$.
By definition,
$\sigma,\partition{J},\alpha,0 \models_\tau \Psi_1$
iff
$\sigma,J_0,\alpha[\cset{\creset{c}{0}}]
\models_\tau
\intvpredupT{c}{\strue}{\always{[2,4]}(x \ge 0)}$
and $J_0$ is left-closed.
Since $J_0=[0,1]$, the latter condition holds.
Moreover, under $\alpha[\cset{\creset{c}{0}}]$, we have $c=0$.
Thus, $\Psi_1$ is satisfied iff
$\intvpredupT{c}{\strue}{\always{[2,4]}(x \ge 0)}$
is satisfied on $J_0$ with $c=0$.

Consider %
$\Psi_2 = \cset{\creset{c}{7},\creset{d}{3}}.(\intvpredT{c}{d}{\strue}{\sfalse}{\always{[2,4]}(x \ge 0)}),
\ipstl{x \ge 0},
\nextop\ast$
on $J_2$.
By definition,
$\sigma,\partition{J},\alpha,2 \models_\tau \Psi_2$ iff
$\sigma,J_2,\alpha[\cset{\creset{c}{7},\creset{d}{3}}]
\models_\tau
\intvpredT{c}{d}{\strue}{\sfalse}{\always{[2,4]}(x \ge 0)}$,
$\sigma,J_2,\alpha \models_\tau \ipstl{x \ge 0}$,
and there is no interval following $J_2$.
Since $J_2$ is the last interval, %
the terminal condition holds.
Thus, $\Psi_2$ is satisfied iff
$\intvpredT{c}{d}{\strue}{\sfalse}{\always{[2,4]}(x \ge 0)}$
is satisfied on $J_2$ with $c=7$ and $d=3$,
and $x \ge 0$ stably holds on $J_2$.
\end{example}

Note that positions range over $\{0,\ldots,|\partition{J}|\}$,
with the terminal position $|\partition{J}|$ used to evaluate $\ast$.
For convenience, $\lty{\ltyVar}$ is also defined at the terminal position $|\partition{J}|$
as explained in Footnote~\ref{footnote:intv_index}.
For example, consider $\partition{J}=\{[0,1),[1,2)\}$.
Then the positions are $0,1,2$, since $|\partition{J}|=2$.
At the terminal position $2$,
$\sigma,\partition{J},\alpha,2 \models_\tau \ast$
and
$\sigma,\partition{J},\alpha,2 \models_\tau \lty{\strue}$ hold,
while
$\sigma,\partition{J},\alpha,2 \not\models_\tau \lty{\sfalse}$.

\paragraph{Properties of Interval Sequence Judgements}

As usual, relations such as equivalence, implication, and reduction hold between interval sequence judgements.
Judgements $\Phi_1$ and $\Phi_2$ are \emph{equivalent}, written
$\Phi_1 \equiv \Phi_2$, if for any signal $\sigma$, partition $\partition{J}$, clock valuation $\alpha$, and a position $k$,
\[
\sigma,\partition{J},\alpha,k\models_\tau\Phi_1
\quad\text{iff}\quad
\sigma,\partition{J},\alpha,k\models_\tau\Phi_2.
\]
Similarly, judgement $\Phi_1$ \emph{implies} $\Phi_2$, written
$\Phi_1 \Rightarrow \Phi_2$, if for any $\sigma,\partition{J},\alpha,k$,
\[
\sigma,\partition{J},\alpha,k\models_\tau\Phi_1
\quad\text{implies}\quad
\sigma,\partition{J},\alpha,k\models_\tau\Phi_2.
\]

Under equivalence, the usual structural laws hold:
conjunction is idempotent and commutative,\footnote{%
Accordingly, we treat judgements as sets and, by abuse of notation,
write $\Phi_1\in(\Phi_1,\Phi_2)$ for $\Phi_1\in\{\Phi_1,\Phi_2\}$.
}
$\nextop$ distributes over conjunction,
and consistent clock resets can be merged.

\begin{proposition}
\label{prop:judg-equiv}
For any judgements $\Psi_1$ and $\Psi_2$, the following hold:
\begin{enumerate}[label=(\arabic*),leftmargin=2em,itemsep=2pt]

\item
$\Psi_1,\Psi_1 \equiv \Psi_1$, \;
$\Psi_1,\Psi_2 \equiv \Psi_2,\Psi_1$, \;
$\nextop(\Psi_1,\Psi_2) \equiv (\nextop\Psi_1),(\nextop\Psi_2)$.

\item
For consistent reset sets $\Lambda_1$ and $\Lambda_2$, if
$\var{\Lambda_1}\cap \fvar{\Lambda_2 . \Psi_2} = \emptyset$ and 
$\var{\Lambda_2}\cap \fvar{\Lambda_1 . \Psi_1} = \emptyset$,
\[
(\Lambda_1.\Psi_1),(\Lambda_2.\Psi_2)
\equiv
(\Lambda_1\cup\Lambda_2).(\Psi_1,\Psi_2),
\quad
\Lambda_1.(\Psi_1,\Lambda_2.\Psi_2)
\equiv
(\Lambda_1\cup\Lambda_2).(\Psi_1,\Psi_2).
\]
\end{enumerate}
\end{proposition}

Interval predicates have equivalent forms in which endpoint information is made explicit.
In particular, predicates can be rewritten into forms with additional clocks
and markers that expose the start and endpoint of the stability interval.
Let $\bothtemp_I \varphi$ denote either $\always{I}\varphi$ or $\eventually{I}\varphi$.

\begin{proposition}
\label{prop:intvpred-equiv}
For fresh clocks $c$ and $d$, the following equivalences hold
under suitable conditions on boundary markers and scope containment
(see \supplementref[our longer report~]{sec:app-judg-equiv} for the details):
\[
\textit{(1)}\;
\ipstl{\bothtemp_I \varphi}
\equiv
\cset{\creset{c}{0}}.
(\intvpredupT{c}{\upLc}{\bothtemp_I \varphi},\lty{\upLc})
\quad\quad\quad
\textit{(2)}\;
\intvpredupT{c}{\upLc}{\bothtemp_I \varphi}\;
\equiv
\cset{\creset{d}{0}}.
(\intvpredT{c}{d}{\upLc}{\neg\dnLc}{\bothtemp_I \varphi},
\lty{\dnLc},
\ipstl{\bothtemp_I \varphi})
\]
\end{proposition}

\begin{figure}[t]
\centering

\begin{subfigure}{0.48\textwidth}
\hspace{-0.02\linewidth}
\begin{minipage}[b]{0.48\textwidth}
\centering
\includegraphics{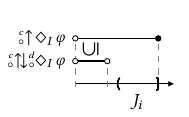}

\end{minipage}
\hspace{-0.02\linewidth}
\begin{minipage}[b]{0.51\textwidth}
\centering
\includegraphics{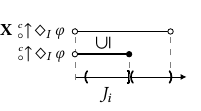}

\end{minipage}
\caption{Subinterval relations}
\end{subfigure}
\hspace{-0.02\linewidth}
\begin{subfigure}{0.48\textwidth}
\begin{minipage}[b]{0.52\textwidth}
\centering
\includegraphics{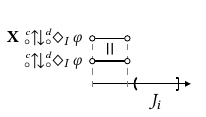}

\end{minipage}
\hspace{-0.07\linewidth}
\begin{minipage}[b]{0.48\textwidth}
\centering
\includegraphics{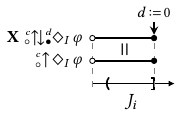}

\end{minipage}
\caption{Equal-interval relations}
\end{subfigure}
\caption{
Illustration of interval relations between scopes of predicates declared at $J_i$.
Each horizontal line denotes the interval represented by the scope of the predicate shown to its left.
}
\label{fig:intvpred-trans}
\Description{Four bar pairs show containment and equality between
intervals represented by clocked predicate scopes. The equality cases
either retain both clock variables on the next interval or reset the
clock that tracks the right endpoint.}
\end{figure}

An interval predicate may imply another.
Such implications arise from the relations between the intervals represented by their scopes
and the partition intervals at which they are declared.
If interval predicates $p$ and $q$ for the same STL formula,
declared at partition intervals
$J_i$ and $J_j$, represent intervals $T_p$ and $T_q$,
respectively, then $q$ implies $p$ when $T_p \subseteq T_q$ and $i \le j$.
\Cref{fig:intvpred-trans} illustrates these interval relations.

By the subinterval relations in \Cref{fig:intvpred-trans}~(a),
$\intvpredupT{c}{\sfalse}{\eventually{I}\varphi}$ implies
$\intvpredT{c}{d}{\sfalse}{\sfalse}{\eventually{I}\varphi}$, and
$\nextop\intvpredupT{c}{\sfalse}{\eventually{I}\varphi}$ implies
$\intvpredupT{c}{\sfalse}{\eventually{I}\varphi}$.
Likewise, by the equal-interval relations in \Cref{fig:intvpred-trans}~(b),
$\nextop\intvpredT{c}{d}{\sfalse}{\sfalse}{\eventually{I}\varphi}$ implies
$\intvpredT{c}{d}{\sfalse}{\sfalse}{\eventually{I}\varphi}$, and
$\nextop(\cset{\creset{d}{0}}.(\intvpredT{c}{d}{\sfalse}{\strue}{\eventually{I}\varphi}))$
implies
$\intvpredupT{c}{\sfalse}{\eventually{I}\varphi}$.
The following lemma generalizes these patterns.

\begin{lemma}
\label{lem:intvpred-trans}
For any interval predicates, the following implications hold:
\[
\begin{aligned}
&\textit{(1)}\;
\textstyle\intvpredupT{c}{\upLc}{\bothtemp_I \varphi} 
\;\Rightarrow\;
\textstyle\intvpredT{c}{d}{\upLc}{\dnLc}{\bothtemp_I \varphi}
\qquad\quad
\textit{(2)}\; 
\textstyle\nextop\intvpredupT{c}{\upLc}{\bothtemp_I \varphi}
\;\Rightarrow\;
\textstyle\intvpredupT{c}{\upLc}{\bothtemp_I \varphi}
\qquad\quad
\textit{(3)}\;
\textstyle\nextop\intvpredT{c}{d}{\upLc}{\dnLc}{\bothtemp_I \varphi}
\;\Rightarrow\;
\textstyle\intvpredT{c}{d}{\upLc}{\dnLc}{\bothtemp_I \varphi}
\\
&\textit{(4)}\;
\textstyle\nextop(\cset{\creset{d}{0}}. (\intvpredT{c}{d}{\upLc}{\neg\dnLc}{\bothtemp_I \varphi}, \lty{\dnLc}))
\Rightarrow\;
\textstyle\intvpredupT{c}{\upLc}{\bothtemp_I \varphi}
\end{aligned}
\]
\end{lemma}

\begin{figure}[t]
\centering
\includegraphics{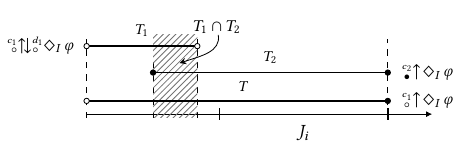}
\caption{
Illustration of the merging of interval predicates declared at $J_i$.
Each horizontal line denotes the interval represented by the scope of the predicate shown to its left or right;
the upper two intervals are denoted by $T_1$ and $T_2$.
The scope of the lower predicate represents their union $T = T_1 \cup T_2$.
}
\label{fig:intvpred-merge}
\Description{Three bars show two overlapping intervals and their union.
Shading marks the intersection; the merged interval retains T1's left
marker and T2's right marker.}
\end{figure}

While \Cref{lem:intvpred-trans} shows implications arising from
subinterval and equal-interval relations,
an interval predicate
$\intvpredT{c_1}{d_1}{m_1}{m'_1}{\bothtemp_I\varphi}$
and an interval predicate
$\intvpredupT{c_2}{m_2}{\bothtemp_I\varphi}$
can also be merged into a single predicate
$\intvpredupT{c_1}{m_1}{\bothtemp_I\varphi}$
when the intervals represented by their scopes overlap or meet without a gap.
As shown in \Cref{fig:intvpred-merge},
the interval represented by the scope of
$\intvpredupT{c_1}{m_1}{\bothtemp_I\varphi}$
is the union of those intervals.

\begin{lemma}
\label{lem:intv-merge}
For any signal $\sigma$, partition $\partition{J}$,
clock valuation $\alpha$, and position $k \in \{0, \cdots, |\partition{J}|\}$,
if the scopes overlap or meet without a gap, i.e.,
$\alpha(d_1) \ltDash{m'_1 \vee m_2} \alpha(c_2) \ltDash{m_1 \vee \neg m_2} \alpha(c_1)$,
then
\[
\sigma,\partition{J},\alpha,k \models_\tau \intvpredT{c_1}{d_1}{m_1}{m'_1}{\bothtemp_I\varphi}, \intvpredupT{c_2}{m_2}{\bothtemp_I\varphi}
\quad\text{iff}\quad
\sigma,\partition{J},\alpha,k \models_\tau \intvpredupT{c_1}{m_1}{\bothtemp_I\varphi}
\]
\end{lemma}

\subsection{Expansion Rules} %

\begin{figure}[b]
\centering
\includegraphics{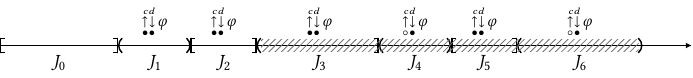}
\caption{
Expansion of
$\protect\intvpredT{c}{d}{\strue}{\strue}{\varphi}$
declared on $(2,3)$ with scope
$\protect\scopetuple{\strue}{0}{2}{\strue}$,
where $\varphi=\always{I}(x\ge0)$ and $I=[5,7]$,
along
$\partition{J}=\{[0,2],(2,3),[3,4],(4,6],(6,7),[7,8],(8,10),\ldots\}$.
The checking interval is $K=[5,9]$. Shading indicates the partition
intervals where $x\ge0$ is stably required.
}
\label{fig:expansion_rules_ex}
\Description{Seven partition intervals show the expansion of a clocked
always predicate from the second interval. Shading on the final four
intervals marks where x must be nonnegative; labels track boundary
markers.}
\end{figure}

We now define \emph{expansion rules} for interval predicates.
An expansion rule decomposes an interval predicate into interval sequence judgements for the current and subsequent partition intervals.

To illustrate the intuition behind expansion rules,
consider the interval predicate
\scalebox{0.93}{$\intvpredT{c}{d}{\strue}{\strue}{\varphi}$}
and the fully stable partition $\partition{J}$ for $\varphi$ shown in
\Cref{fig:expansion_rules_ex},
where
$\varphi=\always{I}(x\ge0)$ and $I=[5,7]$.
Suppose that
\scalebox{0.93}{$\intvpredT{c}{d}{\strue}{\strue}{\varphi}$}
is declared on $J_1$ with $c=2$ and $d=0$.
Then, the scope is
$\scopetuple{\strue}{0}{2}{\strue}$. By definition,
$x\ge0$ must hold on
$K=[5,9]$.
We now illustrate how this interval predicate is expanded along the partition.

On $J_1$ and $J_2$, there is no overlap with $K$.
Hence, no additional judgement is required on either interval.
Instead, the same interval predicate
\scalebox{0.93}{$\intvpredT{c}{d}{\strue}{\strue}{\varphi}$}
is required on the next interval together with the clock condition
$c\le\inf I$,
indicating that $K$ has not yet been reached.

On $J_3$, the overlap with $K$ is $[5,6]$.
Since $x\ge0$ holds on this nonempty intersection and is stable on $J_3$,
it holds throughout $J_3$, requiring $\ipstl{x\ge0}$.
The clock conditions $c\ge\inf I$ and $d<\sup I$ are required on $J_4$
to indicate that $K$ has been reached but not completely checked.
The predicate
$\cset{\creset{c}{\inf I}}.(\intvpredT{c}{d}{\sfalse}{\strue}{\varphi})$
is also required on $J_4$ to represent the remaining portion of $K$.
The left marker $\sfalse$ and reset $\creset{c}{\inf I}$
indicate that its left endpoint coincides with that of $J_4$.

The same pattern repeats on both $J_4$ and $J_5$:
$\ipstl{x\ge0}$ is required on the current interval, and the remaining
obligation is carried to the next interval with the corresponding boundary markers.

Finally, on $J_6$, $K$ still overlaps the partition interval, but no longer overlaps $J_7$.
Hence,
$\ipstl{x\ge0}$
is required on $J_6$,
but no interval predicate is required on $J_7$.
Instead, the clock conditions
$c>\inf I$
and
$d>\sup I$
are required on $J_7$,
indicating that $K$ has been completely checked.

Using the same approach, we analyze the expansion patterns of each interval predicate and 
define the corresponding rules, shown in \Cref{fig:unclocked-rule,fig:clocked-rule}.
Each expansion rule has the form
$
p \to \Psi_1 \mid \cdots \mid \Psi_n,
$
where $p$ is an interval predicate and
$\Psi_1,\ldots,\Psi_n$ are %
expansion branches.

For example, the expansion pattern of
\scalebox{0.93}{$\intvpredT{c}{d}{\strue}{\strue}{\varphi}$}
is captured by the rule
$\scalebox{0.93}{$\alwaysUpDownPred$} \to \Psi_1 \mid \cdots \mid \Psi_5$
in \Cref{fig:clocked-rule}.
$\Psi_1$ corresponds to the case where $K$ has not yet been reached.
$\Psi_4$ and $\Psi_5$ describe the cases where $K$ overlaps the current interval,
with and without a remaining portion of $K$.
$\Psi_2$ and $\Psi_3$, which do not appear in the example, describe 
the cases where the current interval is the last partition interval,
before reaching $K$, and overlapping $K$, respectively.

\begin{figure}[t]
\centering
$\begin{aligned}
\ipstl{\varphi \wedge \varphi'}   
    &\;\longrightarrow\;  \ipstl{\varphi}, \ipstl{\varphi'}  
&&&&&
\ipstl{\varphi \until{} \varphi'}  
    &\;\longrightarrow\;   \ipstl{\varphi}, \ipstl{\varphi'}  
    \;\mid\;   \ipstl{\varphi}, \nextop \ipstl{\varphi \until{} \varphi'} 
\\
\ipstl{\varphi \vee \varphi'}  
    &\;\longrightarrow\;   \ipstl{\varphi} \;\mid\; \ipstl{\varphi'}  
&&&&&
\ipstl{\varphi \release{} \varphi'}  
    &\;\longrightarrow\;   \ipstl{\varphi} 
    \;\mid\; \ipstl{\varphi'}, \nextop \ipstl{\varphi \release{} \varphi'}  
    \;\mid\;  \ipstl{\varphi'}, \nextop \ast  
\end{aligned}$
\caption{Expansion rules for unclocked interval predicates}
\label{fig:unclocked-rule}
\Description{Four unclocked expansion rules cover conjunction,
disjunction, until, and release. Vertical bars separate alternatives;
each branch specifies current requirements and any continuation on the
next interval.}
\end{figure}

\begin{figure}[t]
\raggedright
\TemporalOpen\par\medskip
\RuleAlwaysUp\par\medskip
\RuleAlwaysUpDown\par\medskip
\RuleFinallyUp\par\medskip
\RuleFinallyUpDown
\caption{Expansion rules for clocked interval predicates}
\label{fig:clocked-rule}
\Description{Five rows show clocked expansion rules for always and
eventually. Branches combine clock guards, resets, boundary markers,
current requirements, and either judgements for the next interval or
terminal judgements; markers control strictness.}
\end{figure}

\paragraph{Correctness} %

Expansion rules are designed to preserve the meaning of the original interval predicate.
That is, for a rule
$p \to \Psi_1 \mid \cdots \mid \Psi_n$,
$p$ is satisfied iff at least one branch $\Psi_i$ is satisfied.
The correctness of the rules is established through soundness and completeness.
Soundness states that if a branch $\Psi_i$ is satisfied,
then the predicate $p$ is also satisfied.
Completeness states conversely that if $p$ is satisfied,
then at least one branch $\Psi_i$ is satisfied.

\begin{figure}[t]
\centering
\hspace{-0.02\linewidth}
\begin{minipage}[b]{0.42\linewidth}
\centering
\includegraphics{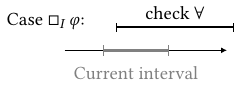}
\end{minipage}
\hspace{-0.02\linewidth}
\begin{minipage}[b]{0.42\linewidth}
\centering
\includegraphics{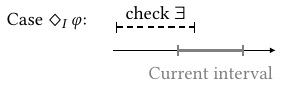}
\end{minipage}

\caption{
Positional conditions for predicate invariants.
For $\always{I}\varphi$, checking must start at or after the start of the current interval.
For $\eventually{I}\varphi$, the earliest witness interval must end at or after that point.
}
\label{fig:predicate-invariant}
\Description{Gray bars mark current intervals. In the left figure, always
checking starts within the current interval. For eventually in the right
figure, the interval searched for a witness starts earlier and overlaps
the current interval.}
\end{figure}

However, the soundness and completeness of expansion rules do not hold in general.
They depend on positional conditions between the partition interval where a predicate is declared
and the checking interval induced by its scope.
For example, consider
$\intvpredupT{c}{\strue}{\always{I}\varphi}$
on $J=[u,v]$ with valuation $\alpha$.
Its scope induces the checking interval
$K=[u-\alpha(c)+\inf I,\;v+\sup I]$.
As illustrated in \Cref{fig:predicate-invariant},
the start of $K$ must not precede that of $J$.
Thus,
$u \le u-\alpha(c)+\inf I$,
i.e., $\alpha(c)\le \inf I$.
Likewise, for
$\intvpredupT{c}{\strue}{\eventually{I}\varphi}$,
the witness interval is
$K=[u-\alpha(c)+\inf I,\;u-\alpha(c)+\sup I]$.
For a witness at or after the start of $J$, it is necessary that
$u \le u-\alpha(c)+\sup I$,
i.e., $\alpha(c)\le \sup I$.

These predicate-specific clock constraints are required
for expansion rules to be sound and complete.
We call them \emph{invariants} of interval predicates
(further details are given \hyperref[par:invariant-preservation]{below}).
When the invariant of an interval predicate holds,
the expansion rule preserves its meaning through its expansion branches.
This correctness is formalized by the following theorem.

\begin{table}[t]
\caption{Clock constraints for invariants of interval predicates:
\invmark{1} position, \invmark{2} no past satisfaction, and \invmark{3} nonemptiness.
Column entries are conjoined; ``-'' means no additional constraint.
$\intvOf{J}{i}=\mathrel{\protect\intv{\ltLc}{u}{u'}{\rtLc}}$ is the current interval;
$k_*<i$ is the last position satisfying $[\varphi]$, with right marker $r_*$ (omit \invmark{2} if no such position exists).}
\label{tab:predicate-invariants}
\centering
\small
\setlength{\tabcolsep}{10pt}
\renewcommand{\arraystretch}{1.05}
\begin{tabular*}{0.98\linewidth}{@{\extracolsep{\fill}}ccccc}
\toprule
Condition
&$\intvpredupT{c}{\upLc}{\always{I}\varphi}$
&$\intvpredT{c}{d}{\upLc}{\dnLc}{\always{I}\varphi}$
&$\intvpredupT{c}{\upLc}{\eventually{I}\varphi}$
&$\intvpredT{c}{d}{\upLc}{\dnLc}{\eventually{I}\varphi}$\\
\midrule
\invmark{1}
&\multicolumn{2}{c}{$\textstyle c\ltDash{\neg\upLc\vee\ltLc}\inf I$}
&\multicolumn{2}{c}{$\textstyle c\ltDash{\neg\upLc\vee\ltLc}\sup I$}\\
\cmidrule(lr){2-3}\cmidrule(lr){4-5}
\invmark{2}
&\multicolumn{2}{c}{-}
&\multicolumn{2}{c}{$\textstyle c\ltDash{\neg\upLc\vee\neg r_*}u+\inf I-\sup\intvOf{J}{k_*}$}\\
\cmidrule(lr){2-3}\cmidrule(lr){4-5}
\invmark{3}&$\text{-}$&$d\ltDash{\upLc\wedge\dnLc}c+|I|$
&$0\ltDash{\upLc\wedge\rtLc}c+|\intvOf{J}{i}|$
&$d\ltDash{\upLc\wedge\dnLc}c$\\
\bottomrule
\end{tabular*}
\Description{Four columns give invariants for the two clocked forms of
always and eventually. Rows cover position, no past satisfaction, and
nonemptiness; boundary markers determine strictness, and dashes indicate
no constraint.}
\end{table}

\begin{theorem}
\label{thm:expansion-rule-correctness}
Let $p \rightarrow \Psi_1 \mid \cdots \mid \Psi_n$
be an expansion rule.
Given a signal $\sigma$ with time horizon $\tau$,
a partition $\partition{J}$ of $\domain{0}{\tau}$,
a clock valuation $\alpha$, and a position $i$,
if the invariant of $p$ %
holds, then: 
\begin{description}[font=\normalfont]
    \item[(Soundness)]
    For every branch $\Psi_j$, $1 \le j \le n$,
    if
    $\sigma,\partition{J},\alpha,i \models_\tau \Psi_j$,
    then
    $\sigma,\partition{J},\alpha,i \models_\tau p$.

    \item[(Completeness)]
    If $\partition{J}$ is fully stable and
    $\sigma,\partition{J},\alpha,i \models_\tau p$,
    then
    $\sigma,\partition{J},\alpha,i \models_\tau \Psi_j$
    for some branch $\Psi_j$. %
    
\end{description}
\end{theorem}

Unlike soundness, completeness requires the stronger assumption
that the partition is \emph{fully stable}.
To see why, consider
$\ipstl{(x \ge 0)\until{}(y \le 0)}$
holding over $J$.
By definition, for every $t \in J$,
there exists a witness point $t'$ where $y \le 0$ holds,
with $x \ge 0$ holding throughout $[t,t']$.
However, the branch
$\ipstl{(x \ge 0)},\ipstl{(y \le 0)}$
requires $y \le 0$ to hold stably over an entire partition interval.
The semantics guarantees $y \le 0$ only at the witness point $t'$,
not over the partition interval containing it.
Thus, the truth at $t'$ must be lifted to the whole interval.
The fully stable condition ensures this lifting.

\paragraph{Invariant Preservation under Expansion.}
\label{par:invariant-preservation}
Besides the positional condition (\invmark{1}), we require that satisfaction
not rely on past intervals (\invmark{2}) and that the interval to be checked
be nonempty (\invmark{3}).
Their conjunction in \Cref{tab:predicate-invariants} defines the \emph{invariant}
of a clocked interval predicate.
For example, the invariant of $\intvpredT{c}{d}{\upLc}{\dnLc}{\always{I}\varphi}$
is $(c\ltDash{\neg\upLc\vee\ltLc}\inf I)\wedge
(d\ltDash{\upLc\wedge\dnLc}c+|I|)$,
while that of $\intvpredupT{c}{\upLc}{\always{I}\varphi}$
is simply $c\ltDash{\neg\upLc\vee\ltLc}\inf I$.
The formal definition of invariant is given in \supplementref{sec:app_pred_inv}.

To apply expansion rules repeatedly, completeness must be witnessed by
a branch that also preserves the invariants of interval predicates.
Such a branch exists under the completeness assumptions.
The following proposition states this fact.

\begin{proposition}
\label{prop:pred_invariant}
Let $p \rightarrow \Psi_1 \mid \cdots \mid \Psi_n$
be an expansion rule.
Given a signal $\sigma$ with time horizon $\tau$,
a fully stable partition $\partition{J}$ of $\domain{0}{\tau}$,
a clock valuation $\alpha$, and a position $i$,
if the invariant of $p$
holds and
$\sigma,\partition{J},\alpha,i \models_\tau p$,
then there exists a branch $\Psi_j$ ($1 \le j \le n$) such that
$\sigma,\partition{J},\alpha,i \models_\tau \Psi_j$
and the invariant of every interval predicate occurring in $\Psi_j$ holds.
\end{proposition}

\section{Constructing Tableaux for STL}
\label{sec:tableau}

In this section, we construct a tableau $\tableau{\varphi}{}$ for an STL
formula $\varphi$ using interval predicates and expansion rules.
The resulting tableau satisfies the following properties:
\begin{enumerate*}[label=(\arabic*)]
\item it is sound and complete,
\item it is finite, and
\item it recognizes every signal satisfying $\varphi$ with respect to
its coarsest fully stable partition.
\end{enumerate*}
We first introduce \emph{closures}, which characterize fully expanded
current-interval judgements and are used to construct tableau states
(\Cref{sec:closure}).
We then establish their soundness and completeness
(\Cref{sec:closure-correct}).
Based on these results, we construct the tableau by repeatedly
expanding the next-interval judgements of each closure
(\Cref{sec:tableau-construct}).
We show that the resulting tableau uses
$O(|\varphi|\cdot r(\varphi))$ clocks and has
$2^{O(|\varphi|\cdot r(\varphi))}$ states
(\Cref{sub:tableau-finite}).
Finally, we present an on-the-fly algorithm that constructs the finite
tableau %
(\Cref{sec:on-the-fly}).

\subsection{Transition Form and Closures}
\label{sec:closure}

Closures are a standard component of tableau constructions for temporal
logics, where they represent fully expanded local information. We adopt
the same idea for STL. A closure is a judgement in which every interval
predicate for the current partition interval has been fully expanded,
together with a $\nextop$-part describing the judgements to be expanded
on the next partition interval. To express this separation, we introduce
\emph{transition forms}.

To define transition forms,
we first define state forms.
A judgement is in state form if it is of the form
$\Phi=C,\Lambda.L$,
where $C$ contains only clock constraints,
$\Lambda$ only clock resets,
and $L$ is a set of judgements containing no $\nextop$ operators,
clock constraints, or clock resets.
Any of $C$, $\Lambda$, and $L$ may be empty.
Transition forms are then defined by extending state forms with a
$\nextop$-prefixed state form.

\begin{definition}
A judgement is in \emph{transition form} if it is of the form
$\Phi=C,\Lambda.(L,\nextop(C',\Lambda'.L'))$,
where $C,\Lambda.L$ and $C',\Lambda'.L'$ are state-form judgements.
The next part $C',\Lambda'.L'$ may be empty.
\end{definition}

Since the next part may be empty,
every state-form judgement is also in transition form.
Moreover, every branch of the expansion rules
(\Cref{fig:unclocked-rule,fig:clocked-rule})
is in transition form.

\begin{example}
The judgements
$(c\ge5,\intvpredupT{c}{\strue}{\always{[2,3]}(x\ge0)},\nextop(c\le3))$
and
$\cset{\creset{c}{0}}.(\intvpredupT{c}{\strue}{\always{[2,3]}(x\ge0)})$
are in transition form with empty reset and next parts, respectively.
In contrast,
$\nextop\nextop(c\ge5)$
and
$\cset{\creset{c}{2}}.(c\ge5)$
are not, because one contains a nested $\nextop$ operator and
the other is not in state form.
\end{example}

Intuitively, a closure is a transition-form judgement
$\Phi=C,\Lambda.(L,\nextop(C',\Lambda'.L'))$
in which every interval predicate $p\in L$ is expanded.
For an expansion rule
$p \rightarrow \Psi_1 \mid \cdots \mid \Psi_n$,
this means that one branch $\Psi_i$ is included in $\Phi$.
Determining this inclusion, however, requires more than simple
syntactic inclusion due to clock resets.
We therefore define syntactic subsumption between judgements.

\begin{definition}\label{def:subsumption}
Let
$\Phi_i=C_i,\Lambda_i.(L_i,\nextop(C_i',\Lambda_i'.L_i'))$
for $i=1,2$
be transition-form judgements.
We say that $\Phi_1$ is \emph{syntactically subsumed} by $\Phi_2$,
written $\Phi_1\sqsubseteq\Phi_2$,
if and only if:
\begin{enumerate}[label=(\alph*)]
  \item
  $C_1 \subseteq C_2$, $\Lambda_1 \subseteq \Lambda_2$, $L_1 \subseteq L_2$, and
  $C'_1 \subseteq C'_2$, $\Lambda'_1 \subseteq \Lambda'_2$, $L'_1 \subseteq L'_2$;
  \item $\vars(\Lambda_2) \cap \fvar{\Lambda_1 . (L_1, \nextop(C_1', \Lambda_1' . L_1'))} = \emptyset$;
  \item $\vars(\Lambda_2') \cap \fvar{\Lambda_1'.L_1'} = \emptyset$.
\end{enumerate}
\end{definition}

\begin{example}
The relation
$(c > 2)
\sqsubseteq
((c > 2), \cset{c:=0}.\nextop(\cset{d:=0}.(\ipstl{x \ge 4})))$
holds because the reset $c:=0$ does not affect the clock constraint
$c>2$, and the clock $d$ is fresh.
In contrast,
$\cset{c:=0}.\nextop(d>4)
\not\sqsubseteq
\cset{c:=0, d:=0}.\nextop(d>4)$,
because $d$ occurs free on the left-hand side.
\end{example}

The relation $\Phi_1 \sqsubseteq \Phi_2$ means that $\Phi_2$
extends $\Phi_1$ with additional judgements, such as clock constraints,
resets, and interval predicates, while preserving the meaning of
$\Phi_1$.
The following lemma formalizes this property.

\begin{lemma}
\label{lem:tnf_subsumption}
For any transition-form judgements $\Phi_1$ and $\Phi_2$,
if $\Phi_1 \sqsubseteq \Phi_2$, then
$\Phi_2 \Rightarrow\Phi_1$.
\end{lemma}

Using syntactic subsumption, we can determine whether a
transition-form judgement is already expanded for an interval predicate
$p$, that is, whether it is \emph{rule-closed} for $p$.

\begin{definition}\label{def:rule-closed-single}
A transition-form judgement $\Phi$ is \emph{rule-closed} for a predicate
$p$ iff the following holds:
if there is an expansion rule
$p \rightarrow \Psi_1 \mid \cdots \mid \Psi_n$,
then some branch $\Psi_i$ satisfies
$\Psi_i \sqsubseteq \Phi$.\footnote{%
For $\Phi=C,\Lambda.(L,\nextop(C',\Lambda'.L'))$ with $p\in L$, both $p$
and its branch are under the current resets $\Lambda$. Thus, $\Psi_i\sqsubseteq\Phi$
means that $\Phi$ subsumes a transition form of $\Lambda.\Psi_i$
with consistent names for bound clocks.}
\end{definition}

\begin{example}\label{ex:rl-closed}
The judgement
$\Phi_1=(c\le3,\ipstl{x\ge0})$
is rule-closed for $\ipstl{x\ge0}$,
since atomic predicates have no expansion rules.
The judgement
$\Phi_2=(c\le3,\intvpredupT{c}{\strue}{\always{[2,3]}(x\ge0)})$
is not rule-closed,
since no branch of the expansion rule for
$\intvpredupT{c}{\strue}{\always{[2,3]}(x\ge0)}$
is subsumed by $\Phi_2$.
\end{example}

A rule-closed extension of $\Phi$ is \emph{minimal} if no strictly
smaller rule-closed extension of $\Phi$ exists.
For $\Phi=\ipstl{x \ge 0}$, $\Phi$ itself is minimal, but
$(c \le 3, \ipstl{x \ge 0})$ is not: removing $c \le 3$ leaves
the already rule-closed judgement $\Phi$.
We now combine rule-closure and minimality to define closures.

\begin{definition}\label{def:closure}
A \emph{closure} of $\Phi$ is a
transition-form judgement $\Phi_c$ such that
\begin{enumerate*}[label=(\roman*)]
\item $\Phi\sqsubseteq\Phi_c$,
\item $\Phi_c$ is rule-closed for every interval predicate in its current part, and
\item $\Phi_c$ is $\sqsubseteq$-minimal among the judgements satisfying (i) and (ii).
\end{enumerate*}
\end{definition}

\begin{example}
Let
$\Phi=\cset{\creset{c}{0}}.
(\intvpredupT{c}{\strue}{\always{[6,12]}(u\ge4)})$.
Then,
$
\Phi_c=
\cset{\creset{c}{0}}.
(\intvpredupT{c}{\strue}{\always{[6,12]}(u\ge4)},\allowbreak
\ipstl{u\ge4},
\nextop(c>6,\ast))$
is a closure of $\Phi$, since
$\Phi\sqsubseteq\Phi_c$,
$\Phi_c$ is rule-closed for
$\intvpredupT{c}{\strue}{\always{[6,12]}(u\ge4)}$,
and minimal.
\end{example}

\subsection{Correctness of Closure}
\label{sec:closure-correct}

In this section, we show that closures are sound and complete,
and that every closure can be obtained by finitely many expansions.
We first define the complexity of interval predicates for %
analyzing closures.
The complexity of an interval predicate $p$, denoted $\comp{p}$, measures the syntactic size by
counting Boolean and temporal operators together with predicate
constructs.

\begin{definition}\label{def:complexity}
For an interval predicate $p$, define $\comp{p}$ recursively as follows, where
$\bothtemp \in \{\always{},\eventually{}\}$,
$\natural \in \{\intvpredupT{c}{\upLc}{},
\intvpredT{c}{d}{\upLc}{\dnLc}{}\}$,
$\star \in \{\vee,\wedge,\until{},\release{}\}$,
and $\bowtie \; \in \{\geq,>,=,<,\leq\}$:
\begin{align*}
\comp{\ipstl{\varphi}} &= \comp{\varphi},
&
\comp{\natural \bothtemp_I\varphi} &= \comp{\varphi} + 1,
&
\comp{f(\vec{x}) \bowtie 0} &= 0,
\\
\comp{\bothtemp_I \varphi} &= \comp{\varphi} + 2,
&
\comp{\varphi \star \varphi'} &= \comp{\varphi} + \comp{\varphi'} + 2,
&
\comp{\mathit{true}} &= 0.
\end{align*}
\end{definition}

The complexity measure strictly decreases under expansion.
In particular, every interval predicate introduced by an expansion rule
has strictly smaller complexity than the predicate being expanded.

\begin{lemma}
\label{lem:complexity-reduced}
Let
$p \rightarrow \Psi_1 \mid \cdots \mid \Psi_n$
be an expansion rule.
For every branch
$\Psi_i=\Lambda.(L,\nextop(C',\Lambda'.L'))$
and every interval predicate $q\in L$,
$\comp{p}>\comp{q}$.
\end{lemma}

\begin{example}
Consider the predicate
$p=\ipstl{(\always{[6,12]}(u\ge4))\release{}(t\le5)}$.
Since $\comp{u\ge4}=\comp{t\le5}=0$ and
$\comp{\always{[6,12]}(u\ge4)}=2$,
we have $\comp{p}=2+0+2=4$.
Expanding $p$ by the release rule can introduce
$\ipstl{\always{[6,12]}(u\ge4)}$, whose complexity is $2<4$.
\end{example}

A closure is obtained by repeatedly expanding interval predicates in the
current part of transition-form judgements.
However, expansion does not preserve transition form in general.
For example, applying the rule
\scalebox{0.96}{$\alwaysUpPred \to \RuleAlwaysUpBranchOne$}
to
$\Phi=C,\Lambda.(L,\alwaysUpPred,\nextop(C',\Lambda'.L'))$
produces a judgement containing two $\nextop$ operators, which is not
in transition form.
To construct closures, we therefore show that every such
judgement can be rewritten as an equivalent transition-form judgement
using the structural laws in \Cref{prop:judg-equiv}.

\begin{lemma}
\label{lem:tnf-existence}
Given a transition-form judgement
$\Phi=C,\Lambda.(L,\nextop(C',\Lambda'.L'))$
and an expansion rule
$p\to\Psi_1\mid\cdots\mid\Psi_n$
for $p\in L$,
for every branch $\Psi_i$, if $\Phi$ and $\Psi_i$ satisfy the
assumptions of \Cref{prop:judg-equiv}, then there exists a
transition-form judgement $\hat{\Phi}$ such that
$\hat{\Phi}\equiv C,\Lambda.(L,\Psi_i,\nextop(C',\Lambda'.L'))$.
\end{lemma}

We can now make precise the notion of rule expansion.
For an expansion rule $p\to\Psi_1\mid\cdots\mid\Psi_n$,
a $j$-th \emph{rule expansion} of a transition-form judgement $\Phi$ for $p$
is a $\sqsubseteq$-minimal transition-form judgement $\Phi'$ with
$\Phi\sqsubseteq\Phi'$ and $\Psi_j\sqsubseteq\Phi'$.
We obtain such an expansion by adding $\Psi_j$ alongside $p$ in $L$ and
restoring transition form as in \Cref{lem:tnf-existence}.
The result contains exactly the components needed to satisfy both subsumptions.
Note that $\Phi'$ is rule-closed for $p$.
So, repeated expansions yield closures.

\paragraph{Soundness of Closures}

We now present the soundness of closures.
Recall that a closure may contain non-atomic interval predicates
introduced by expansion rules.
Soundness means that these predicates are semantically implied by the
atomic interval predicates in the closure.

\begin{theorem}
\label{thm:closure-soundness}
Let $\Phi = C,\Lambda.(L,\nextop(C',\Lambda'.L'))$ be a closure,
and let $A \subseteq L$ be the set obtained from $L$ by removing all non-atomic interval predicates.
Given a signal $\sigma$ with time horizon $\tau$,
a partition $\partition{J}$ of $\domain{0}{\tau}$,
a clock valuation $\alpha$, and a position $i$,
if the invariant of every interval predicate in $L$ is satisfied
and $\sigma,\partition{J},\alpha,i \models_\tau C,\Lambda.(A,\nextop(C',\Lambda'.L'))$,
then $\sigma,\partition{J},\alpha,i \models_\tau C,\Lambda.L$.
\end{theorem}

\begin{proof}[Proof sketch]
The proof is by induction on the complexity of interval predicates.
We show that every $p\in L$ is semantically implied by the remaining judgements of $\Phi$.
The base case is immediate since $\comp{p}=0$ implies $p\in A$.
For the inductive step, let
$p\rightarrow\Psi_1\mid\cdots\mid\Psi_n$ be a rule.
Since $\Phi$ is rule-closed for $p$, there exists a branch
$\Psi_i\sqsubseteq\Phi$.
By \Cref{lem:complexity-reduced}, every interval predicate $q$ appearing in the current part of $\Psi_i$
satisfies $\comp{q}<\comp{p}$.
Hence, each such $q$ is semantically implied by the induction hypothesis.
Together with the remaining judgements inherited from $\Phi$,
this implies that $\Psi_i$ is satisfied.
Finally, \Cref{thm:expansion-rule-correctness}~(1)
implies that $p$ is satisfied.
\end{proof}

\paragraph{Termination of Expansions}

Since a rule expansion yields a judgement that is rule-closed for
the expanded predicate $p$, we define the $p$-expansion relation
$\lexpansion_p$ as follows.

\begin{definition}\label{def:expansion-levels}
For transition-form judgements $\Phi$ and $\Phi'$,
$\lexpansion_p(\Phi,\Phi')$ holds iff:
\begin{enumerate*}[label=(\roman*)]
  \item $\Phi$ is rule-closed for $p$ and $\Phi'=\Phi$; or
  \item for some $j$, $\Phi'$ is a $j$-th rule expansion of $\Phi$ for $p$.
\end{enumerate*}
\end{definition}

Recall from \Cref{prop:pred_invariant} that if the invariant of an
interval predicate $p$ holds, then the invariants of all interval
predicates introduced by some expansion branch also hold.
If $\lexpansion_p(\Phi_1,\Phi_2)$, then
$\Phi_2$ contains all interval predicates occurring in $\Phi_1$,
together with those introduced by the expansion rule.
Hence, some $p$-expansion preserves the invariants of interval
predicates.

\begin{proposition}
\label{prop:expansion-invariant}
Given a signal $\sigma$ with time horizon $\tau$,
a fully stable partition $\partition{J}$ of $\domain{0}{\tau}$,
a clock valuation $\alpha$, and a position $i$,
if all predicate invariants in $\Phi_1$ hold and
$\sigma,\partition{J},\alpha,i\models_\tau\Phi_1$,
then there exists $\Phi_2$ such that
$\lexpansion_p(\Phi_1,\Phi_2)$,
$\sigma,\partition{J},\alpha,i\models_\tau\Phi_2$,
and all predicate invariants in $\Phi_2$ hold.
\end{proposition}

Building on the $p$-expansion relation, we define the $i$-th expansion
of a judgement as the set of all judgements obtained by $i$ successive expansions.
Let $\lexpansion_p(\Phi):=\{\Phi' \mid \lexpansion_p(\Phi,\Phi')\}$.

\begin{definition}
Given a judgement $\Phi$ in transition form,
the $i$-th \emph{expansion} of $\Phi$,
denoted $\expansion_i(\Phi)$,
is defined inductively as follows:
\begin{enumerate*}[label=(\alph*)]
\item
$\expansion_0(\Phi):=\{\Phi\}$, and
\item
for any $i\ge0$,
$\expansion_{i+1}(\Phi)
=
\{
\Phi'
\mid
\Psi=\allowbreak C,\Lambda.(L,\nextop(C',\Lambda'.L'))
\in\expansion_i(\Phi),
p\in L,
\Phi'\in\lexpansion_p(\Psi)
\}$.
\end{enumerate*}
\end{definition}

To show termination, we associate each $i$-th expansion with a ranking.
The ranking %
is defined as
the maximal complexity of the interval predicates newly introduced in
that expansion.

\begin{definition}%
\label{def:expansion-complexity}
Given a judgement $\Phi$ in transition form,
the ranking of the $i$-th expansion of $\Phi$, denoted $\ranklvl(\Phi, i)$, 
is defined as follows:\footnote{We adopt the convention that $\max \emptyset = 0$.}
\begin{enumerate}[label=(\roman*)]
  \item $\ranklvl(\Phi,0) := \max \{ \comp{p} \mid p \in \curlabels_0(\Phi) \}$, and
  \item for $i \ge 1$,
  $\ranklvl(\Phi,i) := \max \{ \comp{p} \mid p \in \curlabels_{i}(\Phi) \setminus \curlabels_{i - 1}(\Phi) \}$,
\end{enumerate}
where for each $i \ge 0$, $\curlabels_i(\Phi) := \bigcup \{ L \mid C, \Lambda.(L, \nextop(C', \Lambda'.L')) \in \expansion_i(\Phi) \}$.
\end{definition}

The ranking decreases because every newly generated interval predicate
has strictly smaller complexity than the predicate from which it is
generated (\Cref{lem:complexity-reduced}). Hence the maximum complexity
among newly generated predicates strictly decreases at each expansion
step.

\begin{lemma}
\label{lem:expansion-ranking-decrease}
For any transition-form judgement $\Phi$ and any $i\ge0$,
if $\ranklvl(\Phi,i)>0$, then
\[
\ranklvl(\Phi,i+1)<\ranklvl(\Phi,i).
\]
\end{lemma}

For any transition-form judgement $\Phi$, as a consequence of
\Cref{lem:expansion-ranking-decrease}, the ranking eventually reaches
$0$. At that point, no further non-atomic interval predicates can be
introduced.

\begin{lemma}
\label{lem:expansion-termination}
There exists $k\in\natset$
such that, for all $i\ge k$, $\ranklvl(\Phi,i)=0$.
\end{lemma}

\paragraph{Completeness of Closures}

We now present the completeness of closures.
Completeness means that every satisfiable transition-form judgement has
a satisfiable closure.
To prove this, we first show that satisfiability is preserved under
$p$-expansion.
Let
$\Phi=C,\Lambda.(L,\nextop(C',\Lambda'.L'))$
be a transition-form judgement, and fix $p\in L$.
If $\Phi$ is satisfiable and the invariants of all interval predicates
occurring in $\Phi$ hold, then there exists
$\Phi'\in\lexpansion_p(\Phi)$ such that $\Phi'$ is satisfiable.

\begin{lemma}
\label{lem:satisfiable-l-expansion}
Given a signal $\sigma$ with time horizon $\tau$,
a fully stable partition $\partition{J}$ of $\domain{0}{\tau}$,
a clock valuation $\alpha$, and a position $i$,
if the invariants of all interval predicates occurring in $\Phi$ hold
and
$\sigma,\partition{J},\alpha,i\models_\tau\Phi$,
then there exists
$\Phi'\in\lexpansion_p(\Phi)$
such that
$\sigma,\partition{J},\alpha,i\models_\tau\Phi'$.
\end{lemma}

\begin{proof}[Proof sketch]
If $\Phi$ is already rule-closed for $p$, the claim is immediate.
Otherwise, since $p$ is satisfied and its invariant holds, by assumption,
\Cref{thm:expansion-rule-correctness}~(2) yields a satisfied expansion branch
$\Psi$ for $p$.
Extending $\Phi$ with the judgements in $\Psi$ yields a satisfied judgement, and
\Cref{lem:tnf-existence} transforms it into an equivalent transition-form judgement
$\hat{\Phi}$.
By construction, $\hat{\Phi}$ is a $p$-expansion of $\Phi$.
Hence, there exists a $p$-expansion of $\Phi$ that is satisfied.
\end{proof}

The completeness of closures follows by repeatedly constructing
satisfiable expansions.
By \Cref{lem:satisfiable-l-expansion,prop:expansion-invariant},
satisfiable expansions can always be chosen repeatedly.
By \Cref{lem:expansion-termination}, this process eventually
terminates.
Hence, a satisfiable closure is obtained.

\begin{theorem}
\label{thm:closure-completeness}
Let $\Phi$ be a transition-form judgement.
Given a signal $\sigma$ with time horizon $\tau$,
a fully stable partition $\partition{J}$ of $\domain{0}{\tau}$,
a clock valuation $\alpha$, and a position $i$,
if the invariants of all predicates occurring in
$\Phi$ hold and
$\sigma,\partition{J},\alpha,i\models_\tau\Phi$,
then there exists a closure $\Phi_c$ of $\Phi$ such that
$\sigma,\partition{J},\alpha,i\models_\tau\Phi_c$.
\end{theorem}

\subsection{Tableau Construction for STL}
\label{sec:tableau-construct}

Using closures, we construct tableaux from STL formulas.
We first give a general definition of a tableau.
Let $\atomset$ be the set of atomic judgements other than clock constraints and
non-atomic interval predicates, and let
$\mathit{CC}$ and $\mathit{CR}$
be the sets of clock constraints and clock resets, respectively.
A tableau is a transition system over
$\atomset$, $\mathit{CC}$, and $\mathit{CR}$,
whose states are labeled by judgements in $\atomset$
and whose transitions are labeled by pairs from
$\mathit{CC}\times\mathit{CR}$.

\begin{definition}
A tableau is a tuple
$\mathcal{T}=(X,X_c,S,T,T_0,S_F,\mathcal L)$,
where
\begin{itemize}
\item $X$ is a set of variables, $X_c$ a set of clock variables, and $S$ a set of states;
\item $T\subseteq S\times(\mathit{CC}\times\mathit{CR})\times S$
      is the transition relation;
\item $T_0\subseteq(\mathit{CC}\times\mathit{CR})\times S$
      is the initial transition relation;
\item $S_F\subseteq S$ is the set of final states;
\item $\mathcal L:S\rightarrow\power{\atomset}$ is the labeling function,
      assigning each state a subset of $\atomset$.
\end{itemize}
\end{definition}

A path is a finite alternating sequence of transition labels and states.

\begin{definition}
A path of a tableau $\mathcal T$ is a finite sequence
$\pi=\eta_0s_0\eta_1s_1\cdots\eta_ns_n$,
where $s_i\in S$, $\eta_i\in\mathit{CC}\times\mathit{CR}$,
and $(s_{i-1},\eta_i,s_i)\in T$ for every $1\le i\le n$.
\end{definition}

A path $\pi$ is \emph{accepting} if $s_n\in S_F$.
For $0\le i\le n$, its $i$-th suffix is
$\pi^i=\eta_is_i\cdots\eta_ns_n$.
A path is a \emph{run} if $(\eta_0,s_0)\in T_0$.
An \emph{accepting run} is an accepting path that is a run.
Each path induces a judgement, called its \emph{trace}, obtained by
recursively unfolding the labels of its states.

\begin{definition}
For a path $\pi = \eta_0 s_0 \cdots \eta_n s_n$ with $\eta_i = (C_i,\Lambda_i)$,
its trace, denoted $\trace{\pi}$, is the judgement defined recursively as follows.
For $0 \le i < n$:
\[
\trace{\eta_n s_n} = C_n, \Lambda_n.\labelcur{s_n},
\quad
\trace{\eta_i s_i \pi^{i+1}} =
C_i, \Lambda_i.(\labelcur{s_i}, \nextop \trace{\pi^{i+1}}).
\]
\end{definition}

\begin{example}
For a path
$\pi=(C_0,\Lambda_0)s_0(C_1,\Lambda_1)\cdots(C_n,\Lambda_n)s_n$,
its trace is
\[
\trace{\pi}
=
C_0,\Lambda_0.
(\labelcur{s_0},
\nextop(
C_1,\Lambda_1.
(\labelcur{s_1},
\nextop(\cdots
\nextop(C_n,\Lambda_n.\labelcur{s_n})))).
\]
\end{example}

Since a trace is a judgement, a tableau can \emph{recognize} a signal
through the traces of its accepting runs.
The following definition formalizes this notion.

\begin{definition}
A tableau
$\tableau{}{}$ recognizes a signal $\sigma$ with respect to a partition $\partition{J}$
if there exists an accepting run
$\pi$ of $\tableau{}{}$ such that, for every clock valuation $\alpha$,
$\sigma,\partition{J},\alpha,0\models_\tau\trace{\pi}$.
\end{definition}

We now construct the tableau for an STL formula $\varphi$.
Interval sequence judgements are used as \emph{generators} of tableau
states.
Given a generator $M_g$, we compute its closure
$\Phi=C,\Lambda.(L,\nextop M_g')$.
The state-form judgement preceding $\nextop$ determines a tableau state, where
the atomic judgements in $L$ label the state, while $(C,\Lambda)$ labels the incoming
transition.
The state-form judgement $M_g'$ becomes the generator for successor states.
Accordingly, a tableau state is represented by a pair
$\langle\Phi,M_g\rangle$.
An initial state is a state whose generator is
$\ipstl{\varphi}$.
A final state is a state whose closure is in state form.

\begin{definition}\label{def:tab-construction}
Given an STL formula $\varphi$,
$\tabG{\varphi}=(X,X_c,S,T,T_0,S_F,\mathcal L)$ is a tableau, where:
\begin{itemize}
  \item $X$ is the set of real-valued variables in $\varphi$, and $X_c$ is a set of clock variables;

  \item $\langle\Phi,M_g\rangle\in S$
  iff $M_g$ is in state form and $\Phi$ is a closure of $M_g$;

  \item
  $(\langle\Phi,M_g\rangle,(C',\Lambda'),
  \langle\Phi',M_g'\rangle)\in T$
  iff
  $\Phi=C,\Lambda.(L,\nextop M_g')$ and
  $\Phi'=C',\Lambda'.(L',\nextop M_g'')$;

  \item
  $((C_0,\Lambda_0),\langle\Phi_0,M_g\rangle)\in T_0$
  iff $M_g=\ipstl{\varphi}$ and $\Phi_0=C_0,\Lambda_0.(L_0,\nextop M_g')$;

  \item
  $\langle\Phi,M_g\rangle\in S_F$
  iff $\Phi$ is in state form;

\item
$\mathcal L(\langle\Phi,M_g\rangle)
=
\{\,l
\mid
\Phi=C,\Lambda.(L,\nextop M),\;
l\in L,\;
l\text{ is }
\ast,\ \lty{\upLc},\text{ or an atomic predicate}
\,\}$.
\end{itemize}
\end{definition}

\begin{figure}[t]
    \includegraphics{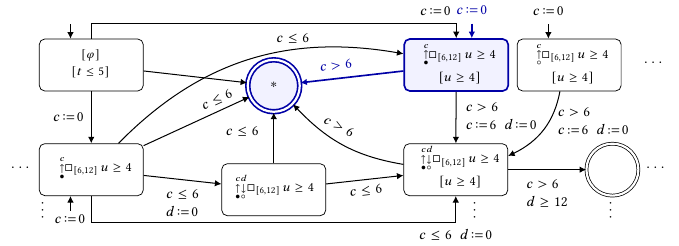}
    \caption{Example of $\tableau{\varphi}{}$, where $\varphi = (\always{[6,12]}u \ge 4) \release{} t \le 5$. Some predicates are omitted for readability. The highlighted rectangle and double circle represent the initial state $s_0$ and final state $s_1$, respectively.}
    \label{fig:ex-tab}
    \Description{Boxes denote tableau states, double circles mark final
    states, and arrows show clock guards and resets. The highlighted path
    leads from the initial state with an always requirement to the final
    state labeled with a star.}
\end{figure}

\begin{example}
Let $\varphi=(\always{[6,12]}(u\ge4))\release{}(t\le5)$.
One closure of $\ipstl{\varphi}$ is
$\Phi_1=\cset{\creset{c}{0}}.(
\ipstl{\varphi},\allowbreak
\ipstl{\always{[6,12]}(u\ge4)},\allowbreak
\intvpredupT{c}{\strue}{\always{[6,12]}u\ge4},\allowbreak
\ipstl{u\ge4},\allowbreak\lty{\strue},\allowbreak\nextop M)$,
where $M=(c>6,\ast)$.
As illustrated in \Cref{fig:ex-tab},
$s_0=\langle\Phi_1,\ipstl{\varphi}\rangle$
is an initial state whose incoming transition is labeled by
$\creset{c}{0}$.
Moreover, since $M$ is its own closure and is in state form, the successor state
$s_1=\langle M,M\rangle$ is final, and there is a transition from $s_0$ to $s_1$ labeled by
$c>6$.
\end{example}

Recall that every path of a tableau induces a trace.
For a path of $\tabG{\varphi}$, the trace consists of the clock constraints,
reset sets, and atomic judgements occurring along the path, excluding
non-atomic interval predicates.
To retain the complete semantic information of paths,
we additionally introduce \emph{full traces}, which also include
non-atomic interval predicates.

\begin{definition}
Given a path
$\pi=\eta_0s_0\cdots\eta_ns_n$
of $\tabG{\varphi}$,
where
$s_i=\langle C_i,\Lambda_i.(L_i,\nextop M_g^{i+1}),M_g^i\rangle$,
the full trace of $\pi$, denoted $\ftrace{\pi}$, is defined recursively by
\[
\ftrace{\eta_ns_n}=C_n,\Lambda_n.L_n,
\qquad
\ftrace{\eta_is_i\pi^{i+1}}
=
C_i,\Lambda_i.(L_i,\nextop\ftrace{\pi^{i+1}}).
\]
\end{definition}

The $\trace{\pi}$ and $\ftrace{\pi}$ are semantically equivalent.
Since $\ftrace{\pi}$ differs from $\trace{\pi}$ only by non-atomic
interval predicates, $\trace{\pi}\sqsubseteq\ftrace{\pi}$. Hence,
by \Cref{lem:tnf_subsumption},
$\ftrace{\pi}$ implies $\trace{\pi}$.
For the converse, backward induction on the suffixes of $\pi$, together
with \Cref{thm:closure-soundness}, recovers the omitted non-atomic
interval predicates at each state. Thus, $\trace{\pi}$ implies
$\ftrace{\pi}$.

\begin{lemma}
\label{lem:trace-equivalence}
For any accepting run $\pi$ of $\tabG{\varphi}$,
$\trace{\pi} \equiv \ftrace{\pi}$.
\end{lemma}

\paragraph{Correctness}

We now present the soundness of $\tabG{\varphi}$.
Soundness means that every signal recognized by the tableau satisfies
$\varphi$.
Since the full trace of an accepting run is constructed from the
closures obtained by expanding $\ipstl{\varphi}$ over a partition,
it captures the semantic meaning of $\varphi$.

\begin{theorem}
\label{thm:tableau-sound}
If a signal $\sigma$ is recognized by $\tabG{\varphi}$ with respect to $\partition{J}$,
then $\sigma, 0 \models_\tau \varphi$.
\end{theorem}

\begin{proof}
Since $\sigma$ is recognized by $\tabG{\varphi}$ with respect to $\partition{J}$,
there exists an accepting run $\pi$ such that
$\sigma,\partition{J},\alpha,0\models_\tau\trace{\pi}$
for any clock valuation $\alpha$.
By \Cref{lem:trace-equivalence},
$\sigma,\partition{J},\alpha,0\models_\tau\ftrace{\pi}$.
Since the initial state of $\pi$ is a closure of $\ipstl{\varphi}$,
$\ipstl{\varphi}\sqsubseteq\ftrace{\pi}$.
Hence, by \Cref{lem:tnf_subsumption},
$\sigma,\partition{J},\alpha,0\models_\tau\ipstl{\varphi}$.
By the semantics of $\ipstl{\varphi}$,
$\sigma,t\models_\tau\varphi$ for every
$t\in\partition{J}(0)$.
Since $0\in\partition{J}(0)$,
$\sigma,0\models_\tau\varphi$.
\end{proof}

Completeness of $\tabG{\varphi}$ means that every signal satisfying
$\varphi$ is recognized by $\tabG{\varphi}$.
To establish this, we first show that every satisfied state-form
judgement has an accepting tableau path starting from one of its
closures.
By \Cref{thm:closure-completeness}, it has a satisfied closure.
If the closure has no
$\nextop$-part, it is already a final state.
Otherwise, its $\nextop$-part becomes the starting point for the
next partition interval, and repeating this construction eventually
reaches a final state.

\begin{lemma}
\label{lem:generator-path-existence}
Given a signal $\sigma$ with time horizon $\tau$,
a fully stable partition $\partition{J}$ of $\domain{0}{\tau}$ for $\varphi$,
a clock valuation $\alpha$, and a position $i$,
let $M_g=C,\Lambda.L$ be a state-form judgement.
If the invariants of all predicates in $L$ hold and
$\sigma,\partition{J},\alpha,i\models_\tau M_g$,
then there exists a closure $\Phi$ of $M_g$ such that the state
$s=\langle\Phi,M_g\rangle$
has an accepting path $\pi$ satisfying
$\sigma,\partition{J},\alpha,i\models_\tau\ftrace{\pi}$.
\end{lemma}

Under a fully stable partition,
$\ipstl{\varphi}$ is satisfied whenever
$\sigma,0\models_\tau\varphi$.
By \Cref{lem:generator-path-existence},
an accepting run whose full trace is satisfied by $\sigma$ exists.
We thus obtain the following completeness theorem.

\begin{theorem}
\label{thm:tableau-complete}
If $\sigma$ has finite variability for $\varphi$ and
$\sigma, 0 \models_\tau \varphi$ then, for every fully stable partition $\partition{J}$ for $\varphi$,
$\sigma$ is recognized by $\tabG{\varphi}$ with respect to $\partition{J}$.
\end{theorem}

\begin{proof}
Let $\partition{J}$ be a fully stable partition of $\domain{0}{\tau}$
for $\varphi$.
Since $0\in\partition{J}(0)$ and
$\sigma,0\models_\tau\varphi$,
the stability of $\partition{J}$ implies
$(\forall t\in\partition{J}(0))\,\sigma,t\models_\tau\varphi$.
Hence, for any clock valuation $\alpha$,
$\sigma,\partition{J},\alpha,0\models_\tau\ipstl{\varphi}$.
Since the invariant of $\ipstl{\varphi}$ is $\true$,
\Cref{lem:generator-path-existence} yields a closure $\Phi$ and an
accepting path $\pi$ from
$s=\langle\Phi,\ipstl{\varphi}\rangle$
such that
$\sigma,\partition{J},\alpha,0\models_\tau\ftrace{\pi}$.
By \Cref{lem:trace-equivalence},
$\sigma,\partition{J},\alpha,0\models_\tau\trace{\pi}$.
Since $s$ is initial, $\pi$ is an accepting run.
Thus,
$\sigma$ is recognized by $\tabG{\varphi}$.
\end{proof}

Hence every signal $\sigma$ with finite variability that satisfies
$\varphi$ is recognized by $\tabG{\varphi}$ with respect to its
coarsest fully stable partition.

\subsection{Finiteness of the Tableau}
\label{sub:tableau-finite}

The tableau construction presented so far is defined over an infinite
state space. This formulation is convenient for establishing the
soundness and completeness of the construction.
In this section, we construct a finite tableau while preserving
soundness and completeness. We first introduce a simplified tableau
that allows interval predicates to be combined. We then show that
accepting runs can be chosen with a bounded number of
interval predicates in each state. These runs can also be chosen
to use a fixed finite set of clocks, leading to a finite tableau.
The resulting tableau uses $O(|\varphi|\cdot r(\varphi))$ clocks
and has at most $2^{O(|\varphi|\cdot r(\varphi))}$ states.
(See \supplementref{sec:app-tab-finiteness} for details.)

Several interval predicates may track the same temporal subformula
over different intervals. When these intervals overlap or meet without a gap, \Cref{lem:intv-merge} allows the predicates to be
combined under its side conditions.
The simplified tableau, denoted $\simptabG{\varphi}$, is obtained
from $\tabG{\varphi}$ by combining these interval predicates,
while preserving the semantics of the tableau. %

\begin{lemma}
\label{lem:simp-trace-equiv}
The accepting runs of $\tabG{\varphi}$ and $\simptabG{\varphi}$ correspond.
\end{lemma}

In $\simptabG{\varphi}$, accepting runs can be chosen with a bounded
number of interval predicates in each state.
The key observation is that, while a timed temporal subformula
remains true, at most one predicate is needed to track
its truth during that time, since additional predicates can be
merged using \Cref{lem:intv-merge}.
Following~\cite{ferrere2019real}, let $r(\varphi)$ denote the resolution
of $\varphi$.\footnote{
The resolution is defined inductively by
$r(\mathit{true})=r(f(\vec{x})\bowtie0)=1$,
$r(\varphi_1\vee\varphi_2)=\max\{r(\varphi_1),\allowbreak r(\varphi_2)\}$,
$r(\neg\varphi)=r(\varphi)$, and
$r(\varphi_1\until{I}\varphi_2)
=\max\{r(\varphi_1),\allowbreak r(\varphi_2),\allowbreak r(I)\}$,
where $r(I)=\lceil{\sup I/(\sup I-\inf I)}\rceil$
if $\sup I<\infty$, and $r(I)=1$ otherwise.}
The number of truth changes of $\bothtemp_I\psi$ over any
interval of length $\sup I$ is $O(r(I))$~\cite{ferrere2019real}.
This also bounds how many predicates are needed simultaneously.
Summing over the $O(|\varphi|)$ subformulas gives
$O(|\varphi|\cdot r(\varphi))$. %

\begin{lemma}\label{lem:bounded-accepting-run}
Accepting runs of $\simptabG{\varphi}$ need only $O(|\varphi|\cdot r(\varphi))$ interval predicates per state.
\end{lemma}

For the accepting runs above,
$O(|\varphi|\cdot r(\varphi))$ clocks suffice in each state,
since each interval predicate uses at most two clocks.
Clock names can be reused once no longer needed.
Thus, an accepting run can be chosen using
$O(|\varphi|\cdot r(\varphi))$ clocks throughout the run.

These runs use states formed from a fixed collection of
$O(|\varphi|\cdot r(\varphi))$ judgements and clock resets.
Let $\tableau{\varphi}{}$ be the subgraph of $\simptabG{\varphi}$
consisting of these states.
Since states are formed from sets of these judgements and resets,
$\tableau{\varphi}{}$ has at most
$2^{O(|\varphi|\cdot r(\varphi))}$ states.
Together with the preceding soundness and completeness results,
this yields the following theorem.

\begin{theorem}
\label{thm:finite-tableau}
For any STL formula $\varphi$, there exists a tableau $\tableau{\varphi}{}$ with
$O(|\varphi|\cdot r(\varphi))$ clocks and
$2^{O(|\varphi|\cdot r(\varphi))}$ states.
Moreover, for every signal $\sigma$ with time horizon $\tau$, the following holds:
\begin{enumerate}
  \item (Soundness) If $\sigma$ is recognized by $\tableau{\varphi}{}$ with respect to $\partition{J}$, then $\sigma,0\models_\tau\varphi$.
  \item (Completeness) If $\sigma$ has finite variability for $\varphi$
  and $\sigma,0\models_\tau\varphi$ then, for any fully stable partition $\partition{J}$, $\sigma$ is recognized by $\tableau{\varphi}{}$ with respect to $\partition{J}$.
\end{enumerate}
\end{theorem}

\begin{algorithm}[t]
\caption{On-the-fly tableau construction}
\label{alg:on-the-fly}
\small
\algsetup{indent=1.5em}
\renewcommand{\algorithmicbody}{\textbf{procedure}\ \textsc{ConstructTableau}($\varphi$)}
\renewcommand{\algorithmiccomment}[1]{\hfill\textcolor{black!55}{$\triangleright$~#1}}
\begin{algorithmic}[1]
\BODY
  \STATE Fix $Q$ and judgement forms according to the bounds \COMMENT{\Cref{sub:tableau-finite}}%
  \STATE $S_0\gets$ initial states from $\mathcal C_Q(\ipstl{\varphi})$, applying the same bounds
  \STATE $S_{-1}\gets\emptyset$; $i\gets0$
  \REPEAT
    \STATE $B\gets$ successors of $S_i\setminus S_{i-1}$ generated using $\mathcal C_Q$ \COMMENT{expand and merge}%
    \STATE Discard from $B$ states exceeding predicate bounds
    \STATE $S_{i+1}\gets S_i\cup B$; add the corresponding transitions
    \STATE $i\gets i+1$
  \UNTIL{$S_i=S_{i-1}$} \COMMENT{fixed point}%
  \STATE \algorithmicreturn\ the constructed tableau
\ENDBODY
\STATE \algorithmicend\ \textbf{procedure}
\end{algorithmic}
\end{algorithm}

\subsection{On-the-Fly Tableau Construction}
\label{sec:on-the-fly}
We now describe how to construct the finite tableau obtained in the
preceding section on the fly (\Cref{alg:on-the-fly}). Starting from
$\ipstl{\varphi}$, the construction generates initial closures and
proceeds breadth-first, expanding their next judgements to obtain
successor closures. Interval predicates are merged during expansion,
and each iteration adds newly discovered states to the tableau.
The construction terminates at a fixed point, when no new states
can be added.

The construction applies the clock and predicate bounds from
\Cref{sub:tableau-finite}. A generated state is discarded if its
number of interval predicates for any timed subformula $\bothtemp_I\psi$
exceeds the corresponding bound of $O(r(I))$.
Let $Q$ be a fixed clock set of the bounded size and $\mathcal C_Q(M)$
the closures generated from $M$ using $Q$ and the judgement forms of
\Cref{sub:tableau-finite}, with merging by \Cref{lem:intv-merge} and
normalization by \Cref{prop:judg-equiv}. For the bounded runs above, it suffices to
merge predicates whose tracked intervals meet without a gap.
In the lemma's notation, their common boundary corresponds to
$\creset{d_1}{0}$ and $\creset{c_2}{0}$ in $\Lambda$.
The side condition is checked from the judgement's constraints, resets,
and markers (\Cref{ex:symbolic-merging}); any condition remaining after
substitution is added as a guard.

\begin{example}\label{ex:symbolic-merging}
Let
$\Phi = \bigl({c_1\ge0},\allowbreak
{\cset{\creset{d_1}{0},\creset{c_2}{0}}}.\bigl(\lty{\strue},\allowbreak
{\intvpredT{c_1}{d_1}{\strue}{\sfalse}{\always{[2,4]}(u\ge0)}},\allowbreak
{\intvpredupT{c_2}{\strue}{\always{[2,4]}(u\ge0)}\bigr)\bigr)}$.
The resets give $d_1=c_2=0$, so the side condition of
\Cref{lem:intv-merge} reduces to $0\le0\le c_1$, which follows from
the guard $c_1\ge0$. After merging, $\Phi$ becomes
${\bigl(c_1\ge0,\bigl(\lty{\strue},
\intvpredupT{c_1}{\strue}{\always{[2,4]}(u\ge0)}\bigr)\bigr)}$;
the resets disappear because $d_1$ and $c_2$ no longer occur.
\end{example}

\section{Constructing Hybrid Automata for STL}
\label{sec:transform}

This section explains how the tableau constructed in the previous
section can be used for robust STL model checking of hybrid automata.
We first recall the reduction from robust to Boolean STL model checking.
We then translate the tableau into a hybrid automaton and establish the
correctness of the translation.
Finally, by composing the translated automaton with the system, we
reduce robust STL model checking to a reachability problem.

\subsection{Background on Robust STL Model Checking}
\label{sec:background_robust}

Given a time horizon $\tau>0$, an STL formula $\varphi$, a hybrid
automaton $H$, and a robustness threshold $\epsilon>0$, the robust STL
model checking problem asks whether
$\robustness{\varphi}{\sigma}{0}{\tau}>\epsilon$
holds for every trajectory $\sigma$ of $H$.
Equivalently, it asks whether there is no trajectory $\sigma$ of $H$
such that
$\robustness{\neg\varphi}{\sigma}{0}{\tau}\ge-\epsilon$.

We use the standard notions of $\epsilon$-weakening and
$\epsilon$-strengthening of STL formulas~\cite{scp2025-stl}, which
respectively relax and strengthen atomic predicates by $\epsilon$ (see
\supplementref{sec:prelim-app}).
The result from~\cite{scp2025-stl} reduces robust STL model
checking to Boolean STL model checking.
In particular, if there exists a trajectory satisfying
$\neg(\varphi^{+\epsilon})$, then there exists a trajectory $\sigma$
such that
$\robustness{\varphi}{\sigma}{0}{\tau}\le\epsilon$.
Conversely, if no trajectory satisfies
$\neg(\varphi^{+\epsilon})$, then
$\robustness{\varphi}{\sigma}{0}{\tau}\ge\epsilon$
for every trajectory $\sigma$.

Finally, note that strict and non-strict atomic predicates have the
same robustness degree. Although $f(\vec{x})>0$ and
$f(\vec{x})\ge0$ may differ under Boolean semantics, they induce the
same robustness value. Therefore, without loss of generality, we assume that all atomic predicates are non-strict.

\subsection{Semantic Gap between Tableaux and Hybrid Automata}

\begin{figure}[t]
\setlength{\fboxsep}{0pt}
\begin{minipage}[t]{0.57\textwidth}
\centering
\includegraphics{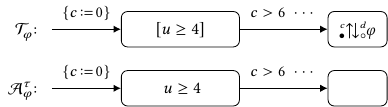}
\caption{Translation from $\tableau{\varphi}{}$ to $\hatableau{\varphi}{\tau}$
}
\label{fig:translation-overview} 
\end{minipage}
\hfill
\begin{minipage}[t]{0.42\textwidth}
\centering
\includegraphics{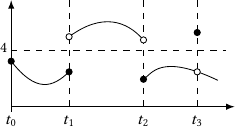}
\caption{A trajectory $\sigma$ of $H$}
\label{fig:boundary-mismatch}
\end{minipage}
\Description{The left figure maps an atomic interval predicate to a mode
invariant, preserving guards and resets. The right figure shows an atomic
proposition holding on a left-open interval but not at its left endpoint.}
\end{figure}

Since a tableau $\tableau{\varphi}{}$ is already a transition
system, a natural translation into a hybrid automaton, which we call
$\hatableau{\varphi}{\tau}$, is to map tableau states to modes
and transitions to jumps, as illustrated in
\Cref{fig:translation-overview}. 
Atomic interval
predicates are translated as mode invariants, so that the stable truth represented by
the tableau is enforced during continuous evolution. Clock constraints
and reset sets on transitions are translated directly into guards and
resets.

However, under this translation, not every signal recognized by
$\tableau{\varphi}{}$ is a trajectory of
$\hatableau{\varphi}{\tau}$.
Consider the signal shown in \Cref{fig:boundary-mismatch}, which
satisfies $u\le4$ over $[t_0,t_1]$ and
$u\ge4$ over $(t_1,t_2)$.
Suppose that the tableau recognizes this signal.
After the translation, the mode with invariant $u\ge4$
requires the signal to satisfy $u\ge4$ at $t_1$.
Since $u < 4$ at $t_1$, the signal is not a trajectory of
$\hatableau{\varphi}{\tau}$.
The same issue also appears at $t_3$, where two consecutive jumps
occur. The first jump, which enters the singleton interval, is valid.
However, after the second jump, the mode corresponding to the left-open
interval requires $u\le4$ to hold at $t_3$. Hence the signal is again not a trajectory.

This example illustrates that discrepancies may arise from the
difference between tableau and hybrid automaton semantics.
\Cref{tab:interval-translation} summarizes the interval types for which
such discrepancies may arise.
A left-closed interval causes no discrepancy, since the
mode invariant can be checked from its left endpoint.
A right-closed interval also causes no discrepancy, since its right
endpoint can be checked by the discrete transition.
Thus, discrepancies arise only for left-open intervals.
Consequently, if an atomic proposition holds on a left-open
interval but not at its left endpoint, the signal may be recognized by
the tableau but cannot be a trajectory of the hybrid automaton.

\begin{table}[h]
\small
\centering
\caption{Correspondence between tableau interval types and hybrid automaton semantics.}
\label{tab:interval-translation}
\begin{tabular}{c c c}
\toprule
Tableau & Hybrid Automaton & Discrepancy? \\
\midrule
$[a,a]$ & discrete transition at $a$ & No \\
$[a,b)$ & $[a,b)$ & No \\
$[a,b]$ & $[a,b)$ followed by discrete transition at $b$ & No \\
$(a,b)$ & $[a,b)$ & Yes \\
$(a,b]$ & $[a,b)$ followed by discrete transition at $b$ & Yes \\
\bottomrule
\end{tabular}
\end{table}

\subsection{Translating Tableaux to Hybrid Automata}

Before formally presenting the translation, we first explain how to avoid the above discrepancy.
In particular, it suffices to require that every atomic
proposition holding over a left-open interval also hold at its left
endpoint. Since all atomic propositions are non-strict, this is ensured
by requiring that, for every signal $\sigma$, the composition
$f\circ\sigma_f$ be right-continuous, where $f$ is the function defining
the atomic proposition. We call such signals \emph{well-behaved}.

\begin{definition}
\label{def:well-behaved-signal}
A signal $\sigma$ is said to be \emph{well-behaved} with respect to an STL formula $\varphi$
if, for every function $f$ appearing in $\varphi$,
$f \circ \sigma_f$ is right-continuous.
\end{definition}

The following lemma gives sufficient conditions for hybrid automata to be well-behaved.\footnote{%
See \Cref{sec:stl-model-checking} for the form of the system automaton used
in model checking.}

\begin{lemma}
\label{lem:well-behaved-suff}
If every flow condition of $H$ is continuous and every function $f$ appearing in $\varphi$ is continuous, 
then every trajectory $\sigma$ of $H$ is well-behaved with respect to $\varphi$.
\end{lemma}

This lemma shows that the well-behaved assumption is mild. Indeed, flow
conditions of hybrid automata are typically continuous, and atomic
propositions are commonly defined by continuous functions. Thus,
the assumption imposes little restriction in typical applications.

We now formally define the translation.
Let $\tableau{\varphi}{}
=(X,X_c,S,T,T_0,S_F,\mathcal L)$.
The hybrid automaton $\hatableau{\varphi}{\tau}$ has the tableau states $S$ as
its modes, with initial and goal modes given by $T_0$ and $S_F$.
The continuous variables are
$X^\sharp=X\cup X_c\cup\{c_g\}$,
where $c_g$ is a fresh global clock.
Clocks in $X_c$ and $c_g$ evolve with rate~1, while variables in $X$
may evolve arbitrarily.
Each atomic interval predicate
$\ipstl{f(\vec{x})\ge0}$ induces the mode invariant
$f(\vec{x})\ge0$.
Tableau transitions induce discrete jumps:
clock constraints become jump guards, reset sets become clock resets,
and $\ast$ becomes the guard $c_g\ge\tau$.

\begin{definition}
\label{def:tableau-ha}
Given a tableau
$\tableau{\varphi}{} = (X, X_c, S, T, T_0, S_F, \mathcal{L})$,
and a time bound $\tau > 0$,
the hybrid automaton
$\hatableau{\varphi}{\tau}
= (Q, G, E, X^\sharp, \init, \inv, \flow, \jump)$
and its goal are defined as follows.
\begin{itemize}
  \item
  $Q = S$, $\goal = S_F$.

  \item
  $G = \{(C,\Lambda)\mid (q,(C,\Lambda),q')\in T\}$.

  \item
  $E = T$.

  \item
  $X^\sharp = X \cup X_c \cup \{c_g\}$,
  where $c_g$ is a fresh global clock.

  \item
  $\init =
  \{ (q_0, v_0) \mid
  ((C_0,\Lambda_0),q_0)\in T_0,\;
  v \in \realset^{X^\sharp},\;
  v \restriction X_c \models C_0,\;
  v_0 = v[\Lambda_0][\cset{\creset{c_g}{0}}]
  \}$.

  \item
  $\inv =
  \{ (q,v) \mid
  q\in Q,\;
  (\ast \not\in \mathcal{L}(q) \Rightarrow v(c_g)<\tau),\;
  \forall \ipstl{f(\vec x)\ge 0}\in\mathcal L(q).\;
  f(v(\vec x))\ge 0
  \}$.

  \item
  $\flow =
  \{
  (\langle q,v\rangle,t,v')\mid
  q\in Q,\;
  t\ge0,\;
  \forall c\in X_c\cup\{c_g\}.\;
  v'(c)=v(c)+t
  \}$.
  \footnote{%
  See \supplementref{sec:app-translate-ha} for the execution convention at $\tau$.}

  \item 
  for each $e \in E$,
  $\jump(e)=
  \{
  (\langle q,v\rangle,\langle q',v'\rangle)
  \mid
  e=(q,(C',\Lambda'),q'),\;
  v\restriction X_c\models C',\;
  v'\restriction(X_c\cup\{c_g\})
  =v[\Lambda']\restriction(X_c\cup\{c_g\}),\;\allowbreak
  \ast\in\mathcal L(q')
  \Rightarrow v'(c_g)\ge\tau
  \}$.
\end{itemize}
\end{definition}

We say that a trajectory of $\hatableau{\varphi}{\tau}$
\emph{reaches a goal} if it reaches a state whose mode belongs to $\goal$.

\begin{figure}
    \includegraphics{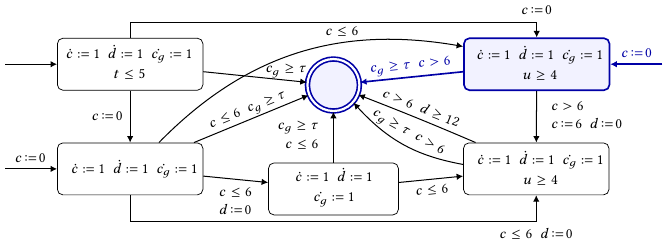}
    \caption{Constructing $\hatableau{\varphi}{\tau}$ from $\tableau{\varphi}{}$, where $\varphi = (\always{[6,12]}u \ge 4) \release{} t \le 5$. The highlighted rectangle and double circle represent modes $q_0$ and $q_1$, respectively; their connecting jump is $e_1$. Some constraints are omitted.}
    \label{fig:ex-ha}
    \Description{Mode boxes list invariants and clock rates; arrows show
    guards and resets. The highlighted jump reaches a goal mode drawn as a
    double circle when c exceeds 6 and the global clock is at least the
    horizon.}
\end{figure}

\begin{example}
\Cref{fig:ex-ha} shows the hybrid automaton obtained from the tableau in
\Cref{fig:ex-tab} after merging equivalent states and transitions.
For example, the initial state $s_0$ and final state $s_1$ in
\Cref{fig:ex-tab} become modes $q_0$ and $q_1$, respectively.
Mode $q_0$ has invariant $u\ge4$ and retains the incoming reset $c:=0$,
while $q_1$ is a goal mode.
The transition from $s_0$ to $s_1$ labeled by $c>6$ becomes jump $e_1$
with guard $c>6,\;c_g\ge\tau$, where the latter condition comes from
the terminal judgement $\ast$.
Finally, the tableau clocks $c,d$ and the global clock $c_g$ evolve at rate~1.
\end{example}

Every trajectory of $\hatableau{\varphi}{\tau}$ naturally induces a
partition by taking its jump times as interval boundaries.
The time spent in each mode determines the length of the corresponding interval
(see \supplementref{sec:app-translate-ha}).
We now establish the correctness of the translation.
Under the well-behaved assumption, every atomic interval predicate is
faithfully interpreted as a mode invariant. Hence every signal
recognized by $\tableau{\varphi}{}$ with respect to a partition becomes
a trajectory of
$\hatableau{\varphi}{\tau}$ that reaches a goal.
Conversely, since a partition induced by a trajectory provides the
interval structure required for tableau recognition, the corresponding
tableau path can be reconstructed.

\begin{lemma}
\label{lemma:translate-correct}
Let $\sigma$ be a signal with time horizon $\tau$.

\begin{enumerate}
\item
If $\sigma$ is well-behaved with respect to $\varphi$ and is
recognized by $\tableau{\varphi}{}$ with respect to a partition
$\partition{J}$, then $\sigma$ is a trajectory of
$\hatableau{\varphi}{\tau}$ that reaches a goal.

\item
If $\sigma$ is a trajectory of
$\hatableau{\varphi}{\tau}$ that reaches a goal, then for every
partition $\partition{J}$ induced by the trajectory,
$\sigma$ is recognized by $\tableau{\varphi}{}$ with respect to
$\partition{J}$.
\end{enumerate}
\end{lemma}

\subsection{STL Model Checking}
\label{sec:stl-model-checking}

We now explain how the tableau construction can be used for STL model
checking of a hybrid automaton $H$.
Given an STL formula $\varphi$ and a time horizon $\tau$, we construct
$\tableau{\neg\varphi}{}$ and translate it into
$\hatableau{\neg\varphi}{\tau}$.
We then compose the translated automaton with the system and reduce STL
model checking to goal reachability in the resulting hybrid automaton.

The system automaton $H$ may exhibit arbitrarily long sequences of
consecutive zero-time jumps.
Since a single tableau transition can synchronize with only one
discrete transition of the system, such sequences cannot in general be
matched by the tableau.
Consequently, the composed automaton
$\composeha{\neg\varphi}{\tau}$ may fail to capture some behaviors of
$H$.

To address this issue, we introduce an
\emph{$n$-bounded zero-time jump extension}.
Given $H$, the extended automaton $\hat H^{(n)}$ adds a single
zero-time jump representing every sequence of at most $n$
consecutive zero-time jumps.
Every trajectory of $\hat H^{(n)}$ is semantically equivalent to a
trajectory of $H$, obtained by expanding each added jump into the
corresponding sequence of zero-time jumps.

We assume that the number of consecutive zero-time jumps is uniformly
bounded over all trajectories of $H$, and choose $n$ at least as large
as this bound. This is a mild assumption in practice, since hybrid
systems are typically designed to avoid arbitrarily long sequences of
instantaneous transitions
\cite{alur1995algorithmic,henzinger2000theory,bae2019bounded,roehm2016stl,kochdumper2024fully,sato2024optimization}.

\begin{proposition}
\label{prop:zero-jump}
Trajectories of $H$ and $\hat H^{(n)}$ are semantically equivalent.
\end{proposition}

By \Cref{prop:zero-jump}, we may replace $H$ by its bounded extension
without changing the signal semantics. Hence, we write $H$ for the bounded extension.
With this convention, STL model checking reduces to a reachability
problem on the composed hybrid automaton.

\begin{theorem}
\label{sec:red_thm}
Let $H$ be a hybrid automaton whose flow condition is continuous, and
let $\varphi$ be an STL formula in which every function $f$ is
continuous.
\begin{enumerate}[label=(\arabic*)]
\item
If there exists a trajectory of
$\composeha{\neg\varphi}{\tau}$
that reaches a goal, then there exists a trajectory $\sigma$ of $H$
such that
$\sigma,0\models_\tau\neg\varphi$.

\item
If no trajectory of
$\composeha{\neg\varphi}{\tau}$
reaches a goal, then every trajectory $\sigma$ of $H$ with finite
variability for $\varphi$ satisfies
$\sigma,0\models_\tau\varphi$.
\end{enumerate}
\end{theorem}

\begin{proof}
By assumption and \Cref{lem:well-behaved-suff}, every trajectory of
$H$ is well-behaved with respect to $\varphi$.

\begin{enumerate}[label=(\arabic*)]
\item
Let $\sigma$ be a trajectory of
$\composeha{\neg\varphi}{\tau}$ that reaches a goal.
Then $\sigma$ is a trajectory of $\hatableau{\neg\varphi}{\tau}$.
By \Cref{lemma:translate-correct},
$\sigma$ is recognized by $\tableau{\neg\varphi}{}$.
Hence, by \Cref{thm:finite-tableau},
$\sigma,0\models_\tau\neg\varphi$.

\item
We prove the contrapositive.
Suppose that there exists a trajectory $\sigma$ of $H$ with finite
variability such that
$\sigma,0\models_\tau\neg\varphi$.
By \Cref{thm:finite-tableau},
$\sigma$ is recognized by $\tableau{\neg\varphi}{}$.
By \Cref{lemma:translate-correct},
there exists a trajectory of
$\hatableau{\neg\varphi}{\tau}$ that reaches a goal, and it is a trajectory of $\composeha{\neg\varphi}{\tau}$.
\qedhere
\end{enumerate}
\end{proof}

\paragraph{Robust STL Model Checking}

As a consequence of \Cref{sec:red_thm} and the reduction in~\cite{scp2025-stl},
reviewed in \Cref{sec:background_robust},
robust STL model checking reduces to determining whether
$\composeha{\neg(\varphi^{+\epsilon})}{\tau}$ can reach a goal.
If such a trajectory exists, it is a counterexample to the robust STL
model checking of $\varphi$.
Otherwise, $\varphi$ is satisfied on every trajectory of $H$ with finite variability, with respect to every
robustness threshold $0<\epsilon'<\epsilon$.

\begin{corollary}
For a hybrid automaton $H$ and an STL formula $\varphi$,
if every flow condition of $H$ is continuous and every function $f$
appearing in $\varphi$ is continuous, then the following hold:
\begin{enumerate}[label=(\arabic*)]
\item
if there exists a trajectory of
$\composeha{\neg(\varphi^{+\epsilon})}{\tau}$
that reaches a goal, then there exists a trajectory $\sigma$ of $H$ such that
$\robustness{\varphi}{\sigma}{0}{\tau}\le\epsilon$;

\item
if no trajectory of
$\composeha{\neg(\varphi^{+\epsilon})}{\tau}$
reaches a goal, then for every trajectory $\sigma$ of $H$ with finite variability,
$\robustness{\varphi}{\sigma}{0}{\tau}\ge\epsilon$.
\end{enumerate}
\end{corollary}

\newcommand{\reduce}{\textsf{Min}}
\newcommand{\none}{\textsf{None}}

\newcommand{\maccs}[1]{\textsf{ACCS0#1}}
\newcommand{\maccu}[1]{\textsf{ACCU0#1}}
\newcommand{\mbat}{\textsf{BAT}}
\newcommand{\mcar}{\textsf{CAR}}
\newcommand{\mwat}{\textsf{WAT}}
\newcommand{\mdisc}[1]{\textsf{DISC0#1}}
\newcommand{\mnav}[1]{\textsf{NAV0#1}}
\newcommand{\mrail}{\textsf{RAIL}}

\newcommand{\mcbc}[1]{\textsf{CBC0#1}}
\newcommand{\mcbf}[1]{\textsf{CBF0#1}}
\newcommand{\mheat}[1]{\textsf{HEAT0#1}}
\newcommand{\mlalo}[1]{\textsf{LALO0#1}}
\newcommand{\mlovo}{\textsf{LOVO}}

\newcommand{\mspa}{\textsf{Spac}}
\newcommand{\moscil}{\textsf{Oscl}}
\newcommand{\mthm}{\textsf{Them}}
\newcommand{\mlane}{\textsf{Lane}}

\newcommand{\mcard}{\textsf{Card}}
\newcommand{\mvand}{\textsf{Vand}}
\newcommand{\mcolumn}{\textsf{Colm}}
\newcommand{\mrobe}[1]{\textsf{ROBO0#1}}
\newcommand{\mlotka}{\textsf{Lotk}}

\newcommand{\spaceex}{\texttt{SE}}
\newcommand{\flowstar}{\texttt{FS}}
\newcommand{\dreach}{\texttt{DH}}

\newcommand{\stlmc}{\texttt{STLmc}}
\newcommand{\ourone}{\texttt{Our1}}
\newcommand{\ourtwo}{\texttt{Our2}}

\newcommand{\timeout}{\texttt{T/O}}

\newcommand{\msub}{\textsf{Sub}}
\newcommand{\mstutter}{\textsf{Stu.}}

\section{Experimental Evaluation}
\label{sec:expr}

We evaluate the proposed automata-based robust STL model checking method
from two perspectives.
First, we compare its model checking performance with the SMT-based approach
STLmc~\cite{cav2022-stl}.
Second, we evaluate the scalability of the proposed tableau construction with
respect to the complexity of STL formulas.

\paragraph{Experimental Setup}
\label{sec:expr-setup}

We use STLmc~\cite{cav2022-stl} as the baseline.%
\footnote{%
STLmc is used as the baseline as it is, to our knowledge, the only publicly available tool
supporting robust STL model checking involving general ODE dynamics.
}
For this purpose, we have implemented a prototype tool that generates
hybrid automata from STL formulas.
We consider benchmark models 
from the STLmc artifact~\cite{cav2022-stl,scp2025-stl}
and 
from the ARCH competition~\cite{Althoff2023,Geretti2023,geretti2022arch,immler2018arch},
covering both linear and nonlinear continuous dynamics.
In total, we use 16 benchmark models:
8 with linear ODE dynamics and 8 with nonlinear ODE dynamics.
For each benchmark, we consider three STL formulas %
that contain multiple timed temporal operators.

For each formula-model pair, we measure the execution time required for
robust STL model checking using STLmc and our automata-based approach,
under identical time bounds and robustness thresholds.
The sizes of the hybrid automata generated by our approach are reported
in \supplementref{tab:sizes_by_formula}.
The resulting automata are analyzed using different reachability engines:
SpaceEx~\cite{spaceex} and SAT-Reach~\cite{Kundu2023} for linear-ODE models,
and dReach~\cite{dReach} and Flow*~\cite{chen2013flow} for nonlinear-ODE models.
SAT-Reach and dReach perform bounded analysis,
whereas SpaceEx and Flow* compute unbounded over-approximations.

We use dReal~\cite{dReal} as the backend solver for STLmc~\cite{cav2022-stl}.
All bounded analysis tools used in our evaluation
(STLmc, SAT-Reach, and dReach)
are run with a bound $N=5$, except for a few nonlinear benchmarks
with particularly complex dynamics, where smaller bounds are used.
For each benchmark, the same time bound~$\tau$ and robustness threshold~$\epsilon$
are used across all tools.
All experiments are conducted on an Intel Xeon 2.8\,GHz machine with 256\,GB memory
and a 1-hour timeout.
Further configuration details and per-instance results are given in
\supplementref{sec:expr-app-stlmc}.

\paragraph{Comparison with STLmc}
\label{sec:expr-stlmc}

\begin{figure*}[t]
  \centering

  \begin{subfigure}[t]{0.32\linewidth}
    \centering
    \includegraphics[width=\linewidth, clip=true, trim=0pt 1.6cm 0pt 0.1cm]{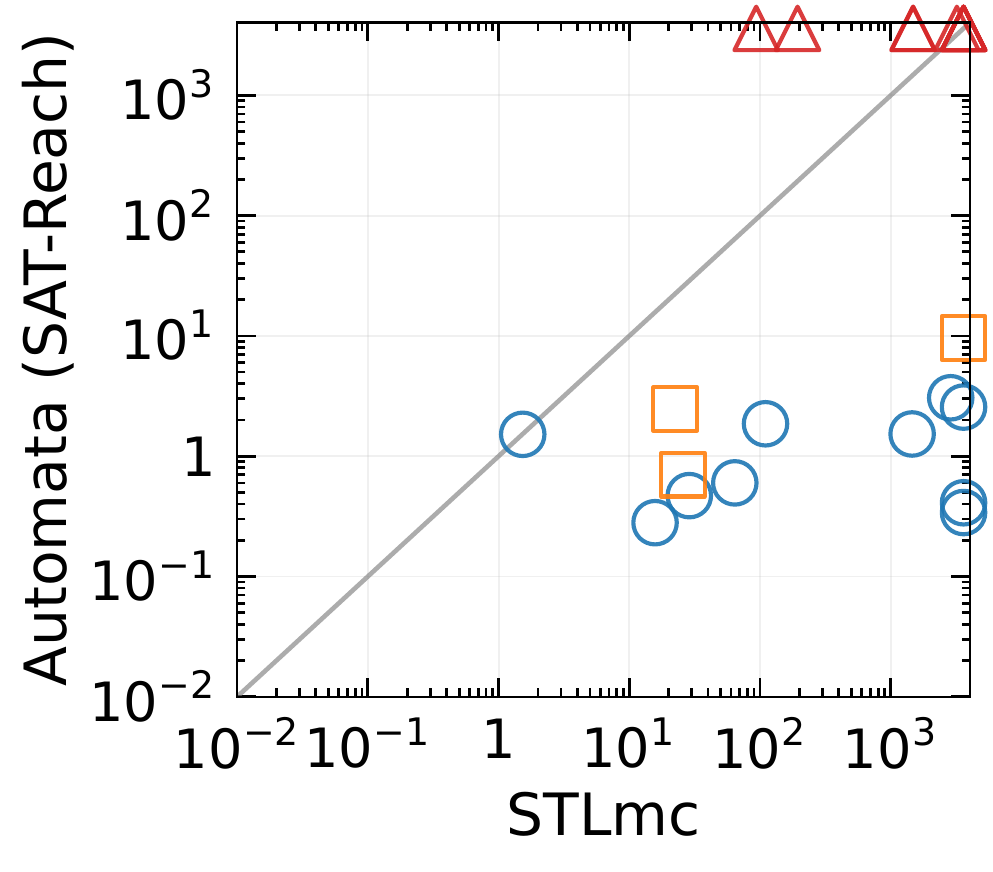}
  \end{subfigure}\hfill
  \begin{subfigure}[t]{0.32\linewidth}
    \centering
    \includegraphics[width=\linewidth, clip=true, trim=0pt 1.6cm 0pt 0.1cm]{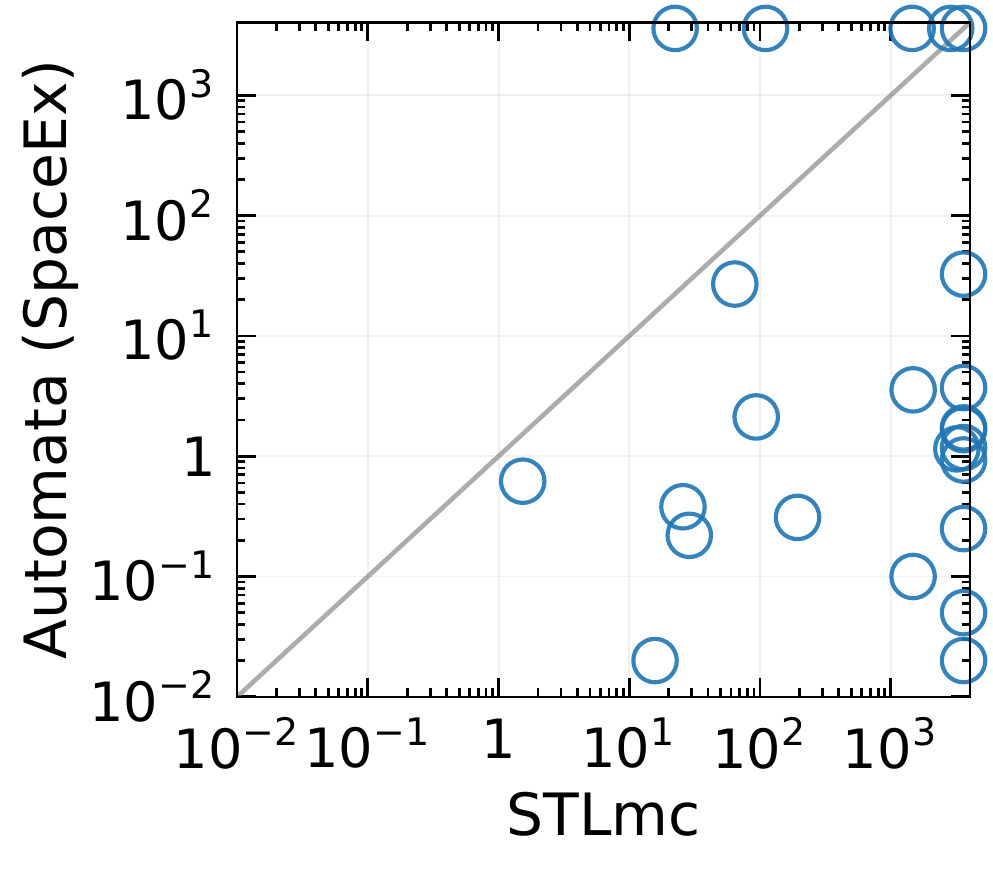}
  \end{subfigure}\hfill
  \begin{subfigure}[t]{0.32\linewidth}
    \centering
    \includegraphics[width=\linewidth, clip=true, trim=0pt 1.6cm 0pt 0.1cm]{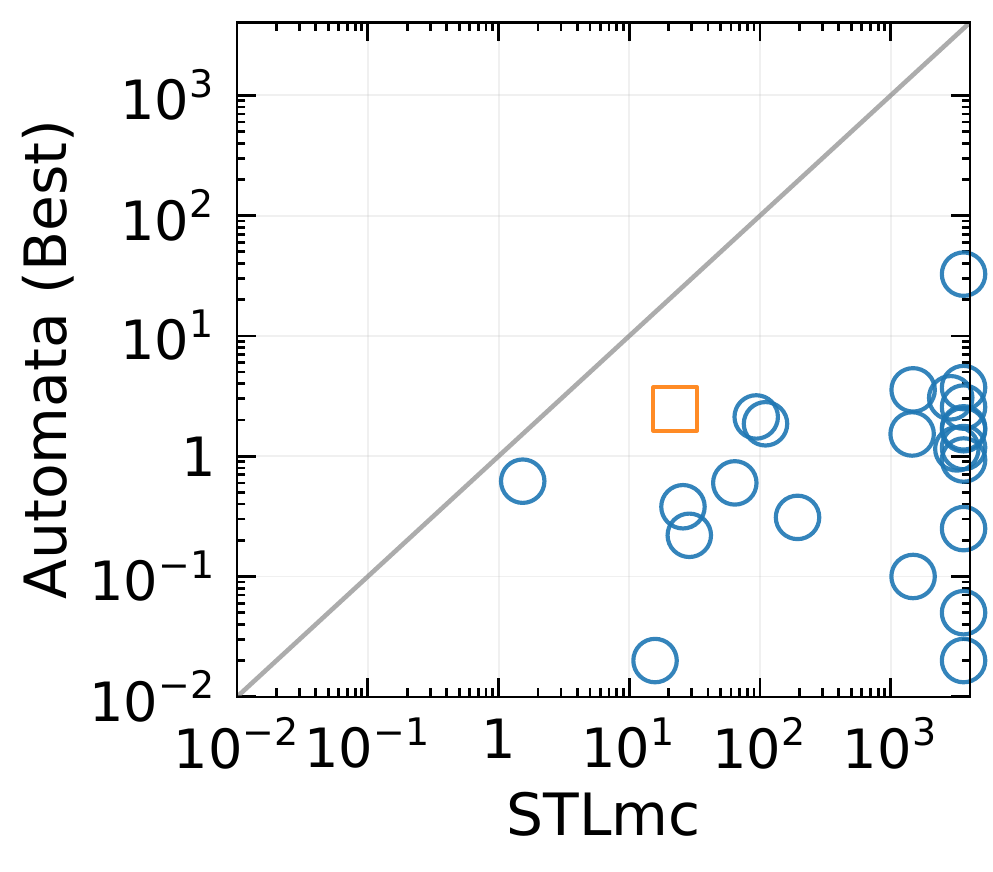}
  \end{subfigure}

  \vspace{1em}

  \begin{subfigure}[t]{0.32\linewidth}
    \centering
    \includegraphics[width=\linewidth, clip=true, trim=0pt 0.4cm 0pt 0.1cm]{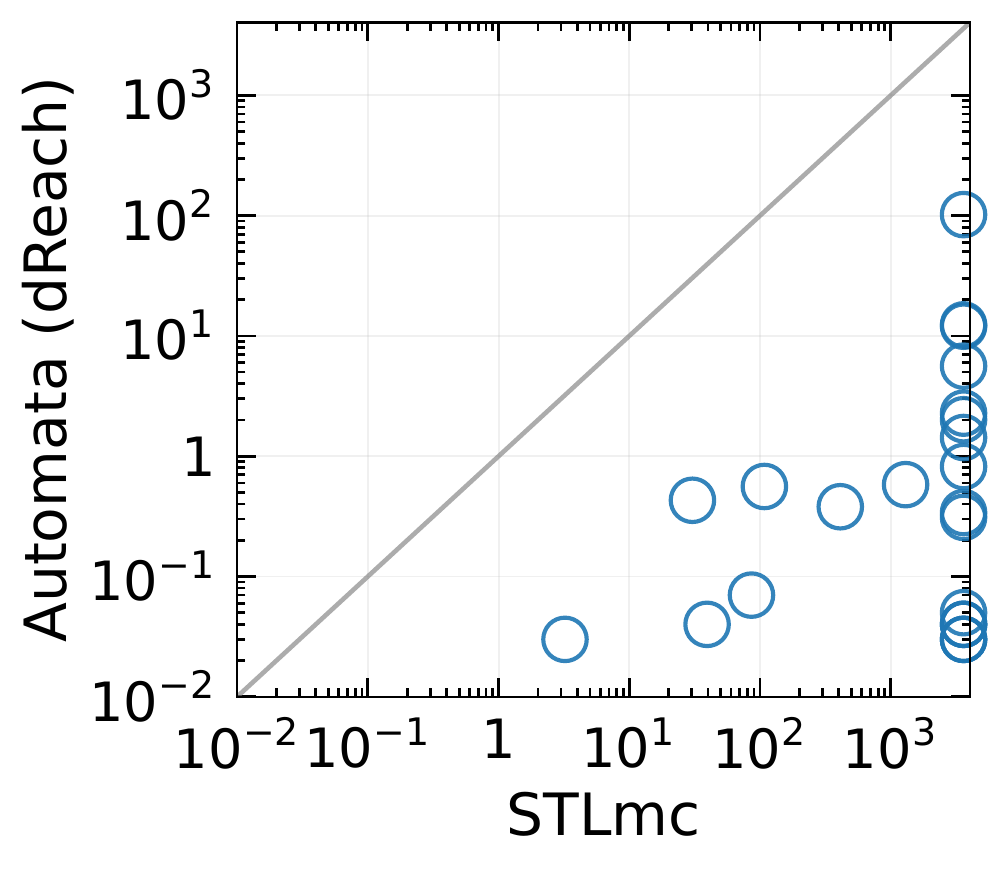}
  \end{subfigure}\hfill
  \begin{subfigure}[t]{0.32\linewidth}
    \centering
    \includegraphics[width=\linewidth, clip=true, trim=0pt 0.4cm 0pt 0.1cm]{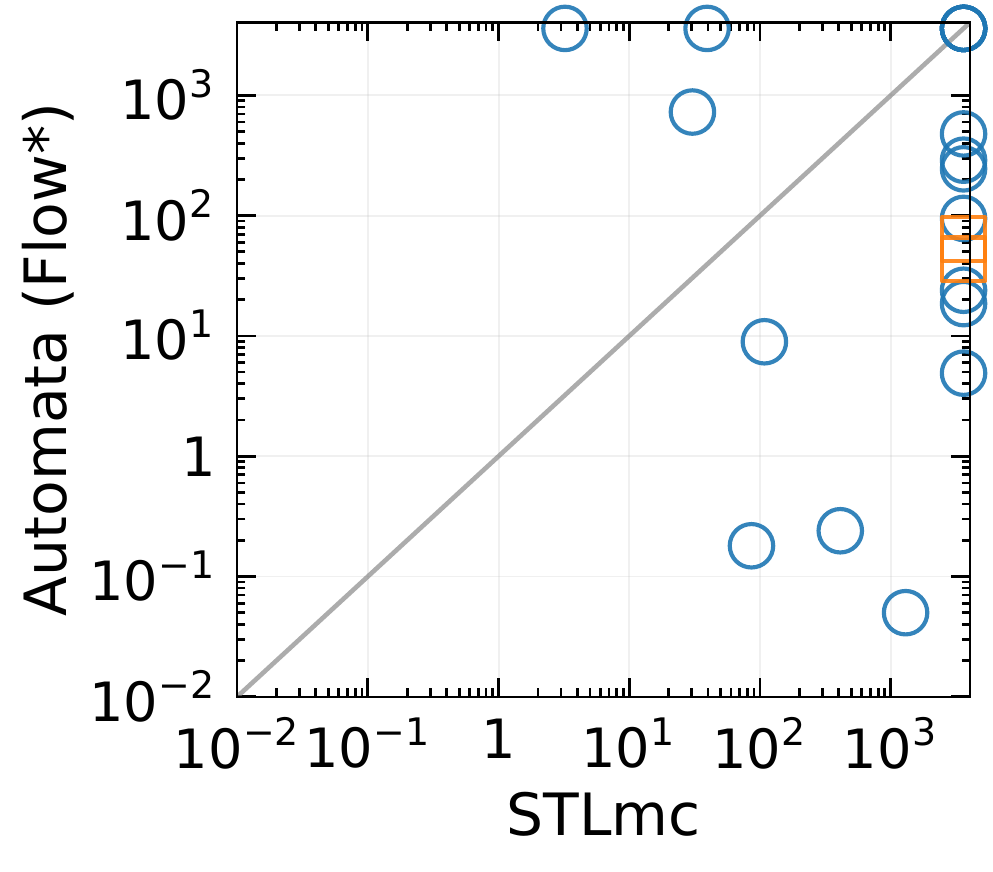}
  \end{subfigure}\hfill
  \begin{subfigure}[t]{0.32\linewidth}
    \centering
    \includegraphics[width=\linewidth, clip=true, trim=0pt 0.4cm 0pt 0.1cm]{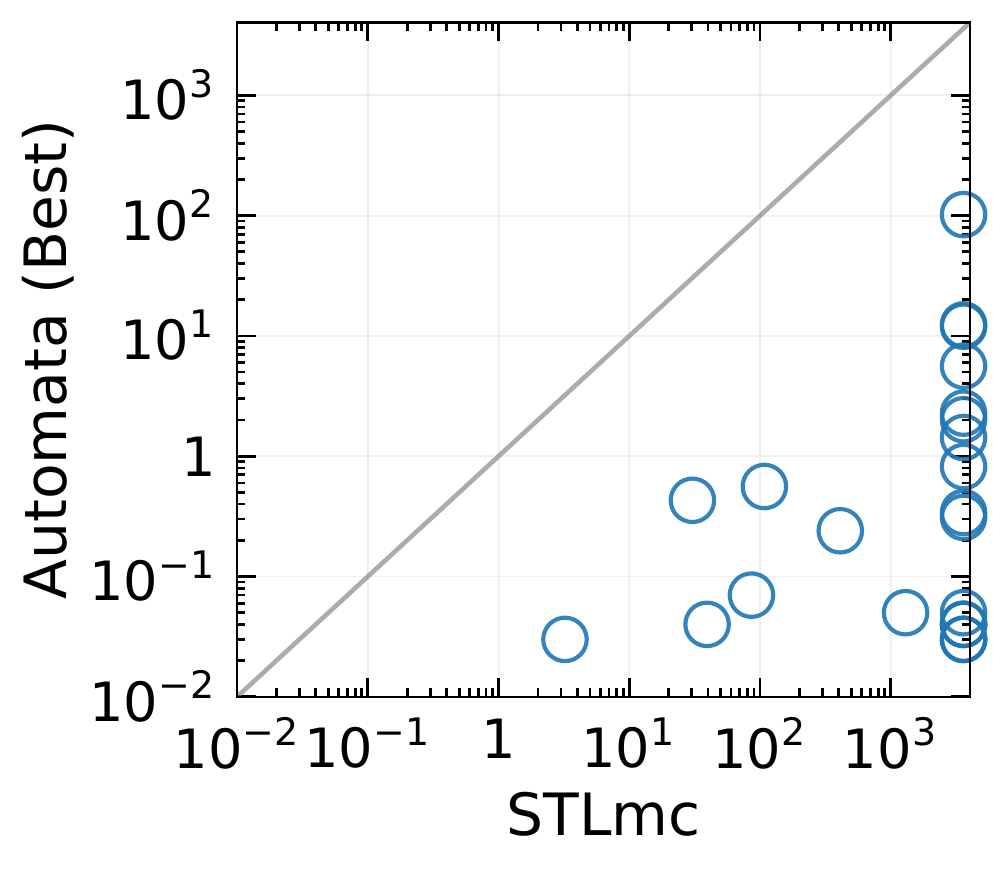}
  \end{subfigure}

\caption{Runtime comparison with STLmc (log--log scale).
Top: linear; bottom: nonlinear. Left/middle: individual backends;
right: best per instance.
Markers indicate outcomes of our reachability backends:
circles denote definitive results or timeouts, triangles solver errors, and
squares UNKNOWN, meaning that the backend could not determine
whether the goal is reachable.
}
  \label{fig:expr1}
\Description{Six scatter plots with logarithmic axes compare runtimes
with STLmc for linear and nonlinear benchmarks, using individual
backends and the best backend per instance. Circles denote definitive
results or timeouts, triangles solver errors, and squares unknown
results.}
\end{figure*}

\Cref{fig:expr1} summarizes the execution time (in seconds).
The left and middle plots compare STLmc with our automata-based approach
using a fixed reachability backend.
For each data point, the x-axis shows the execution time of STLmc,
and the y-axis shows that of our approach on the same %
instance.
Both axes use a logarithmic scale, and the diagonal indicates equal performance.
Circles denote definitive results ($\top$ or $\bot$) or timeouts,
squares denote unknown outcomes,
and triangles denote solver errors;
timeouts are shown at the plot boundary at the time limit.
(See \supplementref{tab:mc-result} for detailed results.)

Overall, our automata-based approach outperforms STLmc on the evaluated benchmarks.
For linear-ODE models, most instances fall below the diagonal, indicating faster execution times,
with only a few cases where STLmc remains competitive.
For nonlinear-ODE models, the advantage is more pronounced: STLmc times out on most instances,
while our approach consistently completes within $10^3$ seconds.
These results show that compiling robust STL properties into hybrid automata and
leveraging an appropriate reachability backend yields substantial performance gains
over SMT-based model checking.

\paragraph{Evaluation of Tableau Construction}
\label{sec:expr-tableau}

We also evaluate whether the proposed tableau construction
scales to nontrivial STL formulas.
To this end, we randomly generate 170 STL formulas
with sizes ranging from 2 to 13 and temporal nesting depth up to~5,
excluding formulas that are equivalent under classical equivalence
relations~\cite{baier2008modelchecking,Fernandez1990,Cortier2009}.
For each generated formula, we use our implementation,
together with tableau minimization based on simulation-equivalence quotients,
to construct the corresponding tableau,
represented as a property hybrid automaton.

\begin{figure}[t]
  \centering
  \includegraphics[width=\textwidth]{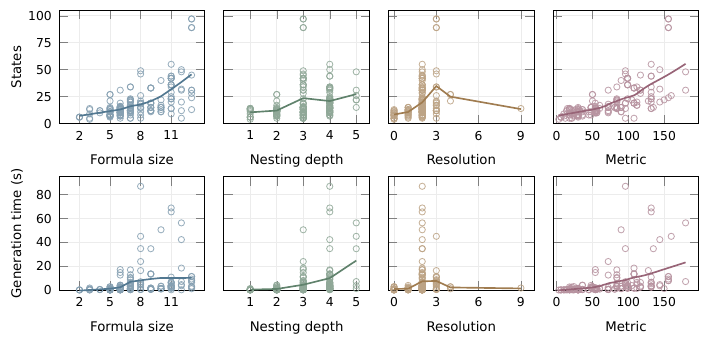}
  \caption{Tableau states after reduction (top) and generation time
  (bottom) for 170 formulas, against formula size,
  temporal nesting depth, interval resolution, and the combined metric.
  Each point represents one formula; points may overlap.
  Lines show local linear fits.}
  \label{fig:expr2-1a}
  \Description{Eight scatter plots in two rows and four columns compare
  reduced tableau states and generation time against formula size,
  temporal nesting depth, interval resolution, and their combined metric.}
\end{figure}

We first analyze tableau size and construction time with respect to
formula size, temporal nesting depth, and interval resolution separately.
The results are shown in the first three columns of \Cref{fig:expr2-1a}.
The individual plots show increasing trends, with considerable variation
among formulas. For example, a smaller formula with higher resolution
can yield more states than a larger formula with lower resolution.
Considering these factors jointly reveals a more consistent overall trend.

To examine tableau size and construction time in relation to these
factors jointly, we summarize them by a complexity metric.
For each generated formula~$\varphi$, it combines formula size,
temporal nesting depth, and interval resolution.
Formally, the complexity metric is defined as
\[
|\varphi|\cdot nd \cdot
\Bigl(\bigl\lceil \frac{\sup(I)}{\sup(I)-\inf(I)} \bigr\rceil + 1\Bigr),
\]
where $|\varphi|$ is the size %
of~$\varphi$,
$nd$ is the nesting depth of temporal operators,
and $I$ is the interval in~$\varphi$ that maximizes
$\bigl\lceil \tfrac{\sup(I)}{\sup(I)-\inf(I)} \bigr\rceil$.
Intuitively, deeper temporal nesting introduces additional dependencies
between subformulas, and higher resolutions require the tableau
to distinguish a larger number of interval predicates.

The last column of \Cref{fig:expr2-1a} plots the same results against
this metric, providing a combined view of the three factors.
Tableau construction succeeds for all 170 formulas within 100 seconds,
and the resulting tableaux contain 4--97 states after reduction.
Despite variation in formula size, temporal nesting depth, and interval
resolution, the resulting tableau sizes remain manageable across these
formulas. These results support the practical feasibility
of the proposed construction.

\section{Related Work}
\label{sec:related}

\paragraph{Tableaux for Temporal Logics.}
Expansion rules and tableaux underlie satisfiability checking and
automata-theoretic analysis for discrete-time temporal logics such as
LTL and LTLf~\cite{gerth1995simple,schwendimann1998,ltlf-ijcai13,
ltlf2014ecai}.
For discrete-time STL, Melani et al.~\cite{stl-tecs25} develop a one-pass
tableau for bounded operators.
Their construction assumes discrete time, whereas we handle STL under
continuous signal semantics.

Closest to our work, Geilen and Dams~\cite{geilen2000onthefly,geilen2003cav}
develop expansion-rule tableaux for a fragment of MITL over timed state
sequences.
Our construction handles the full syntax of STL with non-singular timing intervals under
continuous signal semantics, using intervals of stable truth to obtain a finite tableau that
translates into a hybrid automaton for robust model checking.

\paragraph{MITL-to-Automata Translations.}
MITL has been translated into timed and signal automata under pointwise
and continuous
semantics.
Under pointwise semantics, such constructions advance over event positions,
using clocks to check timing constraints \cite{mitlautomata}.
As discussed in \Cref{sec:intro}, continuous semantics
constructions \cite{alur1996benefits,ferrere2019real,
brihaye2017m,brihaye2017signal,geilen2003cav}
may introduce transitions at timing bounds even when no
relevant subformula changes truth value.
We instead advance across intervals of stable truth.

The signal-based construction of Brihaye et
al.~\cite{brihaye2017signal} is particularly close because it also
represents signals by intervals of stable values.
There, intervals encode unbounded signals into timed automata; in our
construction, they are the successive steps of a tableau over a finite
signal, across which temporal obligations are unfolded.

\paragraph{Interval-Based Reasoning.}
Interval Temporal Logic and Duration Calculus treat intervals as
first-class semantic objects.
Interval predicates also appear in refinement-based reasoning for
real-time programs, where they describe semantic properties of intervals
and streams~\cite{dongol2012algebra}.
Our object logic remains point-based STL: interval predicates are finite
syntactic judgements used to decompose formula satisfaction over intervals
of stable truth.

\paragraph{Automata-Based Verification of Hybrid Systems.}
HyLTL~\cite{hyltlbresolin2013} translates temporal properties of hybrid
traces into hybrid automata and verifies them by composition with the
system.
Our reduction has the same general form, but HyLTL extends untimed LTL
with flow constraints and targets infinite traces.
Our construction handles metric STL operators over finite signals and
provides guarantees up to a robustness threshold.
Other reachability-based methods verify STL over computed reachable
sets~\cite{roehm2016stl,kochdumper2024fully}.
We instead compile the property into a hybrid automaton, allowing its
composition with the system to be analyzed by different reachability
tools.

\paragraph{Constraint-Based STL Model Checking.}
SMT- and optimization-based approaches encode bounded STL model checking
as constraint
solving~\cite{bae2019bounded,cav2022-stl,scp2025-stl,
sato2024optimization}.
Our approach instead reduces robust STL model checking to reachability of
a composed hybrid automaton and can use analyzers designed for different
classes of dynamics.
We incorporate robustness through $\epsilon$-weakening of atomic
predicates~\cite{scp2025-stl} and compare against SMT-based robust STL
model checking in our evaluation.

\paragraph{Runtime Verification and Falsification.}
STL monitoring evaluates satisfaction or robustness on individual signals,
including online monitoring of partial signals~\cite{maler2004monitoring,
donze2013efficient,deshmukh2017robust}.
Falsification searches for violating signals through simulation and
optimization~\cite{annpureddy2011s,sankaranarayanan2012falsification}.
A counterexample establishes a violation, but an unsuccessful search does
not in general establish satisfaction for all system behaviors.
Our approach instead targets robust model checking over all trajectories
of a hybrid automaton within a bounded horizon.

\section{Concluding Remarks}
\label{sec:conclusion}

Temporal reasoning in discrete time advances by unfolding obligations from
the current instant to the next, but continuous-time signals provide no
distinguished successor through which this unfolding can proceed.
This work shows that intervals of stable truth can play that role.
Interval predicates and their expansion rules carry STL obligations across
a finite partition of a signal, recovering the stepwise decomposition
underlying discrete-time tableaux.
This yields a finite tableau that translates into a hybrid automaton,
thereby reducing robust STL model checking to reachability analysis with
different backends.
Our evaluation with four reachability analyzers on linear and nonlinear
systems shows that this backend-independent reduction can substantially
outperform state-of-the-art SMT-based robust STL model checking.

Several directions remain for extending this foundation.
The current approach targets finite signals over a bounded horizon and
relies on a well-behavedness condition when translating tableaux into
hybrid automata.
Extending the framework to infinite signals, together with suitable
B\"uchi-style acceptance and fairness conditions, is a natural next step.
Further work could integrate the construction more closely with
individual reachability backends and improve the automation and usability
of the prototype.
Beyond STL, interval predicates provide a reusable principle for
automata-based reasoning whenever temporal progress can be organized by
intervals of semantic stability.
Indeed, when restricted to propositional atoms, our construction already
yields a tableau for MITL over finite signals.
Applying this principle to other temporal logics and automata-based
verification methods is another promising direction.

\begin{acks}
This work was partially supported by the
\grantsponsor{iitp}{Institute of Information \& Communications Technology Planning \& Evaluation (IITP)}{https://www.iitp.kr/}
grant \mbox{(No.~\grantnum{iitp}{RS-2024-00439856})} and the
\grantsponsor{nrf}{National Research Foundation of Korea (NRF)}{https://www.nrf.re.kr/}
grant \mbox{(No.~\grantnum{nrf}{RS-2021-NR060080})},
both funded by the Korean government (MSIT).
\end{acks}

\section*{Disclosure of Use of AI in Research}
We used OpenAI Codex to assist with the Lean formalization of the paper's main results.
We also used it to check the manuscript's grammar and identify typos.

\section*{Data Availability Statement}
The Figshare artifact~\cite{stl-tableaux-artifact} contains our prototype
for translating STL to hybrid automata and the experimental results in \Cref{sec:expr}.
The Lean formalization is available on GitHub~\cite{stl-tableaux-lean}.

\bibliographystyle{ACM-Reference-Format}
  \raggedbottom
\bibliography{kbib-stl}

\clearpage
\appendix

\section{More Preliminaries}
\label{sec:prelim-app}

\subsection{Intervals, Partitions, and Markers}
\label{sec:prelim-app-marker}

Let $I \subseteq \realset$ be an interval with endpoints $a < b$. 
Then $I$ is \emph{open}, written $(a, b)$, if $a, b \notin I$;
\emph{closed}, written $[a, b]$, if $a, b \in I$;
\emph{left-closed and right-open}, written $[a, b)$, if $a \in I$ but $b \notin I$;
and \emph{left-open and right-closed}, written $(a, b]$, if $a \notin I$ but $b \in I$.

The following property for interval operations is particularly useful for STL~\cite{bae2019bounded}.

\begin{lemma}
\label{lem:interval_operation}
For intervals $I,K \subseteq \realset_{\ge 0}$,
$t \in K - I$ iff there exists $t' \ge t$ such that $t' \in K$ and $t' \in t + I$.
\end{lemma}

The least upper bound and the greatest lower bound of the set 
$\{g(a) \mid a \in A\}$
are denoted by 
$\sup_{a \in A} g(a)$ and $\inf_{a \in A} g(a)$, respectively.

Consider  
two partitions 
$\mathcal{P} = \{J_i\}_{i=0}^{N-1}$
and $\mathcal{Q} = \{K_j\}_{j=0}^{M-1}$
of the same interval. %
A partition
$\mathcal{Q}$ 
is called 
\emph{finer} than $\mathcal{P}$,
denoted by $\mathcal{Q} \sqsubseteq \mathcal{P}$,
if every interval $K_j$ in $\mathcal{Q}$ is a subset 
of some interval $J_i \in \mathcal{P}$.
Conversely, 
a partition $\mathcal{P}$ is called 
\emph{coarser} than $\mathcal{Q}$,
denoted by $\mathcal{P} \sqsupseteq \mathcal{Q}$,
when $\mathcal{Q}$ is finer than $\mathcal{P}$.
The coarsest partition that is finer than both $\mathcal{P}$ and $\mathcal{Q}$,
denoted by $\mathcal{P} \sqcap \mathcal{Q}$, always exists~\cite{bae2019bounded}.

The following lemmas present several basic properties of intervals, under our marker notation, 
including non-emptiness, inclusion, and intersection.

\begin{lemma}
  \label{lemma:interval_non_empty}
  Let $I = \intv{\upLc}{\intvLe{}}{\intvRe{}}{\dnLc}$ be an interval.
  Then, $I \neq \emptyset$ iff $\intvLe{} \ltDash{\upLc \wedge \dnLc} \intvRe{}$.
\end{lemma}

\begin{proof}
($\Rightarrow$)
Assume $I \neq \emptyset$.
Then there exists $t \in I$ such that $\intvLe{} \le t \le \intvRe{}$.
If $\intvLe{}<\intvRe{}$, then $\intvLe{} \ltDash{\upLc \wedge \dnLc} \intvRe{}$ holds trivially.
If $\intvLe{}=\intvRe{}$, this forces $\upLc=\dnLc=\strue$.
Hence $\intvLe{} \ltDash{\upLc \wedge \dnLc} \intvRe{}$.

\noindent
($\Leftarrow$)
Assume $\intvLe{} \ltDash{\upLc \wedge \dnLc} \intvRe{}$.
If $\intvLe{}<\intvRe{}$, then $(\intvLe{},\intvRe{})\subseteq I$.
If $\intvLe{}=\intvRe{}$, then $\upLc=\dnLc=\strue$, hence $\intvLe{}\in I$.
In either case, $I\neq\emptyset$.
\qedhere
\end{proof}

\begin{lemma}
  \label{lemma:interval}
  For nonempty intervals $I_i=\intv{\intvLty{i}}{\intvLe{i}}{\intvRe{i}}{\intvRty{i}}$ ($i=1,2$), the following hold:
  \begin{enumerate}[label=(\arabic*)]
    \item $I_1\subseteq I_2$ iff
    $\intvLe{2}\ltDash{\neg\intvLty{1}\vee\intvLty{2}}\intvLe{1}$ and
    $\intvRe{1}\ltDash{\neg\intvRty{1}\vee\intvRty{2}}\intvRe{2}$.
    \item $I_1\cap I_2=\emptyset$ iff
    $\intvRe{1}\ltDash{\neg\intvRty{1}\vee\neg\intvLty{2}}\intvLe{2}$ or
    $\intvRe{2}\ltDash{\neg\intvLty{1}\vee\neg\intvRty{2}}\intvLe{1}$.
    \item $I_1\cap I_2\neq\emptyset$ if
    $\intvLe{1}\ltDash{\intvLty{1}\vee\neg\intvLty{2}}\intvLe{2}$ and
    $\intvLe{2}\ltDash{\intvRty{1}\wedge\intvLty{2}}\intvRe{1}$.
  \end{enumerate}
\end{lemma}

\begin{proof}
We derive the inclusion and intersection conditions from interval membership.
\begin{proofrules}[topsep=2pt,itemsep=2pt]
\item[(1)]
($\Rightarrow$)
Assume $I_1\subseteq I_2$.
Then necessarily $\intvLe{2}\le\intvLe{1}$.
If $\intvLe{2}=\intvLe{1}$ and $\intvLty{1}=\strue$, then
$\intvLe{1}\in I_1$ implies $\intvLe{1}\in I_2$, hence $\intvLty{2}=\strue$.
Thus $\intvLe{2}\ltDash{\neg\intvLty{1}\vee\intvLty{2}}\intvLe{1}$.
The same argument applies to the right endpoints, yielding
$\intvRe{1}\ltDash{\neg\intvRty{1}\vee\intvRty{2}}\intvRe{2}$.

($\Leftarrow$)
Assume $\intvLe{2}\ltDash{\neg\intvLty{1}\vee\intvLty{2}}\intvLe{1}$ and
$\intvRe{1}\ltDash{\neg\intvRty{1}\vee\intvRty{2}}\intvRe{2}$.
Let $t\in I_1$.
Then $\intvLe{1}\le t\le\intvRe{1}$.
If $\intvLe{2}<\intvLe{1}$, then $\intvLe{2}\le t$ holds trivially.
If $\intvLe{2}=\intvLe{1}$, then $\intvLe{2}\le t$ follows from $\intvLe{1}\le t$.
Moreover, $t\le\intvRe{1}\le\intvRe{2}$.
If $t$ equals an endpoint of $I_2$, it equals the corresponding closed endpoint
of $I_1$, and the marked comparison forces that endpoint of $I_2$ to be closed.
Hence $t\in I_2$, and therefore $I_1\subseteq I_2$.

\item[(2)]
Equivalently, $I_1\cap I_2\neq\emptyset$ iff both comparisons below hold:
\begin{equation*}
\intvLe{2}\ltDash{\intvRty{1}\wedge\intvLty{2}}\intvRe{1}
\quad\text{and}\quad
\intvLe{1}\ltDash{\intvLty{1}\wedge\intvRty{2}}\intvRe{2}.
\tag{O}\label{eq:interval-overlap-bounds}
\end{equation*}
($\Rightarrow$)
For $t\in I_1\cap I_2$, $\intvLe{2}\ltDash{\intvLty{2}}t\ltDash{\intvRty{1}}\intvRe{1}$
and its symmetric counterpart give \eqref{eq:interval-overlap-bounds}.

($\Leftarrow$)
Let $\intvLe{}=\max_i\intvLe{i}$, $\intvRe{}=\min_j\intvRe{j}$,
$\upLc=\bigwedge_{\intvLe{i}=\intvLe{}}\intvLty{i}$, and
$\dnLc=\bigwedge_{\intvRe{j}=\intvRe{}}\intvRty{j}$.
Then $I_1\cap I_2=\intv{\upLc}{\intvLe{}}{\intvRe{}}{\dnLc}$.
For $\intvLe{i}=\intvLe{}$ and $\intvRe{j}=\intvRe{}$,
$\intvLe{}\ltDash{\intvLty{i}\wedge\intvRty{j}}\intvRe{}$ follows from
\eqref{eq:interval-overlap-bounds} if $i\ne j$, and from $I_i\neq\emptyset$
by \Cref{lemma:interval_non_empty} if $i=j$.
This gives $\intvLe{}\ltDash{\upLc\wedge\dnLc}\intvRe{}$,
so $I_1\cap I_2\neq\emptyset$ by \Cref{lemma:interval_non_empty}.

\item[(3)]
Assume $\intvLe{1}\ltDash{\intvLty{1}\vee\neg\intvLty{2}}\intvLe{2}$ and
$\intvLe{2}\ltDash{\intvRty{1}\wedge\intvLty{2}}\intvRe{1}$.
If $\intvLty{2}=\strue$, then $\intvLe{2}\in I_2$.
The two inequalities imply
$\intvLe{1}\ltDash{\intvLty{1}}\intvLe{2}\ltDash{\intvRty{1}}\intvRe{1}$,
hence $\intvLe{2}\in I_1$ and therefore $\intvLe{2}\in I_1\cap I_2$.
If $\intvLty{2}=\sfalse$, then $\intvLe{2}<\intvRe{1}$ holds strictly.
Since $I_2$ is nonempty, there exists $t>\intvLe{2}$ with $t\in I_2$.
Choosing $t$ sufficiently close to $\intvLe{2}$, we have
$\intvLe{1}\le\intvLe{2}<t<\intvRe{1}$, hence $t\in I_1$ as well.
Therefore $t\in I_1\cap I_2$.\qedhere
\end{proofrules}
\end{proof}

\subsection{Robust Semantics of STL}

The robust semantics of STL \cite{donze2013efficient,donze2010robust}
quantifies 
how strongly an STL formula $\varphi$ is satisfied.
In contrast, the Boolean semantics \cite{maler2004monitoring} %
defines the satisfaction relation $\sigma, t \models_\tau \varphi$.

\begin{definition}
Given 
an STL formula $\varphi$, 
a signal $\sigma: \domain{0}{\tau} \to {\realset}^X$,
and 
a time $t \in \domain{0}{\tau}$, 
the \emph{robustness degree}
$\robustness{\varphi}{\sigma}{t}{\tau} \in \bar{\realset}$
is defined inductively  by:
\begin{align*}
\robustness{\mathit{true}}{\sigma}{t}{\tau} 
    &= \ttrue
&
\robustness{f(\vec{x}) \geq 0}{\sigma}{t}{\tau}
    &= f(\sigma(t))
\\
\robustness{\neg \varphi}{\sigma}{t}{\tau}
    &= -\robustness{\varphi}{\sigma}{t}{\tau}
&
\robustness{\varphi \wedge \varphi'}{\sigma}{t}{\tau}  &= 
    \min (\robustness{\varphi}{\sigma}{t}{\tau}, \robustness{\varphi'}{\sigma}{t}{\tau})
\\
\robustness{\varphi \until{I} \varphi'}{\sigma}{t}{\tau} &= 
    \textstyle\sup_{t' \in (t+I) \cap \domain{0}{\tau}} \min (\robustness{\varphi'}{\sigma}{t'}{\tau}, \inf_{t'' \in [t,t']} \robustness{\varphi}{\sigma}{t''}{\tau})
    \span\span
\end{align*} 
\end{definition}

The notions of 
$\epsilon$-weakening and
$\epsilon$-strengthening for STL
are defined as follows \cite{scp2025-stl}. 
As expected, 
the robustness degree of $\varphi^{- \epsilon}$ (resp., $\varphi^{+ \epsilon}$ )  is 
greater (resp., smaller) than that of $\varphi$ by $\epsilon$.

\begin{definition}\label{def:ep-str}
For an STL formula $\varphi$ and a robustness threshold $\epsilon > 0$, 
the  $\epsilon$-weakening $\varphi^{- \epsilon}$ and $\epsilon$-strengthening $\varphi^{+ \epsilon}$ 
are defined inductively
as follows:
\begin{align*}
(\neg \varphi)^{- \epsilon} &\equiv \neg (\varphi^{+ \epsilon})
&
(\varphi_1 \wedge \varphi_2)^{- \epsilon} &\equiv \varphi_1^{- \epsilon} \wedge \varphi_2^{- \epsilon} 
&
(\varphi_1 \until{I} \varphi_2)^{- \epsilon} &\equiv \varphi_1^{- \epsilon} \until{I} \varphi_2^{- \epsilon}
\\
(\neg \varphi)^{+ \epsilon} &\equiv \neg (\varphi^{- \epsilon})
&
(\varphi_1 \wedge \varphi_2)^{+ \epsilon} &\equiv \varphi_1^{+ \epsilon} \wedge \varphi_2^{+ \epsilon} 
&
(\varphi_1 \until{I} \varphi_2)^{+ \epsilon} &\equiv \varphi_1^{+ \epsilon} \until{I} \varphi_2^{+ \epsilon} 
\end{align*}
where
$\mathit{true}^{-\epsilon}\equiv\mathit{true}^{+\epsilon}\equiv\mathit{true}$,
$(f(\vec{x}) \geq 0)^{-\epsilon} \equiv f(\vec{x}) \geq -\epsilon$
and
$(f(\vec{x}) \geq 0)^{+\epsilon} \equiv f(\vec{x}) \geq \epsilon$.
\end{definition}

\subsection{Hybrid Automata}
\label{sec:prelim-app-ha}

We follow the standard approach to parallel composition~\cite{henzinger2000theory,lynch2003hybrid}.

\begin{definition}
Let
$H_i=(Q_i,G_i,E_i,X_i,\mathit{init}_i,\mathit{inv}_i,
\mathit{flow}_i,\mathit{jump}_i)\,(i=1,2)$
be hybrid automata.
Their \emph{parallel composition} is
$H_1\parallel H_2
=
(Q_1\times Q_2,G,E,X_1\cup X_2,\mathit{init},\mathit{inv},\mathit{flow},\mathit{jump})$,
where:
\begin{itemize}

\item
$G$ is the set of labels $g$, represented by nonempty partial
maps on $\{1,2\}$ with $g(i)\in G_i$ whenever $g$ has a label for $H_i$;

\item
$((q_1,q_2),g,(q_1',q_2'))\in E$ iff $g\in G$ and, for each $i=1,2$,
$q_i,q_i'\in Q_i$ and either
(a) $g$ has a label for $H_i$ and $(q_i,g(i),q_i')\in E_i$, or
(b) $g$ has no label for $H_i$ and $q_i=q_i'$;

\item
$\langle(q_1,q_2),v\rangle\in\mathit{init}$
iff
$\langle q_i,v|_{X_i}\rangle\in\mathit{init}_i$
for $i=1,2$;

\item
$\langle(q_1,q_2),v\rangle\in\mathit{inv}$
iff
$\langle q_i,v|_{X_i}\rangle\in\mathit{inv}_i$
for $i=1,2$;

\item
$(\langle(q_1,q_2),v\rangle,t,v_t)\in\mathit{flow}$
iff
$(\langle q_i,v|_{X_i}\rangle,t,v_t|_{X_i})\in\mathit{flow}_i$
for $i=1,2$;

\item
for $e=((q_1,q_2),g,(q_1',q_2'))\in E$,
$(\langle(q_1,q_2),v\rangle,\langle(q_1',q_2'),v'\rangle)
\in\mathit{jump}(e)$ iff, for each $i=1,2$, either
(a) $g$ has a label for $H_i$ and
$(\langle q_i,v|_{X_i}\rangle,\langle q_i',v'|_{X_i}\rangle)
\in\mathit{jump}_i((q_i,g(i),q_i'))$, or
(b) $g$ has no label for $H_i$ and $v|_{X_i}=v'|_{X_i}$.
\end{itemize}
\end{definition}

\begingroup
\setlength{\intextsep}{6pt}
\begin{table}[!htbp]
\centering
\footnotesize
\caption{Sizes of our composed and property automata in the model checking experiments.}
\label{tab:sizes_by_formula}
\begin{minipage}[t]{0.48\textwidth}
\centering
\textbf{Linear models}\\[2pt]
\begin{tabular}{c@{\hspace{2pt}}cc@{\hspace{3pt}}cc@{\hspace{3pt}}cc}
\toprule
\multirow{2}{*}{Model}
  & \multicolumn{2}{c}{$f_1$} & \multicolumn{2}{c}{$f_2$} & \multicolumn{2}{c}{$f_3$} \\
\cmidrule(lr){2-3}\cmidrule(lr){4-5}\cmidrule(lr){6-7}
  & M/J & V/E & M/J & V/E & M/J & V/E \\
\midrule
\textsf{bat} & 27/204 & 10/37 & 25/134 & 9/20 & 19/78 & 7/14 \\
\textsf{lane} & 4/4 & 13/53 & 4/5 & 4/5 & 9/19 & 9/19 \\
\textsf{oscl} & 24/141 & 8/24 & 21/85 & 7/15 & 25/115 & 7/14 \\
\textsf{rail02} & 49/277 & 13/33 & 45/272 & 12/32 & 13/30 & 4/4 \\
\textsf{spac01} & 9/20 & 9/20 & 11/25 & 11/25 & 14/43 & 14/43 \\
\textsf{spac02} & 25/160 & 9/20 & 15/83 & 9/34 & 14/51 & 7/14 \\
\textsf{thm02} & 23/213 & 9/34 & 6/13 & 5/15 & 16/100 & 6/11 \\
\textsf{wat01} & 13/86 & 4/5 & 21/242 & 6/13 & 21/210 & 6/11 \\
\bottomrule
\end{tabular}
\end{minipage}
\hfill
\begin{minipage}[t]{0.48\textwidth}
\centering
\textbf{Nonlinear models}\\[2pt]
\begin{tabular}{c@{\hspace{2pt}}cc@{\hspace{3pt}}cc@{\hspace{3pt}}cc}
\toprule
\multirow{2}{*}{Model}
  & \multicolumn{2}{c}{$f_1$} & \multicolumn{2}{c}{$f_2$} & \multicolumn{2}{c}{$f_3$} \\
\cmidrule(lr){2-3}\cmidrule(lr){4-5}\cmidrule(lr){6-7}
  & M/J & V/E & M/J & V/E & M/J & V/E \\
\midrule
\textsf{car02} & 13/90 & 4/4 & 17/146 & 5/6 & 25/319 & 7/14 \\
\textsf{card} & 7/13 & 7/13 & 16/75 & 16/75 & 14/37 & 14/37 \\
\textsf{colm} & 13/53 & 13/53 & 12/41 & 12/41 & 7/13 & 7/13 \\
\textsf{lovo} & 7/18 & 5/8 & 6/13 & 4/5 & 7/12 & 5/6 \\
\textsf{robe02} & 3/2 & 3/2 & 11/38 & 11/38 & 10/37 & 10/37 \\
\textsf{robe03} & 3/2 & 3/2 & 14/37 & 14/37 & 11/38 & 11/38 \\
\textsf{vand} & 8/14 & 8/14 & 4/5 & 4/5 & 11/42 & 11/42 \\
\textsf{wat02} & 6/8 & 4/4 & 15/64 & 8/14 & 37/293 & 19/68 \\
\bottomrule
\end{tabular}
\end{minipage}
\end{table}

\endgroup
\clearpage
\suppressfloats[t]

\section{Details of the Model Checking Experiments}
\label{sec:expr-app-stlmc}

We have developed a prototype implementation of the proposed automata-based
robust STL model checking method.
The model checking prototype implements tableau construction in Python,
including the standard simulation-equivalence reduction method
\cite{baier2008modelchecking}.
Hybrid automata are specified using the STLmc input language
and parsed through its Python API~\cite{cav2022-stl}.
Given a hybrid system and an STL formula,
our tool constructs the composed hybrid automaton
$\composeha{\neg (\varphi^{+\epsilon})}{\tau}$ together with the associated
property automaton $\hatableau{\neg (\varphi^{+\epsilon})}{\tau}$,
and invokes existing reachability analysis tools to perform
robust model checking.

The sizes of the composed hybrid automata and their corresponding
property automata generated by our approach
are summarized in \Cref{tab:sizes_by_formula}.
For each benchmark and formula, we report the size of the composed hybrid automaton
as M/J (number of modes and jumps),
and the size of the property automaton as V/E
(number of vertices and edges).
The left part of the table corresponds to linear-ODE benchmarks,
and the right part to nonlinear-ODE benchmarks.

STLmc is executed using dReal~\cite{dReal}
as the backend solver with the \texttt{2-step} algorithm
and 5 cores in parallel.
The bounded analysis tools SAT-Reach, dReach, and STLmc
are run with a bound $N=5$ for all benchmarks,
except for a few nonlinear systems with particularly complex dynamics,
where we use $N=1$ (\texttt{robe02}, \texttt{robe03}) or $N=2$ (\texttt{colm}).
These models are highlighted in \Cref{tab:mc-result}.
Time bounds~$\tau$ and robustness thresholds~$\epsilon$
are chosen based on the characteristics of each benchmark model.
Flow* is executed with adaptive step sizes in the range $[0.001,\,0.5]$
by default and a fixed Taylor order of 6, 8, 12, or 15 per instance,
with tighter settings for stiff systems.
dReal is used with precision $\delta = 0.001$.
SpaceEx and SAT-Reach are run in the \texttt{stc} scenario
with octagonal directions,
with SAT-Reach using a time step of $0.5$.

\Cref{tab:mc-result} reports detailed per-instance results for the comparison
between STLmc and our automata-based approach across all benchmarks and formulas.
For each model, results are grouped by dynamics type (linear or nonlinear) and
by analysis method.
For each of the STL formulas $f_1$, $f_2$, and $f_3$, the table reports the
size of the composed automaton (M/J), the execution time in seconds, and
the analysis outcome.
Outcomes are denoted by $\top$ if the unsafe region is proven unreachable,
$\bot$ if it is reachable, $*$ if the result is unknown, $-$ if the analysis
times out, and $X$ if the analysis fails due to an error.
For our approach, multiple reachability backends are shown for each benchmark,
illustrating how different engines behave on the same generated hybrid automaton,
whereas STLmc results are reported using dReal as its backend solver.

\begin{table}
\caption{Per-instance model checking runtimes and outcomes.
M/J: number of modes/jumps in the composed automaton; Time: seconds;
R: $\top$ indicates that the unsafe region is unreachable,
$\bot$ indicates reachable,
$*$ denotes unknown,
 $-$ denotes timeouts,
 and
 $X$ denotes solver errors.
}
\label{tab:mc-result}
\scriptsize
\centering
\begin{tabular}{
@{\hspace{0.1cm}}c @{\hspace{0.1cm}}c @{\hspace{0.1cm}}c
@{\hspace{0.1cm}}c @{\hspace{0.1cm}}r @{\hspace{0.1cm}}c
@{\hspace{0.1cm}}c @{\hspace{0.1cm}}r @{\hspace{0.1cm}}c
@{\hspace{0.1cm}}c @{\hspace{0.1cm}}r @{\hspace{0.1cm}}c
}
\toprule
\multirow{2}{*}{Dyn.} & 
\multirow{2}{*}{Model} & 
\multirow{2}{*}{Method (Solver)} & 
\multicolumn{3}{c}{$f_1$} & 
\multicolumn{3}{c}{$f_2$} & 
\multicolumn{3}{c}{$f_3$} \\
\cmidrule(r){4-6} \cmidrule(r){7-9} \cmidrule(r){10-12}
& & & M/J & Time & R
& M/J & Time & R
& M/J & Time & R \\
\midrule
\multirow{24}{*}{\rotatebox[origin=c]{90}{linear}}
&\multirow{3}{*}{\texttt{bat}}
& Autm (SR) & 27/204 & 2.47 & * & 25/134 & 1.53 & $\top$ & 19/78 & 2.56 & $\top$ \\
& & Autm (SX) & 27/204 & \texttt{T/O} & - & 25/134 & \texttt{T/O} & - & 19/78 & \texttt{T/O} & - \\
& & STLmc (DRL) & -/- & 22.40 & $\bot$ & -/- & 1455.05 & $\bot$ & -/- & \texttt{T/O} & - \\
\cmidrule(r){3-12}&\multirow{3}{*}{\texttt{lane}}
& Autm (SR) & 4/4 & 1.52 & $\top$ & 4/5 & 0.47 & $\top$ & 9/19 & 9.61 & * \\
& & Autm (SX) & 4/4 & 0.62 & $\top$ & 4/5 & 0.22 & $\top$ & 9/19 & 32.63 & $\bot$ \\
& & STLmc (DRL) & -/- & 1.53 & $\top$ & -/- & 28.75 & $\top$ & -/- & \texttt{T/O} & - \\
\cmidrule(r){3-12}&\multirow{3}{*}{\texttt{oscl}}
& Autm (SR) & 24/141 & - & X & 21/85 & - & X & 25/115 & - & X \\
& & Autm (SX) & 24/141 & 0.31 & $\top$ & 21/85 & 2.12 & $\top$ & 25/115 & 3.56 & $\bot$ \\
& & STLmc (DRL) & -/- & 193.43 & $\top$ & -/- & 93.43 & $\bot$ & -/- & 1481.28 & $\top$ \\
\cmidrule(r){3-12}&\multirow{3}{*}{\texttt{rail02}}
& Autm (SR) & 49/277 & - & X & 45/272 & - & X & 13/30 & 0.70 & * \\
& & Autm (SX) & 49/277 & 3.72 & $\bot$ & 45/272 & 1.15 & $\bot$ & 13/30 & 0.38 & $\bot$ \\
& & STLmc (DRL) & -/- & \texttt{T/O} & - & -/- & 3199.26 & $\bot$ & -/- & 25.72 & $\bot$ \\
\cmidrule(r){3-12}&\multirow{3}{*}{\texttt{spac01}}
& Autm (SR) & 9/20 & - & X & 11/25 & - & X & 14/43 & - & X \\
& & Autm (SX) & 9/20 & 0.93 & $\top$ & 11/25 & 1.72 & $\top$ & 14/43 & 1.66 & $\top$ \\
& & STLmc (DRL) & -/- & \texttt{T/O} & - & -/- & \texttt{T/O} & - & -/- & \texttt{T/O} & - \\
\cmidrule(r){3-12}&\multirow{3}{*}{\texttt{spac02}}
& Autm (SR) & 25/160 & - & X & 15/83 & - & X & 14/51 & - & X \\
& & Autm (SX) & 25/160 & 1.18 & $\bot$ & 15/83 & 0.10 & $\top$ & 14/51 & 0.25 & $\top$ \\
& & STLmc (DRL) & -/- & \texttt{T/O} & - & -/- & 1480.14 & $\top$ & -/- & \texttt{T/O} & - \\
\cmidrule(r){3-12}&\multirow{3}{*}{\texttt{thm02}}
& Autm (SR) & 23/213 & 3.06 & $\top$ & 6/13 & 0.28 & $\top$ & 16/100 & 1.86 & $\top$ \\
& & Autm (SX) & 23/213 & \texttt{T/O} & - & 6/13 & 0.02 & $\top$ & 16/100 & \texttt{T/O} & - \\
& & STLmc (DRL) & -/- & 2871.23 & $\top$ & -/- & 15.73 & $\top$ & -/- & 110.06 & $\bot$ \\
\cmidrule(r){3-12}&\multirow{3}{*}{\texttt{wat01}}
& Autm (SR) & 13/86 & 0.60 & $\top$ & 21/242 & 0.34 & $\top$ & 21/210 & 0.41 & $\top$ \\
& & Autm (SX) & 13/86 & 27.00 & $\bot$ & 21/242 & 0.05 & $\top$ & 21/210 & 0.02 & $\top$ \\
& & STLmc (DRL) & -/- & 64.09 & $\top$ & -/- & \texttt{T/O} & - & -/- & \texttt{T/O} & - \\
\midrule
\multirow{24}{*}{\rotatebox[origin=c]{90}{nonlinear}}
&\multirow{3}{*}{\texttt{car02}}
& Autm (DRH) & 13/90 & 0.34 & $\bot$ & 17/146 & 0.56 & $\bot$ & 25/319 & 101.94 & $\bot$ \\
& & Autm (Flow*) & 13/90 & 4.90 & $\bot$ & 17/146 & 8.94 & $\bot$ & 25/319 & 477.71 & $\bot$ \\
& & STLmc (DRL) & -/- & \texttt{T/O} & - & -/- & 107.88 & $\bot$ & -/- & \texttt{T/O} & - \\
\cmidrule(r){3-12}&\multirow{3}{*}{\texttt{card}}
& Autm (DRH) & 7/13 & 12.36 & $\bot$ & 16/75 & 12.04 & $\top$ & 14/37 & 2.01 & $\top$ \\
& & Autm (Flow*) & 7/13 & \texttt{T/O} & - & 16/75 & 94.80 & $\top$ & 14/37 & \texttt{T/O} & - \\
& & STLmc (DRL) & -/- & \texttt{T/O} & - & -/- & \texttt{T/O} & - & -/- & \texttt{T/O} & - \\
\cmidrule(r){3-12}&\multirow{3}{*}{\textcolor{red}{\texttt{colm}}}
& Autm (DRH) & 13/53 & 0.05 & $\top$ & 12/41 & 0.03 & $\top$ & 7/13 & 0.04 & $\top$ \\
& & Autm (Flow*) & 13/53 & \texttt{T/O} & - & 12/41 & 244.85 & $\top$ & 7/13 & 289.04 & $\top$ \\
& & STLmc (DRL) & -/- & \texttt{T/O} & - & -/- & \texttt{T/O} & - & -/- & \texttt{T/O} & - \\
\cmidrule(r){3-12}&\multirow{3}{*}{\texttt{lovo}}
& Autm (DRH) & 7/18 & 0.82 & $\top$ & 6/13 & 0.58 & $\top$ & 7/12 & 0.43 & $\top$ \\
& & Autm (Flow*) & 7/18 & \texttt{T/O} & - & 6/13 & 0.05 & $\top$ & 7/12 & 726.44 & $\top$ \\
& & STLmc (DRL) & -/- & \texttt{T/O} & - & -/- & 1297.04 & $\top$ & -/- & 30.38 & $\top$ \\
\cmidrule(r){3-12}&\multirow{3}{*}{\textcolor{red}{\texttt{robe02}}}
& Autm (DRH) & 3/2 & 0.03 & $\top$ & 11/38 & 0.03 & $\top$ & 10/37 & 0.03 & $\top$ \\
& & Autm (Flow*) & 3/2 & \texttt{T/O} & - & 11/38 & \texttt{T/O} & - & 10/37 & \texttt{T/O} & - \\
& & STLmc (DRL) & -/- & 3.22 & $\top$ & -/- & \texttt{T/O} & - & -/- & \texttt{T/O} & - \\
\cmidrule(r){3-12}&\multirow{3}{*}{\textcolor{red}{\texttt{robe03}}}
& Autm (DRH) & 3/2 & 0.04 & $\top$ & 14/37 & 0.04 & $\top$ & 11/38 & 0.04 & $\top$ \\
& & Autm (Flow*) & 3/2 & \texttt{T/O} & - & 14/37 & \texttt{T/O} & - & 11/38 & \texttt{T/O} & - \\
& & STLmc (DRL) & -/- & 39.32 & $\top$ & -/- & \texttt{T/O} & - & -/- & \texttt{T/O} & - \\
\cmidrule(r){3-12}&\multirow{3}{*}{\texttt{vand}}
& Autm (DRH) & 8/14 & 5.61 & $\bot$ & 4/5 & 2.28 & $\bot$ & 11/42 & 1.43 & $\top$ \\
& & Autm (Flow*) & 8/14 & 63.95 & * & 4/5 & 43.35 & * & 11/42 & 23.96 & $\top$ \\
& & STLmc (DRL) & -/- & \texttt{T/O} & - & -/- & \texttt{T/O} & - & -/- & \texttt{T/O} & - \\
\cmidrule(r){3-12}&\multirow{3}{*}{\texttt{wat02}}
& Autm (DRH) & 6/8 & 0.07 & $\top$ & 15/64 & 0.38 & $\bot$ & 37/293 & 0.31 & $\bot$ \\
& & Autm (Flow*) & 6/8 & 0.18 & $\top$ & 15/64 & 0.24 & $\bot$ & 37/293 & 18.55 & $\bot$ \\
& & STLmc (DRL) & -/- & 85.94 & $\top$ & -/- & 410.56 & $\top$ & -/- & \texttt{T/O} & - \\
\bottomrule
\end{tabular}
\end{table}

Beyond execution time, \Cref{tab:mc-result} also highlights differences in the
verification guarantees provided by the underlying tools.
STLmc, SAT-Reach, and dReach perform bounded analysis and can establish correctness
only up to a bound~$N$; consequently, a $\top$ outcome certifies
safety only within that bound.
In contrast, SpaceEx and Flow* perform unbounded reachability analysis via
fixpoint-based over-approximation.
When they return $\top$, the unsafe region is proven unreachable for all
execution lengths.
This distinction is reflected in \Cref{tab:mc-result}, where SpaceEx or Flow*
sometimes yields definitive results even when bounded tools report unknown
or require very small bounds.
This demonstrates that our %
approach enables flexible
trade-offs between performance and precision.

\clearpage

\section{Interval Predicates, Expansion Rules, and Closures}
\label{sec:app-rule}

\Crefrange{sec:app-clock}{sec:app_pred_inv} establish the properties of
judgements and expansion rules from \Cref{sec:rule}.
\Crefrange{sec:app-subsumption}{sec:app_closure_invariant} then develop
transition form, subsumption, and closure correctness for \Cref{sec:tableau}.

\subsection{Details on Clock Valuations and Resets}
\label{sec:app-clock}

Formally, the free variables are defined as follows.

\begin{definition}\label{def:judgement-free-variables}
  The set $\fvar{\Psi}$ of free clock variables of a judgement $\Psi$ is defined inductively by:
\begin{alignat*}{6}
  &\fvar{\ast} = \emptyset
\qquad\qquad
  \fvar{c \bowtie a} = \{ c \}
\qquad\;\;
   \fvar{\Lambda.\, \Psi} = \fvar{\Psi} \setminus \var{\Lambda} \\
&\fvar{\ipstl{\varphi}} = \emptyset
\qquad\qquad
\fvar{\intvpredupT{c}{\upLc}{\varphi}} = \{ c \}
\qquad\qquad
\fvar{\intvpredT{c}{d}{\upLc}{\dnLc}{\varphi}} = \{ c, d \}\\
&\fvar{\Psi_1, \Psi_2} = \fvar{\Psi_1} \cup \fvar{\Psi_2}
\qquad
\fvar{\nextop \Psi} = \fvar{\Psi}
\qquad
\fvar{\lty{\ltyVar}} = \emptyset
 \end{alignat*}
\end{definition}

\Cref{lem:clock-independence} states that two clock-reset sets $\Lambda_1$ and $\Lambda_2$
commute whenever they assign the same values to their shared clocks. In this case,
applying $\Lambda_1$ and $\Lambda_2$ sequentially has exactly the same effect as
applying their union at once.

\begin{lemma}
\label{lem:clock-independence}
For any clock valuation $\alpha$ and consistent resets $\Lambda_1$ and $\Lambda_2$,
$\alpha[\Lambda_1][\Lambda_2] = \alpha[\Lambda_1 \cup \Lambda_2]$.
\end{lemma}

\begin{proof}
Let $c$ be a clock variable such that $c \in \var{\Lambda_1} \cup \var{\Lambda_2}$.
There are two cases for $c$: (i) $c \in \var{\Lambda_1} \cap \var{\Lambda_2}$ and (ii) otherwise.
Suppose $c \in \var{\Lambda_1} \cap \var{\Lambda_2}$.
By assumption $\alpha[\Lambda_1](c) = \alpha[\Lambda_2](c)$. Thus
$\alpha[\Lambda_1][\Lambda_2](c) = \alpha[\Lambda_2](c)$.
Since $\Lambda_2 \subseteq \Lambda_1 \cup \Lambda_2$,
$\alpha[\Lambda_1][\Lambda_2](c) = \alpha[\Lambda_1 \cup \Lambda_2](c)$.
Now suppose $c \not\in \var{\Lambda_1} \cap \var{\Lambda_2}$.
Then, either $c \in \var{\Lambda_1 \setminus \Lambda_2}$ or
$c \in \var{\Lambda_2 \setminus \Lambda_1}$.
For either case, the claim holds.
\end{proof}

The satisfaction of an interval sequence judgement depends only on the clock valuations of the clocks that actually occur in the judgement.
Therefore, modifying the values of irrelevant clocks does not affect its semantics.

\begin{lemma}
\label{lem:irrelevant-reset}
For any signal $\sigma$, partition $\partition{J}$, judgement $\Psi$, position $i$,
and clock valuations $\alpha_1, \alpha_2$,
if $\alpha_1(c) = \alpha_2(c)$ for every $c \in \fvar{\Psi}$, then
$\sigma,\partition{J},\alpha_1, i \models_\tau \Psi$ iff
$\sigma,\partition{J},\alpha_2, i \models_\tau \Psi$.
\end{lemma}

\begin{proof}
  By induction on $\Psi$. In the $p$ and $\nextop\Phi$ cases, neither side holds at $i=|\partition{J}|$.
  \begin{proofrules}[topsep=2pt,itemsep=2pt]
    \item \emph{Case $\Psi = c \bowtie a$.}
    Since $\fvar{c \bowtie a} = \{ c \}$ and by assumption $\alpha_1(c) = \alpha_2(c)$,
    $\sigma,\partition{J},\alpha_1, i \models_\tau c \bowtie a$ iff
    $\sigma,\partition{J},\alpha_2, i \models_\tau c \bowtie a$.

    \item \emph{Case $\Psi = \ast$ and $\Psi = \lty{\ltyVar}$.}
    By definition, the satisfaction of the judgements $\ast$ and $\lty{\ltyVar}$
    depends only on $\partition{J}$. Thus the claim holds.
    \item \emph{Case $\Psi = p$.}
    $p$ is either
    $\ipstl{\varphi}$ or $\ipred \bothtemp_I \varphi$, where $\bothtemp \in \{ \always{}, \eventually{} \}$ and
      $\ipred \in \{ \intvpredupT{c}{\upLc}{}, \intvpredT{c}{d}{\upLc}{\dnLc}{} \}$.
    \begin{proofsubcases}[topsep=1pt,itemsep=1pt]
      \item \emph{Case $p = \ipstl{\varphi}$.}
      The scopes $\scope{\intvOf{J}{i}}{\alpha_1}{\ipstl{\varphi}}$ and
      $\scope{\intvOf{J}{i}}{\alpha_2}{\ipstl{\varphi}}$ coincide and represent $\intvOf{J}{i}$.
      By definition,
      $\sigma, \partition{J}, \alpha_1, i \models_\tau \Psi$ iff
      $(\forall t \in \intvOf{J}{i})\, \sigma, t \models_\tau \varphi$ iff
      $\sigma, \partition{J}, \alpha_2, i \models_\tau \Psi$.

      \item \emph{Case $p = \ipred \bothtemp_I \varphi$.}
      By assumption, for every $c \in \fvar{\ipred \bothtemp_I \varphi}$, $\alpha_1(c) = \alpha_2(c)$.
      This implies $\scope{\intvOf{J}{i}}{\alpha_1}{\ipred \bothtemp_I \varphi} =
      \scope{\intvOf{J}{i}}{\alpha_2}{\ipred \bothtemp_I \varphi}$.
      Since the satisfaction depends on $\sigma$ and their scopes,
      we have $\sigma, \partition{J}, \alpha_1, i \models_\tau \ipred \bothtemp_I \varphi$ iff
      $\sigma, \partition{J}, \alpha_2, i \models_\tau \ipred \bothtemp_I \varphi$.
    \end{proofsubcases}

    \item \emph{Case $\Psi = \Lambda \, . \, \Phi$.}
    By definition, $\fvar{\Lambda . \Phi} = \fvar{\Phi} \setminus \var{\Lambda}$.
    Consider $\alpha_1[\Lambda]$ and $\alpha_2[\Lambda]$.
    Then the assumption implies
    $\alpha_1[\Lambda](c) = \alpha_2[\Lambda](c)$ for every $c \in \fvar{\Phi}$.
    By induction hypothesis,
    $\sigma, \partition{J}, \alpha_1[\Lambda], i \models_\tau \Phi$ iff
    $\sigma, \partition{J}, \alpha_2[\Lambda], i \models_\tau \Phi$.

    \item \emph{Case $\Psi = \nextop \Phi$.}
    By definition, $\fvar{\nextop \Phi} = \fvar{\Phi}$.
    By assumption, $\alpha_1(c) = \alpha_2(c)$ for every $c \in \fvar{\Phi}$.
    Note that, for any constant $a$ and every $c\in\fvar{\Phi}$,
    $(\alpha_1 + a)(c) = (\alpha_2 + a)(c)$.
    By induction hypothesis,
    $\sigma, \partition{J}, \alpha_1 + |\intvOf{J}{i}|, i + 1 \models_\tau \Phi$ iff
    $\sigma, \partition{J}, \alpha_2 + |\intvOf{J}{i}|, i + 1 \models_\tau \Phi$.

    \item \emph{Case $\Psi = \Phi_1, \Phi_2$.}
    By definition,
    $\fvar{\Phi_1,\Phi_2} = \fvar{\Phi_1} \cup \fvar{\Phi_2}$.
    For each $j\in\{1,2\}$, the valuations agree on $\fvar{\Phi_j}$,
    so the induction hypothesis gives
    $\sigma, \partition{J}, \alpha_1, i \models_\tau \Phi_j$ iff
    $\sigma, \partition{J}, \alpha_2, i \models_\tau \Phi_j$.
    Thus, the claim holds.
  \qedhere
  \end{proofrules}
\end{proof}

\begin{example}
Let $\Psi = (c > 1)$, so $\fvar{\Psi} = \{c\}$.
Consider two reset sets $\Lambda_1 = \cset{\creset{c}{2}}$ and
$\Lambda_2 = \cset{\creset{d}{5}}$ with $d \neq c$.
Since $\var{\Lambda_1} \cap \var{\Lambda_2} = \emptyset$,
the resets are disjoint and thus consistent.
Let $\alpha$ be a clock assignment.
By \Cref{lem:clock-independence},
$\alpha[\Lambda_1][\Lambda_2] = \alpha[\Lambda_1 \cup \Lambda_2]$.
Let $\alpha_1 = \alpha[\Lambda_1][\Lambda_2]$ and $\alpha_2 = \alpha[\Lambda_1 \cup \Lambda_2]$.
Because $c$ is the only clock occurring in $\Psi$,
the condition of \Cref{lem:irrelevant-reset} holds.
Hence, for any signal $\sigma$, partition $\partition{J}$, and position $i$,
$\sigma,\partition{J},\alpha_1,i \models_\tau c > 1$
iff
$\sigma,\partition{J},\alpha_2,i \models_\tau c > 1$.
\end{example}

Extending a reset set may introduce additional clock resets.
However, if these additional resets affect only clocks irrelevant to the resulting judgement, then the valuations of all relevant clocks remain unchanged.

\begin{lemma}\label{lem:reset-agreement}
Given reset sets $\Lambda_1$ and $\Lambda_2$ such that $\Lambda_1 \subseteq \Lambda_2$
and a clock valuation $\alpha$,
if $\vars(\Lambda_2)\cap \fvar{\Lambda_1 . \Psi} = \emptyset$, then
$\alpha[\Lambda_1](x)=\alpha[\Lambda_2](x)$ for every $x\in \fvar{\Psi}$.
\end{lemma}

\begin{proof}
Let $x \in \fvar{\Psi}$.
If $x \in \vars(\Lambda_2)$, then by
$\vars(\Lambda_2)\cap \fvar{\Lambda_1 . \Psi} = \emptyset$, we have
$x \in \vars(\Lambda_1)$. Since $\Lambda_1 \subseteq \Lambda_2$,
both reset sets assign the same value to $x$, and hence
$\alpha[\Lambda_1](x)=\alpha[\Lambda_2](x)$.
Suppose $x \notin \vars(\Lambda_2)$. Then $x \notin \vars(\Lambda_1)$. Therefore,
$\alpha[\Lambda_1](x)=\alpha(x)=\alpha[\Lambda_2](x)$.
\end{proof}

\subsection{Proofs of Judgement Equivalence and Implications}
\label{sec:app-judg-equiv}

Up to equivalence, judgement conjunction is associative by the definition of
satisfaction, and idempotent and commutative by \Cref{prop:judg-equiv}. Thus a conjunction can be
treated as the finite set of its conjuncts (e.g., writing
$\Phi_1,\Phi_2$ as $\{\Phi_1,\Phi_2\}$ and expressions such as
$\Phi_1 \in (\Phi_1,\Phi_2)$ by abuse of notation).
The proposition also shows how to merge reset-annotated judgements in parallel
or nested form: when the reset sets touch disjoint free variables and agree on
shared clocks, $(\Lambda_1.\Psi_1)$ and $(\Lambda_2.\Psi_2)$ combine into a
judgement with reset $\Lambda_1 \cup \Lambda_2$ and body $\Psi_1,\Psi_2$.

\restatetheorem{prop:judg-equiv}

\begin{proof}
For each equivalence of the form $\Phi_1 \equiv \Phi_2$, we show that,
for any signal $\sigma$, partition $\partition{J}$, clock valuation $\alpha$,
and position $i \in \{0,\ldots,|\partition{J}|\}$,
$\sigma, \partition{J}, \alpha, i \models_\tau \Phi_1$ iff $\sigma, \partition{J}, \alpha, i \models_\tau \Phi_2$.
\begin{proofrules}[topsep=2pt,itemsep=2pt]
\item \emph{Cases $\Psi_1, \Psi_1 \equiv \Psi_1$ and
$\Psi_1, \Psi_2 \equiv \Psi_2, \Psi_1$.}
Both follow directly from the definition.

\item \emph{Case $\nextop (\Psi_1, \Psi_2) \equiv (\nextop \Psi_1), (\nextop \Psi_2)$.}
If $i = |\partition{J}|$, both sides are false.
Suppose $i < |\partition{J}|$.
By definition,
$\sigma, \partition{J}, \alpha, i \models_\tau \nextop (\Psi_1, \Psi_2)$ iff
$\sigma, \partition{J}, \alpha + |\intvOf{J}{i}|, i+1 \models_\tau \Psi_1,\Psi_2$ iff
$\sigma, \partition{J}, \alpha, i \models_\tau \nextop \Psi_1$ and
$\sigma, \partition{J}, \alpha, i \models_\tau \nextop \Psi_2$ iff
$\sigma, \partition{J}, \alpha, i \models_\tau (\nextop \Psi_1), (\nextop \Psi_2)$.
\end{proofrules}

For the two reset cases, we first derive a common equivalence.
By the hypotheses and the definition of free variables,
$\var{\Lambda_1\cup\Lambda_2}\cap\fvar{\Lambda_j.\Psi_j}=\emptyset$ for $j=1,2$.
Since $\Lambda_j\subseteq\Lambda_1\cup\Lambda_2$, \Cref{lem:reset-agreement}
gives agreement on $\fvar{\Psi_j}$, so \Cref{lem:irrelevant-reset} yields
\begin{equation*}
\sigma,\partition{J},\alpha[\Lambda_j],i\models_\tau\Psi_j
\quad\text{iff}\quad
\sigma,\partition{J},\alpha[\Lambda_1\cup\Lambda_2],i\models_\tau\Psi_j
\qquad (j=1,2).
\tag{R}\label{eq:judg-equiv-reset-agreement}
\end{equation*}
\begin{proofrules}[topsep=2pt,itemsep=2pt]
\item \emph{Case $\Phi_1 = (\Lambda_1.\, \Psi_1), (\Lambda_2.\, \Psi_2)$ and
$\Phi_2 = (\Lambda_1 \cup \Lambda_2) . (\Psi_1, \Psi_2)$.}
\begin{align*}
&\sigma, \partition{J}, \alpha, i \models_\tau (\Lambda_1 . \Psi_1), (\Lambda_2 . \Psi_2) &\\
\text{iff}\;\;\;
&\sigma, \partition{J}, \alpha[\Lambda_1], i \models_\tau \Psi_1 \;\; \text{and} \;\;
\sigma, \partition{J}, \alpha[\Lambda_2], i \models_\tau \Psi_2 \\
\text{iff}\;\;\;
&\sigma, \partition{J}, \alpha[\Lambda_1 \cup \Lambda_2], i \models_\tau \Psi_1,\Psi_2
\quad &&\text{(by \eqref{eq:judg-equiv-reset-agreement})\quad\phantom{\qedsymbol}}
\end{align*}

\item \emph{Case $\Phi_1 = \Lambda_1. (\Psi_1, \Lambda_2.\, \Psi_2)$ and
$\Phi_2 = (\Lambda_1 \cup \Lambda_2) . (\Psi_1, \Psi_2)$.}
\begin{align*}
&\sigma, \partition{J}, \alpha, i \models_\tau \Lambda_1. (\Psi_1, \Lambda_2.\, \Psi_2) & \\
\text{iff}\;\;\;
&\sigma, \partition{J}, \alpha[\Lambda_1], i \models_\tau \Psi_1 \;\; \text{and} \;\; \sigma, \partition{J}, \alpha[\Lambda_1][\Lambda_2], i \models_\tau \Psi_2 & \\
\text{iff}\;\;\;
&\sigma, \partition{J}, \alpha[\Lambda_1\cup\Lambda_2], i \models_\tau \Psi_1 \;\; \text{and} \;\;
\sigma, \partition{J}, \alpha[\Lambda_1][\Lambda_2], i \models_\tau \Psi_2
\quad &&\text{(by \eqref{eq:judg-equiv-reset-agreement} with $j=1$)} \\
\text{iff}\;\;\;
&\sigma, \partition{J}, \alpha[\Lambda_1 \cup \Lambda_2], i \models_\tau \Psi_1,\Psi_2
\quad &&\text{(by \Cref{lem:clock-independence})\quad\qedhere}
\end{align*}
\end{proofrules}
\end{proof}

The equivalences in \Cref{prop:intvpred-equiv} are understood under the following
implicit side conditions at a current partition interval $\intvOf{J}{i}$
with clock valuation $\alpha$:
in (1), $\upLc$ matches the left-endpoint type of $\intvOf{J}{i}$;
in (2), $\dnLc$ matches that type and
$\alpha(c)\gtDash{\neg\dnLc\vee\upLc}0$,
so the interval specified by the scope of
$\intvpredupT{c}{\upLc}{\bothtemp_I\varphi}$ starts no later than the current interval,
with endpoint markers taken into account.

\restatetheorem{prop:intvpred-equiv}

\begin{proof}
Fix $\sigma$, $\partition{J}$, $\alpha$, and a position
$i\in\{0,\ldots,|\partition{J}|\}$.
At $i=|\partition{J}|$, both sides of each equivalence are false.
Otherwise, write
$\intvOf{J}{i}=\intv{\ltLc}{u}{v}{\rtLc}$ and use the
endpoint-marker conventions stated above.

\begin{proofrules}[topsep=2pt,itemsep=2pt]
\item \emph{(1)}
Let $\beta=\alpha[\cset{\creset{c}{0}}]$.
Since $\beta(c)=0$ and $\upLc=\ltLc$, the type judgement holds and
\[
\scope{\intvOf{J}{i}}{\beta}{\intvpredupT{c}{\upLc}{\bothtemp_I\varphi}}
=\pair{\ltLc}{u}{v}{\rtLc}
=\scope{\intvOf{J}{i}}{\alpha}{\ipstl{\bothtemp_I\varphi}}.
\]
Thus the scopes coincide, and the equivalence follows from the
satisfaction definitions.

\item \emph{(2)}
We show that the interval for
$\intvpredupT{c}{\upLc}{\bothtemp_I\varphi}$ is the union of those for
$\intvpredT{c}{d}{\upLc}{\neg\dnLc}{\bothtemp_I\varphi}$ and
$\ipstl{\bothtemp_I\varphi}$ after resetting $d$ to $0$.
Since $d$ is fresh and $\dnLc=\ltLc$, the type judgement holds, and their scopes are
\[
\begin{gathered}
\scope{\intvOf{J}{i}}{\alpha}{\intvpredupT{c}{\upLc}{\bothtemp_I\varphi}}
=\pair{\upLc}{u-\alpha(c)}{v}{\rtLc},\\
\scope{\intvOf{J}{i}}{\alpha[\cset{\creset{d}{0}}]}{\intvpredT{c}{d}{\upLc}{\neg\dnLc}{\bothtemp_I\varphi}}
=\pair{\upLc}{u-\alpha(c)}{u}{\neg\ltLc},\quad
\scope{\intvOf{J}{i}}{\alpha[\cset{\creset{d}{0}}]}{\ipstl{\bothtemp_I\varphi}}
=\pair{\ltLc}{u}{v}{\rtLc}.
\end{gathered}
\]
For $\eventually{I}\varphi$, the intervals represented by the scopes satisfy
\[
\annotatedInterval{\intv{\upLc}{u-\alpha(c)}{v}{\rtLc}}
  {\intvpredupT{c}{\upLc}{\eventually{I}\varphi}}
=\annotatedInterval{\intv{\upLc}{u-\alpha(c)}{u}{\neg\ltLc}}
  {\intvpredT{c}{d}{\upLc}{\neg\dnLc}{\eventually{I}\varphi}}
\ \cup\ \annotatedInterval{\intvOf{J}{i}}{\ipstl{\eventually{I}\varphi}}.
\]
For $\always{I}\varphi$, the checking intervals instead satisfy
\begin{align*}
\annotatedInterval{\intv{\upLc}{u-\alpha(c)+\inf I}{v+\sup I}{\rtLc}}
  {\intvpredupT{c}{\upLc}{\always{I}\varphi}}
&=\annotatedInterval{\intv{\upLc}{u-\alpha(c)+\inf I}{u+\sup I}{\neg\ltLc}}
  {\intvpredT{c}{d}{\upLc}{\neg\dnLc}{\always{I}\varphi}}\\
&\quad\cup\ \annotatedInterval{\intv{\ltLc}{u+\inf I}{v+\sup I}{\rtLc}}
  {\ipstl{\always{I}\varphi}}.
\end{align*}
Since $I$ is non-singular, $u+\inf I<u+\sup I$,
so the two checking intervals overlap.
Intersecting with $\domain{0}{\tau}$ preserves unions; the satisfaction
definitions therefore give the required conjunction.\qedhere
\end{proofrules}
\end{proof}

\restatetheorem{lem:intvpred-trans}

\begin{proof}
At $k=|\partition{J}|$, every premise is false. Otherwise write
$\intvOf{J}{k}=\intv{\intvLty{k}}{\intvLe{k}}{\intvRe{k}}{\intvRty{k}}$.
In (2)--(4), the premise forces $k+1<|\partition{J}|$, so
$\intvLe{k+1}=\intvRe{k}=\intvLe{k}+|\intvOf{J}{k}|$ and
\[
\intvLe{k+1}-(\alpha(x)+|\intvOf{J}{k}|)
=\intvLe{k}-\alpha(x)\qquad(x\in\{c,d\}).
\]
We first consider $\eventually{I}\varphi$. The displays give the intervals
specified by the predicates' scopes: $T_1$ for the premise and $T_2$ for
the conclusion. It suffices to show $T_2\subseteq T_1$.
For nonempty $T_2$, the marked endpoint comparisons below also ensure
$T_1\ne\emptyset$, allowing \Cref{lemma:interval}~(1); the empty case is immediate.

\begin{proofrules}[topsep=3pt,itemsep=3pt]
\setlength{\abovedisplayskip}{3pt}
\setlength{\belowdisplayskip}{3pt}
\setlength{\abovedisplayshortskip}{3pt}
\setlength{\belowdisplayshortskip}{3pt}
\item[(1)] By \Cref{lemma:interval_non_empty} and $\alpha(d)\ge0$,
$\intvLe{k}-\alpha(d)\ltDash{\neg\dnLc\vee\intvRty{k}}\intvRe{k}$.
The marked left endpoints coincide, so
\[
\annotatedInterval{\intv{\upLc}{\intvLe{k}-\alpha(c)}{\intvRe{k}}{\intvRty{k}}}
  {T_1:\,\intvpredupT{c}{\upLc}{\eventually{I}\varphi}}
\quad\supseteq\quad
\annotatedInterval{\intv{\upLc}{\intvLe{k}-\alpha(c)}{\intvLe{k}-\alpha(d)}{\dnLc}}
  {T_2:\,\intvpredT{c}{d}{\upLc}{\dnLc}{\eventually{I}\varphi}}.
\]
\item[(2)] The clock identity gives the same marked left endpoint.
By \Cref{lemma:interval_non_empty} for the next interval,
$\intvRe{k}\ltDash{\neg\intvRty{k}\vee\intvRty{k+1}}\intvRe{k+1}$,
and hence
\[
\annotatedInterval{\intv{\upLc}{\intvLe{k}-\alpha(c)}{\intvRe{k+1}}{\intvRty{k+1}}}
  {T_1:\,\nextop\intvpredupT{c}{\upLc}{\eventually{I}\varphi}}
\quad\supseteq\quad
\annotatedInterval{\intv{\upLc}{\intvLe{k}-\alpha(c)}{\intvRe{k}}{\intvRty{k}}}
  {T_2:\,\intvpredupT{c}{\upLc}{\eventually{I}\varphi}}.
\]
\item[(3)] The clock identity gives equal endpoints, with unchanged markers:
\[
\annotatedInterval{\intv{\upLc}{\intvLe{k}-\alpha(c)}{\intvLe{k}-\alpha(d)}{\dnLc}}
  {T_1:\,\nextop\intvpredT{c}{d}{\upLc}{\dnLc}{\eventually{I}\varphi}}
\quad=\quad
\annotatedInterval{\intv{\upLc}{\intvLe{k}-\alpha(c)}{\intvLe{k}-\alpha(d)}{\dnLc}}
  {T_2:\,\intvpredT{c}{d}{\upLc}{\dnLc}{\eventually{I}\varphi}}.
\]
\item[(4)] The fresh clock $d$ is distinct from $c$, so resetting $d$ to $0$ leaves
the advanced value of $c$ unchanged. Moreover, $\lty{\dnLc}$ gives
$\dnLc=\intvLty{k+1}$, hence $\neg\dnLc=\intvRty{k}$. Thus
\[
\annotatedInterval{\intv{\upLc}{\intvLe{k}-\alpha(c)}{\intvLe{k+1}}{\neg\dnLc}}
  {T_1:\,\nextop(\cset{\creset{d}{0}}.
    (\intvpredT{c}{d}{\upLc}{\neg\dnLc}{\eventually{I}\varphi},\lty{\dnLc}))}
\quad=\quad
\annotatedInterval{\intv{\upLc}{\intvLe{k}-\alpha(c)}{\intvRe{k}}{\intvRty{k}}}
  {T_2:\,\intvpredupT{c}{\upLc}{\eventually{I}\varphi}},
\]
using the identity above and $\intvLe{k+1}=\intvRe{k}$.
\end{proofrules}

For $\always{I}\varphi$, add $\inf I$ and $\sup I$ to the left and right
endpoints in the displays, respectively (right-open at $\infty$).
The same endpoint comparisons and markers give $T_2\subseteq T_1$ for
these checking intervals. Since $\varphi$ holds throughout
$T_1\cap\domain{0}{\tau}$, it also holds throughout
$T_2\cap\domain{0}{\tau}$, proving all four implications.
\end{proof}

\restatetheorem{lem:intv-merge}

\begin{proof}
At $k=|\partition{J}|$, both sides are false, since there is no current
interval. Otherwise write
$\intvOf{J}{k}=\intv{\upLc}{u}{v}{\dnLc}$.
Since $\intvOf{J}{k}$ is nonempty, \Cref{lemma:interval_non_empty}
gives $u\ltDash{\upLc\wedge\dnLc}v$, hence $u\ltDash{\dnLc}v$.
With $\alpha(d_1)\ge0$ and the clock inequalities, we obtain
\begin{equation*}
u-\alpha(c_1)\ltDash{m_1\vee\neg m_2}
\tikz[baseline=(join.base)]{%
  \node[inner xsep=2pt,inner ysep=0pt,outer sep=0pt]
    (join) {$\displaystyle u-\alpha(c_2)\ltDash{m'_1\vee m_2}u-\alpha(d_1)$};
  \draw[black!50,line width=0.3pt]
    ([yshift=0.75pt]join.north west) rectangle
    ([yshift=-2.25pt]join.south east);
}
\ltDash{\dnLc}v.
\tag{M}\label{eq:intvmerge-comparisons}
\end{equation*}
\begin{proofrules}[topsep=3pt,itemsep=3pt]
\item \emph{Case $\eventually{I}\varphi$.}
We first show $T_1\cup T_2=T$ for the following intervals:
\[
\annotatedInterval{\intv{m_1}{u-\alpha(c_1)}{u-\alpha(d_1)}{m'_1}}
  {T_1:\,\intvpredT{c_1}{d_1}{m_1}{m'_1}{\eventually{I}\varphi}},\qquad
\annotatedInterval{\intv{m_2}{u-\alpha(c_2)}{v}{\dnLc}}
  {T_2:\,\intvpredupT{c_2}{m_2}{\eventually{I}\varphi}},\qquad
\annotatedInterval{\intv{m_1}{u-\alpha(c_1)}{v}{\dnLc}}
  {T:\,\intvpredupT{c_1}{m_1}{\eventually{I}\varphi}}.
\]
For $T_1\cup T_2\subseteq T$, the outer comparisons in \eqref{eq:intvmerge-comparisons}
bound $T_1$ on the right and $T_2$ on the left. The other marked endpoints
already match those of $T$.

For $T\subseteq T_1\cup T_2$, suppose $t\in T\setminus(T_1\cup T_2)$. Then
$u-\alpha(d_1)\ltDash{\neg m'_1}t\ltDash{\neg m_2}u-\alpha(c_2)$.
By \Cref{lemma:interval_non_empty}, this implies
$u-\alpha(d_1)\ltDash{\neg m'_1\wedge\neg m_2}u-\alpha(c_2)$,
contradicting the boxed comparison in \eqref{eq:intvmerge-comparisons}.
\begin{proofbranches}[topsep=1pt,itemsep=1pt]
\item[($\Rightarrow$)]
If both predicates hold, each $t\in T_1\cup T_2$ has a witness $t'\ge t$ with
$t'\in(t+I)\cap\domain{0}{\tau}$ and $\sigma,t'\models_\tau\varphi$.
Since $T_1\cup T_2=T$, the merged predicate holds.
\item[($\Leftarrow$)]
If the merged predicate holds, it provides a witness $t'$ for each $t\in T_1\cup T_2$.
\end{proofbranches}

\item \emph{Case $\always{I}\varphi$.}
Denote the checking intervals as follows (right-open at $\infty$):
\[
\begin{gathered}
\annotatedInterval{\intv{m_1}{u-\alpha(c_1)+\linf{I}}{u-\alpha(d_1)+\rsup{I}}{m'_1}}
  {T_1:\,\intvpredT{c_1}{d_1}{m_1}{m'_1}{\always{I}\varphi}},\qquad
\annotatedInterval{\intv{m_2}{u-\alpha(c_2)+\linf{I}}{v+\rsup{I}}{\dnLc}}
  {T_2:\,\intvpredupT{c_2}{m_2}{\always{I}\varphi}},\\
\annotatedInterval{\intv{m_1}{u-\alpha(c_1)+\linf{I}}{v+\rsup{I}}{\dnLc}}
  {T:\,\intvpredupT{c_1}{m_1}{\always{I}\varphi}}.
\end{gathered}
\]
\begin{proofbranches}[topsep=1pt,itemsep=1pt]
\item[($\Rightarrow$)]
We first claim $T\subseteq T_1\cup T_2$. We distinguish the following cases.
\begin{proofsubcases}[topsep=1pt,itemsep=1pt]
\item \emph{$\rsup{I}<\infty$.}
By \eqref{eq:intvmerge-comparisons} and non-singularity of $I$,
the checking intervals are nonempty and
$u-\alpha(c_2)+\linf{I}<u-\alpha(d_1)+\rsup{I}$.
By \Cref{lemma:interval}~(3), $T_1\cap T_2\ne\emptyset$;
the outer comparisons then give $T\subseteq T_1\cup T_2$.
\item \emph{$\rsup{I}=\infty$.}
$T_1=T$, since both are right-open at $\infty$ with the same marked left
endpoint.
\end{proofsubcases}
Now, assume both predicates hold. Then, $\sigma,t\models_\tau\varphi$ for every $t\in T\cap\domain{0}{\tau}$.
\item[($\Leftarrow$)]
We show $T_1\cup T_2\subseteq T$.
\begin{proofsubcases}[topsep=1pt,itemsep=1pt]
\item \emph{$\rsup{I}<\infty$.}
By \eqref{eq:intvmerge-comparisons},
$u-\alpha(d_1)+\rsup{I}\ltDash{\neg m'_1\vee\dnLc}v+\rsup{I}$ and
$u-\alpha(c_1)+\linf{I}\ltDash{m_1\vee\neg m_2}u-\alpha(c_2)+\linf{I}$.
Thus $T_1\subseteq T$ and $T_2\subseteq T$ by \Cref{lemma:interval}~(1).
\item \emph{$\rsup{I}=\infty$.}
$T_1=T$ as above, and $T_2\subseteq T$ by \eqref{eq:intvmerge-comparisons}.
\end{proofsubcases}
If the merged predicate holds,
then $\sigma,t\models_\tau\varphi$ for every
$t\in (T_1\cup T_2)\cap\domain{0}{\tau}$.\qedhere
\end{proofbranches}
\end{proofrules}
\end{proof}

\subsection{Proofs of Correctness of Expansion Rules}

\restatetheorem{thm:expansion-rule-correctness}

\begin{proof}
The theorem follows directly from
\Cref{thm:expansion-rule-soundness,thm:expansion-rule-completeness}.
\end{proof}

\begin{restatable}{theorem}{rulesoundness}
\label{thm:expansion-rule-soundness}
Given a signal $\sigma$ with time horizon $\tau$,
a partition $\partition{J}$ of $\domain{0}{\tau}$,
a clock valuation $\alpha$, and a position $i$,
let $p \rightarrow \Psi_1 \mid \cdots \mid \Psi_n$
be an expansion rule.
If
$\alpha$ satisfies conditions \invmark{1} and \invmark{3} of the invariant
$\predInv_{\intvOf{J}{i}}(p)$,
then for every branch $\Psi_j$ ($1 \le j \le n$),
$\sigma, \partition{J}, \alpha, i \models_\tau \Psi_j$ implies
$\sigma, \partition{J}, \alpha, i \models_\tau p$.
\end{restatable}

\begin{proof}
For each branch $\Psi$ of the rule for $p$,
we show $\sigma, \partition{J}, \alpha, i \models_\tau p$, assuming
conditions \invmark{1} and \invmark{3} of the invariant of $p$
in \Cref{tab:predicate-invariants} hold
and $\sigma, \partition{J}, \alpha, i \models_\tau \Psi$.
Write
$\intvOf{J}{k} = \intv{\intvLty{k}}{\intvLe{k}}{\intvRe{k}}{\intvRty{k}}$.
\begin{proofrules}[itemsep=2pt]
\item $\ipstl{\varphi \wedge \varphi'} \rightarrow \ipstl{\varphi},\ipstl{\varphi'}$

Suppose $\sigma, \partition{J}, \alpha, i \models_\tau \ipstl{\varphi},\ipstl{\varphi'}$.
By definition, $\textstyle\sigma, t \models_\tau \varphi$ and $\sigma, t \models_\tau \varphi'$ for all $t \in \intvOf{J}{i}$.
This implies $\sigma, t \models_\tau \varphi \wedge \varphi'$ for all $t \in \intvOf{J}{i}$.
By definition,
$\sigma, \partition{J}, \alpha, i \models_\tau \ipstl{\varphi \wedge \varphi'}$.

\item $\ipstl{\varphi \vee \varphi'} \rightarrow \ipstl{\varphi} \; | \; \ipstl{\varphi'}$

Suppose $\sigma, \partition{J}, \alpha, i \models_\tau \ipstl{\varphi}$.
By definition, $\sigma, t \models_\tau \varphi$ for all $t \in \intvOf{J}{i}$.
Thus, $\sigma, t \models_\tau \varphi \vee \varphi'$ for all $t \in \intvOf{J}{i}$.
Then, $\sigma, \partition{J}, \alpha, i \models_\tau \ipstl{\varphi \vee \varphi'}$.
Similarly, we obtain $\sigma, \partition{J}, \alpha, i \models_\tau \ipstl{\varphi \vee \varphi'}$ for $\varphi'$.

\item $\ipstl{\varphi \until{} \varphi'} \rightarrow
\ipstl{\varphi}, \ipstl{\varphi'} \mid
\ipstl{\varphi},\nextop{\ipstl{\varphi \until{} \varphi'}}$

By definition,
$\sigma,\partition{J},\alpha,i\models_\tau
\ipstl{\varphi\until{}\varphi'}$
iff for all $t\in\intvOf{J}{i}$,
there exists $t'\in\domain{t}{\tau}$ such that
$\sigma,t'\models_\tau\varphi'$
and
$\sigma,t''\models_\tau\varphi$
for all $t''\in[t,t']$.

\begin{proofsubcases}
\setlength{\abovedisplayskip}{5pt}
\setlength{\belowdisplayskip}{1pt}

\item Suppose
$\sigma,\partition{J},\alpha,i
\models_\tau
\ipstl{\varphi},\ipstl{\varphi'}$.
Then
$\sigma,t\models_\tau\varphi$
and
$\sigma,t\models_\tau\varphi'$
for all $t\in\intvOf{J}{i}$.
Fix $t\in\intvOf{J}{i}$.
Taking $t'=t$, we have
$t'\in\domain{t}{\tau}$,
$\sigma,t'\models_\tau\varphi'$,
and
$\sigma,t''\models_\tau\varphi$
for all $t''\in[t,t']$.
Thus,
$\sigma,t\models_\tau\varphi\until{}\varphi'$.
Since $t$ was arbitrary,
$\sigma,\partition{J},\alpha,i
\models_\tau
\ipstl{\varphi\until{}\varphi'}$.

\item Suppose
$\sigma,\partition{J},\alpha,i
\models_\tau
\ipstl{\varphi},
\nextop\ipstl{\varphi\until{}\varphi'}$.
By definition,
$\sigma,t\models_\tau\varphi$
for all $t\in\intvOf{J}{i}$.
Also,
$\sigma,\partition{J},\alpha+|\intvOf{J}{i}|,i+1
\models_\tau
\ipstl{\varphi\until{}\varphi'}$.
Hence, for all $u\in\intvOf{J}{i+1}$,
there exists $u'\in\domain{u}{\tau}$ such that
$\sigma,u'\models_\tau\varphi'$
and
$\sigma,u''\models_\tau\varphi$
for all $u''\in[u,u']$.
Moreover, by the semantics of $\until{}$,
$\sigma,u\models_\tau\varphi$
for all $u\in\intvOf{J}{i+1}$.
Fix $t\in\intvOf{J}{i}$ and choose
$u\in\intvOf{J}{i+1}$.
Let $u'$ be the witness for $u$ obtained above.
Then $u'\in\domain{t}{\tau}$ and
$\sigma,u'\models_\tau\varphi'$.
For any $u''\in[t,u']$,
if $u''\in\intvOf{J}{i}$, then
$\sigma,u''\models_\tau\varphi$.
Otherwise,
$u''\in\intvOf{J}{i+1}$ or
$u''\in[u,u']$.
In either case,
$\sigma,u''\models_\tau\varphi$.
Thus,
$\sigma,u''\models_\tau\varphi$
for all $u''\in[t,u']$.
Therefore,
$\sigma,t\models_\tau\varphi\until{}\varphi'$.
Since $t$ was arbitrary,
$\sigma,\partition{J},\alpha,i
\models_\tau
\ipstl{\varphi\until{}\varphi'}$.
\end{proofsubcases}

\item $\ipstl{\varphi \release{} \varphi'} \rightarrow \;
\ipstl{\varphi} \mid
\ipstl{\varphi'},\nextop{\ipstl{\varphi \release{} \varphi'}} \mid
\ipstl{\varphi'},\nextop\ast$

By definition, $\varphi \release{} \varphi' = \neg(\neg\varphi \until{} \neg \varphi')$.
Thus,
$\sigma,\partition{J},\alpha,i\models_\tau
\ipstl{\varphi\release{}\varphi'}$
iff for all $t\in\intvOf{J}{i}$ and all $t'\in\domain{t}{\tau}$,
either
$\sigma,t'\models_\tau\varphi'$
or there exists $t''\in[t,t']$ such that
$\sigma,t''\models_\tau\varphi$.

\begin{proofsubcases}
\setlength{\abovedisplayskip}{5pt}
\setlength{\belowdisplayskip}{1pt}

\item Suppose
$\sigma,\partition{J},\alpha,i
\models_\tau \ipstl{\varphi}$.
Then $\sigma,t\models_\tau\varphi$
for all $t\in\intvOf{J}{i}$.
Let $t\in\intvOf{J}{i}$ and
$t'\in\domain{t}{\tau}$.
Since $t\in[t,t']$,
there exists $t''\in[t,t']$ such that
$\sigma,t''\models_\tau\varphi$.
Therefore,
$\sigma,t\models_\tau\varphi\release{}\varphi'$.
Since $t$ was arbitrary,
$\sigma,\partition{J},\alpha,i
\models_\tau
\ipstl{\varphi\release{}\varphi'}$.

\item Suppose
$\sigma,\partition{J},\alpha,i
\models_\tau
\ipstl{\varphi'},\nextop\ipstl{\varphi\release{}\varphi'}$.
By definition,
$\sigma,t\models_\tau\varphi'$
for all $t\in\intvOf{J}{i}$.
Also,
$\sigma,\partition{J},\alpha+|\intvOf{J}{i}|,i+1
\models_\tau
\ipstl{\varphi\release{}\varphi'}$.
Hence, for all $u\in\intvOf{J}{i+1}$
and all $u'\in\domain{u}{\tau}$, either
$\sigma,u'\models_\tau\varphi'$
or there exists $u''\in[u,u']$ such that
$\sigma,u''\models_\tau\varphi$.
Let $t\in\intvOf{J}{i}$ and let
$t'\in\domain{t}{\tau}$.
If $t'\in\intvOf{J}{i}$, then
$\sigma,t'\models_\tau\varphi'$.
Otherwise,
choose
$u\in\intvOf{J}{i+1}$ such that $u\le t'$.
By the assumption on the next interval, either
$\sigma,t'\models_\tau\varphi'$
or there exists
$u''\in[u,t']$
such that
$\sigma,u''\models_\tau\varphi$.
In the latter case, since
$t\le u\le u''\le t'$,
we have $u''\in[t,t']$.
Thus, either
$\sigma,t'\models_\tau\varphi'$
or there exists $u''\in[t,t']$ such that
$\sigma,u''\models_\tau\varphi$.
Therefore,
$\sigma,t\models_\tau\varphi\release{}\varphi'$.
Since $t$ was arbitrary,
$\sigma,\partition{J},\alpha,i
\models_\tau
\ipstl{\varphi\release{}\varphi'}$.

\item Suppose
$\sigma,\partition{J},\alpha,i
\models_\tau \ipstl{\varphi'},\nextop\ast$.
Then $\sigma,t\models_\tau\varphi'$
for all $t\in\intvOf{J}{i}$, and
$\intvOf{J}{i}$ is the last interval of $\partition{J}$.
Thus, for any $t\in\intvOf{J}{i}$,
$\domain{t}{\tau}\subseteq\intvOf{J}{i}$.
Hence, for any $t\in\intvOf{J}{i}$ and any
$t'\in\domain{t}{\tau}$,
$\sigma,t'\models_\tau\varphi'$.
Therefore,
$\sigma,\partition{J},\alpha,i
\models_\tau
\ipstl{\varphi\release{}\varphi'}$.
\end{proofsubcases}

\end{proofrules}

The remaining rules use the intervals represented by the predicates' scopes.
\begin{proofrules}[itemsep=2pt]

\item $\ipstl{\bothtemp_I \varphi} \rightarrow \cset{\creset{c}{0}} \bindot (\intvpredupT{c}{\upLc}{\bothtemp_I \varphi}, \lty{\upLc})$, where $\bothtemp \in \{\always{}, \eventually{}\}$

Assume $\sigma, \partition{J}, \alpha, i \models_\tau \cset{\creset{c}{0}} \bindot (\intvpredupT{c}{\upLc}{\bothtemp_I \varphi}, \lty{\upLc})$.
The type judgement in the branch ensures that $\upLc$ matches the
current interval's left marker. Hence \Cref{prop:intvpred-equiv}~(1) applies.
Thus, $\sigma, \partition{J}, \alpha, i \models_\tau \ipstl{\bothtemp_I \varphi}$.

\end{proofrules}

For each clocked rule below, $T$ denotes the shifted checking interval
for $\always{I}\varphi$ or the interval specified by the scope for
$\eventually{I}\varphi$.
By definition, it suffices to show $\sigma,t\models_\tau\varphi$
or $\sigma,t\models_\tau\eventually{I}\varphi$, respectively,
for every $t\in T$ with $t<\tau$.

\begin{proofrules}[itemsep=2pt]
\item $\alwaysUpPred \; \rightarrow \; \Psi_1 \mid \Psi_2 \mid \Psi_3 \mid \Psi_4$, where
{\small
\setlength{\abovedisplayskip}{3pt}
\setlength{\belowdisplayskip}{3pt}
\setlength{\abovedisplayshortskip}{3pt}
\setlength{\belowdisplayshortskip}{3pt}
\[
\begin{aligned}
\Psi_1 &= \RuleAlwaysUpBranchOne\\
\Psi_2 &= \RuleAlwaysUpBranchTwo\\
\Psi_3 &= \RuleAlwaysUpBranchThree\\
\Psi_4 &= \RuleAlwaysUpBranchFour
\end{aligned}
\]
}

Here
$\textstyle
T = \intv{\upLc}{\intvLe{i} - \alpha(c) + \inf I}{\intvRe{i} + \sup I}{\intvRty{i}}$.

\begin{proofbranches}
\setlength{\abovedisplayskip}{5pt}
\setlength{\belowdisplayskip}{1pt}

\item[$\Psi_1$.]
Assume $\sigma, \partition{J}, \alpha, i \models_\tau \nextop(c \le \inf I, \ast)$.
Since
$\sigma, \partition{J}, \alpha, i \models_\tau \nextop \ast$,
$\intvOf{J}{i}$ is the last interval of $\partition{J}$, and thus
$\intvOf{J}{i} = \intvro{\intvLty{i}}{\intvLe{i}}{\tau}$.
Moreover,
$\sigma, \partition{J}, \alpha, i \models_\tau \nextop(c \le \inf I)$
implies
$\alpha(c) + |\intvOf{J}{i}| \le \inf I$.
Let $K = \intvro{\upLc}{\intvLe{i} - \alpha(c)}{\tau}$.
Then $|K| = \tau - (\intvLe{i} - \alpha(c)) = \alpha(c) + |\intvOf{J}{i}| \le \inf I$.
Hence, by \Cref{lem:always-interval-vacuous},
$\sigma, t \models_\tau \always{I}\varphi$
for all $t \in K$.
By the semantics of $\always{I}$, $\sigma, s \models_\tau \varphi$ for all
$s \in \intvro{\upLc}{\intvLe{i} - \alpha(c) + \inf I}{\tau}$,
equivalently, for all
$s \in \intvro{\upLc}{\intvLe{i} - \alpha(c) + \inf I}{\tau + \sup I} \cap \domain{0}{\tau}$.
Therefore,
$\sigma, \partition{J}, \alpha, i \models_\tau \intvpredupT{c}{\upLc}{\always{I}\varphi}$.

\item[$\Psi_2$.]
Assume
$\sigma, \partition{J}, \alpha, i \models_\tau \ipstl{\varphi}, \nextop \ast$.
Since
$\sigma, \partition{J}, \alpha, i \models_\tau \nextop \ast$,
$\intvOf{J}{i}$ is the last interval of $\partition{J}$, and hence
$\intvOf{J}{i}=\intvro{\intvLty{i}}{\intvLe{i}}{\tau}$.
Moreover,
$\sigma, \partition{J}, \alpha, i \models_\tau \ipstl{\varphi}$
implies
$\sigma,t\models_\tau\varphi$ for all
$t\in\intvOf{J}{i}$.
Let
$K=\intvro{\upLc}{\intvLe{i}-\alpha(c)+\inf I}{\tau}$.
By \invmark{1},
$\alpha(c)\ltDash{\neg\upLc\vee\intvLty{i}}\inf I$,
and thus
$\intvLe{i}
\ltDash{\neg\upLc\vee\intvLty{i}}
\intvLe{i}-\alpha(c)+\inf I$.
By \Cref{lemma:interval}~(1), $K\subseteq\intvOf{J}{i}$.
Hence
$\sigma,t\models_\tau\varphi$ for all $t\in K$.
Equivalently,
$\sigma,t\models_\tau\varphi$ for all
$t\in
\intvro{\upLc}{\intvLe{i}-\alpha(c)+\inf I}{\tau+\sup I}
\cap\domain{0}{\tau}$.
Therefore, by definition,
$\sigma,\partition{J},\alpha,i
\models_\tau
\intvpredupT{c}{\upLc}{\always{I}\varphi}$.

\item[$\Psi_3$.]
By assumption, $\sigma, \partition{J}, \alpha, i \models_\tau \allowbreak \nextop(\cset{\creset{d}{0}} . (\intvpredT{c}{d}{\upLc}{\neg \dnLc}{\always{I} \varphi}, \lty{\dnLc}))$.
\newline
By \Cref{lem:intvpred-trans}~(4), $\sigma, \partition{J}, \alpha, i \models_\tau \intvpredupT{c}{\upLc}{\always{I} \varphi}$.

\item[$\Psi_4$.]
By assumption,
$\sigma,\partition{J},\alpha,i\models_\tau \allowbreak \ipstl{\varphi},
\nextop(\cset{\creset{c}{\inf I},\creset{d}{0}}.{}$\linebreak[0] $(\intvpredT{c}{d}{\dnLc}{\neg\dnLc}{\always{I}\varphi},
\lty{\dnLc}))$.\newline
Hence,
$\sigma,t\models_\tau\varphi$ for all $t\in\intvOf{J}{i}$, and
$\sigma,t\models_\tau\varphi$ for all
\[
\textstyle
t\in
\intv{\intvLty{i+1}}{\intvLe{i+1}}{\intvLe{i+1}+\sup I}{\neg\intvLty{i+1}}
\cap\domain{0}{\tau}.
\]
Therefore,
$\sigma,t\models_\tau \varphi$ for all
\[
\textstyle
t\in\intv{\intvLty{i}}{\intvLe{i}}{\intvRe{i}+\sup I}{\intvRty{i}}
\cap\domain{0}{\tau},
\]
since $\neg\intvLty{i+1}=\intvRty{i}$ and $\intvLe{i+1}=\intvRe{i}$.
By \invmark{1},
$\alpha(c)\ltDash{\neg\upLc\vee\intvLty{i}}\inf I$,
and hence
$\intvLe{i}
\ltDash{\neg\upLc\vee\intvLty{i}}
\intvLe{i}-\alpha(c)+\inf I$.
Thus, by \Cref{lemma:interval}~(1),
$T\subseteq\intv{\intvLty{i}}{\intvLe{i}}{\intvRe{i}+\sup I}{\intvRty{i}}$.
Since
$\sigma,t\models_\tau\varphi$
for all
$t\in
\intv{\intvLty{i}}{\intvLe{i}}
{\intvRe{i}+\sup I}{\intvRty{i}}$
and $T$ is included in this interval,
$\sigma,t\models_\tau\varphi$ for all $t\in T$ such that $t < \tau$.
Thus,
$\sigma,\partition{J},\alpha,i
\models_\tau
\intvpredupT{c}{\upLc}{\always{I}\varphi}$.

\end{proofbranches}

\item $\alwaysUpDownPred \; \rightarrow \; \Psi_1 \mid \Psi_2 \mid \Psi_3 \mid \Psi_4 \mid \Psi_5$, where
{\small
\setlength{\abovedisplayskip}{3pt}
\setlength{\belowdisplayskip}{3pt}
\setlength{\abovedisplayshortskip}{3pt}
\setlength{\belowdisplayshortskip}{3pt}
\[
\begin{aligned}
\Psi_1 &= \RuleAlwaysUpDownBranchOne\\
\Psi_2 &= \RuleAlwaysUpDownBranchTwo\\
\Psi_3 &= \RuleAlwaysUpDownBranchThree\\
\Psi_4 &= \RuleAlwaysUpDownBranchFour\\
\Psi_5 &= \RuleAlwaysUpDownBranchFive
\end{aligned}
\]
}

Here
$\textstyle
T = \intv{\upLc}{\intvLe{i} - \alpha(c) + \inf I}{\intvLe{i} - \alpha(d) + \sup I}{\dnLc}$.

\begin{proofbranches}
\setlength{\abovedisplayskip}{5pt}
\setlength{\belowdisplayskip}{1pt}

\item[$\Psi_1$.]
Assume $\sigma, \partition{J}, \alpha, i \models_\tau \allowbreak \nextop(\intvpredT{c}{d}{\upLc}{\dnLc}{\always{I} \varphi})$.
By \Cref{lem:intvpred-trans}~(3), $\sigma, \partition{J}, \alpha, i \models_\tau \intvpredT{c}{d}{\upLc}{\dnLc}{\always{I} \varphi}$.

\item[$\Psi_2,\Psi_3$.]
Assume $\sigma, \partition{J}, \alpha, i \models_\tau \nextop(c \le \inf I, \ast)$.
By the soundness of the $\alwaysUpPred$ rule under \invmark{1},
we have $\sigma, \partition{J}, \alpha, i \models_\tau \intvpredupT{c}{\upLc}{\always{I} \varphi}$.
By \Cref{lem:intvpred-trans}~(1),
$\intvpredupT{c}{\upLc}{\always{I} \varphi} \models \intvpredT{c}{d}{\upLc}{\dnLc}{\always{I} \varphi}$.
\newline
Thus,~\mbox{$\sigma, \partition{J}, \alpha, i \models_\tau \intvpredT{c}{d}{\upLc}{\dnLc}{\always{I} \varphi}$}.
The branch $\Psi_3$ is similar.

\item[$\Psi_4$.]
Assume $\sigma, \partition{J}, \alpha, i \models_\tau \allowbreak \ipstl{\varphi},\nextop(\cset{\creset{c}{\inf I}} . (\intvpredT{c}{d}{\newLc}{\dnLc}{\always{I} \varphi}, \lty{\newLc}))$.
By definition,
$\sigma, t \models_\tau \varphi$ for all $t \in \intvOf{J}{i}$, and
$\sigma, t \models_\tau \varphi$ for all \[
\textstyle
t \in \intv{\intvLty{i + 1}}{\intvLe{i + 1}}{\intvLe{i + 1} - (\alpha(d) + |\intvOf{J}{i}|) + \sup I}{\dnLc} \cap \domain{0}{\tau}.
\]
Since $\intvLe{i + 1} = \intvLe{i} + |\intvOf{J}{i}|$ and $\intvLty{i + 1}= \neg \intvRty{i}$,
$\sigma, t \models_\tau \varphi$ for all
\[
\textstyle
t \in \intv{\intvLty{i}}{\intvLe{i}}{\intvLe{i} - \alpha(d) + \sup I}{\dnLc} \cap \domain{0}{\tau}.
\]
By \invmark{1}, $\alpha(c) \ltDash{\neg \upLc \vee \intvLty{i}} \inf I$.
Then, $\intvLe{i} \ltDash{\neg \upLc \vee \intvLty{i}} \intvLe{i} - \alpha(c) + \inf I$.

By \Cref{lemma:interval}~(1), $T \subseteq \intv{\intvLty{i}}{\intvLe{i}}{\intvLe{i} - \alpha(d) + \sup I}{\dnLc}$.
Hence, $\sigma, t \models_\tau \varphi$ for all $t \in T$ such that $t < \tau$.
Thus, $\sigma, \partition{J}, \alpha, i \models_\tau \intvpredT{c}{d}{\upLc}{\dnLc}{\always{I} \varphi}$.

\item[$\Psi_5$.]
Assume
$\sigma,\partition{J},\alpha,i
\models_\tau
\ipstl{\varphi},
\nextop (\lty{\newLc}, d\gtDash{\neg\dnLc\vee\neg\newLc}\sup I)$.
Then
$\sigma,t\models_\tau\varphi$
for all $t\in\intvOf{J}{i}$, and
$\alpha(d)+|\intvOf{J}{i}|
\gtDash{\neg\dnLc\vee\neg\intvLty{i+1}}
\sup I$.
Since
$\intvRe{i}=\intvLe{i}+|\intvOf{J}{i}|$ and $\neg \intvLty{i+1} = \intvRty{i}$, we have
$\intvLe{i}-\alpha(d)+\sup I
\allowbreak\ltDash{\neg\dnLc\vee\intvRty{i}}\intvRe{i}$.
By \invmark{1},
$\alpha(c)\ltDash{\neg\upLc\vee\intvLty{i}}\inf I$.
Hence
$\intvLe{i}
\ltDash{\neg\upLc\vee\intvLty{i}}
\intvLe{i}-\alpha(c)+\inf I$.
Therefore, by \Cref{lemma:interval}~(1),
\[
\textstyle
T=
\intv{\upLc}{\intvLe{i}-\alpha(c)+\inf I}
{\intvLe{i}-\alpha(d)+\sup I}{\dnLc}
\subseteq
\intvOf{J}{i}.
\]
Since
$\sigma,t\models_\tau\varphi$
for all $t\in\intvOf{J}{i}$,
we conclude that
$\sigma,t\models_\tau\varphi$
for all $t\in T$.
\newline
Thus,~\mbox{$\sigma,\partition{J},\alpha,i
\models_\tau
\intvpredT{c}{d}{\upLc}{\dnLc}{\always{I}\varphi}$}.
\end{proofbranches}

\end{proofrules}

For either eventually rule, whenever a branch contains $\ipstl{\varphi}$,
its satisfaction gives $\sigma,t\models_\tau\varphi$ for all
$t\in\intvOf{J}{i}$. By \Cref{lem:interval-eventually},
$\sigma,t\models_\tau\eventually{I}\varphi$ for all
$t\in\intvOf{J}{i}-I$.

\begin{proofrules}[itemsep=2pt]
\item $\finallyUpPred \; \rightarrow \; \Psi_1 \mid \Psi_2 \mid \Psi_3$, where
{\small
\setlength{\abovedisplayskip}{3pt}
\setlength{\belowdisplayskip}{3pt}
\setlength{\abovedisplayshortskip}{3pt}
\setlength{\belowdisplayshortskip}{3pt}
\[
\begin{aligned}
\Psi_1 &= \RuleFinallyUpBranchOne\\
\Psi_2 &= \RuleFinallyUpBranchTwo\\
\Psi_3 &= \RuleFinallyUpBranchThree
\end{aligned}
\]
}

Here
$\textstyle
T = \intv{\upLc}{\intvLe{i} - \alpha(c)}{\intvRe{i}}{\intvRty{i}}$.

\begin{proofbranches}
\setlength{\abovedisplayskip}{5pt}
\setlength{\belowdisplayskip}{1pt}

\item[$\Psi_1$.]
By assumption, $\sigma, \partition{J}, \alpha, i \models_\tau \allowbreak \nextop(\cset{\creset{d}{0}} . (\intvpredT{c}{d}{\upLc}{\neg \dnLc}{\eventually{I} \varphi}, \lty{\dnLc}))$.
\newline
By \Cref{lem:intvpred-trans}~(4), $\sigma, \partition{J}, \alpha, i \models_\tau \intvpredupT{c}{\upLc}{\eventually{I} \varphi}$.

\item[$\Psi_2$.]
The common cover argument gives
$\sigma,t\models_\tau\eventually{I}\varphi$ for all $t\in\intvOf{J}{i}-I$.
Also, by assumption,
$\sigma,\partition{J},\alpha,i\models_\tau
\allowbreak \nextop(\cset{\creset{c}{\inf I},\creset{d}{0}}.{}$\linebreak[0] $(\intvpredT{c}{d}{\newLc}{\neg\newLc}{\eventually{I}\varphi},
\lty{\newLc}))$.
By definition,
$\sigma,t\models_\tau\eventually{I}\varphi$
for all $t<\tau$ such that
$t\in
\intv{\neg\intvRty{i}}{\intvRe{i}-\inf I}
{\intvRe{i}}{\intvRty{i}}$.
Since
$\intvOf{J}{i}-I
=
\intv{\intvLty{i}}{\intvLe{i}-\sup I}
{\intvRe{i}-\inf I}{\intvRty{i}}$,
together with the previous observation,
$\sigma,t\models_\tau\eventually{I}\varphi$
for all $t<\tau$ such that
$t\in
\intv{\intvLty{i}}{\intvLe{i}-\sup I}
{\intvRe{i}}{\intvRty{i}}$.
By \invmark{1},
$\alpha(c)\ltDash{\neg\upLc\vee\intvLty{i}}\sup I$,
and hence
$\intvLe{i}-\alpha(c)
\gtDash{\neg\upLc\vee\intvLty{i}}
\intvLe{i}-\sup I$.
By \Cref{lemma:interval}~(1),
$T\subseteq
\intv{\intvLty{i}}{\intvLe{i}-\sup I}
{\intvRe{i}}{\intvRty{i}}$.
Hence
$(\forall t\in T)\,
t<\tau,
\sigma,t\models_\tau\eventually{I}\varphi$.
Therefore,
$\sigma,\partition{J},\alpha,i
\models_\tau
\intvpredupT{c}{\upLc}{\eventually{I}\varphi}$.

\item[$\Psi_3$.]
Assume
$\sigma,\partition{J},\alpha,i
\models_\tau
\ipstl{\varphi},\inf I = 0$.
The common cover argument gives
$\sigma,t\models_\tau\eventually{I}\varphi$ for all $t\in\intvOf{J}{i}-I$.
By \invmark{1},
$\intvLe{i}-\sup I
\ltDash{\neg\upLc\vee\intvLty{i}}
\intvLe{i}-\alpha(c)$.
Since
$\intvOf{J}{i}-I
=
\intv{\intvLty{i}}{\intvLe{i}-\sup I}
{\intvRe{i}}{\intvRty{i}}$,
by \Cref{lemma:interval}~(1),
$T\subseteq\intvOf{J}{i}-I$.
Thus,
$(\forall t\in T)\,
t<\tau,
\sigma,t\models_\tau\eventually{I}\varphi$.
\newline
Therefore,~\mbox{$\sigma,\partition{J},\alpha,i
\models_\tau
\intvpredupT{c}{\upLc}{\eventually{I}\varphi}$}.
\end{proofbranches}

\item $\finallyUpDownPred \; \rightarrow \; \Psi_1 \mid \Psi_2 \mid \Psi_3$, where
{\small
\setlength{\abovedisplayskip}{3pt}
\setlength{\belowdisplayskip}{3pt}
\setlength{\abovedisplayshortskip}{3pt}
\setlength{\belowdisplayshortskip}{3pt}
\[
\begin{aligned}
\Psi_1 &= \RuleFinallyUpDownBranchOne\\
\Psi_2 &= \RuleFinallyUpDownBranchTwo\\
\Psi_3 &= \RuleFinallyUpDownBranchThree
\end{aligned}
\]
}

Here
$\textstyle
T = \intv{\upLc}{\intvLe{i} - \alpha(c)}{\intvLe{i} - \alpha(d)}{\dnLc}$.

\begin{proofbranches}
\setlength{\abovedisplayskip}{5pt}
\setlength{\belowdisplayskip}{1pt}

\item[$\Psi_1$.]
By assumption,
$\sigma, \partition{J}, \alpha, i \models_\tau \allowbreak \nextop(\intvpredT{c}{d}{\upLc}{\dnLc}{\eventually{I} \varphi})$.
By \Cref{lem:intvpred-trans}~(3), $\sigma, \partition{J}, \alpha, i \models_\tau \intvpredT{c}{d}{\upLc}{\dnLc}{\eventually{I} \varphi}$.

\item[$\Psi_2$.]
The common cover argument gives
$\sigma,t\models_\tau\eventually{I}\varphi$ for all $t\in\intvOf{J}{i}-I$.
Also, by assumption,
$\sigma,\partition{J},\alpha,i\models_\tau
\allowbreak \nextop(\cset{\creset{c}{\inf I}}.{}$\linebreak[0] $(\intvpredT{c}{d}{\newLc}{\dnLc}{\eventually{I}\varphi},
\lty{\newLc}))$.
By definition,
$\sigma,t\models_\tau\eventually{I}\varphi$
for all $t<\tau$ such that
\[
\textstyle
t\in
\intv{\intvLty{i + 1}}{\intvLe{i + 1}-\inf I}
{\intvLe{i + 1} - (\alpha(d) + |\intvOf{J}{i}|)}{\dnLc}.
\]
Since $\intvLe{i + 1} = \intvRe{i}$, $\intvLty{i + 1} = \neg \intvRty{i}$, and $\intvLe{i + 1} = \intvLe{i} + |\intvOf{J}{i}|$, we have
$\sigma,t\models_\tau\eventually{I}\varphi$
for all $t<\tau$ such that
$t\in
\intv{\neg \intvRty{i}}{\intvRe{i}-\inf I}
{\intvLe{i} - \alpha(d)}{\dnLc}$.
Since
$\intvOf{J}{i}-I
=
\intv{\intvLty{i}}{\intvLe{i}-\sup I}
{\intvRe{i}-\inf I}{\intvRty{i}}$,
together with the previous observation,
$\sigma,t\models_\tau\eventually{I}\varphi$
for all $t<\tau$ such that
\[
\textstyle
t\in
\intv{\intvLty{i}}{\intvLe{i}-\sup I}
{\intvLe{i} - \alpha(d)}{\dnLc}.
\]
By \invmark{1},
$\alpha(c)\ltDash{\neg\upLc\vee\intvLty{i}}\sup I$,
and hence
$\intvLe{i}-\alpha(c)
\gtDash{\neg\upLc\vee\intvLty{i}}
\intvLe{i}-\sup I$.
By \Cref{lemma:interval}~(1),
$T\subseteq
\intv{\intvLty{i}}{\intvLe{i}-\sup I}
{\intvLe{i} - \alpha(d)}{\dnLc}$.
Hence
$(\forall t\in T)\,
t<\tau,
\sigma,t\models_\tau\eventually{I}\varphi$.
\newline
Therefore,~\mbox{$\sigma,\partition{J},\alpha,i
\models_\tau
\intvpredT{c}{d}{\upLc}{\dnLc}{\eventually{I}\varphi}$}.

\item[$\Psi_3$.]
By assumption,
$\sigma, \partition{J},\alpha, i \models_\tau \nextop(d \gtDash{\neg \dnLc \vee \neg \newLc} \inf I, \lty{\newLc})$.

By definition, $\alpha(d) + |\intvOf{J}{i}| \gtDash{\neg \dnLc \vee \neg \intvLty{i + 1}} \inf I$.
By \invmark{1}, $\alpha(c) \ltDash{\neg \upLc \vee \intvLty{i}} \sup I$.
Since $\intvRe{i} = \intvLe{i} + |\intvOf{J}{i}|$ and $\neg \intvLty{i + 1} = \intvRty{i}$, we have
$\intvRe{i} - \inf I \gtDash{\neg \dnLc \vee \intvRty{i}} \intvLe{i} - \alpha(d)$ and
$\intvLe{i} - \sup I \ltDash{\neg \upLc \vee \intvLty{i}} \intvLe{i} - \alpha(c)$.
Since $\intvOf{J}{i} - I = \intv{\intvLty{i}}{\intvLe{i} - \sup I}{\intvRe{i} - \inf I}{\intvRty{i}}$,
by \Cref{lemma:interval}~(1), $T \subseteq \intvOf{J}{i} - I$.
Since the branch contains $\ipstl{\varphi}$, the common cover argument gives
$\sigma,t\models_\tau\eventually{I}\varphi$ for all $t\in\intvOf{J}{i}-I$.
Thus, $(\forall t \in T) \, t < \tau, \sigma, t \models_\tau \eventually{I}\varphi$.
Therefore,
$\sigma,\partition{J},\alpha,i
\models_\tau
\intvpredT{c}{d}{\upLc}{\dnLc}{\eventually{I}\varphi}$.
\qedhere
\end{proofbranches}
\end{proofrules}
\end{proof}

\begin{restatable}{theorem}{rulecompleteness}
\label{thm:expansion-rule-completeness}
Given a signal $\sigma$ with time horizon $\tau$,
a fully stable partition $\partition{J}$ of $\domain{0}{\tau}$,
a clock valuation $\alpha$, and a position $i$,
let
$p \rightarrow \Psi_1 \mid \cdots \mid \Psi_n$
be an expansion rule.
If
$\alpha$ satisfies the invariant
$\predInv_{\intvOf{J}{i}}(p)$
and
$\sigma, \partition{J}, \alpha, i \models_\tau p$,
then there exists a branch $\Psi_j$ ($1 \le j \le n$)
such that
$\sigma, \partition{J}, \alpha, i \models_\tau \Psi_j$.
\end{restatable}

\begin{proof}
For each rule, we prove that there exists a branch $\Psi$ such that
$\sigma, \partition{J}, \alpha, i \models_\tau \Psi$,
assuming the invariant of $p$ holds and
$\sigma, \partition{J}, \alpha, i \models_\tau p$.
Here conditions \invmark{1}, \invmark{2}, and \invmark{3} are as specified
in \Cref{tab:predicate-invariants}.

\begin{proofrules}[itemsep=2pt]
\item $\ipstl{\varphi \wedge \varphi'} \rightarrow \ipstl{\varphi},\ipstl{\varphi'}$

Suppose $\sigma, \partition{J}, \alpha, i \models_\tau \ipstl{\varphi \wedge \varphi'}$.
Then, $\sigma, t \models_\tau \varphi$ and $\sigma, t \models_\tau \varphi'$ for all $t \in \intvOf{J}{i}$.
Thus, by definition,
$\sigma, \partition{J}, \alpha, i \models_\tau \ipstl{\varphi},\ipstl{\varphi'}$.

\item $\ipstl{\varphi \vee \varphi'} \rightarrow
\ipstl{\varphi} \mid \ipstl{\varphi'}$

Suppose $\sigma,\partition{J},\alpha,i \models_\tau \ipstl{\varphi \vee \varphi'}$.
Then $\sigma,t\models_\tau \varphi\vee\varphi'$ for all $t\in\intvOf{J}{i}$.
Choose any $u\in\intvOf{J}{i}$ and a disjunct
$\psi\in\{\varphi,\varphi'\}$ such that $\sigma,u\models_\tau\psi$.
By stability, $\sigma,t\models_\tau\psi$ for all $t\in\intvOf{J}{i}$.
Hence $\sigma,\partition{J},\alpha,i\models_\tau\ipstl{\psi}$,
so the corresponding branch is satisfied.

\item $\ipstl{\varphi \until{} \varphi'} \rightarrow
\ipstl{\varphi},\ipstl{\varphi'}
\mid
\ipstl{\varphi},\nextop{\ipstl{\varphi \until{} \varphi'}}$

Suppose
$\sigma,\partition{J},\alpha,i
\models_\tau
\ipstl{\varphi\until{}\varphi'}$.
Then, for all $t\in\intvOf{J}{i}$,
there exists $t'\in\domain{t}{\tau}$ such that
$\sigma,t'\models_\tau\varphi'$
and
$\sigma,t''\models_\tau\varphi$
for all $t''\in[t,t']$.
Choose any $u\in\intvOf{J}{i}$.
Let $u'$ be a witness for $u$.
Let $k\ge i$ be such that
$u'\in\intvOf{J}{k}$.
Since
$\sigma,u''\models_\tau\varphi$
for all $u''\in[u,u']$ and $\sigma,u'\models_\tau\varphi'$,
by the stability of $\partition{J}$,
we obtain
$\sigma,\partition{J},\alpha,j
\models_\tau
\ipstl{\varphi}$
for every $i\le j\le k$,
and
$\sigma,\partition{J},\alpha,k
\models_\tau
\ipstl{\varphi'}$.

\begin{proofsubcases}
\item If $k=i$, then
$\sigma,\partition{J},\alpha,i
\models_\tau
\ipstl{\varphi},\ipstl{\varphi'}$.

\item Otherwise, let
$v\in\intvOf{J}{i+1}$ and choose any
$v'\in\intvOf{J}{k}$ such that
$v\le v'$.
Then
$v'\in\domain{v}{\tau}$ and
$\sigma,v'\models_\tau\varphi'$.
Also, since
$\sigma,\partition{J},\alpha,j
\models_\tau
\ipstl{\varphi}$
for every $i\le j\le k$,
we have
$\sigma,s\models_\tau\varphi$
for all $s\in[v,v']$.
Hence
$\sigma,v\models_\tau\varphi\until{}\varphi'$.
Since $v$ was arbitrary,
$\sigma,\partition{J},\alpha+|\intvOf{J}{i}|,i+1
\models_\tau
\ipstl{\varphi\until{}\varphi'}$.
Therefore,
$\sigma,\partition{J},\alpha,i
\models_\tau
\ipstl{\varphi},
\nextop\ipstl{\varphi\until{}\varphi'}$.
\end{proofsubcases}

\item $\ipstl{\varphi \release{} \varphi'} \rightarrow \; 
\ipstl{\varphi} \mid
\ipstl{\varphi'},\nextop{\ipstl{\varphi \release{} \varphi'}} \mid 
\ipstl{\varphi'},\nextop\ast$

Suppose
$\sigma,\partition{J},\alpha,i
\models_\tau
\ipstl{\varphi\release{}\varphi'}$.
By definition, for all $t\in\intvOf{J}{i}$ and all
$t'\in\domain{t}{\tau}$,
either
$\sigma,t'\models_\tau\varphi'$
or there exists
$t''\in[t,t']$
such that
$\sigma,t''\models_\tau\varphi$.

We distinguish whether $\intvOf{J}{i}$ is the last interval.

\begin{proofsubcases}

\item Suppose $\intvOf{J}{i}$ is not the last interval.

\begin{proofsubcases}

\item If there exists
$v\in\intvOf{J}{i}$
such that
$\sigma,v\models_\tau\varphi$,
then, by the stability of $\partition{J}$ for $\varphi$,
$\sigma,t\models_\tau\varphi$
for all $t\in\intvOf{J}{i}$.
Thus, for every $t\in\intvOf{J}{i}$ and $t' \in \domain{t}{\tau}$,
$t$ itself is a witness for $\varphi$.
Hence
$\sigma,\partition{J},\alpha,i
\models_\tau
\ipstl{\varphi}$.

\item Otherwise, suppose $\varphi$ does not hold in
$\intvOf{J}{i}$.
We verify the current and next judgements separately.
\begin{proofsubcases}
\item For any $t\in\intvOf{J}{i}$, taking $t'=t$ in the release
definition gives $\sigma,t\models_\tau\varphi'$ or
$\sigma,t\models_\tau\varphi$. The latter is excluded by assumption.
Hence $\sigma,\partition{J},\alpha,i\models_\tau\ipstl{\varphi'}$.

\item Let $v\in\intvOf{J}{i+1}$ and $v'\in\domain{v}{\tau}$,
and choose any $u\in\intvOf{J}{i}$.
Since $u\le v\le v'$, we have $v'\in\domain{u}{\tau}$.
By the release definition, either $\sigma,v'\models_\tau\varphi'$
or some $w\in[u,v']$ satisfies $\sigma,w\models_\tau\varphi$.

In the latter case, $w\notin\intvOf{J}{i}$ by assumption.
If $w\in[v,v']$, it already witnesses the release condition.
Otherwise, $w<v$, so $w\in\intvOf{J}{i+1}$; stability then gives
$\sigma,s\models_\tau\varphi$ for all $s\in\intvOf{J}{i+1}$,
and $v$ itself is a witness.
Thus, for every $v$ and $v'$ as above, either $\varphi'$ holds at $v'$
or $\varphi$ has a witness in $[v,v']$.
Hence $\sigma,v\models_\tau\varphi\release{}\varphi'$ for every
$v\in\intvOf{J}{i+1}$, proving the next judgement.
\end{proofsubcases}
Therefore,
$\sigma,\partition{J},\alpha,i
\models_\tau
\ipstl{\varphi'},
\nextop{\ipstl{\varphi\release{}\varphi'}}$.
\end{proofsubcases}

\item Suppose $\intvOf{J}{i}$ is the last interval.
If there exists
$u\in\intvOf{J}{i}$
such that
$\sigma,u\models_\tau\varphi$,
then, by the stability of $\partition{J}$ for $\varphi$,
$\sigma,\partition{J},\alpha,i
\models_\tau
\ipstl{\varphi}$.
Note that this case is already captured by the first branch.
Otherwise, $\varphi$ does not hold in
$\intvOf{J}{i}$.
For any $t\in\intvOf{J}{i}$,
taking $t'=t$ in the release definition gives
$\sigma,t\models_\tau\varphi'$.
Hence
$\sigma,\partition{J},\alpha,i
\models_\tau
\ipstl{\varphi'}$.
Moreover, since $\intvOf{J}{i}$ is the last interval,
$\sigma,\partition{J},\alpha,i
\models_\tau
\nextop\ast$.
Therefore,
$\sigma,\partition{J},\alpha,i
\models_\tau
\ipstl{\varphi'},\nextop\ast$.

\end{proofsubcases}

\item $\ipstl{\bothtemp_I \varphi} \rightarrow \cset{\creset{c}{0}} . (\intvpredupT{c}{\upLc}{\bothtemp_I \varphi}, \lty{\upLc})$, where $\bothtemp \in \{ \always{}, \eventually{} \}$

Assume $\sigma, \partition{J}, \alpha, i \models_\tau \ipstl{\bothtemp_I \varphi}$.
Choose $\upLc=\intvLty{i}$, the left marker of the current interval.
By \Cref{prop:intvpred-equiv}~(1), the corresponding opening branch is satisfied.
Thus, $\sigma, \partition{J}, \alpha, i \models_\tau \cset{\creset{c}{0}} \bindot (\intvpredupT{c}{\upLc}{\bothtemp_I \varphi}, \lty{\upLc})$.

\end{proofrules}

For the clocked rules below, whenever a successor interval exists,
$\intvRe{i}=\intvLe{i+1}$,
$\intvRty{i}=\neg\intvLty{i+1}$, and
$\intvLe{i}=\intvLe{i+1}-|\intvOf{J}{i}|$.
In each branch, choose its next type marker to match the next position.
For either always rule, let $K$ be its shifted checking interval.
By assumption, $\sigma,t\models_\tau\varphi$ for all
$t\in K\cap\domain{0}{\tau}$.
Thus, whenever $K\cap\intvOf{J}{i}\ne\varnothing$, stability gives
$\sigma,\partition{J},\alpha,i\models_\tau\ipstl{\varphi}$.

\begin{proofrules}[itemsep=2pt]

\item $\alwaysUpPred \; \rightarrow \; \Psi_1 \mid \Psi_2 \mid \Psi_3 \mid \Psi_4$, where
{\small
\setlength{\abovedisplayskip}{3pt}
\setlength{\belowdisplayskip}{3pt}
\setlength{\abovedisplayshortskip}{3pt}
\setlength{\belowdisplayshortskip}{3pt}
\[
\begin{aligned}
\Psi_1 &= \RuleAlwaysUpBranchOne\\
\Psi_2 &= \RuleAlwaysUpBranchTwo\\
\Psi_3 &= \RuleAlwaysUpBranchThree\\
\Psi_4 &= \RuleAlwaysUpBranchFour
\end{aligned}
\]
}

By assumption,
$\sigma,\partition{J},\alpha,i
\models_\tau
\intvpredupT{c}{\upLc}{\always{I}\varphi}$.
Here
$K=
\intv{\upLc}{\intvLe{i}-\alpha(c)+\inf I}
{\intvRe{i}+\sup I}{\intvRty{i}}$
is its shifted checking interval.
By \invmark{1},
$\alpha(c)\ltDash{\neg\upLc\vee\intvLty{i}}\inf I$.
We now consider two cases depending on whether
$\intvOf{J}{i}$ is the last interval.

\begin{proofsubcases}
\setlength{\abovedisplayskip}{5pt}
\setlength{\belowdisplayskip}{1pt}

\item Suppose $\intvOf{J}{i}$ is the last interval.
Then
$\intvOf{J}{i}=\intvro{\intvLty{i}}{\intvLe{i}}{\tau}$ and
\[
\textstyle
K\cap\domain{0}{\tau}
=
\intvro{\upLc}{\intvLe{i}-\alpha(c)+\inf I}{\tau}.
\]
Trivially if $K\cap\domain{0}{\tau}=\emptyset$, and otherwise by
\Cref{lemma:interval}~(1),
$K\cap\domain{0}{\tau}\subseteq\intvOf{J}{i}$.

\begin{proofbranches}
\item[$\Psi_1$.] If $K\cap\domain{0}{\tau}=\emptyset$, then by
\Cref{lemma:interval_non_empty},
$\tau\le \intvLe{i}-\alpha(c)+\inf I$.
Since $\tau=\intvLe{i}+|\intvOf{J}{i}|$,
$\alpha(c)+|\intvOf{J}{i}|\le\inf I$.
Together with the fact that $\intvOf{J}{i}$ is the last interval,
$\sigma,\partition{J},\alpha,i
\models_\tau
\nextop(c\leq\inf I,\ast)$.
Thus,
$\sigma,\partition{J},\alpha,i\models_\tau\Psi_1$.

\item[$\Psi_2$.] If $K\cap\domain{0}{\tau}\neq\emptyset$, then by
\Cref{lemma:interval_non_empty},
$\tau> \intvLe{i}-\alpha(c)+\inf I$.
Thus
$\alpha(c)+|\intvOf{J}{i}|>\inf I$,
and hence
$\sigma,\partition{J},\alpha,i
\models_\tau
\nextop(c>\inf I,\ast)$.
Since $K\cap\domain{0}{\tau}\subseteq\intvOf{J}{i}$ is nonempty,
stability gives
$\sigma,\partition{J},\alpha,i\models_\tau\ipstl{\varphi}$.
Thus, we have
$\sigma,\partition{J},\alpha,i\models_\tau\Psi_2$.
\end{proofbranches}

\item Suppose $\intvOf{J}{i}$ is not the last interval.
\newline
We distinguish whether $K$ intersects the \mbox{current interval}.

\begin{proofbranches}
\item[$\Psi_3$.] If $K\cap\intvOf{J}{i}=\emptyset$, then by
\Cref{lemma:interval}~(2),
$\intvRe{i}
\ltDash{\neg\intvRty{i}\vee\neg\upLc}
\intvLe{i}-\alpha(c)+\inf I$.
Hence
$\alpha(c)+|\intvOf{J}{i}|
\ltDash{\intvLty{i+1}\vee\neg\upLc}
\inf I$,
which establishes the clock constraint.
Also, using
$\intvLe{i}=\intvLe{i+1}-|\intvOf{J}{i}|$,
we rewrite
$K$ as
\[
\textstyle
\intv{\upLc}
{\intvLe{i+1}-(\alpha(c)+|\intvOf{J}{i}|)+\inf I}
{\intvLe{i+1}+\sup I}
{\neg\intvLty{i+1}}.
\]
This is precisely the shifted checking interval of
$\intvpredT{c}{d}{\upLc}{\neg\intvLty{i+1}}{\always{I}\varphi}$
with respect to $\intvOf{J}{i+1}$ and
$(\alpha+|\intvOf{J}{i}|)[\cset{\creset{d}{0}}]$.
Since
$\sigma,t\models_\tau\varphi$
for all $t\in K\cap\domain{0}{\tau}$,
the successor predicate is satisfied. Hence
$\sigma,\partition{J},\alpha,i\models_\tau\Psi_3$.

\item[$\Psi_4$.] If $K\cap\intvOf{J}{i}\neq\emptyset$, then by
\Cref{lemma:interval}~(2),
$\intvRe{i}
\gtDash{\intvRty{i}\wedge\upLc}
\intvLe{i}-\alpha(c)+\inf I$.
Hence
$\alpha(c)+|\intvOf{J}{i}|
\gtDash{\neg\intvLty{i+1}\wedge\upLc}
\inf I$,
which establishes the clock constraint.
The common stability argument gives
$\sigma,\partition{J},\alpha,i\models_\tau\ipstl{\varphi}$.
The remaining part is
\[
\textstyle
K\setminus\intvOf{J}{i}
=
\intv{\intvLty{i+1}}{\intvLe{i+1}}
{\intvLe{i+1}+\sup I}{\neg\intvLty{i+1}}.
\]
\noindent This residual is nonempty by \Cref{lemma:interval_non_empty},
since $\sup I>0$, and is precisely the shifted checking interval of
$\intvpredT{c}{d}{\intvLty{i+1}}{\neg\intvLty{i+1}}{\always{I}\varphi}$
with respect to $\intvOf{J}{i+1}$ and
$(\alpha+|\intvOf{J}{i}|)\allowbreak[\cset{\creset{c}{\inf I},\creset{d}{0}}]$.
Since
$(K\setminus\intvOf{J}{i})\cap\domain{0}{\tau}
\subseteq K\cap\domain{0}{\tau}$,
by assumption,
$\sigma,t\models_\tau\varphi$
for all
$t\in (K\setminus\intvOf{J}{i})\cap\domain{0}{\tau}$.
Thus the successor predicate is satisfied, and
$\sigma,\partition{J},\alpha,i\models_\tau\Psi_4$.
\end{proofbranches}
\end{proofsubcases}

\item $\alwaysUpDownPred \; \rightarrow \; \Psi_1 \mid \Psi_2 \mid \Psi_3 \mid \Psi_4 \mid \Psi_5$, where
{\small
\setlength{\abovedisplayskip}{3pt}
\setlength{\belowdisplayskip}{3pt}
\setlength{\abovedisplayshortskip}{3pt}
\setlength{\belowdisplayshortskip}{3pt}
\[
\begin{aligned}
\Psi_1 &= \RuleAlwaysUpDownBranchOne\\
\Psi_2 &= \RuleAlwaysUpDownBranchTwo\\
\Psi_3 &= \RuleAlwaysUpDownBranchThree\\
\Psi_4 &= \RuleAlwaysUpDownBranchFour\\
\Psi_5 &= \RuleAlwaysUpDownBranchFive
\end{aligned}
\]
}

By assumption,
$\sigma,\partition{J},\alpha,i
\models_\tau
\intvpredT{c}{d}{\upLc}{\dnLc}{\always{I}\varphi}$.
Here
$K=
\intv{\upLc}{\intvLe{i}-\alpha(c)+\inf I}
{\intvLe{i}-\alpha(d)+\sup I}{\dnLc}$
is its shifted checking interval.
By \invmark{1},
$\alpha(c)\ltDash{\neg\upLc\vee\intvLty{i}}\inf I$.
Hence $K$ is placed to the right of $\intvOf{J}{i}$.
We consider two cases depending on whether
$\intvOf{J}{i}$ is the last interval.

\begin{proofsubcases}
\setlength{\abovedisplayskip}{5pt}
\setlength{\belowdisplayskip}{1pt}

\item Suppose $\intvOf{J}{i}$ is the last interval.
We distinguish whether $K$ intersects the \mbox{current interval}.

\begin{proofbranches}

\item[$\Psi_2$.] If $K\cap\intvOf{J}{i}=\emptyset$, then
$K\cap\domain{0}{\tau}=\emptyset$.
By \Cref{lemma:interval}~(2),
$\tau\le \intvLe{i}-\alpha(c)+\inf I$.
Since $\tau=\intvLe{i}+|\intvOf{J}{i}|$,
$\alpha(c)+|\intvOf{J}{i}|\le\inf I$.
Hence
$\sigma,\partition{J},\alpha,i
\models_\tau
\nextop(c\leq\inf I,\ast)$.
Thus,
$\sigma,\partition{J},\alpha,i\models_\tau\Psi_2$.

\item[$\Psi_3$.] If $K\cap\intvOf{J}{i}\neq\emptyset$, then
$K\cap\domain{0}{\tau}\neq\emptyset$.
By \Cref{lemma:interval}~(2),
$\tau>\intvLe{i}-\alpha(c)+\inf I$.
Thus
$\alpha(c)+|\intvOf{J}{i}|>\inf I$,
and hence
$\sigma,\partition{J},\alpha,i
\models_\tau
\nextop(c>\inf I,\ast)$.
The common stability argument gives
$\sigma,\partition{J},\alpha,i\models_\tau\ipstl{\varphi}$.
Hence,
$\sigma,\partition{J},\alpha,i\models_\tau\Psi_3$.
\end{proofbranches}

\item Suppose $\intvOf{J}{i}$ is not the last interval.
We distinguish whether
$K\cap\intvOf{J}{i}=\emptyset$.

\begin{proofsubcases}

\item If $K\cap\intvOf{J}{i}=\emptyset$, then by
\Cref{lemma:interval}~(2),
$\intvRe{i}
\ltDash{\neg\intvRty{i}\vee\neg\upLc}
\intvLe{i}-\alpha(c)+\inf I$.
Since
$\intvRe{i}=\intvLe{i}+|\intvOf{J}{i}|$
and
$\neg\intvRty{i}=\intvLty{i+1}$,
we obtain
$\alpha(c)+|\intvOf{J}{i}|
\ltDash{\intvLty{i+1}\vee\neg\upLc}
\inf I$,
which establishes the clock constraint.
Using
$\intvLe{i}=\intvLe{i+1}-|\intvOf{J}{i}|$,
we rewrite
$K$ as
\[
\textstyle
\intv{\upLc}
{\intvLe{i+1}-(\alpha(c)+|\intvOf{J}{i}|)+\inf I}
{\intvLe{i+1}-(\alpha(d)+|\intvOf{J}{i}|)+\sup I}
{\dnLc}
\]
This is precisely the shifted checking interval of
$\intvpredT{c}{d}{\upLc}{\dnLc}{\always{I}\varphi}$
with respect to $\intvOf{J}{i+1}$ and
$\alpha+|\intvOf{J}{i}|$.
Since
$\sigma,t\models_\tau\varphi$
for all $t\in K\cap\domain{0}{\tau}$,
the successor predicate is satisfied. Hence
$\sigma,\partition{J},\alpha,i\models_\tau\Psi_1$.

\item If $K\cap\intvOf{J}{i}\neq\emptyset$, then by
\Cref{lemma:interval}~(2),
$\intvRe{i}
\gtDash{\intvRty{i}\wedge\upLc}
\intvLe{i}-\alpha(c)+\inf I$.
Since
$\intvRe{i}=\intvLe{i}+|\intvOf{J}{i}|$
and
$\intvRty{i}=\neg\intvLty{i+1}$,
we obtain
$\alpha(c)+|\intvOf{J}{i}|
\gtDash{\neg\intvLty{i+1}\wedge\upLc}
\inf I$,
which establishes the clock constraint.
The common stability argument gives
$\sigma,\partition{J},\alpha,i\models_\tau\ipstl{\varphi}$.
Now consider the remaining part
$K\setminus\intvOf{J}{i}$.
Since $K$ is placed to the right of $\intvOf{J}{i}$,
$K\setminus\intvOf{J}{i}
=
\intv{\neg\intvRty{i}}{\intvRe{i}}
{\intvLe{i}-\alpha(d)+\sup I}{\dnLc}$.
Using
$\intvRe{i}=\intvLe{i+1}$ and
$\intvRty{i}=\neg\intvLty{i+1}$,
this becomes
$\intv{\intvLty{i+1}}{\intvLe{i+1}}
{\intvLe{i+1}-(\alpha(d)+|\intvOf{J}{i}|)+\sup I}
{\dnLc}$.

\begin{proofbranches}

\item[$\Psi_5$.] If $K\setminus\intvOf{J}{i}=\emptyset$ then,
by \Cref{lemma:interval_non_empty},
$\alpha(d)+|\intvOf{J}{i}|
\gtDash{\neg\dnLc\vee\neg\intvLty{i+1}}
\sup I$,
which establishes the clock constraint.
Hence
$\sigma,\partition{J},\alpha,i
\models_\tau
\Psi_5$.

\item[$\Psi_4$.] If $K\setminus\intvOf{J}{i}\neq\emptyset$ then,
by \Cref{lemma:interval_non_empty},
$\alpha(d)+|\intvOf{J}{i}|
\ltDash{\dnLc\wedge\intvLty{i+1}}
\sup I$,
which establishes the clock constraint.
Moreover,
$K\setminus\intvOf{J}{i}$
is precisely the shifted checking interval of
$\intvpredT{c}{d}{\intvLty{i+1}}{\dnLc}{\always{I}\varphi}$
with respect to $\intvOf{J}{i+1}$ and
$(\alpha+|\intvOf{J}{i}|)[\cset{\creset{c}{\inf I}}]$.
Since
$(K\setminus\intvOf{J}{i})\cap\domain{0}{\tau}
\subseteq K\cap\domain{0}{\tau}$,
by assumption,
$\sigma,t\models_\tau\varphi$
for all
$t\in (K\setminus\intvOf{J}{i})\cap\domain{0}{\tau}$.
Thus the successor predicate is satisfied, and
$\sigma,\partition{J},\alpha,i\models_\tau\Psi_4$.
\end{proofbranches}
\end{proofsubcases}
\end{proofsubcases}

\end{proofrules}

For the two eventually rules below, let $T$ be the interval specified
by the scope of $p$. By assumption,
$\sigma,t\models_\tau\eventually{I}\varphi$ for every $t\in T$
and, by \invmark{3}, $T\ne\varnothing$.
By the cover argument of \Cref{lem:interval-eventually-stable},
full stability supplies a cover of the entire interval $T$ by sets
$\intvOf{J}{m}-I$ from intervals on which $\varphi$ holds throughout.
By \invmark{2}, no past true cover intersects $T$.
In both rules, we select a consuming branch when $[\varphi]$ holds on
$\intvOf{J}{i}$ and $T\cap(\intvOf{J}{i}-I)\ne\varnothing$;
otherwise we propagate.

\begin{proofrules}[itemsep=2pt]
\item $\finallyUpPred \; \rightarrow \; \Psi_1 \mid \Psi_2 \mid \Psi_3$, where
{\small
\setlength{\abovedisplayskip}{3pt}
\setlength{\belowdisplayskip}{3pt}
\setlength{\abovedisplayshortskip}{3pt}
\setlength{\belowdisplayshortskip}{3pt}
\[
\begin{aligned}
\Psi_1 &= \RuleFinallyUpBranchOne\\
\Psi_2 &= \RuleFinallyUpBranchTwo\\
\Psi_3 &= \RuleFinallyUpBranchThree
\end{aligned}
\]
}

By assumption,
$\sigma,\partition{J},\alpha,i\models_\tau\finallyUpPred$.
Its scope specifies the interval
$T=\intv{\upLc}{\intvLe{i}-\alpha(c)}
{\intvRe{i}}{\intvRty{i}}$.

\begin{proofsubcases}
\item Suppose $[\varphi]$ holds on $\intvOf{J}{i}$ and
$T\cap(\intvOf{J}{i}-I)\ne\varnothing$.
By \invmark{1}, the current cover begins
no later than $T$.
If $\inf I=0$, then $T\subseteq\intvOf{J}{i}-I$,
and $[\varphi],\inf I=0$ satisfies $\Psi_3$.

If $\inf I>0$, the nonempty residual is
\[
\textstyle
R=T\setminus(\intvOf{J}{i}-I)
=\intv{\neg\intvRty{i}}{\intvRe{i}-\inf I}
{\intvRe{i}}{\intvRty{i}}.
\]
By \invmark{2} and the definition of $R$, its satisfaction requires
a future covering interval, so a successor exists.
The intersection with the current cover gives
$\alpha(c)+|\intvOf{J}{i}|
\gtDash{\upLc\wedge\neg\intvLty{i+1}}\inf I$.
At the next position, the scope after the resets $c:=\inf I,d:=0$ specifies the interval
\[
\textstyle
\intv{\intvLty{i+1}}{\intvLe{i+1}-\inf I}
{\intvLe{i+1}}{\neg\intvLty{i+1}}=R.
\]
Because $R\subseteq T$, the successor predicate is satisfied.
Hence $\sigma,\partition{J},\alpha,i\models_\tau\Psi_2$.

\item Otherwise, $[\varphi]$ is false on the current interval or
$T\cap(\intvOf{J}{i}-I)=\varnothing$.
By \invmark{2} and the case assumption, all covering intervals are future.
Nonemptiness of $T$ implies existence of such an interval and a successor.
Their covers start no earlier than $\intvLe{i+1}-\sup I$, with that
boundary included only if the next interval is left-closed. Consequently
\[
\begin{aligned}
&\intvLe{i+1}-\sup I
\ltDash{\intvLty{i+1}\vee\neg\upLc}\intvLe{i}-\alpha(c)\\
\Longleftrightarrow\quad
&\alpha(c)+|\intvOf{J}{i}|
\ltDash{\intvLty{i+1}\vee\neg\upLc}\sup I.
\end{aligned}
\]
This is the clock constraint of $\Psi_1$.
Under time elapse and the reset $d:=0$, its successor scope specifies
\[
\textstyle
\intv{\upLc}
{\intvLe{i+1}-(\alpha(c)+|\intvOf{J}{i}|)}
{\intvLe{i+1}}{\neg\intvLty{i+1}}=T.
\]
The successor predicate is satisfied by assumption, so $\Psi_1$ is satisfied.
\end{proofsubcases}

\item $\finallyUpDownPred \; \rightarrow \; \Psi_1 \mid \Psi_2 \mid \Psi_3$, where
{\small
\setlength{\abovedisplayskip}{3pt}
\setlength{\belowdisplayskip}{3pt}
\setlength{\abovedisplayshortskip}{3pt}
\setlength{\belowdisplayshortskip}{3pt}
\[
\begin{aligned}
\Psi_1 &= \RuleFinallyUpDownBranchOne\\
\Psi_2 &= \RuleFinallyUpDownBranchTwo\\
\Psi_3 &= \RuleFinallyUpDownBranchThree
\end{aligned}
\]
}

By assumption,
$\sigma,\partition{J},\alpha,i\models_\tau\finallyUpDownPred$.
Here
$T=\intv{\upLc}{\intvLe{i}-\alpha(c)}
{\intvLe{i}-\alpha(d)}{\dnLc}$
is the interval specified by its scope.

\begin{proofsubcases}
\setlength{\abovedisplayskip}{5pt}
\setlength{\belowdisplayskip}{1pt}

\item Suppose $[\varphi]$ holds on $\intvOf{J}{i}$ and
$T\cap(\intvOf{J}{i}-I)\ne\varnothing$.
Then $\sigma,t\models_\tau\varphi$ for all $t\in\intvOf{J}{i}$.
Hence
$\sigma,\partition{J},\alpha,i\models_\tau\ipstl{\varphi}$.
Let
$R=T\setminus(\intvOf{J}{i}-I)$.
Since
$\intvOf{J}{i}-I
=
\intv{\intvLty{i}}{\intvLe{i}-\sup I}
{\intvRe{i}-\inf I}{\intvRty{i}}$,
\invmark{1} places the start of the current cover no later
than that of $T$, respecting endpoint inclusion. Hence
$R=
\intv{\neg\intvRty{i}}{\intvRe{i}-\inf I}
{\intvLe{i}-\alpha(d)}{\dnLc}$.
We distinguish whether $R$ is empty.

\begin{proofbranches}
\item[$\Psi_3$.] If $R=\emptyset$, then by \Cref{lemma:interval_non_empty},
$\alpha(d)+|\intvOf{J}{i}|
\gtDash{\neg\dnLc\vee\neg\intvLty{i+1}}\inf I$,
which establishes the clock constraint.
If the current interval is last, the next clock constraint and type judgement
are evaluated at the terminal position as in
Footnote~\ref{footnote:intv_index}; no successor predicate is introduced.
Since $R=\emptyset$, $T$ is covered by $\intvOf{J}{i}-I$.
Hence
$\sigma,\partition{J},\alpha,i\models_\tau\Psi_3$.

\item[$\Psi_2$.] If $R\neq\emptyset$, \invmark{2} and the definition of $R$
ensure that its satisfaction requires a future covering interval,
so a successor exists.
Then by \Cref{lemma:interval_non_empty},
$\alpha(d)+|\intvOf{J}{i}|
\ltDash{\dnLc\wedge\intvLty{i+1}}\inf I$,
which establishes the clock constraint.
By the condition for selecting the consuming branch,
$(\intvOf{J}{i}-I)\cap T\neq\emptyset$.
By \Cref{lemma:interval}~(2),
$\alpha(c)+|\intvOf{J}{i}|
\gtDash{\neg\intvLty{i+1}\wedge\upLc}\inf I$,
which establishes the clock constraint.
Using
$\intvRe{i}=\intvLe{i+1}$
and
$\intvRty{i}=\neg\intvLty{i+1}$,
the region $R$ is rewritten as
\[
\textstyle
\intv{\intvLty{i+1}}{\intvLe{i+1}-\inf I}
{\intvLe{i+1}-(\alpha(d)+|\intvOf{J}{i}|)}{\dnLc}.
\]
This is precisely the time region of
$\intvpredT{c}{d}{\intvLty{i+1}}{\dnLc}{\eventually{I}\varphi}$
with respect to $\intvOf{J}{i+1}$ and
$(\alpha+|\intvOf{J}{i}|)[\cset{\creset{c}{\inf I}}]$.
Since
$\sigma,t\models_\tau\eventually{I}\varphi$
for all $t\in T$,
the same holds for all $t\in R$.
Thus the successor predicate is satisfied, and
$\sigma,\partition{J},\alpha,i\models_\tau\Psi_2$.
\end{proofbranches}

\item Otherwise, $[\varphi]$ is false on the current interval or
$T\cap(\intvOf{J}{i}-I)=\varnothing$.
By \invmark{2} and the case assumption, all covering intervals are future.
Nonemptiness of $T$ therefore
gives a successor, to which we propagate $T$.
Since all witness intervals occur after $\intvOf{J}{i}$,
the earliest possible region of the form $\intvOf{J}{k}-I$
starts no earlier than
$\intvLe{i+1}-\sup I$.
Hence
$\intvLe{i+1}-\sup I
\ltDash{\intvLty{i+1}\vee\neg\upLc}
\intvLe{i}-\alpha(c)$.
Then
$\alpha(c)+|\intvOf{J}{i}|
\ltDash{\intvLty{i+1}\vee\neg\upLc}\sup I$,
which establishes the clock constraint.
By rewriting $T$ with
$\intvLe{i}=\intvLe{i+1}-|\intvOf{J}{i}|$,
we obtain
$\intv{\upLc}{\intvLe{i+1}-(\alpha(c)+|\intvOf{J}{i}|)}
{\intvLe{i+1}-(\alpha(d)+|\intvOf{J}{i}|)}{\dnLc}$,
which is the time region of
$\intvpredT{c}{d}{\upLc}{\dnLc}{\eventually{I}\varphi}$
with respect to $\intvOf{J}{i+1}$ and
$\alpha+|\intvOf{J}{i}|$.
Since $\eventually{I}\varphi$ holds on $T$, the successor predicate is satisfied.
Therefore,
$\sigma,\partition{J},\alpha,i\models_\tau\Psi_1$.
\qedhere
\end{proofsubcases}
\end{proofrules}
\end{proof}

\subsection{Auxiliary Lemmas for Expansion Rules}

Near the time horizon,
$\always{I}\varphi$ becomes vacuously satisfied
when the future interval required by $I$ lies entirely beyond $\tau$.

\begin{lemma}\label{lem:always-interval-vacuous}
For any right-open interval $K$ with right endpoint $\tau$, and any closed interval $I \subseteq \realset_{\ge 0}$,
if $|K| \le \inf I$ then, $(\forall t \in K)\, \sigma, t \models_\tau \always{I} \varphi$.
\end{lemma}

\begin{proof}
Suppose $|K| \le \inf I$.
For any $t\in K$, we have $t+\inf I \ge \tau$, and hence
$(t+I)\cap[0,\tau)=\emptyset$.
Thus, $\always{I}\varphi$ is vacuously satisfied at every $t\in K$.
\end{proof}

The interval where $\varphi$ is satisfied
characterizes where the eventuality formula $\eventually{I} \varphi$ holds.
If $\varphi$ holds throughout an interval $K$,
then $\eventually{I} \varphi$ holds throughout $(K-I)\cap\domain{0}{\tau}$.

\begin{lemma}\label{lem:interval-eventually}
For any intervals $K \subseteq \domain{0}{\tau}$ and $I\subseteq\realset_{\ge0}$,
if
$(\forall t\in K)\,
\sigma,t\models_\tau\varphi$,
then
$(\forall t\in (K-I)\cap\domain{0}{\tau})\,
\sigma,t\models_\tau\eventually{I}\varphi$.
\end{lemma}

\begin{proof}
Let $t\in (K-I)\cap\domain{0}{\tau}$.
By \Cref{lem:interval_operation},
there exists $t'\in K\cap(t+I)$.
By assumption,
$\sigma,t'\models_\tau\varphi$.
Hence
$\sigma,t\models_\tau\eventually{I}\varphi$.
\end{proof}

By a similar intuition,
the satisfaction region of
$\eventually{I}\varphi$
can be determined from stable partition intervals satisfying $\varphi$.
That is,
$\eventually{I}\varphi$
holds over regions covered by shifted intervals
$K-I$, where $K$ is a stable partition interval satisfying $\varphi$.

\begin{lemma}\label{lem:interval-eventually-stable}
Given a signal $\sigma$ with time horizon $\tau$,
an interval $K\subseteq\domain{0}{\tau}$,
and a partition $\partition{J}$ of $\domain{0}{\tau}$
that is stable for $\varphi$,
$(\forall t\in K)\,\sigma,t\models_\tau\eventually{I}\varphi$
iff there exist intervals
$\intvOf{J}{i_1},\dots,\intvOf{J}{i_k}$
such that
$K\subseteq
\bigcup_{m\in\{i_1,\dots,i_k\}}(\intvOf{J}{m}-I)$
and
$(\forall t\in \intvOf{J}{m})\,\sigma,t\models_\tau\varphi$
for every $m\in\{i_1,\dots,i_k\}$.
\end{lemma}

\begin{proof}
($\Rightarrow$)
Assume
$(\forall t\in K)\,
\sigma,t\models_\tau\eventually{I}\varphi$.
For each $t\in K$,
there exists
$t'\in (t+I) \cap \domain{0}{\tau}$
such that
$\sigma,t'\models_\tau\varphi$.
Since $\partition{J}$ is a partition of $\domain{0}{\tau}$,
$t'\in\intvOf{J}{m}$
for some $m$.
By stability,
$(\forall s\in\intvOf{J}{m})\,
\sigma,s\models_\tau\varphi$.
Moreover, by \Cref{lem:interval_operation},
$t'\in\intvOf{J}{m}\cap(t+I)$ implies
$t\in\intvOf{J}{m}-I$.
Thus every $t\in K$ belongs to
$\intvOf{J}{m}-I$
for some interval
$\intvOf{J}{m}$
on which $\varphi$ is stably true.
Collecting all such intervals yields
$\intvOf{J}{i_1},\dots,\intvOf{J}{i_k}$
such that
$K\subseteq
\bigcup_{m\in\{i_1,\dots,i_k\}}
(\intvOf{J}{m}-I)$
and
$(\forall t\in\intvOf{J}{m})\,
\sigma,t\models_\tau\varphi$
for every
$m\in\{i_1,\dots,i_k\}$.

($\Leftarrow$)
Suppose
$K\subseteq
\bigcup_{m\in\{i_1,\dots,i_k\}}
(\intvOf{J}{m}-I)$
for intervals
$\intvOf{J}{i_1},\dots,\intvOf{J}{i_k}$
on which $\varphi$ is stably true.
Pick any $t\in K$.
Then
$t\in\intvOf{J}{m}-I$
for some
$m\in\{i_1,\dots,i_k\}$.
By \Cref{lem:interval_operation},
there exists
$t'\in\intvOf{J}{m}\cap(t+I)$.
Since
$(\forall s\in\intvOf{J}{m})\,
\sigma,s\models_\tau\varphi$,
we have
$\sigma,t'\models_\tau\varphi$.
Hence
$\sigma,t\models_\tau\eventually{I}\varphi$.
\end{proof}

\subsection{Invariants of Interval Predicates}
\label{sec:app_pred_inv}

Expansion rules decompose interval predicates into conditions on the current and subsequent intervals.
A further condition (\invmark{2}) is that the satisfaction of predicates must not rely on past intervals.
For example, $\intvpredT{c}{d}{\strue}{\strue}{\eventually{[1,3]}\varphi}$
on $J=[3,4)$ with $\alpha(c)=3$, $\alpha(d)=2$ has scope
$\scopetuple{\strue}{0}{1}{\strue}$.
If $\varphi$ holds only on the past interval $[1,3)$, the predicate
is nevertheless satisfied, since shifting $[0,1]$ by $1\in[1,3]$
yields $[1,2]\subseteq[1,3)$.
Such reliance on the past must therefore be excluded.

Finally, the interval to be checked must be nonempty (\invmark{3}): the interval specified by the scope for
$\eventually{I}\varphi$, or the shifted checking interval for
$\always{I}\varphi$. Otherwise, the predicate is vacuously satisfied.

For $\intvpredupT{c}{\upLc}{\always{I}\varphi}$, the shifted checking interval
has positive length $|\intvOf{J}{i}|+c+|I|$, so condition \invmark{3} is already satisfied.
For $\always{I}\varphi$, condition \invmark{2} is ensured by condition \invmark{1},
since checking starts no earlier than the current interval.

The following definition makes the clock constraints in
\Cref{tab:predicate-invariants} explicit, including their dependence
on the signal and endpoint markers.
Fix a signal $\sigma$, a partition $\partition{J}$, and a position $i$.
For each eventuality $\eventually{I}\varphi$, let
\[
k_*:=\max\{k<i\mid
(\forall t\in\intvOf{J}{k})\,\sigma,t\models_\tau\varphi\}
\]
be the last position satisfying $[\varphi]$, with right marker $r_*=\intvRty{k_*}$.
If no such position exists, condition \invmark{2} is taken to be $\true$.
The dependence on $\sigma$ is implicit.

\begin{definition}
The invariant of an interval predicate $p$
with respect to an interval
$\intvOf{J}{i}=\intv{\ltLc}{u}{u'}{\rtLc}$,
denoted $\predInv_{\intvOf{J}{i}}(p)$,
is the following clock constraint:
\begingroup
\setlength{\jot}{0pt}
\setlength{\abovedisplayskip}{4pt}
\setlength{\belowdisplayskip}{4pt}
\[
\begin{aligned}
\predInv_{\intvOf{J}{i}}(\intvpredupT{c}{\upLc}{\always{I}\varphi})
&=c \ltDash{\neg\upLc \vee \ltLc} \inf I,\\
\predInv_{\intvOf{J}{i}}(\intvpredT{c}{d}{\upLc}{\dnLc}{\always{I}\varphi})
&=(c \ltDash{\neg\upLc \vee \ltLc} \inf I)
\wedge(d \ltDash{\upLc \wedge \dnLc} c + |I|),\\
\predInv_{\intvOf{J}{i}}(\intvpredupT{c}{\upLc}{\eventually{I}\varphi})
&=(c \ltDash{\neg\upLc \vee \ltLc} \sup I)
\wedge(c \ltDash{\neg\upLc\vee\neg r_*} u+\inf I-\sup\intvOf{J}{k_*})
\wedge(0 \ltDash{\upLc \wedge \rtLc} c + |\intvOf{J}{i}|),\\
\predInv_{\intvOf{J}{i}}(\intvpredT{c}{d}{\upLc}{\dnLc}{\eventually{I}\varphi})
&=(c \ltDash{\neg\upLc \vee \ltLc} \sup I)
\wedge(c \ltDash{\neg\upLc\vee\neg r_*} u+\inf I-\sup\intvOf{J}{k_*})
\wedge(d \ltDash{\upLc \wedge \dnLc} c),\\
\predInv_{\intvOf{J}{i}}(\ipstl{\varphi})
&=
\true.
\end{aligned}
\]
\endgroup
\end{definition}

Condition \invmark{2} places the left endpoint of the eventuality scope
to the right of every cover $\intvOf{J}{k}-I$ contributed by a past
interval on which $\varphi$ holds throughout.
It suffices to consider $k_*$ because the right endpoints of these covers
are ordered by the partition positions.
Under a clock valuation $\alpha$, its bound is equivalent to placing
$u-\alpha(c)$ at or after $\sup\intvOf{J}{k_*}-\inf I$.
Equality is allowed unless both the left endpoint specified by the scope
and the right endpoint of the past cover are included, as recorded by
$\neg\upLc\vee\neg r_*$.

For condition \invmark{3}, an interval with endpoints $x,y$ and
markers $m,n$ is nonempty iff $x\ltDash{m\wedge n}y$.
The one-clock scope has endpoint difference
$\alpha(c)+|\intvOf{J}{i}|$, while the two-clock scope has endpoint
difference $\alpha(c)-\alpha(d)$.
Shifting the checking endpoints for $\always{I}\varphi$ adds $|I|$
to these differences, yielding the stated constraints for condition \invmark{3}.
These tests precede intersection with $\domain{0}{\tau}$;
when the endpoints coincide, both must be included.

The invariant of a set of interval predicates is the conjunction of their invariants:
\[
\predInv_{\intvOf{J}{i}}(L) = \bigwedge_{l \in L} \predInv_{\intvOf{J}{i}}(l),
\qquad
\predInv_{\intvOf{J}{i}}(\emptyset) = \true.
\]

Under the completeness assumptions, a satisfied branch can be chosen
that also preserves the invariants of its interval predicates.
The following proposition formalizes this property.

\restatetheorem{prop:pred_invariant}

\begin{proof}
Let $p$ be the predicate being expanded and $\alpha$ the clock valuation before expansion.
Choose the same satisfied branch as in the completeness proof of
\Cref{thm:expansion-rule-correctness}, using its branch numbering throughout.
It suffices to verify conditions \invmark{1}, \invmark{2}, and \invmark{3}
in \Cref{tab:predicate-invariants} for its interval predicates
after the relevant time elapse and clock resets.

Let $\intvOf{J}{i}=\intv{\intvLty{i}}{\intvLe{i}}{\intvRe{i}}{\intvRty{i}}$
and, whenever a successor predicate occurs,
$\intvOf{J}{i+1}=\intv{\intvLty{i+1}}{\intvLe{i+1}}{\intvRe{i+1}}{\intvRty{i+1}}$.
The following rules introduce no clocked interval predicates, so their
invariants are $\true$:
\begin{proofrules}[itemsep=2pt]
\item $\ipstl{\varphi\wedge\varphi'}\rightarrow\ipstl{\varphi},\ipstl{\varphi'}$.
\item $\ipstl{\varphi\vee\varphi'}\rightarrow\ipstl{\varphi}\mid\ipstl{\varphi'}$.
\item $\ipstl{\varphi\until{}\varphi'}\rightarrow
\ipstl{\varphi},\ipstl{\varphi'}\mid
\ipstl{\varphi},\nextop{\ipstl{\varphi\until{}\varphi'}}$.
\item $\ipstl{\varphi\release{}\varphi'}\rightarrow
\ipstl{\varphi}\mid
\ipstl{\varphi'},\nextop{\ipstl{\varphi\release{}\varphi'}}\mid
\ipstl{\varphi'},\nextop\ast$.
\end{proofrules}

The rules for $\ipstl{\always{I}\varphi}$ and
$\ipstl{\eventually{I}\varphi}$ introduce a clock $c$, reset to $0$.
The invariant of the generated predicate is checked below.

\begin{proofrules}[itemsep=2pt]
\item
$\ipstl{\bothtemp_I\varphi}\rightarrow
\cset{\creset{c}{0}}\bindot
(\intvpredupT{c}{\upLc}{\bothtemp_I\varphi},\lty{\upLc})$,
where $\bothtemp\in\{\always{},\eventually{}\}$.

There is no $\nextop$ in this branch, so the valuation is
$\alpha[\cset{\creset{c}{0}}]$.
The type judgement gives $\upLc=\intvLty{i}$.

\invmark{1} The condition reduces to $c\le\inf I$ for always
and $c\le\sup I$ for eventually; both follow from
$\alpha[\cset{\creset{c}{0}}](c)=0\le\inf I\le\sup I$.

\invmark{2} For eventually, if a past true interval exists,
$\intvLe{i}+\inf I-\sup\intvOf{J}{k_*}\ge\inf I\ge0$.
If this bound is zero, the right endpoint of that past interval
coincides with $\intvLe{i}$, and disjointness gives
$\neg\upLc\vee\neg\intvRty{k_*}=\true$.
Thus the bound holds at $c=0$; it is vacuous if no such interval exists.

\invmark{3} The scope specifies the nonempty interval $\intvOf{J}{i}$,
giving $0\ltDash{\upLc\wedge\intvRty{i}}c+|\intvOf{J}{i}|$
after the reset. For one-clock always, no additional constraint is needed.
\end{proofrules}

For each selected branch below, write
$\Psi=\Lambda.(L,\nextop(C',\Lambda'.L'))$ and put
$\alpha'=(\alpha[\Lambda]+|\intvOf{J}{i}|)[\Lambda']$.
It suffices to show
\[
\alpha[\Lambda]\models\predInv_{\intvOf{J}{i}}(L),
\qquad
\alpha'\models\predInv_{\intvOf{J}{i+1}}(L').
\]
Since $L$ contains no clocked interval predicates,
$\predInv_{\intvOf{J}{i}}(L)=\true$, so the first claim is immediate;
only the next invariants need checking.
Branches with no successor clocked predicate have invariant $\true$.
For always, condition \invmark{2} is trivially true.
For eventually, let $T$ be the interval specified by the scope of $p$;
by \invmark{3} for $p$, $T\ne\varnothing$.

\begin{proofrules}[itemsep=2pt]
\item $\alwaysUpPred\rightarrow\Psi_1\mid\Psi_2\mid\Psi_3\mid\Psi_4$.

\begin{proofbranches}
\item[$\Psi_1,\Psi_2$.] Immediate: no successor clocked predicate occurs.

\item[$\Psi_3$.]
\invmark{1} The type judgement and reset give
$\dnLc=\intvLty{i+1}$ and
$\alpha'=(\alpha+|\intvOf{J}{i}|)[\cset{\creset{d}{0}}]$.
The clock constraint therefore yields
$\alpha'\models c\ltDash{\neg\upLc\vee\intvLty{i+1}}\inf I$,
since resetting $d$ leaves $c$ unchanged.
\invmark{3} $\alpha'(d)=0$ and
$\alpha'(c)+|I|=\alpha(c)+|\intvOf{J}{i}|+|I|>0$.

\item[$\Psi_4$.]
The type judgement gives $\dnLc=\intvLty{i+1}$, and
$\alpha'=(\alpha+|\intvOf{J}{i}|)
[\cset{\creset{c}{\inf I},\creset{d}{0}}]$.
Hence
\[
\text{\invmark{1}}\quad\alpha'(c)=\inf I\le\inf I,
\qquad
\text{\invmark{3}}\quad\alpha'(d)=0<\alpha'(c)+|I|.
\]
\end{proofbranches}

\item $\alwaysUpDownPred\rightarrow\Psi_1\mid\Psi_2\mid\Psi_3\mid\Psi_4\mid\Psi_5$.

\begin{proofbranches}
\item[$\Psi_2,\Psi_3,\Psi_5$.] Immediate: no successor clocked predicate occurs.

\item[$\Psi_1$.]
\invmark{1} With $\newLc=\intvLty{i+1}$ and
$\alpha'=\alpha+|\intvOf{J}{i}|$, the clock constraint gives
$\alpha'\models c\ltDash{\neg\upLc\vee\intvLty{i+1}}\inf I$.
\invmark{3} The endpoint markers are unchanged, and
$\alpha'(d)-\alpha'(c)=\alpha(d)-\alpha(c)$,
so \invmark{3} for~$p$~is~preserved.

\item[$\Psi_4$.]
\invmark{1} With $\newLc=\intvLty{i+1}$,
$\alpha'=(\alpha+|\intvOf{J}{i}|)[\cset{\creset{c}{\inf I}}]$
gives $\alpha'(c)=\inf I\le\inf I$.
\invmark{3} The clock constraint and reset give
$\alpha'(d)\ltDash{\dnLc\wedge\intvLty{i+1}}\sup I
=\alpha'(c)+|I|$,
including the closed-singleton~case.
\end{proofbranches}

\item $\finallyUpPred\rightarrow\Psi_1\mid\Psi_2\mid\Psi_3$.

\begin{proofbranches}
\item[$\Psi_3$.] Immediate: no successor clocked predicate occurs.

\item[$\Psi_1$.]
\invmark{1} With $\dnLc=\intvLty{i+1}$ and
$\alpha'=(\alpha+|\intvOf{J}{i}|)[\cset{\creset{d}{0}}]$,
\newline
the clock constraint gives
\mbox{$\alpha'\models c\ltDash{\neg\upLc\vee\intvLty{i+1}}\sup I$}.

\invmark{2} By the completeness selection, either
$[\varphi]$ is false on $\intvOf{J}{i}$ or
$T\cap(\intvOf{J}{i}-I)=\varnothing$:
\begin{proofsubcases}
\item If $[\varphi]$ is false, the last past true interval and its
right marker are unchanged. Both $c$ and its bound increase by
$|\intvOf{J}{i}|$, since
\[
\alpha'(c)=\alpha(c)+|\intvOf{J}{i}|,\qquad
\intvLe{i+1}+\inf I-\sup\intvOf{J}{k_*}
=\intvLe{i}+\inf I-\sup\intvOf{J}{k_*}+|\intvOf{J}{i}|.
\]
Thus \invmark{2} is preserved, or remains vacuous.
\item Otherwise, $[\varphi]$ holds and the current cover is disjoint
from $T$. By \invmark{1} for $p$ and \Cref{lemma:interval}~(2),
\[
\alpha'(c)=\alpha(c)+|\intvOf{J}{i}|
\ltDash{\neg\upLc\vee\neg\intvRty{i}}\inf I.
\]
The last past true interval at $i+1$ is $\intvOf{J}{i}$, and
$\intvLe{i+1}+\inf I-\sup\intvOf{J}{i}=\inf I$,
which is exactly the successor bound.
\end{proofsubcases}
\invmark{3} The successor scope specifies the same interval $T\ne\varnothing$.

\item[$\Psi_2$.]
The type judgement gives $\dnLc=\intvLty{i+1}$, and
$\alpha'=(\alpha+|\intvOf{J}{i}|)
[\cset{\creset{c}{\inf I},\creset{d}{0}}]$.
Thus
\[
\text{\invmark{1}}\quad\alpha'(c)=\inf I\le\sup I,
\qquad
\text{\invmark{3}}\quad\alpha'(d)=0<\alpha'(c)=\inf I,
\]
where $\inf I>0$ is required by the branch.
\invmark{2} Since $[\varphi]$ holds on $\intvOf{J}{i}$,
the last past true interval at $i+1$ is $\intvOf{J}{i}$.
The new left marker is
$\intvLty{i+1}=\neg\intvRty{i}$, so
$\neg\intvLty{i+1}\vee\neg\intvRty{i}=\true$.
Hence $\alpha'(c)=\inf I$ satisfies the successor bound with equality.
\end{proofbranches}

\item $\finallyUpDownPred\rightarrow\Psi_1\mid\Psi_2\mid\Psi_3$.

\begin{proofbranches}
\item[$\Psi_3$.] Immediate: no successor clocked predicate occurs.

\item[$\Psi_1$.]
\invmark{1} With $\newLc=\intvLty{i+1}$ and
$\alpha'=\alpha+|\intvOf{J}{i}|$, the clock constraint gives
$\alpha'\models c\ltDash{\neg\upLc\vee\intvLty{i+1}}\sup I$.
\invmark{2} The argument for $\finallyUpPred$, branch $\Psi_1$, applies unchanged.
\invmark{3} The successor scope specifies the same interval $T\ne\varnothing$.

\item[$\Psi_2$.]
\invmark{1} With $\newLc=\intvLty{i+1}$,
$\alpha'=(\alpha+|\intvOf{J}{i}|)[\cset{\creset{c}{\inf I}}]$
gives $\alpha'(c)=\inf I\le\sup I$.
\invmark{2} The argument for $\finallyUpPred$, branch $\Psi_2$, applies unchanged.
\invmark{3} The clock constraint and reset give
$\alpha'(d)\ltDash{\dnLc\wedge\intvLty{i+1}}\inf I=\alpha'(c)$,
which is exactly condition \invmark{3} for the successor predicate.
\qedhere
\end{proofbranches}
\end{proofrules}
\end{proof}

\subsection{Transition-Form Judgements and Subsumption}
\label{sec:app-subsumption}

\restatetheorem{lem:tnf_subsumption}

\begin{proof}
Assume $\Phi_1 \sqsubseteq \Phi_2$ and
$\sigma,\partition{J},\alpha,i \models_\tau \Phi_2$.
Let
$\Phi_r=C_r,\Lambda_r.(L_r,\nextop(C_r',\Lambda_r'.L_r'))$
for $r=1,2$.
By definition,
$\sigma,\partition{J},\alpha,i \models_\tau C_2$ and
$\sigma,\partition{J},\alpha[\Lambda_2],i \models_\tau L_2$.
When a next part is absent, its components and free variables below
are omitted.
Also, by $\Phi_1 \sqsubseteq \Phi_2$, we have
$C_1 \subseteq C_2$, $\Lambda_1 \subseteq \Lambda_2$,
$L_1 \subseteq L_2$, $C_1' \subseteq C_2'$,
$\Lambda_1' \subseteq \Lambda_2'$, $L_1' \subseteq L_2'$,
$\vars(\Lambda_2)\cap
\fvar{\Lambda_1.(L_1,\nextop(C_1',\Lambda_1'.L_1'))}
=\emptyset$, and
$\vars(\Lambda_2')\cap\fvar{\Lambda_1'.L_1'}=\emptyset$.
To show
$\sigma,\partition{J},\alpha,i\models_\tau\Phi_1$,
we first verify its current part, at any position including the terminal position.
\begin{proofrules}[topsep=3pt,itemsep=3pt]
\item
($\sigma,\partition{J},\alpha,i\models_\tau C_1$.)
Since $C_1\subseteq C_2$ and
$\sigma,\partition{J},\alpha,i\models_\tau C_2$,
the claim follows.

\item
($\sigma,\partition{J},\alpha[\Lambda_1],i\models_\tau L_1$.)
Since $L_1\subseteq L_2$ and
$\sigma,\partition{J},\alpha[\Lambda_2],i\models_\tau L_2$,
we have
$\sigma,\partition{J},\alpha[\Lambda_2],i\models_\tau L_1$.
Moreover,
$\vars(\Lambda_2)\cap\fvar{\Lambda_1.L_1}=\emptyset$.
By \Cref{lem:reset-agreement},
$\alpha[\Lambda_1]$ and $\alpha[\Lambda_2]$ agree on
$\fvar{L_1}$.
Hence, by \Cref{lem:irrelevant-reset},
$\sigma,\partition{J},\alpha[\Lambda_1],i\models_\tau L_1$.

\end{proofrules}

If $\Phi_1$ has no next part, this proves the claim.
Otherwise, subsumption ensures that $\Phi_2$ also has a next part.
Since $\Phi_2$ is satisfied, $i<|\partition{J}|$, so $J_i$ exists.
The semantics of its next part gives
$\sigma,\partition{J},\alpha[\Lambda_2]+|\intvOf{J}{i}|,i+1
\models_\tau C_2'$ and
$\sigma,\partition{J},
(\alpha[\Lambda_2]+|\intvOf{J}{i}|)[\Lambda_2'],i+1
\models_\tau L_2'$.
It remains to verify the next part of $\Phi_1$.
\begin{proofrules}[topsep=3pt,itemsep=3pt]

\item
($\sigma,\partition{J},
\alpha[\Lambda_1]+|\intvOf{J}{i}|,i+1
\models_\tau C_1'$.)
Since $C_1'\subseteq C_2'$ and
$\sigma,\partition{J},
\alpha[\Lambda_2]+|\intvOf{J}{i}|,i+1
\models_\tau C_2'$,
we have
$\sigma,\partition{J},
\alpha[\Lambda_2]+|\intvOf{J}{i}|,i+1
\models_\tau C_1'$.
Moreover,
$\vars(\Lambda_2)\cap\fvar{\Lambda_1.\nextop C_1'}=\emptyset$.
By \Cref{lem:reset-agreement},
$\alpha[\Lambda_1]$ and $\alpha[\Lambda_2]$ agree on
$\fvar{C_1'}$.
Thus
$\alpha[\Lambda_1]+|\intvOf{J}{i}|$
and
$\alpha[\Lambda_2]+|\intvOf{J}{i}|$
also agree on $\fvar{C_1'}$.
Hence, by \Cref{lem:irrelevant-reset},
$\sigma,\partition{J},
\alpha[\Lambda_1]+|\intvOf{J}{i}|,i+1
\models_\tau C_1'$.

\item
($\sigma,\partition{J},
(\alpha[\Lambda_1]+|\intvOf{J}{i}|)[\Lambda_1'],i+1
\models_\tau L_1'$.)
Since $L_1'\subseteq L_2'$ and
$\sigma,\partition{J},
(\alpha[\Lambda_2]+|\intvOf{J}{i}|)[\Lambda_2'],i+1
\models_\tau L_2'$,
we have
$\sigma,\partition{J},
(\alpha[\Lambda_2]+|\intvOf{J}{i}|)[\Lambda_2'],i+1
\models_\tau L_1'$.
Since
$\vars(\Lambda_2')\cap\fvar{\Lambda_1'.L_1'}=\emptyset$,
by \Cref{lem:reset-agreement},
$(\alpha[\Lambda_2]+|\intvOf{J}{i}|)[\Lambda_1']$
and
$(\alpha[\Lambda_2]+|\intvOf{J}{i}|)[\Lambda_2']$
agree on $\fvar{L_1'}$.
Hence, by \Cref{lem:irrelevant-reset},
$\sigma,\partition{J},
(\alpha[\Lambda_2]+|\intvOf{J}{i}|)[\Lambda_1'],i+1
\models_\tau L_1'$.
Moreover,
$\vars(\Lambda_2)\cap
\fvar{\Lambda_1.\nextop(\Lambda_1'.L_1')}
=\emptyset$.
By \Cref{lem:reset-agreement},
$\alpha[\Lambda_1]$ and $\alpha[\Lambda_2]$ agree on
$\fvar{\Lambda_1'.L_1'}$.
Thus
$\alpha[\Lambda_1]+|\intvOf{J}{i}|$
and
$\alpha[\Lambda_2]+|\intvOf{J}{i}|$
also agree on $\fvar{\Lambda_1'.L_1'}$.
Hence, by \Cref{lem:irrelevant-reset},
$\sigma,\partition{J},
(\alpha[\Lambda_1]+|\intvOf{J}{i}|)[\Lambda_1'],i+1
\models_\tau L_1'$.
\end{proofrules}

Therefore,
$\sigma,\partition{J},\alpha,i\models_\tau\Phi_1$.
\end{proof}

\begin{lemma}
\label{lem:type-subsumption}
Let $\Phi_i=C_i,\Lambda_i.(L_i,\nextop(C_i',\Lambda_i'.L_i'))$ $(i=1,2)$
be transition-form judgements, and let $T_1,T_1'$ and $T_2,T_2'$ be the sets of judgements of the form
$\lty{\cdot}$ occurring in $L_1,L_1'$ and $L_2,L_2'$, respectively.
If $\Phi_1\sqsubseteq\Phi_2$, then the following hold:
\begin{itemize}
  \item
$C_1,\Lambda_1.(T_1,\nextop(C_1',\Lambda_1'.L_1')) \sqsubseteq
C_2,\Lambda_2.(T_2,\nextop(C_2',\Lambda_2'.L_2'))$
  \item
$C_1,\Lambda_1.(T_1,\nextop(C_1',\Lambda_1'.T_1')) \sqsubseteq
C_2,\Lambda_2.(T_2,\nextop(C_2',\Lambda_2'.T_2'))$
\end{itemize}
\end{lemma}

\begin{proof}
Assume $\Phi_1\sqsubseteq\Phi_2$.
By definition of subsumption, we have
$C_1\subseteq C_2$,
$\Lambda_1\subseteq\Lambda_2$,
$L_1\subseteq L_2$,
$C_1'\subseteq C_2'$,
$\Lambda_1'\subseteq\Lambda_2'$,
$L_1'\subseteq L_2'$,
$\vars(\Lambda_2)\cap
\fvar{\Lambda_1.(L_1,\nextop(C_1',\Lambda_1'.L_1'))}=\emptyset$,
and
$\vars(\Lambda_2')\cap\fvar{\Lambda_1'.L_1'}=\emptyset$.
Since $T_1,T_1'$ and $T_2,T_2'$ are the sets of judgements of the form
$\lty{\cdot}$ occurring in $L_1,L_1'$ and $L_2,L_2'$, respectively,
$L_1\subseteq L_2$ and $L_1'\subseteq L_2'$ imply
$T_1\subseteq T_2$ and $T_1'\subseteq T_2'$.
Moreover, since $T_1\subseteq L_1$ and $T_1'\subseteq L_1'$,
we have
$\fvar{\Lambda_1.(T_1,\nextop(C_1',\Lambda_1'.L_1'))}
\subseteq
\fvar{\Lambda_1.(L_1,\nextop(C_1',\Lambda_1'.L_1'))}$
and
$\fvar{\Lambda_1.(T_1,\nextop(C_1',\Lambda_1'.T_1'))}
\subseteq
\fvar{\Lambda_1.(L_1,\nextop(C_1',\Lambda_1'.L_1'))}$.
Hence, we have:
\[
\vars(\Lambda_2)\cap
\fvar{\Lambda_1.(T_1,\nextop(C_1',\Lambda_1'.L_1'))}=\emptyset,\\
\vars(\Lambda_2)\cap
\fvar{\Lambda_1.(T_1,\nextop(C_1',\Lambda_1'.T_1'))}=\emptyset.
\]

Also, since $T_1'\subseteq L_1'$,
$\fvar{\Lambda_1'.T_1'}\subseteq\fvar{\Lambda_1'.L_1'}$.
Thus
$\vars(\Lambda_2')\cap\fvar{\Lambda_1'.T_1'}=\emptyset$.
Thus, by definition of $\sqsubseteq$,
$C_1,\Lambda_1.(T_1,\nextop(C_1',\Lambda_1'.L_1'))
\sqsubseteq
C_2,\Lambda_2.(T_2,\nextop(C_2',\Lambda_2'.L_2'))$
and
$C_1,\Lambda_1.(T_1,\nextop(C_1',\Lambda_1'.T_1'))
\sqsubseteq
C_2,\Lambda_2.(T_2,\nextop(C_2',\Lambda_2'.T_2'))$.
\end{proof}

\begin{lemma}\label{lem:subsumption-fvar}
Let $\Phi_i=C_i,\Lambda_i.(L_i,\nextop(C_i',\Lambda_i'.L_i'))$ $(i=1,2)$
be transition-form judgements. If $\Phi_1 \sqsubseteq \Phi_2$, then
$\fvar{L_1, \nextop(C_1',\Lambda_1'.L_1')} \subseteq \fvar{L_2, \nextop(C_2',\Lambda_2'.L_2')}$.
\end{lemma}

\begin{proof}
Assume $\Phi_1 \sqsubseteq \Phi_2$, and let
$x \in \fvar{L_1, \nextop(C_1',\Lambda_1'.L_1')}$.
If $x \in \fvar{L_1}$ or $x \in \fvar{C_1'}$, then the claim follows immediately from
$L_1 \subseteq L_2$ and $C_1' \subseteq C_2'$.
Otherwise, $x \in \fvar{L_1'}\setminus\vars(\Lambda_1')$.
If $x \notin \vars(\Lambda_2')$, then $x$ remains free in $\Lambda_2'.L_2'$ because
$L_1' \subseteq L_2'$, hence
$x \in \fvar{\nextop(C_2',\Lambda_2'.L_2')}$.
If instead $x \in \vars(\Lambda_2')$, then by \Cref{def:subsumption}~(c),
$\vars(\Lambda_2') \cap \fvar{L_1'} \subseteq \vars(\Lambda_1')$.
Thus $x \in \vars(\Lambda_1')$. This contradicts $x \notin \vars(\Lambda_1')$.
Hence this case is impossible.
Therefore,
$\fvar{L_1, \nextop(C_1',\Lambda_1'.L_1')} \subseteq
\fvar{L_2, \nextop(C_2',\Lambda_2'.L_2')}$.
\end{proof}

The following proposition shows that the subsumption relation induces a partial
order on transition forms.

\begin{proposition}\label{prop:subsumption-poset}
The relation $\sqsubseteq$ is a partial order on transition-form judgements.
\end{proposition}

\begin{proof}
Let $\Phi_i = C_i,\Lambda_i.(L_i, \nextop(C_i',\Lambda_i'.L_i'))$ for $i=1,2,3$.
We show the relation $\sqsubseteq$ is reflexive, antisymmetric, and transitive.
\begin{proofrules}[topsep=3pt,itemsep=3pt]
  \item (Reflexivity) Every component is contained in itself.
Moreover, the clocks bound by each reset set are disjoint from
the free clocks of the judgement under that reset.
Thus conditions (b) and (c) hold.
Hence $\Phi_1 \sqsubseteq \Phi_1$.
  \item (Antisymmetry) Assume $\Phi_1 \sqsubseteq \Phi_2$ and $\Phi_2 \sqsubseteq \Phi_1$.
Then, by \Cref{def:subsumption}~(a),
$C_1 = C_2$, $\Lambda_1 = \Lambda_2$, $L_1 = L_2$,
$C_1' = C_2'$, $\Lambda_1' = \Lambda_2'$, and $L_1' = L_2'$.
Thus $\Phi_1 = \Phi_2$.
  \item (Transitivity) Assume $\Phi_1 \sqsubseteq \Phi_2$ and $\Phi_2 \sqsubseteq \Phi_3$. We show the conditions (a), (b), (c) in \Cref{def:subsumption} hold
  for $\Phi_1 \sqsubseteq \Phi_3$.
  \begin{proofbranches}[topsep=2pt,itemsep=2pt]
    \item[(a)] From the assumptions $\Phi_1 \sqsubseteq \Phi_2$ and $\Phi_2 \sqsubseteq \Phi_3$, we obtain
$C_1 \subseteq C_3$, $\Lambda_1 \subseteq \Lambda_3$, $L_1 \subseteq L_3$,
$C_1' \subseteq C_3'$, $\Lambda_1' \subseteq \Lambda_3'$, and $L_1' \subseteq L_3'$.
    \item[(b)] Let $x \in \vars(\Lambda_3)\cap \fvar{L_1, \nextop(C_1',\Lambda_1'.L_1')}$.
Note that, by \Cref{lem:subsumption-fvar},
\[
\fvar{L_1, \nextop(C_1',\Lambda_1'.L_1')} \subseteq \fvar{L_2, \nextop(C_2',\Lambda_2'.L_2')}.
\]
This implies that $x \in \vars(\Lambda_3)\cap \fvar{L_2, \nextop(C_2',\Lambda_2'.L_2')}$.
Since $\Phi_2 \sqsubseteq \Phi_3$, we obtain $x \in \vars(\Lambda_2)$.
By \Cref{def:subsumption}~(b) for $\Phi_1 \sqsubseteq \Phi_2$, we conclude
$x \in \vars(\Lambda_1)$.
Thus,
$\vars(\Lambda_3)\cap \fvar{L_1, \nextop(C_1',\Lambda_1'.L_1')}
\subseteq \vars(\Lambda_1)$.
    \item[(c)]
    Let $x \in \vars(\Lambda_3')\cap \fvar{L_1'}$.
Since $L_1' \subseteq L_2'$, we have $x \in \fvar{L_2'}$.
Thus, from $\Phi_2 \sqsubseteq \Phi_3$, we obtain $x \in \vars(\Lambda_2')$.
Since $x \in \var{\Lambda_2'} \cap \fvar{L_1'}$ and $\Phi_1 \sqsubseteq \Phi_2$, we conclude
$x \in \vars(\Lambda_1')$.
Hence,
$\vars(\Lambda_3')\cap \fvar{L_1'} \subseteq \vars(\Lambda_1')$.
\qedhere
  \end{proofbranches}
\end{proofrules}
\end{proof}

\paragraph{Restoring Transition Form}
We prove the following lemma by explicitly rewriting each expansion
branch using the structural laws of \Cref{prop:judg-equiv}.

\restatetheorem{lem:tnf-existence}

\begin{proof}
{\raggedbottom
\allowdisplaybreaks[4]
Let $\Phi=C,\Lambda.(L,\nextop(C',\Lambda'.L'))$ be the given
transition-form judgement, and let
$\Psi=\Lambda_\Psi.(L_\Psi,\nextop(C_\Psi',\Lambda_\Psi'.L_\Psi'))$
be a selected branch. Write
\[
\tilde{\Phi}=C,\Lambda.(L,\Psi,\nextop(C',\Lambda'.L')).
\]
We construct a transition-form judgement $\hat{\Phi}\equiv\tilde{\Phi}$
for each branch.

Bound-clock names are chosen consistently, preserving free clocks
and inherited bindings; comparisons below a fixed judgement use its
bound-clock names.
By the assumed side conditions of \Cref{prop:judg-equiv}, the reset
sets to be merged are consistent and satisfy the no-capture conditions
\[
\begin{aligned}
\vars(\Lambda_\Psi)\cap
\fvar{\Lambda.(L,\nextop(C',\Lambda'.L'))}&=\emptyset,\\
\vars(\Lambda')\cap\fvar{\Lambda_\Psi'.L_\Psi'}&=\emptyset,\\
\vars(\Lambda_\Psi')\cap\fvar{\Lambda'.L'}&=\emptyset.
\end{aligned}
\]
An absent next part is omitted: retain the other next part if present,
and introduce none if both are absent.
In the one-clock cases, the fresh next-interval marker $\dnLc$ in the
branch list is written $\newLc$ in the calculations below.

We denote each structural rule as follows.
\begin{proofrules}[topsep=3pt,itemsep=3pt]
    \item \comm : $\Psi_1,\Psi_2 \equiv \Psi_2,\Psi_1$
    \item \xmerge : $\nextop(\Psi_1,\Psi_2) \equiv (\nextop\Psi_1),(\nextop\Psi_2)$
    \item \cmerge : $(\Lambda_1.\Psi_1),(\Lambda_2.\Psi_2) \equiv (\Lambda_1\cup\Lambda_2).(\Psi_1,\Psi_2)$
    \item \cnestedmerge : $\Lambda_1.(\Psi_1,\Lambda_2.\Psi_2) \equiv (\Lambda_1\cup\Lambda_2).(\Psi_1,\Psi_2)$
\end{proofrules}

The following rules introduce no clocks or resets. Adding a branch
therefore preserves transition form after combining any next judgements
by \xmerge; all existing components are retained.

\begin{proofrules}[itemsep=2pt]

\item $\ipstl{\varphi \wedge \varphi'} \rightarrow \ipstl{\varphi},\ipstl{\varphi'}$
\item $\ipstl{\varphi \vee \varphi'} \rightarrow \ipstl{\varphi} \mid \ipstl{\varphi'}$
\item $\ipstl{\varphi \until{} \varphi'} \rightarrow \; \ipstl{\varphi}, \ipstl{\varphi'} \mid \ipstl{\varphi},\nextop{\ipstl{\varphi \until{} \varphi'}}$
\item $\ipstl{\varphi \release{} \varphi'} \rightarrow \; \ipstl{\varphi} \mid \ipstl{\varphi'},\nextop{\ipstl{\varphi \release{} \varphi'}} \mid \ipstl{\varphi'},\nextop\ast$
\end{proofrules}

\newcommand{\tnfstrut}{\vphantom{\textstyle
\nextop(c\ltDash{\neg\upLc\vee\newLc}\inf I,
\cset{\creset{d}{0}}.(\intvpredT{c}{d}{\upLc}{\neg\newLc}{\always{I}\varphi},\lty{\newLc}))}}
\newcommand{\tnfboxsep}{2pt}
\newcommand{\tnfmark}[1]{\begingroup\setlength{\fboxsep}{\tnfboxsep}%
\setlength{\fboxrule}{0.5pt}%
\fcolorbox{black!40}{white}{$\textstyle\tnfstrut#1$}\endgroup}
\noindent Boxes mark the terms used by the next rewriting step;
the last box marks the merged result.

\begin{proofrules}[itemsep=2pt]

\item \begin{minipage}[t]{\linewidth}
$\ipstl{\bothtemp_I \varphi} \rightarrow \cset{\creset{c}{0}} \bindot (\intvpredupT{c}{\upLc}{\bothtemp_I \varphi}, \lty{\upLc})$, where $\bothtemp \in \{\always{}, \eventually{}\}$
\par\addvspace{\smallskipamount}
\centering
\begin{tikzpicture}[term/.style={inner sep=1pt,font=\small},
 equivalence/.style={inner sep=0pt},
 reason/.style={font=\scriptsize,inner sep=2pt}]
\node[term] (t0) {$\textstyle \tilde{\Phi} =C, \Lambda . (\tnfmark{L, \cset{\creset{c}{0}} . (\intvpredupT{c}{\upLc}{\bothtemp_I \varphi}, \lty{\upLc})}, \nextop (C', \Lambda' . L'))$};
\node[equivalence,below=5pt of t0] (e1) {\rotatebox{90}{$\equiv$}};
\node[reason,right=3pt of e1] {\comm};
\node[term,below=5pt of e1] (t1) {$\textstyle C, \tnfmark{\Lambda . (\cset{\creset{c}{0}} . (\intvpredupT{c}{\upLc}{\bothtemp_I \varphi}, \lty{\upLc}), L,\nextop (C',\Lambda' . L'))}$};
\node[equivalence,below=5pt of t1] (e2) {\rotatebox{90}{$\equiv$}};
\node[reason,right=3pt of e2] {\cnestedmerge};
\node[term,below=5pt of e2] (t2) {$\textstyle C, \tnfmark{\Lambda \cup \cset{\creset{c}{0}} . (\intvpredupT{c}{\upLc}{\bothtemp_I \varphi}, \lty{\upLc}, L, \nextop (C',\Lambda' . L'))} = \hat{\Phi}$};
\end{tikzpicture}
\par
\end{minipage}
\item $\alwaysUpPred \; \rightarrow \; \Psi_1 \mid \Psi_2 \mid \Psi_3 \mid \Psi_4$

\begin{proofbranches}[beginpenalty=10000,style=standard,leftmargin=0pt,labelwidth=0pt,labelsep=0pt,itemindent=0pt,itemsep=8pt]
\setlength{\abovedisplayskip}{5pt}
\setlength{\belowdisplayskip}{1pt}

\item[]\begin{minipage}[t]{0.48\linewidth}
\raggedright
{\small $\textstyle\Psi_1 = \RuleAlwaysUpBranchOne$}\par\addvspace{\smallskipamount}
\centering

\begin{tikzpicture}[baseline=(t0.base),term/.style={inner sep=1pt,font=\small},
 equivalence/.style={inner sep=0pt},
 reason/.style={font=\scriptsize,inner sep=2pt}]
\node[term] (t0) {$\textstyle \tilde{\Phi} =C, \Lambda  .  (L, \tnfmark{\nextop(c \leq \linf{I},\ast)}, \tnfmark{\nextop (C', \Lambda'  .  L')})$};
\node[equivalence,below=5pt of t0] (e1) {\rotatebox{90}{$\equiv$}};
\node[reason,right=3pt of e1] {\xmerge};
\node[term,below=5pt of e1] (t1) {$\textstyle C, \Lambda  .  (L, \nextop(c \leq \linf{I},\tnfmark{\ast},\tnfmark{C'}, \Lambda'  .  L'))$};
\node[equivalence,below=5pt of t1] (e2) {\rotatebox{90}{$\equiv$}};
\node[reason,right=3pt of e2] {\comm};
\node[term,below=5pt of e2] (t2) {$\textstyle C, \Lambda  .  (L, \nextop(c \leq \linf{I}, C', \tnfmark{\ast},\tnfmark{\Lambda'  .  L'}))$};
\node[equivalence,below=5pt of t2] (e3) {\rotatebox{90}{$\equiv$}};
\node[reason,right=3pt of e3] {\cmerge};
\node[term,below=5pt of e3] (t3) {$\textstyle C, \Lambda  .  (L, \nextop(c \leq \linf{I}, C',\tnfmark{\Lambda'  .  (\ast, L')})) = \hat{\Phi}$};
\end{tikzpicture}

\end{minipage}\hfill%
\begin{minipage}[t]{0.48\linewidth}
\raggedright
{\small $\textstyle\Psi_2 = \RuleAlwaysUpBranchTwo$}\par\addvspace{\smallskipamount}
\centering

\begin{tikzpicture}[baseline=(t0.base),term/.style={inner sep=1pt,font=\small},
 equivalence/.style={inner sep=0pt},
 reason/.style={font=\scriptsize,inner sep=2pt}]
\node[term] (t0) {$\textstyle \tilde{\Phi} =C, \Lambda  .  (L, \ipstl{\varphi}, \tnfmark{\nextop(c > \inf I, \ast)}, \tnfmark{\nextop (C', \Lambda'  .  L')})$};
\node[equivalence,below=5pt of t0] (e1) {\rotatebox{90}{$\equiv$}};
\node[reason,right=3pt of e1] {\xmerge};
\node[term,below=5pt of e1] (t1) {$\textstyle C, \Lambda  .  (L, \ipstl{\varphi}, \nextop(c > \inf I,\tnfmark{\ast},\tnfmark{C'}, \Lambda'  .  L'))$};
\node[equivalence,below=5pt of t1] (e2) {\rotatebox{90}{$\equiv$}};
\node[reason,right=3pt of e2] {\comm};
\node[term,below=5pt of e2] (t2) {$\textstyle C, \Lambda  .  (L, \ipstl{\varphi}, \nextop(c > \inf I, C', \tnfmark{\ast},\tnfmark{\Lambda'  .  L'}))$};
\node[equivalence,below=5pt of t2] (e3) {\rotatebox{90}{$\equiv$}};
\node[reason,right=3pt of e3] {\cmerge};
\node[term,below=5pt of e3] (t3) {$\textstyle C, \Lambda  .  (L, \ipstl{\varphi}, \nextop(c > \inf I, C',\tnfmark{\Lambda'  .  (\ast, L')})) = \hat{\Phi}$};
\end{tikzpicture}

\end{minipage}

\item[]\begin{minipage}[t]{\linewidth}
\raggedright
{\small $\textstyle\Psi_3 = \RuleAlwaysUpBranchThree$}\par\addvspace{\smallskipamount}
\centering
\begin{tikzpicture}[baseline=(a.base),
 term/.style={inner sep=1pt,font=\small},
 equivalence/.style={inner sep=0pt},
 reason/.style={font=\scriptsize,inner sep=2pt}
]
\node[term] (a) {$\textstyle
\tilde\Phi=C,\Lambda.\bigl(L,
\tnfmark{\nextop(c\ltDash{\neg\upLc\vee\newLc}\inf I,
\cset{\creset{d}{0}}.(\intvpredT{c}{d}{\upLc}{\neg\newLc}{\always{I}\varphi},\lty{\newLc}))},
\tnfmark{\nextop(C',\Lambda'.L')}\bigr)$};
\node[equivalence,below=5pt of a] (ab) {\rotatebox{90}{$\equiv$}};
\node[reason,right=3pt of ab] {\xmerge};
\node[term,below=5pt of ab] (b) {$\textstyle
C,\Lambda.\bigl(L,\nextop(c\ltDash{\neg\upLc\vee\newLc}\inf I,
\tnfmark{\cset{\creset{d}{0}}.(\intvpredT{c}{d}{\upLc}{\neg\newLc}{\always{I}\varphi},\lty{\newLc})},
\tnfmark{C'},\Lambda'.L')\bigr)$};
\node[equivalence,below=5pt of b] (bc) {\rotatebox{90}{$\equiv$}};
\node[reason,right=3pt of bc] {\comm};
\node[term,below=5pt of bc] (c) {$\textstyle
C,\Lambda.\bigl(L,\nextop(c\ltDash{\neg\upLc\vee\newLc}\inf I,C',
\tnfmark{\cset{\creset{d}{0}}.(\intvpredT{c}{d}{\upLc}{\neg\newLc}{\always{I}\varphi},\lty{\newLc})},
\tnfmark{\Lambda'.L'} )\bigr)$};
\node[equivalence,below=5pt of c] (cd) {\rotatebox{90}{$\equiv$}};
\node[reason,right=3pt of cd] {\cmerge};
\node[term,below=5pt of cd] (d) {$\textstyle
C,\Lambda.\bigl(L,\nextop(c\ltDash{\neg\upLc\vee\newLc}\inf I,C',
\tnfmark{(\Lambda'\cup\cset{\creset{d}{0}}).
(\intvpredT{c}{d}{\upLc}{\neg\newLc}{\always{I}\varphi},\lty{\newLc},L')} )\bigr)
=\hat\Phi$};
\end{tikzpicture}
\end{minipage}
\item[]\begin{minipage}[t]{\linewidth}
\raggedright
{\small $\textstyle\Psi_4 = \RuleAlwaysUpBranchFour$}\par\addvspace{\smallskipamount}
\centering

\begin{tikzpicture}[baseline=(t0.base),term/.style={inner sep=1pt,font=\small},
 equivalence/.style={inner sep=0pt},
 reason/.style={font=\scriptsize,inner sep=2pt}]
\node[term] (t0) {$\textstyle \tilde{\Phi} =C, \Lambda  .  (L,\ipstl{\varphi},\tnfmark{\nextop(c \gtDash{\upLc \wedge \neg \newLc} \inf I, \cset{\creset{c}{\inf I}, \creset{d}{0}}  .  (\intvpredT{c}{d}{\newLc}{\neg{\newLc}}{\always{I} \varphi}, \lty{\newLc}))}, \tnfmark{\nextop (C', \Lambda'  .  L')})$};
\node[equivalence,below=5pt of t0] (e1) {\rotatebox{90}{$\equiv$}};
\node[reason,right=3pt of e1] {\xmerge};
\node[term,below=5pt of e1] (t1) {$\textstyle C, \Lambda  .  (L,\ipstl{\varphi},\nextop(c \gtDash{\upLc \wedge \neg \newLc} \inf I,\tnfmark{\cset{\creset{c}{\inf I}, \creset{d}{0}}  .  (\intvpredT{c}{d}{\newLc}{\neg{\newLc}}{\always{I} \varphi}, \lty{\newLc})},\tnfmark{C'}, \Lambda'  .  L'))$};
\node[equivalence,below=5pt of t1] (e2) {\rotatebox{90}{$\equiv$}};
\node[reason,right=3pt of e2] {\comm};
\node[term,below=5pt of e2] (t2) {$\textstyle C, \Lambda  .  (L,\ipstl{\varphi},\nextop( c \gtDash{\upLc \wedge \neg \newLc} \inf I, C',\tnfmark{\cset{\creset{c}{\inf I}, \creset{d}{0}}  .  (\intvpredT{c}{d}{\newLc}{\neg{\newLc}}{\always{I} \varphi}, \lty{\newLc})},\tnfmark{\Lambda'  .  L'}))$};
\node[equivalence,below=5pt of t2] (e3) {\rotatebox{90}{$\equiv$}};
\node[reason,right=3pt of e3] {\cmerge};
\node[term,below=5pt of e3] (t3) {$\textstyle C, \Lambda  .  (L,\ipstl{\varphi},\nextop( c \gtDash{\upLc \wedge \neg \newLc} \inf I, C',\tnfmark{(\cset{\creset{c}{\inf I}, \creset{d}{0}}\cup\Lambda')  .  (\intvpredT{c}{d}{\newLc}{\neg{\newLc}}{\always{I} \varphi}, \lty{\newLc}, L')})) = \hat{\Phi}$};
\end{tikzpicture}

\end{minipage}

\end{proofbranches}

\item $\alwaysUpDownPred \; \rightarrow \; \Psi_1 \mid \Psi_2 \mid \Psi_3 \mid \Psi_4 \mid \Psi_5$

\begin{proofbranches}[beginpenalty=10000,style=standard,leftmargin=0pt,labelwidth=0pt,labelsep=0pt,itemindent=0pt,itemsep=8pt]
\setlength{\abovedisplayskip}{5pt}
\setlength{\belowdisplayskip}{1pt}

\item[]\begin{minipage}[t]{\linewidth}
\raggedright
{\small $\textstyle\Psi_1 = \RuleAlwaysUpDownBranchOne$}\par\addvspace{\smallskipamount}
\centering

\begin{tikzpicture}[baseline=(t0.base),term/.style={inner sep=1pt,font=\small},
 equivalence/.style={inner sep=0pt},
 reason/.style={font=\scriptsize,inner sep=2pt}]
\node[term] (t0) {$\textstyle \tilde{\Phi} =C, \Lambda  .  (L, \tnfmark{\nextop(c \ltDash{\neg \upLc \vee \newLc} \inf I, \lty{\newLc}, \intvpredT{c}{d}{\upLc}{\dnLc}{\always{I} \varphi})}, \tnfmark{\nextop (C', \Lambda'  .  L')})$};
\node[equivalence,below=5pt of t0] (e1) {\rotatebox{90}{$\equiv$}};
\node[reason,right=3pt of e1] {\xmerge};
\node[term,below=5pt of e1] (t1) {$\textstyle C, \Lambda  .  (L, \nextop(c \ltDash{\neg \upLc \vee \newLc} \inf I, \tnfmark{\lty{\newLc}}, \tnfmark{\intvpredT{c}{d}{\upLc}{\dnLc}{\always{I} \varphi}},\tnfmark{C'}, \Lambda'  .  L'))$};
\node[equivalence,below=5pt of t1] (e2) {\rotatebox{90}{$\equiv$}};
\node[reason,right=3pt of e2] {\comm};
\node[term,below=5pt of e2] (t2) {$\textstyle C, \Lambda  .  (L, \nextop(c \ltDash{\neg \upLc \vee \newLc} \inf I, C', \tnfmark{\lty{\newLc}}, \tnfmark{\intvpredT{c}{d}{\upLc}{\dnLc}{\always{I} \varphi}},\tnfmark{\Lambda'  .  L'}))$};
\node[equivalence,below=5pt of t2] (e3) {\rotatebox{90}{$\equiv$}};
\node[reason,right=3pt of e3] {\cmerge};
\node[term,below=5pt of e3] (t3) {$\textstyle C, \Lambda  .  (L, \nextop(c \ltDash{\neg \upLc \vee \newLc} \inf I, C',\tnfmark{\Lambda'  .  (\lty{\newLc}, \intvpredT{c}{d}{\upLc}{\dnLc}{\always{I} \varphi}, L')})) = \hat{\Phi}$};
\end{tikzpicture}

\end{minipage}

\item[]\begin{minipage}[t]{0.48\linewidth}
\raggedright
{\small $\textstyle\Psi_2 = \RuleAlwaysUpDownBranchTwo$}\par\addvspace{\smallskipamount}
\centering

\begin{tikzpicture}[baseline=(t0.base),term/.style={inner sep=1pt,font=\small},
 equivalence/.style={inner sep=0pt},
 reason/.style={font=\scriptsize,inner sep=2pt}]
\node[term] (t0) {$\textstyle \tilde{\Phi} =C, \Lambda  .  (L, \tnfmark{\nextop(c \le \inf I, \ast)}, \tnfmark{\nextop (C', \Lambda'  .  L')})$};
\node[equivalence,below=5pt of t0] (e1) {\rotatebox{90}{$\equiv$}};
\node[reason,right=3pt of e1] {\xmerge};
\node[term,below=5pt of e1] (t1) {$\textstyle C, \Lambda  .  (L, \tnfmark{\nextop(c \le \inf I, \ast, C', \Lambda'  .  L')})$};
\node[equivalence,below=5pt of t1] (e2) {\rotatebox{90}{$\equiv$}};
\node[reason,right=3pt of e2] {\comm, \cmerge};
\node[term,below=5pt of e2] (t2) {$\textstyle C, \Lambda  .  (L, \nextop(c \le \inf I, C',\tnfmark{\Lambda'  .  (\ast, L')})) = \hat{\Phi}$};
\end{tikzpicture}

\end{minipage}\hfill%
\begin{minipage}[t]{0.48\linewidth}
\raggedright
{\small $\textstyle\Psi_3 = \RuleAlwaysUpDownBranchThree$}\par\addvspace{\smallskipamount}
\centering

\begin{tikzpicture}[baseline=(t0.base),term/.style={inner sep=1pt,font=\small},
 equivalence/.style={inner sep=0pt},
 reason/.style={font=\scriptsize,inner sep=2pt}]
\node[term] (t0) {$\textstyle \tilde{\Phi} =C, \Lambda  .  (L,\ipstl{\varphi}, \tnfmark{\nextop(c > \inf I, \ast)},\tnfmark{\nextop (C', \Lambda'  .  L')})$};
\node[equivalence,below=5pt of t0] (e1) {\rotatebox{90}{$\equiv$}};
\node[reason,right=3pt of e1] {\xmerge};
\node[term,below=5pt of e1] (t1) {$\textstyle C, \Lambda  .  (L, \ipstl{\varphi}, \nextop(c > \inf I,\tnfmark{\ast},\tnfmark{C'}, \Lambda'  .  L'))$};
\node[equivalence,below=5pt of t1] (e2) {\rotatebox{90}{$\equiv$}};
\node[reason,right=3pt of e2] {\comm};
\node[term,below=5pt of e2] (t2) {$\textstyle C, \Lambda  .  (L, \ipstl{\varphi}, \nextop(c > \inf I, C', \tnfmark{\ast},\tnfmark{\Lambda'  .  L'}))$};
\node[equivalence,below=5pt of t2] (e3) {\rotatebox{90}{$\equiv$}};
\node[reason,right=3pt of e3] {\cmerge};
\node[term,below=5pt of e3] (t3) {$\textstyle C, \Lambda  .  (L, \ipstl{\varphi}, \nextop(c > \inf I, C',\tnfmark{\Lambda'  .  (\ast, L')})) = \hat{\Phi}$};
\end{tikzpicture}

\end{minipage}

\item[]\begin{minipage}[t]{\linewidth}
\raggedright
{\small $\textstyle\Psi_4 = \RuleAlwaysUpDownBranchFour$}\par\addvspace{\smallskipamount}
\centering

\begingroup
\renewcommand{\tnfboxsep}{1pt}
\begin{tikzpicture}[baseline=(t0.base),term/.style={inner sep=0pt,font=\small},
 equivalence/.style={inner sep=0pt},
 reason/.style={font=\scriptsize,inner sep=2pt}]
\node[term] (t0) {$\textstyle\textstyle  \tilde{\Phi} =C, \Lambda  .  (L, \ipstl{\varphi},    \tnfmark{\nextop(c \gtDash{\upLc \wedge \neg \newLc} \inf I, d \ltDash{\dnLc \wedge \newLc} \sup I, \cset{\creset{c}{\inf I}}  .  (\intvpredT{c}{d}{\newLc}{\dnLc}{\always{I} \varphi}, \lty{\newLc}))}, \tnfmark{\nextop (C', \Lambda'  .  L')})$};
\node[equivalence,below=5pt of t0] (e1) {\rotatebox{90}{$\equiv$}};
\node[reason,right=3pt of e1] {\xmerge};
\node[term,below=5pt of e1] (t1) {$\textstyle\textstyle  C, \Lambda  .  (L, \ipstl{\varphi},    \nextop(c \gtDash{\upLc \wedge \neg \newLc} \inf I, d \ltDash{\dnLc \wedge \newLc} \sup I,\tnfmark{\cset{\creset{c}{\inf I}}  .  (\intvpredT{c}{d}{\newLc}{\dnLc}{\always{I} \varphi}, \lty{\newLc})},\tnfmark{C'}, \Lambda'  .  L'))$};
\node[equivalence,below=5pt of t1] (e2) {\rotatebox{90}{$\equiv$}};
\node[reason,right=3pt of e2] {\comm};
\node[term,below=5pt of e2] (t2) {$\textstyle\textstyle  C, \Lambda  .  (L, \ipstl{\varphi},    \nextop(c \gtDash{\upLc \wedge \neg \newLc} \inf I, d \ltDash{\dnLc \wedge \newLc} \sup I, C',\tnfmark{\cset{\creset{c}{\inf I}}  .  (\intvpredT{c}{d}{\newLc}{\dnLc}{\always{I} \varphi}, \lty{\newLc})},\tnfmark{\Lambda'  .  L'}))$};
\node[equivalence,below=5pt of t2] (e3) {\rotatebox{90}{$\equiv$}};
\node[reason,right=3pt of e3] {\cmerge};
\node[term,below=5pt of e3] (t3) {$\textstyle\textstyle  C, \Lambda  .  (L, \ipstl{\varphi},    \nextop(c \gtDash{\upLc \wedge \neg \newLc} \inf I, d \ltDash{\dnLc \wedge \newLc} \sup I, C',\tnfmark{(\cset{\creset{c}{\inf I}}\cup\Lambda')  .  (\intvpredT{c}{d}{\newLc}{\dnLc}{\always{I} \varphi}, \lty{\newLc}, L')})) = \hat{\Phi}$};
\end{tikzpicture}
\endgroup

\end{minipage}

\item[]\begin{minipage}[t]{\linewidth}
\raggedright
{\small $\textstyle\Psi_5 = \RuleAlwaysUpDownBranchFive$}\par\addvspace{\smallskipamount}
\centering

\begin{tikzpicture}[baseline=(t0.base),term/.style={inner sep=1pt,font=\small},
 equivalence/.style={inner sep=0pt},
 reason/.style={font=\scriptsize,inner sep=2pt}]
\node[term] (t0) {$\textstyle \tilde{\Phi} =C, \Lambda  .  (L, \ipstl{\varphi}, \tnfmark{\nextop(\lty{\newLc}, c \gtDash{\upLc \wedge \neg \newLc} \inf I, d \gtDash{\neg \dnLc \vee \neg \newLc} \sup I)}, \tnfmark{\nextop (C', \Lambda'  .  L')})$};
\node[equivalence,below=5pt of t0] (e1) {\rotatebox{90}{$\equiv$}};
\node[reason,right=3pt of e1] {\xmerge, \comm, \cmerge};
\node[term,below=5pt of e1] (t1) {$\textstyle C, \Lambda  .  (L, \ipstl{\varphi}, \nextop(c \gtDash{\upLc \wedge \neg \newLc} \inf I, d \gtDash{\neg \dnLc \vee \neg \newLc} \sup I, C',\tnfmark{\Lambda'  .  (\lty{\newLc}, L')})) = \hat{\Phi}$};
\end{tikzpicture}

\end{minipage}
\end{proofbranches}

\item $\finallyUpPred \; \rightarrow \; \Psi_1 \mid \Psi_2 \mid \Psi_3$

The branches requiring $\inf I>0$ and $\inf I=0$ are considered
only when their respective conditions hold.

\begin{proofbranches}[beginpenalty=10000,style=standard,leftmargin=0pt,labelwidth=0pt,labelsep=0pt,itemindent=0pt,itemsep=8pt]
\setlength{\abovedisplayskip}{5pt}
\setlength{\belowdisplayskip}{1pt}

\item[]\begin{minipage}[t]{\linewidth}
\raggedright
{\small $\textstyle\Psi_1 = \RuleFinallyUpBranchOne$}\par\addvspace{\smallskipamount}
\centering

\begin{tikzpicture}[baseline=(t0.base),term/.style={inner sep=1pt,font=\small},
 equivalence/.style={inner sep=0pt},
 reason/.style={font=\scriptsize,inner sep=2pt}]
\node[term] (t0) {$\textstyle \tilde{\Phi} =C, \Lambda  .  (L, \tnfmark{\nextop(c \ltDash{\neg \upLc \vee \newLc} \sup I, \cset{\creset{d}{0}}  .  (\intvpredT{c}{d}{\upLc}{\neg{\newLc}}{\eventually{I} \varphi}, \lty{\newLc}))}, \tnfmark{\nextop(C', \Lambda'  .  L')})$};
\node[equivalence,below=5pt of t0] (e1) {\rotatebox{90}{$\equiv$}};
\node[reason,right=3pt of e1] {\xmerge};
\node[term,below=5pt of e1] (t1) {$\textstyle C, \Lambda  .  (L, \nextop(c \ltDash{\neg \upLc \vee \newLc} \sup I,\tnfmark{\cset{\creset{d}{0}}  .  (\intvpredT{c}{d}{\upLc}{\neg{\newLc}}{\eventually{I} \varphi}, \lty{\newLc})},\tnfmark{C'}, \Lambda'  .  L'))$};
\node[equivalence,below=5pt of t1] (e2) {\rotatebox{90}{$\equiv$}};
\node[reason,right=3pt of e2] {\comm};
\node[term,below=5pt of e2] (t2) {$\textstyle C, \Lambda  .  (L, \nextop(c \ltDash{\neg \upLc \vee \newLc} \sup I, C',\tnfmark{\cset{\creset{d}{0}}  .  (\intvpredT{c}{d}{\upLc}{\neg{\newLc}}{\eventually{I} \varphi}, \lty{\newLc})},\tnfmark{\Lambda'  .  L'}))$};
\node[equivalence,below=5pt of t2] (e3) {\rotatebox{90}{$\equiv$}};
\node[reason,right=3pt of e3] {\cmerge};
\node[term,below=5pt of e3] (t3) {$\textstyle C, \Lambda  .  (L, \nextop(c \ltDash{\neg \upLc \vee \newLc} \sup I, C',\tnfmark{(\cset{\creset{d}{0}}\cup\Lambda')  .  (\intvpredT{c}{d}{\upLc}{\neg{\newLc}}{\eventually{I} \varphi}, \lty{\newLc}, L')})) = \hat{\Phi}$};
\end{tikzpicture}

\end{minipage}

\item[]\begin{minipage}[t]{\linewidth}
\raggedright
{\small $\textstyle\Psi_2 = \RuleFinallyUpBranchTwo$}\par\addvspace{\smallskipamount}
\centering

\begin{tikzpicture}[baseline=(t0.base),term/.style={inner sep=1pt,font=\small},
 equivalence/.style={inner sep=0pt},
 reason/.style={font=\scriptsize,inner sep=2pt}]
\node[term] (t0) {$\textstyle \tilde{\Phi} =C, \Lambda  .  (L, \ipstl{\varphi}, \tnfmark{\nextop(c \gtDash{\upLc \wedge \neg \newLc} \inf I, \cset{\creset{c}{\inf I}, \creset{d}{0}}  .  (\intvpredT{c}{d}{\newLc}{\neg{\newLc}}{\eventually{I} \varphi}, \lty{\newLc}))}, \tnfmark{\nextop(C', \Lambda'  .  L')})$};
\node[equivalence,below=5pt of t0] (e1) {\rotatebox{90}{$\equiv$}};
\node[reason,right=3pt of e1] {\xmerge};
\node[term,below=5pt of e1] (t1) {$\textstyle C, \Lambda  .  (L, \ipstl{\varphi}, \nextop(c \gtDash{\upLc \wedge \neg \newLc} \inf I,\tnfmark{\cset{\creset{c}{\inf I}, \creset{d}{0}}  .  (\intvpredT{c}{d}{\newLc}{\neg{\newLc}}{\eventually{I} \varphi}, \lty{\newLc})},\tnfmark{C'}, \Lambda'  .  L'))$};
\node[equivalence,below=5pt of t1] (e2) {\rotatebox{90}{$\equiv$}};
\node[reason,right=3pt of e2] {\comm};
\node[term,below=5pt of e2] (t2) {$\textstyle C, \Lambda  .  (L, \ipstl{\varphi}, \nextop(c \gtDash{\upLc \wedge \neg \newLc} \inf I, C',\tnfmark{\cset{\creset{c}{\inf I}, \creset{d}{0}}  .  (\intvpredT{c}{d}{\newLc}{\neg{\newLc}}{\eventually{I} \varphi}, \lty{\newLc})},\tnfmark{\Lambda'  .  L'}))$};
\node[equivalence,below=5pt of t2] (e3) {\rotatebox{90}{$\equiv$}};
\node[reason,right=3pt of e3] {\cmerge};
\node[term,below=5pt of e3] (t3) {$\textstyle C, \Lambda  .  (L, \ipstl{\varphi}, \nextop(c \gtDash{\upLc \wedge \neg \newLc} \inf I, C',\tnfmark{(\cset{\creset{c}{\inf I}, \creset{d}{0}}\cup\Lambda')  .  (\intvpredT{c}{d}{\newLc}{\neg{\newLc}}{\eventually{I} \varphi}, \lty{\newLc}, L')})) = \hat{\Phi}$};
\end{tikzpicture}

\end{minipage}

\item[]\begin{minipage}[t]{\linewidth}
\raggedright
{\small $\textstyle\Psi_3 = \RuleFinallyUpBranchThree$}\par\addvspace{\smallskipamount}
\centering

\begin{tikzpicture}[baseline=(t0.base),term/.style={inner sep=1pt,font=\small},
 equivalence/.style={inner sep=0pt},
 reason/.style={font=\scriptsize,inner sep=2pt}]
\node[term] (t0) {$\textstyle \tilde{\Phi} =C, \Lambda  .  (L, \ipstl{\varphi}, \nextop(C',\Lambda'  .  L')) = \hat{\Phi}$};
\end{tikzpicture}

\end{minipage}
\end{proofbranches}

\item $\finallyUpDownPred \; \rightarrow \; \Psi_1 \mid \Psi_2 \mid \Psi_3$

\begin{proofbranches}[beginpenalty=10000,style=standard,leftmargin=0pt,labelwidth=0pt,labelsep=0pt,itemindent=0pt,itemsep=8pt]
\setlength{\abovedisplayskip}{5pt}
\setlength{\belowdisplayskip}{1pt}

\item[]\begin{minipage}[t]{\linewidth}
\raggedright
{\small $\textstyle\Psi_1 = \RuleFinallyUpDownBranchOne$}\par\addvspace{\smallskipamount}
\centering

\begin{tikzpicture}[baseline=(t0.base),term/.style={inner sep=1pt,font=\small},
 equivalence/.style={inner sep=0pt},
 reason/.style={font=\scriptsize,inner sep=2pt}]
\node[term] (t0) {$\textstyle \tilde{\Phi} =C, \Lambda  .  (L, \tnfmark{\nextop(c \ltDash{\neg \upLc \vee \newLc} \sup I, \lty{\newLc}, \intvpredT{c}{d}{\upLc}{\dnLc}{\eventually{I} \varphi})}, \tnfmark{\nextop(C', \Lambda'  .  L')})$};
\node[equivalence,below=5pt of t0] (e1) {\rotatebox{90}{$\equiv$}};
\node[reason,right=3pt of e1] {\xmerge};
\node[term,below=5pt of e1] (t1) {$\textstyle C, \Lambda  .  (L, \nextop(c \ltDash{\neg \upLc \vee \newLc} \sup I, \tnfmark{\lty{\newLc}}, \tnfmark{\intvpredT{c}{d}{\upLc}{\dnLc}{\eventually{I} \varphi}},\tnfmark{C'}, \Lambda'  .  L'))$};
\node[equivalence,below=5pt of t1] (e2) {\rotatebox{90}{$\equiv$}};
\node[reason,right=3pt of e2] {\comm};
\node[term,below=5pt of e2] (t2) {$\textstyle C, \Lambda  .  (L, \nextop(c \ltDash{\neg \upLc \vee \newLc} \sup I, C', \tnfmark{\lty{\newLc}}, \tnfmark{\intvpredT{c}{d}{\upLc}{\dnLc}{\eventually{I} \varphi}},\tnfmark{\Lambda'  .  L'}))$};
\node[equivalence,below=5pt of t2] (e3) {\rotatebox{90}{$\equiv$}};
\node[reason,right=3pt of e3] {\cmerge};
\node[term,below=5pt of e3] (t3) {$\textstyle C, \Lambda  .  (L, \nextop(c \ltDash{\neg \upLc \vee \newLc} \sup I, C',\tnfmark{\Lambda'  .  (\lty{\newLc}, \intvpredT{c}{d}{\upLc}{\dnLc}{\eventually{I} \varphi}, L')})) = \hat{\Phi}$};
\end{tikzpicture}

\end{minipage}

\item[]\begin{minipage}[t]{\linewidth}
\raggedright
{\small $\textstyle\Psi_2 = \RuleFinallyUpDownBranchTwo$}\par\addvspace{\smallskipamount}
\centering

\begingroup
\renewcommand{\tnfboxsep}{1pt}
\begin{tikzpicture}[baseline=(t0.base),term/.style={inner sep=0pt,font=\small},
 equivalence/.style={inner sep=0pt},
 reason/.style={font=\scriptsize,inner sep=2pt}]
\node[term] (t0) {$\textstyle\textstyle  \tilde{\Phi} =C, \Lambda  .  (L, \ipstl{\varphi},    \tnfmark{\nextop(c \gtDash{\upLc \wedge \neg \newLc} \inf I, d \ltDash{\dnLc \wedge \newLc} \inf I, \cset{\creset{c}{\inf I}}  .  (\intvpredT{c}{d}{\newLc}{\dnLc}{\eventually{I} \varphi}, \lty{\newLc}))}, \tnfmark{\nextop(C', \Lambda'  .  L')})$};
\node[equivalence,below=5pt of t0] (e1) {\rotatebox{90}{$\equiv$}};
\node[reason,right=3pt of e1] {\xmerge};
\node[term,below=5pt of e1] (t1) {$\textstyle\textstyle  C, \Lambda  .  (L, \ipstl{\varphi},    \nextop(c \gtDash{\upLc \wedge \neg \newLc} \inf I, d \ltDash{\dnLc \wedge \newLc} \inf I,\tnfmark{\cset{\creset{c}{\inf I}}  .  (\intvpredT{c}{d}{\newLc}{\dnLc}{\eventually{I} \varphi}, \lty{\newLc})},\tnfmark{C'}, \Lambda'  .  L'))$};
\node[equivalence,below=5pt of t1] (e2) {\rotatebox{90}{$\equiv$}};
\node[reason,right=3pt of e2] {\comm};
\node[term,below=5pt of e2] (t2) {$\textstyle\textstyle  C, \Lambda  .  (L, \ipstl{\varphi},    \nextop(c \gtDash{\upLc \wedge \neg \newLc} \inf I, d \ltDash{\dnLc \wedge \newLc} \inf I, C',\tnfmark{\cset{\creset{c}{\inf I}}  .  (\intvpredT{c}{d}{\newLc}{\dnLc}{\eventually{I} \varphi}, \lty{\newLc})},\tnfmark{\Lambda'  .  L'}))$};
\node[equivalence,below=5pt of t2] (e3) {\rotatebox{90}{$\equiv$}};
\node[reason,right=3pt of e3] {\cmerge};
\node[term,below=5pt of e3] (t3) {$\textstyle\textstyle  C, \Lambda  .  (L, \ipstl{\varphi},    \nextop(c \gtDash{\upLc \wedge \neg \newLc} \inf I, d \ltDash{\dnLc \wedge \newLc} \inf I, C',\tnfmark{(\cset{\creset{c}{\inf I}}\cup\Lambda')  .  (\intvpredT{c}{d}{\newLc}{\dnLc}{\eventually{I} \varphi}, \lty{\newLc}, L')})) = \hat{\Phi}$};
\end{tikzpicture}
\endgroup

\end{minipage}

\item[]\begin{minipage}[t]{\linewidth}
\raggedright
{\small $\textstyle\Psi_3 = \RuleFinallyUpDownBranchThree$}\par\addvspace{\smallskipamount}
\centering

\begin{tikzpicture}[baseline=(t0.base),term/.style={inner sep=1pt,font=\small},
 equivalence/.style={inner sep=0pt},
 reason/.style={font=\scriptsize,inner sep=2pt}]
\node[term] (t0) {$\textstyle \tilde{\Phi} =C, \Lambda  .  (L, \ipstl{\varphi}, \tnfmark{\nextop(d \gtDash{\neg \dnLc \vee \neg \newLc} \inf I, \lty{\newLc})}, \tnfmark{\nextop(C', \Lambda'  .  L')})$};
\node[equivalence,below=5pt of t0] (e1) {\rotatebox{90}{$\equiv$}};
\node[reason,right=3pt of e1] {\xmerge, \comm, \cmerge};
\node[term,below=5pt of e1] (t1) {$\textstyle C, \Lambda  .  (L, \ipstl{\varphi}, \nextop(d \gtDash{\neg \dnLc \vee \neg \newLc} \inf I, C',\tnfmark{\Lambda'  .  (\lty{\newLc}, L')})) = \hat{\Phi}$};
\end{tikzpicture}

\end{minipage}
\end{proofbranches}
\end{proofrules}

In each case, the final judgement $\hat{\Phi}$ is in transition form,
and every rewriting step uses a structural law under the stated
side conditions. Hence $\hat{\Phi}\equiv\tilde{\Phi}$, as required.

The construction also preserves subsumption. For the comparisons in
\Cref{def:rule-closed-single} and in rule expansions, the branch under
the source resets $\Lambda$ has transition form
\[
\Psi_\Lambda=(\Lambda\cup\Lambda_\Psi).
 (L_\Psi,\nextop(C_\Psi',\Lambda_\Psi'.L_\Psi'))
\equiv\Lambda.\Psi.
\]
This equivalence follows from the nested reset law when a fresh
current reset is introduced, and is immediate otherwise.
Every component of $\Phi$ and $\Psi_\Lambda$ is included in the
corresponding component of $\hat{\Phi}$.
For $\Phi$, additional current resets are fresh, and additional next
resets do not capture free clocks of $\Lambda'.L'$.
For $\Psi_\Lambda$, the current reset sets are identical, and existing
next resets do not capture free clocks of $\Lambda_\Psi'.L_\Psi'$.
Thus the component inclusions and both no-capture conditions of
\Cref{def:subsumption} hold, giving
$\Phi\sqsubseteq\hat{\Phi}$ and $\Psi_\Lambda\sqsubseteq\hat{\Phi}$.
\qedhere
}
\end{proof}

\subsection{Correctness of Closures}
\label{sec:app-closure-correctness}

Recall the complexity measure in \Cref{def:complexity}.

\restatetheorem{lem:complexity-reduced}

\begin{proof}
Fix a branch $\Psi_i$ and an interval predicate $q$ in its current part.
Using the notation of \Cref{def:complexity}, the rules in
\Cref{fig:unclocked-rule,fig:clocked-rule} give the following cases.
\begin{proofrules}[topsep=3pt,itemsep=3pt]
\item \emph{Case $p=\ipstl{\varphi\star\varphi'}$.}
Here $q$ is either $\ipstl{\varphi}$ or $\ipstl{\varphi'}$, so
$\comp{q}<\comp{\varphi}+\comp{\varphi'}+2=\comp{p}$.
\item \emph{Case $p=\ipstl{\bothtemp_I\varphi}$.}
The branch introduces only $q=\intvpredupT{c}{\upLc}{\bothtemp_I\varphi}$
as a current interval predicate. Thus
$\comp{q}=\comp{\varphi}+1<\comp{\varphi}+2=\comp{p}$.
\item \emph{Case $p=\natural\bothtemp_I\varphi$.}
If the branch contains a current interval predicate, it is $q=\ipstl{\varphi}$.
Hence $\comp{q}=\comp{\varphi}<\comp{\varphi}+1=\comp{p}$.\qedhere
\end{proofrules}
\end{proof}

\paragraph{Termination of Closure}
\label{par:closure-termination}
The ranking of \Cref{def:expansion-complexity} concerns first appearances
across expansion levels, so the termination argument must account for
clock renaming in these comparisons.
In \Cref{def:expansion-levels}, all compatible choices of the branch's
bound-clock names from an unbounded supply are considered.
Fresh branch clocks and clocks bound by the source's existing next resets
may be renamed to avoid capture, leaving the source's current part and all
free occurrences unchanged. Subsumption and minimality are tested
with these representatives fixed. For the ranking, current interval
predicates in $\curlabels_i(\Phi)$ are identified up to bijective clock
renaming that fixes the clocks of $\Phi$.

\begin{lemma}\label{lem:closure-search-termination}
Let $k$ be the first level with $\ranklvl(\Phi,k)=0$. There exists
$N\ge k$ such that $\Phi_c\in\expansion_N(\Phi)$ for every closure
$\Phi_c$ of $\Phi$.
\end{lemma}

\begin{proof}
If $k=0$, every current interval predicate has complexity zero and no
expansion rule, so $\Phi$ is its only closure and $N=0$ suffices.
For $k>0$, consider a sequence $(\Phi_i)$ starting at
$\Phi_k\in\expansion_k(\Phi)$, and let $L_i$ be the current label of
$\Phi_i$. For each consecutive pair, $\lexpansion_{p_i}(\Phi_i,\Phi_{i+1})$
holds for an interval predicate $p_i\in L_i$ for which $\Phi_i$ is not
rule-closed. Let $m$ be the maximum complexity of the interval predicates
in $L_k$. For $d\in\{0,\ldots,m\}$,
let $I_d=\{i\mid\comp{p_i}\ge d\}$, with $i$ ranging over the consecutive
pairs. We prove that $I_d$ is finite by downward induction on $d$.

Every $\Phi_j$ with $j>i$ is rule-closed for $p_i$, since its witnessing
branch remains subsumed, so the $p_i$ are pairwise distinct. As in the
proof of \Cref{lem:expansion-ranking-decrease}, any interval predicate
$q\in L_{i+1}\setminus L_i$ comes from the rule branch for
$p_i$, so $\comp{q}<\comp{p_i}$ by \Cref{lem:complexity-reduced}.
In particular, no current predicate has complexity greater than $m$.
\begin{proofrules}[topsep=3pt,itemsep=3pt]
\item \emph{Base case ($d=m$).}
For $i\in I_m$, $p_i\in L_k$. Since $L_k$ is finite and the $p_i$ are
distinct, $I_m$ is finite.
\item \emph{Inductive case ($d+1$ to $d$).}
By the induction hypothesis, $I_{d+1}$ is finite. For
$i\in I_d\setminus I_{d+1}$, either
\begin{enumerate*}[label=(\alph*)]
\item $p_i\in L_k$ or
\item $p_i\in L_{j+1}\setminus L_j$ for some $k\le j<i$.
\end{enumerate*}
In (b), $\comp{p_j}>\comp{p_i}=d$ by \Cref{lem:complexity-reduced},
so $j\in I_{d+1}$. Thus the $p_i$ in (a) lie in the finite set $L_k$,
and those in (b) lie in finitely many finite sets $L_{j+1}$.
Since the $p_i$ are distinct, $I_d\setminus I_{d+1}$, and hence $I_d$,
is finite.
\end{proofrules}
Since $I_0$ contains every index of a consecutive pair, the sequence is finite.

We identify finite expansion sequences up to clock renaming fixing the
clocks of $\Phi$. Any finite sequence of $p$-expansions has finitely many
extensions by one judgement under this identification, since the possible
predicates, rule branches, and relative choices of clock names are finite.
Hence $\expansion_k(\Phi)$ has finitely many judgements up to renaming.
Since no sequence of the kind considered in the induction is infinite,
finite branching gives a common length bound $D$.

Let $N=k+D$. It remains to show that every closure of $\Phi$ belongs to
$\expansion_\ell(\Phi)$ for some $\ell\le N$. First fix a closure $\Phi_c$
of $\Phi$. Here minimality in \Cref{def:closure} is taken among
transition-form judgements satisfying conditions (i) and (ii).
Starting with $\Phi_0=\Phi\sqsubseteq\Phi_c$, choose at
$\Phi_i$ a current interval predicate $p_i$ of $\Phi_c$ that also occurs
in the current part of $\Phi_i$ and for which $\Phi_i$ is not rule-closed.
If $\Phi\sqsubseteq\Phi_i\sqsubseteq\Phi_c$ and $\Phi_i\ne\Phi_c$, such
a $p_i$ exists by minimality of $\Phi_c$ and inclusion of current labels.
Since $\Phi_c$ is rule-closed for $p_i$, there exists $\Phi_{i+1}$
such that
\[
  \Phi_{i+1}\in\lexpansion_{p_i}(\Phi_i)
  \quad\land\quad \Phi_{i+1}\sqsubseteq\Phi_c.
\]
Thus the sequence can be extended whenever $\Phi_i\ne\Phi_c$. Suppose
$\Phi_c\notin\expansion_\ell(\Phi)$ for all $\ell\le N$. Then $\Phi_i\ne\Phi_c$
at each level $i\le N$, so successive extensions give
$(\Phi_0,\ldots,\Phi_{N+1})$ with $N+1-k=D+1$ expansions after $\Phi_k$,
a contradiction. Since $k>0$, $\Phi_c$ retains a current interval predicate
of $\Phi$, so identity expansions give $\Phi_c\in\expansion_N(\Phi)$.
\end{proof}

\restatetheorem{thm:closure-soundness}

\begin{proof}
We prove the following stronger statement.
Let $\Phi = C,\Lambda.(L,\nextop(C',\Lambda'.L'))$ be a closure,
and let $A \subseteq L$ be the set obtained from $L$ by removing all non-atomic interval predicates.
Given a signal $\sigma$ with time horizon $\tau$,
a partition $\partition{J}$ of $\domain{0}{\tau}$,
a clock valuation $\alpha$, and a position $i$,
if $\alpha[\Lambda]$ satisfies conditions \invmark{1} and \invmark{3}
of $\predInv_{\intvOf{J}{i}}(L)$ and $\sigma,\partition{J},\alpha,i \models_\tau C,\Lambda.(A,\nextop(C',\Lambda'.L'))$,
then $\sigma,\partition{J},\alpha,i \models_\tau C,\Lambda.L$.
Condition \invmark{2} is not needed, since rule soundness requires only
conditions \invmark{1} and \invmark{3}.
Let $\Phi_a = C,\Lambda.(A,\nextop(C',\Lambda'.L'))$
and assume $\sigma,\partition{J},\alpha,i \models_\tau \Phi_a$.
We show that $\sigma,\partition{J},\alpha,i \models_\tau C,\Lambda.L$.
It therefore suffices to prove
$\sigma,\partition{J},\alpha[\Lambda],i \models_\tau L$.
Non-interval judgements in $L$ remain in $A$ and are already satisfied.
We show $\sigma,\partition{J},\alpha[\Lambda],i \models_\tau l$ for every interval predicate $l \in L$ by induction on $\comp{l}$.

If $\comp{l}=0$, then $l \in A$, and the claim follows immediately from
$\sigma,\partition{J},\alpha,i \models_\tau \Phi_a$.
Now suppose $\comp{l}=n+1$ and the claim holds for every interval
predicate of complexity at most $n$.
Since $\Phi$ is a closure and $l \in L$,
$\Phi$ is rule-closed for $l$.
By \Cref{def:rule-closed-single}, there is a transition form
$\Psi=\Lambda_\Psi.(L_\Psi,\nextop(C_\Psi',\Lambda_\Psi'.L_\Psi'))$
of a branch for $l$ under $\Lambda$ with $\Psi\sqsubseteq\Phi$.
Its current reset set is contained
in $\Lambda$. At $\alpha[\Lambda]$, the inherited resets act as the
identity, so this representative and the underlying rule branch
have the same satisfaction.
Also, by assumption, $\alpha[\Lambda]$ satisfies conditions \invmark{1} and \invmark{3} of $\predInv_{\intvOf{J}{i}}(L)$.
Since this invariant is a conjunction over $L$, the same conjuncts
of $\predInv_{\intvOf{J}{i}}(l)$ hold under $\alpha[\Lambda]$.

Let $A_\Psi \subseteq L_\Psi$ be the set obtained from $L_\Psi$ by removing all non-atomic interval predicates, and define
$\Psi_a = \Lambda_\Psi.(A_\Psi,\nextop(C_\Psi',\Lambda_\Psi'.L_\Psi'))$.
Since $L_\Psi\subseteq L$, removing non-atomic interval predicates
on both sides gives $A_\Psi\subseteq A$. All other component inclusions
are unchanged, and the free-clock sets in the no-capture conditions
can only shrink. Thus $\Psi_a\sqsubseteq\Phi_a$ directly by
\Cref{def:subsumption}.
Hence, by \Cref{lem:tnf_subsumption}, $\sigma,\partition{J},\alpha,i \models_\tau \Psi_a$.
Since $\Psi_a \sqsubseteq \Phi_a$, by definition of subsumption, we obtain
$\vars(\Lambda) \cap \fvar{\Psi_a} = \emptyset$ and $\Lambda_\Psi \subseteq \Lambda$.
By \Cref{lem:reset-agreement}, this implies
$\alpha[\Lambda_\Psi](x) = \alpha[\Lambda](x)$ for every $x \in \fvar{A_\Psi,\nextop(C_\Psi',\Lambda_\Psi'.L_\Psi')}$.
By definition, $\sigma,\partition{J},\alpha,i \models_\tau \Psi_a$ iff
$\sigma,\partition{J},\alpha[\Lambda_\Psi],i \models_\tau A_\Psi,\nextop(C_\Psi',\Lambda_\Psi'.L_\Psi')$.
Thus, by \Cref{lem:irrelevant-reset},
$\sigma,\partition{J},\alpha[\Lambda],i \models_\tau A_\Psi,\nextop(C_\Psi',\Lambda_\Psi'.L_\Psi')$.
Since $\Lambda_\Psi \subseteq \Lambda$, we have $\alpha[\Lambda] = \alpha[\Lambda][\Lambda_\Psi]$.
Thus, $\sigma,\partition{J},\alpha[\Lambda],i \models_\tau \Lambda_\Psi.(A_\Psi,\nextop(C_\Psi',\Lambda_\Psi'.L_\Psi'))$.
I.e., $\sigma,\partition{J},\alpha[\Lambda],i \models_\tau \Psi_a$.

Now let $l_\Psi \in L_\Psi \setminus A_\Psi$.
Since $l_\Psi$ appears in the branch $\Psi$ of a rule for $l$,
\Cref{lem:complexity-reduced} gives
$\comp{l_\Psi} < \comp{l}$.
By the induction hypothesis,
$\sigma,\partition{J},\alpha[\Lambda],i \models_\tau l_\Psi$.
Since $\alpha[\Lambda] = \alpha[\Lambda][\Lambda_\Psi]$, we obtain
$\sigma,\partition{J},\alpha[\Lambda][\Lambda_\Psi],i \models_\tau l_\Psi$,
i.e., $\sigma,\partition{J},\alpha[\Lambda],i \models_\tau \Lambda_\Psi.l_\Psi$.
Since $l_\Psi$ was arbitrary, every non-atomic interval predicate in
$L_\Psi \setminus A_\Psi$ is satisfied under $\alpha[\Lambda]$ after
applying $\Lambda_\Psi$.
Together with
$\sigma,\partition{J},\alpha[\Lambda],i \models_\tau \Psi_a$,
which gives the satisfaction of $A_\Psi$ and the next part,
this implies
$\sigma,\partition{J},\alpha[\Lambda],i \models_\tau \Psi$.
Since $\alpha[\Lambda][\Lambda_\Psi]=\alpha[\Lambda]$, this also
establishes satisfaction of the underlying rule branch under
$\alpha[\Lambda]$, with the next reset judgements and their bound clock occurrences unchanged.
Also recall that $\alpha[\Lambda]$ satisfies conditions \invmark{1} and \invmark{3} of $\predInv_{\intvOf{J}{i}}(l)$.
Hence, by \Cref{thm:expansion-rule-soundness},
$\sigma,\partition{J},\alpha[\Lambda],i \models_\tau l$.
\end{proof}

\restatetheorem{lem:satisfiable-l-expansion}

\begin{proof}
We show the following.
Let
$\Phi=C,\Lambda.(L,\nextop(C',\Lambda'.L'))$
be a transition-form judgement, and let $l\in L$.
Given a signal $\sigma$ with time horizon $\tau$,
a fully stable partition $\partition{J}$ of $\domain{0}{\tau}$,
a clock valuation $\alpha$, and a position $i$,
if
$\sigma,\partition{J},\alpha,i\models_\tau\Phi$
and the current and next predicate invariants of $\Phi$ are satisfied by
$\alpha[\Lambda]$
and
$(\alpha[\Lambda]+|\intvOf{J}{i}|)[\Lambda']$,
respectively,
then there exists an $l$-expansion $\Phi'$ of $\Phi$ such that
$\sigma,\partition{J},\alpha,i\models_\tau\Phi'$.

Assume the stated premises.
If $\Phi$ is rule-closed for $l$, then $\Phi\in\lexpansion_l(\Phi)$, and the claim follows immediately from the assumption.
Suppose $\Phi$ is not rule-closed for $l$.
Since $l\in L$, we obtain $\sigma,\partition{J},\alpha[\Lambda],i\models_\tau l$.
Moreover, since the current predicate invariants of $\Phi$ are satisfied by $\alpha[\Lambda]$,
the predicate invariant of $l$ is satisfied under $\alpha[\Lambda]$.

Since $\alpha[\Lambda]$ satisfies the predicate invariant of $l$ and
$\sigma,\partition{J},\alpha[\Lambda],i\models_\tau l$, by
\Cref{thm:expansion-rule-completeness}, choose the branch $\Psi$
constructed in its proof such that
$\sigma,\partition{J},\alpha[\Lambda],i\models_\tau\Psi$.
Let $\Phi^\flat = C,\Lambda.(L,\Psi,\nextop(C',\Lambda'.L'))$.
Combining satisfaction of $\Phi$ under $\alpha$ with satisfaction of
$\Psi$ under $\alpha[\Lambda]$ gives
\[
\sigma,\partition{J},\alpha,i\models_\tau\Phi^\flat.
\]
Write
$\Psi=C_\Psi,\Lambda_\Psi.
(L_\Psi,\nextop(C_\Psi',\Lambda_\Psi'.L_\Psi'))$.
Let $\widetilde\Phi^\flat$ be obtained by capture-avoiding renaming
of the bound clocks in $\Lambda_\Psi,\Lambda',\Lambda_\Psi'$ to
pairwise distinct names fresh for $\Phi^\flat$, changing only each
clock reset and the occurrences of the clock bound by it. Such names exist since
$\Phi^\flat$ contains finitely many clocks.
The enclosing $\Lambda$, free occurrences, and inherited bindings
are unchanged, so $\widetilde\Phi^\flat\equiv\Phi^\flat$ with the
same satisfaction under $\alpha$.
The reset sets to be merged assign values to disjoint sets of clocks and are therefore consistent,
and the merges introduce no capture. Thus the side conditions of
\Cref{prop:judg-equiv} hold.

Let $\widetilde\Phi$ and $\widetilde\Psi$ be the source and branch
representatives induced by this same renaming.
By \Cref{lem:tnf-existence} and its construction, there is a
transition-form judgement $\hat\Phi\equiv\widetilde\Phi^\flat
\equiv\Phi^\flat$ with
$\widetilde\Phi\sqsubseteq\hat\Phi$ and
$\widetilde\Psi\sqsubseteq\hat\Phi$, the latter under the inherited
current resets. Interpreting these comparisons up to the same
capture-avoiding bound-clock renaming, as in the proof of
\Cref{lem:tnf-existence}, gives
$\Phi\sqsubseteq\hat\Phi$ and $\widetilde\Psi\sqsubseteq\hat\Phi$
under the inherited current resets.
In particular, $\sigma,\partition{J},\alpha,i\models_\tau\hat\Phi$.

Choose a $\sqsubseteq$-minimal element $\Phi'$ of
\[
\{\Theta\mid
\Phi\sqsubseteq\Theta\sqsubseteq\hat\Phi,
\ \widetilde\Psi\sqsubseteq\Theta,
\ \Theta\text{ is in transition form}\}.
\]
This set contains $\hat\Phi$ and is finite with its clock names fixed,
since candidates consist of subsets of its finitely many components.
Any strictly smaller judgement below $\Phi'$ that subsumes both
$\Phi$ and $\widetilde\Psi$ would, by transitivity, belong to this same set.
Hence $\Phi'\in\lexpansion_l(\Phi)$.
Finally, $\Phi'\sqsubseteq\hat\Phi$ and
\Cref{lem:tnf_subsumption} give $\hat\Phi\Rightarrow\Phi'$, so
$\sigma,\partition{J},\alpha,i\models_\tau\Phi'$.
This proves the claim.
\end{proof}

\restatetheorem{thm:closure-completeness}

\begin{proof}
We show the following.
Let
$\Phi=C,\Lambda.(L,\nextop(C',\Lambda'.L'))$
be a transition-form judgement.
Given a signal $\sigma$ with time horizon $\tau$,
a fully stable partition $\partition{J}$ of $\domain{0}{\tau}$,
a clock valuation $\alpha$, and a position $i$,
if the current and next predicate invariants of $\Phi$
are satisfied by $\alpha[\Lambda]$ and
$(\alpha[\Lambda] + |\intvOf{J}{i}|)[\Lambda']$,
respectively, and
$\sigma,\partition{J},\alpha,i\models_\tau\Phi$,
then there exists a closure $\Phi_c$ of $\Phi$ such that
$\sigma,\partition{J},\alpha,i\models_\tau\Phi_c$
and its current and next predicate invariants are satisfied under
its corresponding reset valuations.

Assume the stated premises.
Starting from $\Phi$, select only current predicates for which the
judgement is not rule-closed.
For each selected predicate, \Cref{prop:expansion-invariant} gives a
branchwise expansion that preserves satisfaction and the current and
next predicate invariants.
If the search reaches the first level of rank zero, its remaining
steps are finite by the proof of \Cref{lem:closure-search-termination}.
Hence the search reaches a rule-closed judgement $\Phi_k$ satisfying
the same conditions after finitely many expansions. Along this finite
sequence, any renamings of clocks bound by next resets can be applied to the retained
components of $\Phi$ as well: the next no-capture condition guarantees
that an occurrence free in the retained source is not renamed.
Thus $\Phi\sqsubseteq\Phi_k$ with clock names fixed.

Choose a $\sqsubseteq$-minimal element $\Phi_c$ of
\[
\left\{\Theta\;\middle|\;
\begin{gathered}
\Phi\sqsubseteq\Theta\sqsubseteq\Phi_k,\quad
\Theta\text{ is in transition form},\\
\Theta\text{ is rule-closed for all interval predicates in its current label}
\end{gathered}
\right\}.
\]
The set contains $\Phi_k$ and is finite with its clock names fixed.
Any strictly smaller rule-closed extension of $\Phi$ below $\Phi_c$
would belong to the same set by transitivity, contradicting minimality.
Thus $\Phi_c$ is a closure of $\Phi$ by \Cref{def:closure},
which requires rule-closure for predicates in the current part.
Finally, $\Phi_c\sqsubseteq\Phi_k$ and \Cref{lem:tnf_subsumption}
imply $\Phi_k\Rightarrow\Phi_c$. Hence
$\sigma,\partition{J},\alpha,i\models_\tau\Phi_c$.
Each predicate of $\Phi_c$ is retained from $\Phi_k$.
By the reset inclusions and no-capture conditions of
$\Phi_c\sqsubseteq\Phi_k$, its free clocks have the same values under
the respective current and next reset valuations.
The satisfied predicate invariants of $\Phi_k$ therefore also
hold for $\Phi_c$.
\qedhere
\end{proof}

\subsection{Closure Invariants}
\label{sec:app_closure_invariant}

\restatetheorem{prop:expansion-invariant}

\begin{proof}
We show the following.
Let $\Phi_1=C_1,\Lambda_1.(L_1,\nextop(C_1',\Lambda_1'.L_1'))$
and let $l\in L_1$.
Given a signal $\sigma$ with time horizon $\tau$,
a fully stable partition $\partition{J}$ of $\domain{0}{\tau}$,
a clock valuation $\alpha$, and a position $i$,
if the current and next predicate invariants of $\Phi_1$
are satisfied by
$\alpha[\Lambda_1]$
and
$(\alpha[\Lambda_1]+|\intvOf{J}{i}|)[\Lambda_1']$,
respectively, and
$\sigma,\partition{J},\alpha,i\models_\tau\Phi_1$,
then there exists
$\Phi_2=C_2,\Lambda_2.(L_2,\nextop(C_2',\Lambda_2'.L_2'))$
such that $\lexpansion_l(\Phi_1,\Phi_2)$,
$\sigma,\partition{J},\alpha,i\models_\tau\Phi_2$,
and the current and next predicate invariants of $\Phi_2$
are satisfied by
$\alpha[\Lambda_2]$
and
$(\alpha[\Lambda_2]+|\intvOf{J}{i}|)[\Lambda_2']$,
respectively.
Assume the premises of the proposition.
If $\Phi_1$ is rule-closed for $l$, choose $\Phi_2=\Phi_1$,
and the claim is immediate.
Suppose $\Phi_1$ is not rule-closed for $l$.
Since $l\in L_1$, its satisfaction and invariant hold under
$\beta=\alpha[\Lambda_1]$.
Choose the satisfied branch
$\Psi=\Lambda_\Psi.(L_\Psi,\nextop(C_\Psi',\Lambda_\Psi'.L_\Psi'))$
from the corresponding case of the rule completeness proof.
By \Cref{prop:pred_invariant}, this same branch can be chosen with
current predicate invariants satisfied under $\beta[\Lambda_\Psi]$
and next predicate invariants satisfied under
$(\beta[\Lambda_\Psi]+|\intvOf{J}{i}|)[\Lambda_\Psi']$.
Here $\Psi$ denotes the rule branch before adding the inherited resets.

As in the proof of \Cref{lem:satisfiable-l-expansion}, normalize
\[
C_1,\Lambda_1.(L_1,\Psi,\nextop(C_1',\Lambda_1'.L_1'))
\]
using fresh bound-clock names and compatible representatives to obtain
a satisfied transition-form judgement
$\widehat\Phi_2=C_2,\Lambda_2.(L_2,\nextop(C_2',\Lambda_2'.L_2'))$.
Then $\Phi_1\sqsubseteq\widehat\Phi_2$ and
$\Psi\sqsubseteq\widehat\Phi_2$, with the latter representative taken
under the inherited current resets.
We verify its invariants before taking the $\sqsubseteq$-minimal judgement selected in
that proof.

Every predicate of $\widehat\Phi_2$ is inherited from $\Phi_1$ or
introduced by $\Psi$. The inherited resets act as the identity at
$\beta$. After the compatible bound-clock renaming, the reset side
conditions and \Cref{lem:reset-agreement} give the same free clock
values for the branch predicates under their branch valuations and
under the normalized current and next valuations
$\alpha[\Lambda_2]$ and
$(\alpha[\Lambda_2]+|\intvOf{J}{i}|)[\Lambda_2']$, respectively.
The additional resets do not affect the free clocks of the inherited
predicates, so their satisfied invariants are preserved as well.
The formula, signal, partition, and position determining the past-cover
bounds are unchanged by renaming. Hence both sets of predicate
invariants hold under these normalized valuations.

Finally, choose the same $\sqsubseteq$-minimal judgement $\Phi_2$ between $\Phi_1$
and $\widehat\Phi_2$ as in the proof of
\Cref{lem:satisfiable-l-expansion}.
That construction gives $\lexpansion_l(\Phi_1,\Phi_2)$ and
$\sigma,\partition{J},\alpha,i\models_\tau\Phi_2$.
Every predicate retained in $\Phi_2$ comes from $\widehat\Phi_2$.
The reset inclusions and no-capture conditions of
$\Phi_2\sqsubseteq\widehat\Phi_2$ make its free clock values agree
under the corresponding current and next reset valuations.
Consequently its predicate invariants also hold.
Here the components $C_2,\Lambda_2,L_2,C_2',\Lambda_2',L_2'$ are
understood to be those of the final judgement.
\end{proof}

\begin{corollary}\label{cor:closure-invariant}
Let $\Phi_c$ be a closure of $\Phi$.
For a fixed signal, partition, clock valuation, and position,
if the clock constraints, resets, and type judgements of $\Phi_c$
are satisfied and conditions \invmark{1} and \invmark{3} of all
predicate invariants in $\Phi$ hold, then the same conditions
hold for $\Phi_c$.
\end{corollary}

\begin{proof}
We extend the proof of \Cref{prop:pred_invariant} to closures
for conditions \invmark{1} and \invmark{3}.

Let
$\Phi_c=C_c,\Lambda_c.(L_c,\nextop(C_c',\Lambda_c'.L_c'))$
be a closure of
$\Phi=C,\Lambda.(L,\nextop(C',\Lambda'.L'))$.
Let $T_c$ and $T_c'$ be the sets of judgements of the form
$\lty{\cdot}$ occurring in $L_c$ and $L_c'$, respectively.
Given a signal $\sigma$ with time horizon $\tau$,
a partition $\partition{J}$ of $\domain{0}{\tau}$,
a clock valuation $\alpha$, and a position $i$,
if conditions \invmark{1} and \invmark{3} of the current and next
predicate invariants of $\Phi$ are satisfied by
$\alpha[\Lambda]$ and
$(\alpha[\Lambda]+|\intvOf{J}{i}|)[\Lambda']$,
respectively, and
$\sigma,\partition{J},\alpha,i
\models_\tau
C_c,\Lambda_c.(T_c,\nextop(C_c',\Lambda_c'.T_c'))$,
then the same conditions of the current and next predicate invariants
of $\Phi_c$ are satisfied by
$\alpha[\Lambda_c]$ and
$(\alpha[\Lambda_c]+|\intvOf{J}{i}|)[\Lambda_c']$,
respectively.

Throughout this proof, preservation refers only to conditions
\invmark{1} and \invmark{3}.
The proof of \Cref{lem:closure-search-termination} gives a finite sequence
$\Phi_0,\ldots,\Phi_k$ such that
$\Phi_0=\Phi$, $\Phi_k=\Phi_c$, and for every $0\le r<k$,
there exists an interval predicate $l_r$ with
$\lexpansion_{l_r}(\Phi_r,\Phi_{r+1})$.
Use the fixed clock representatives of that construction throughout. Then
$\Phi_r\sqsubseteq\Phi_{r+1}$ for every $0\le r<k$.

Write
$\Phi_r=C_r,\Lambda_r.(L_r,\nextop(C_r',\Lambda_r'.L_r'))$.
Let $T_r$ and $T_r'$ be the sets of judgements of the form
$\lty{\cdot}$ occurring in $L_r$ and $L_r'$, respectively.
By repeated applications of \Cref{lem:type-subsumption},
for every $0\le r\le k$, we obtain
$C_r,\Lambda_r.(T_r,\nextop(C_r',\Lambda_r'.T_r'))
\sqsubseteq
C_k,\Lambda_k.(T_k,\nextop(C_k',\Lambda_k'.T_k'))$.
Since $\sigma,\partition{J},\alpha,i\models_\tau
C_k,\Lambda_k.(T_k,\nextop(C_k',\Lambda_k'.T_k'))$,
by \Cref{lem:tnf_subsumption},
for every $0\le r\le k$,
$\sigma,\partition{J},\alpha,i \models_\tau
C_r,\Lambda_r.(T_r,\nextop(C_r',\Lambda_r'.T_r'))$.

We now prove by induction on $r$ that the current and next predicate
invariants of $\Phi_r$ are satisfied by
$\alpha[\Lambda_r]$
and
$(\alpha[\Lambda_r]+|\intvOf{J}{i}|)[\Lambda_r']$,
respectively.
The base case $r=0$ holds by assumption since $\Phi_0=\Phi$.
Suppose the claim holds for some $0\le r<k$.
An identity expansion preserves the claim immediately.
Since
$\lexpansion_{l_r}(\Phi_r,\Phi_{r+1})$
and
$\sigma,\partition{J},\alpha,i
\models_\tau
C_{r+1},\Lambda_{r+1}.
(T_{r+1},\nextop(C_{r+1}',\Lambda_{r+1}'.T_{r+1}'))$,
consider a rule branch realizing this expansion.
The calculations for conditions \invmark{1} and \invmark{3} in the proof of
\Cref{prop:pred_invariant} use only the input conjuncts, branch guards,
and resets, not the branch selection used in the completeness proof or formula satisfaction.
For a predicate propagated without resets, the interval specified by
its scope or its shifted checking interval remains nonempty;
one-clock always propagation has endpoint difference
$\alpha(c)+|\intvOf{J}{i}|+|I|>0$;
the initial resets and the resets $d:=0$ or $c:=\inf I$ in the remaining
branches establish condition \invmark{3}, using the clock constraint on
$d$ in the two-clock cases.
These calculations therefore apply to this branch even if it was not
selected by completeness. The inherited and introduced conjuncts are
combined using the same reset-agreement and no-capture argument as above.
Thus
the current and next predicate invariants of $\Phi_{r+1}$
are satisfied by
$\alpha[\Lambda_{r+1}]$
and
$(\alpha[\Lambda_{r+1}]+|\intvOf{J}{i}|)[\Lambda_{r+1}']$,
respectively.
Therefore, the current and next predicate invariants of
$\Phi_k=\Phi_c$ are satisfied.
\qedhere
\end{proof}

\restatetheorem{lem:expansion-ranking-decrease}

\begin{proof}
If $\curlabels_{i+1}(\Phi)\setminus\curlabels_i(\Phi)$ contains no interval
predicate, the rank at level $i+1$ is $0$ and the claim follows. Otherwise
take any interval predicate $q$ in this difference.
By the definition of $\expansion_{i+1}(\Phi)$,
some step from level $i$ selects a current predicate $p$ and adds $q$.
We show that $p$ must itself be new at level $i$, with membership in
$\curlabels_i(\Phi)$ tested under the clock-renaming convention in
\hyperref[par:closure-termination]{the termination argument above}.

Let $B$ be the source of the step adding $q$, $B'$ its result, and
$\Psi_j$ its selected branch under the current resets of $B$.
Both $B$ and $\Psi_j$ are subsumed by $B'$. Taking the union of
their corresponding components yields a transition-form judgement
subsumed by $B'$ and subsuming both: reset consistency and
no-capture follow from the two subsumptions into $B'$. By the
$\sqsubseteq$-minimality of $B'$ in \Cref{def:expansion-levels}~(ii),
this union is $B'$ itself. Since $q$ is not in the current part of
$B$, it is in that of $\Psi_j$, so
\[
\comp{q}<\comp{p}
\qquad\text{(\Cref{lem:complexity-reduced}).}
\]

Suppose $i>0$ and $p\in\curlabels_{i-1}(\Phi)$. Choose a judgement
$A\in\expansion_{i-1}(\Phi)$ with a current predicate that renames to
$p$. Apply the same renaming to the sequence from $\Phi$ to $A$.
It fixes $\Phi$ and preserves each expansion, since subsumption and
branchwise minimality are preserved by consistent renaming. Thus we
may take $p$ itself to occur in $A$. There are two cases.
\begin{proofrules}
\item \emph{$q$ is unclocked.}
Use the same branch under the current resets of $A$. Any free clock of
the branch is a clock of $p$; only bound clocks need to be renamed.
Choose the branch's bound clocks fresh and rename any conflicting
clocks bound by the next resets of $A$.
The branch can then be added to $A$ without changing $q$.
\item \emph{$q$ is clocked.}
The only expansion that introduces a clocked interval predicate in the
current part is that of $\ipstl{\bothtemp_I\varphi}$ with a reset $c:=0$
(\Cref{fig:clocked-rule}). To obtain the same $q$, use the same $c$ at $A$.
The branch adds no next part, so only two checks are needed:
\begin{proofsubcases}
\item \emph{No capture.}
Along expansions, the free clocks of $\Lambda.(L,\nextop(C',\Lambda'.L'))$
do not change: subsumption prevents capture, every free clock of an added
branch comes from the selected current predicate under the same resets,
and renaming clocks bound by next resets preserves free clocks. Thus $c:=0$ cannot
capture a free clock in $A$ if it did not in $B$.
\item \emph{Reset consistency.}
All new current resets assign $0$. Any conflicting assignment in $A$
would therefore come from $\Phi$ and would also have blocked the step
from $B$.
\end{proofsubcases}
\end{proofrules}

In either case, after the indicated renaming, the union of $A$ and the
branch is $\sqsubseteq$-minimal among judgements subsuming both, so it
is a $p$-expansion by \Cref{def:expansion-levels}~(ii). It contains $q$,
so $q\in\curlabels_i(\Phi)$, a contradiction. Thus $p$ is new
at level $i$ (or belongs to the initial current part if $i=0$), and
\[
\comp{q}<\comp{p}\le\ranklvl(\Phi,i).
\]
Taking the maximum over the new predicates at level $i+1$ proves the
claim.
\end{proof}

\restatetheorem{lem:expansion-termination}

\begin{proof}
For each $i \in \natset$, let $a_i = \ranklvl(\Phi,i) \in \natset$.
If $a_i=0$, no new current interval predicate can occur at level $i+1$
up to the permitted clock renamings: the proof of
\Cref{lem:expansion-ranking-decrease} would give
$\comp{q}<\comp{p}\le a_i$ for any such predicate $q$.
Hence $a_{i+1}=0$. While $a_i>0$, the same lemma gives
$a_{i+1}<a_i$. Therefore the rank reaches $0$ in finitely many steps
and stays there.
\end{proof}

\clearpage
\section{Constructing Tableaux for STL}
\label{sec:app-tableau}

We first prove the trace and path properties of \Cref{sec:tableau-construct},
then develop the finiteness argument of \Cref{sub:tableau-finite}.

\subsection{Traces and Accepting Paths}

\restatetheorem{lem:trace-equivalence}

\begin{proof}
Fix any signal $\sigma$ with time horizon $\tau$,
a partition $\partition{J}$ of
$\domain{0}{\tau}$,
a clock valuation $\alpha$,
and a position $i$.
We show that
$\sigma,\partition{J},\alpha,i\models_\tau\trace{\pi}$
iff
$\sigma,\partition{J},\alpha,i\models_\tau\ftrace{\pi}$.

\noindent
($\Rightarrow$)
Assume
$\sigma,\partition{J},\alpha,i\models_\tau\trace{\pi}$.
By \Cref{lem:trace-ftrace-implication},
$\sigma,\partition{J},\alpha,i\models_\tau\ftrace{\pi}$.

\noindent
($\Leftarrow$)
Assume
$\sigma,\partition{J},\alpha,i\models_\tau\ftrace{\pi}$.
Since
$\trace{\pi}\sqsubseteq\ftrace{\pi}$,
by \Cref{lem:tnf_subsumption},
$\sigma,\partition{J},\alpha,i\models_\tau\trace{\pi}$.
\qedhere
\end{proof}

Traces capture the essential semantic information of tableau runs:
clock constraints, resets, and interval type judgements.
This information is sufficient to guarantee that conditions \invmark{1} and \invmark{3} of the predicate invariants of the closures of states along the path are satisfied.
Condition \invmark{2} is not required for this soundness argument.
The following lemma formalizes this.

\begin{lemma}\label{lem:trace-invariant}
Let $\pi$ be an accepting run of $\tabG{\varphi}$.
Given a signal $\sigma$ with time horizon $\tau$,
a partition $\partition{J}$ of $\domain{0}{\tau}$,
a clock valuation $\alpha$, and a position $i$,
if $\sigma,\partition{J},\alpha,i\models_\tau\trace{\pi}$,
then conditions \invmark{1} and \invmark{3} of the predicate
invariants of the closure represented by each state of $\pi$
hold along the trace.
\end{lemma}

\begin{proof}
The invariant claims in this proof concern only conditions
\invmark{1} and \invmark{3}.
Let $\pi=\eta_0s_0\cdots\eta_ns_n$.
Write $s_j=\langle\Phi_j,M_g^j\rangle$, where
$\Phi_j=C_j,\Lambda_j.(L_j,\nextop M_g^{j+1})$ for $0\le j<n$
and $\Phi_n=C_n,\Lambda_n.L_n$.
Write $M_g^j=C_g^j,\Gamma_j.G_j$ for each generator.
Thus $\Gamma_{j+1}$ is the reset set in the next part of $\Phi_j$;
it need not equal the reset set $\Lambda_{j+1}$ of its closure.
Assume $\sigma,\partition{J},\alpha,i\models_\tau\trace{\pi}$.
Let $\alpha_0=\alpha$ and $\alpha_{j+1}=\alpha_j[\Lambda_j]+|\intvOf{J}{i+j}|$ for $0\le j<n$.
If a next part contains no interval predicates, its invariant claim is
vacuous and the corresponding successor valuation is omitted.
At the terminal position there is no interval on which to evaluate a
predicate; invariant claims for predicate-free labels are likewise vacuous.
By definition of $\trace{\pi}$, for every $0\le j<n$,
the clock constraints, resets, and type judgements of $\Phi_j$
are satisfied at position $i+j$ under $\alpha_j$.
Similarly, the clock constraints, resets, and type judgements of
$\Phi_n$ are satisfied at position $i+n$ under $\alpha_n$.
We prove by induction on $j$ that:
\begin{proofrules}[topsep=3pt,itemsep=2pt]
\item
if $0\le j<n$, then the current and next predicate invariants of
$\Phi_j$
are satisfied by
$\alpha_j[\Lambda_j]$
and
$\alpha_{j+1}[\Gamma_{j+1}]$,
respectively;
\item
if $j=n$, then the predicate invariants of
$\Phi_n$
are satisfied by
$\alpha_n[\Lambda_n]$.
\end{proofrules}

For $j=0$, the generator of $s_0$ is $\ipstl{\varphi}$.
Since $\predInv_{\intvOf{J}{i}}(\ipstl{\varphi})=\true$, the predicate invariants of
the generator are satisfied.
Moreover, $\Phi_0$ is a closure of $\ipstl{\varphi}$, and the clock
constraints, resets, and type judgements of $\Phi_0$ are satisfied by
the trace as observed above.
Hence, by \Cref{cor:closure-invariant}, the current predicate invariants
of $\Phi_0$ are satisfied by $\alpha_0[\Lambda_0]$.
If $n>0$, its next predicate invariants are also satisfied by
$\alpha_1[\Gamma_1]$.
If $n=0$, there is no next part, and the claim is proved.

Suppose the claim holds for $0\le j<n$.
Then the next predicate invariants of $\Phi_j$ are satisfied by
$\alpha_{j+1}[\Gamma_{j+1}]$.
These are the predicate invariants of the label $G_{j+1}$ in the
generator $M_g^{j+1}=C_g^{j+1},\Gamma_{j+1}.G_{j+1}$.
They therefore supply exactly the generator invariant premise of
\Cref{cor:closure-invariant} at incoming valuation $\alpha_{j+1}$.
If $j+1<n$, then $\Phi_{j+1}$ is a closure of $M_g^{j+1}$.
Also, the clock constraints, resets, and type judgements of
$\Phi_{j+1}$ are satisfied at position $i+j+1$ under $\alpha_{j+1}$.
Therefore, by \Cref{cor:closure-invariant}, the current and next
predicate invariants of $\Phi_{j+1}$ are satisfied by
$\alpha_{j+1}[\Lambda_{j+1}]$
and
$\alpha_{j+2}[\Gamma_{j+2}]$,
respectively.
If $j+1=n$, then $\Phi_n$ is a closure of $M_g^n$.
The predicate invariants of $G_n$ are satisfied by $\alpha_n[\Gamma_n]$.
Moreover, the clock constraints, resets, and type judgements of
$\Phi_n$ are satisfied at position $i+n$ under $\alpha_n$.
Hence, by \Cref{cor:closure-invariant}, the predicate invariants of
$\Phi_n$ are satisfied by
$\alpha_n[\Lambda_n]$.

Thus, by induction, the predicate invariants of the closure of
every state of $\pi$ are satisfied along the trace.
\end{proof}

Traces already contain enough information
to recover the full semantic meaning of tableau runs.
Although they only retain clock constraints, resets, and judgements
obtained by removing non-atomic interval predicates, the omitted
non-atomic judgements can be reconstructed using closure invariants and
closure soundness.
Consequently, every trace induced by an accepting run semantically
implies the corresponding full trace.

\begin{lemma}\label{lem:trace-ftrace-implication}
Let $\pi=\eta_0s_0\cdots\eta_ns_n$ be an accepting run of $\tabG{\varphi}$.
For each $0\le j<n$, let $s_j= \langle C_j,\Lambda_j.(L_j,\nextop M_g^{j+1}), M_g^j \rangle$, and let
$s_n = \langle C_n,\Lambda_n.L_n,M_g^n\rangle$.
Given a signal $\sigma$ with time horizon $\tau$,
a partition $\partition{J}$ of $\domain{0}{\tau}$,
a clock valuation $\alpha$, and a position $i$,
suppose that
$\sigma,\partition{J},\alpha,i\models_\tau\trace{\pi}$.
Then, for every $0\le j\le n$,
if $\sigma,\partition{J},\alpha_j,i+j \models_\tau \trace{\pi^j}$,
then $\sigma,\partition{J},\alpha_j,i+j \models_\tau \ftrace{\pi^j}$
and $\sigma,\partition{J},\alpha_j,i+j \models_\tau M_g^j$, where $\alpha_0=\alpha$ and
$\alpha_{j+1} = \alpha_j[\Lambda_j] + |\intvOf{J}{i+j}|$ for $0\le j<n$.
\end{lemma}

\begin{proof}
We use the version of \Cref{thm:closure-soundness} requiring only
conditions \invmark{1} and \invmark{3}, as supplied by
\Cref{lem:trace-invariant}.
We prove the claim by backward induction on $j$.

For each $0\le j\le n$, let $A_j$ be the set of judgements in $L_j$
obtained by removing non-atomic interval predicates.

\begin{proofrules}[topsep=3pt,itemsep=4pt]
\item
(Base case $j=n$)
Assume $\sigma,\partition{J},\alpha_n,i+n \models_\tau \trace{\pi^n}$.
The final state is in state form, so
$\trace{\pi^n}=C_n,\Lambda_n.A_n$ by definition.
Hence $\sigma,\partition{J},\alpha_n,i+n \models_\tau C_n,\Lambda_n.A_n$.
By \Cref{lem:trace-invariant}, conditions \invmark{1} and \invmark{3}
of the predicate invariants of $L_n$ hold under $\alpha_n[\Lambda_n]$.
Since $C_n,\Lambda_n.L_n$ is a closure of $M_g^n$, by \Cref{thm:closure-soundness},
$\sigma,\partition{J},\alpha_n,i+n \models_\tau C_n,\Lambda_n.L_n$.
Hence $\sigma,\partition{J},\alpha_n,i+n \models_\tau \ftrace{\pi^n}$.
Since $M_g^n \sqsubseteq C_n,\Lambda_n.L_n$, by \Cref{lem:tnf_subsumption},
$\sigma,\partition{J},\alpha_n,i+n \models_\tau M_g^n$.

\item
(Inductive case)
Suppose the claim holds for $j+1>0$. We show the claim for $j$.
Assume $\sigma,\partition{J},\alpha_j,i+j \models_\tau \trace{\pi^j}$.
Since $\trace{\pi^j} = C_j,\Lambda_j. (A_j,\nextop\trace{\pi^{j+1}})$,
we have $\sigma,\partition{J},\alpha_{j+1},i+j+1 \models_\tau \trace{\pi^{j+1}}$.
By the induction hypothesis,
$\sigma,\partition{J},\alpha_{j+1},i+j+1 \models_\tau \ftrace{\pi^{j+1}}$
and $\sigma,\partition{J},\alpha_{j+1},i+j+1 \models_\tau M_g^{j+1}$.
Also, by assumption, $\sigma,\partition{J},\alpha_j,i+j \models_\tau C_j,\Lambda_j.A_j$.
Hence, $\sigma,\partition{J},\alpha_j,i+j \models_\tau C_j,\Lambda_j.(A_j,\nextop M_g^{j+1})$.
By \Cref{lem:trace-invariant}, conditions \invmark{1} and \invmark{3}
of the current predicate invariants of $L_j$ hold under
$\alpha_j[\Lambda_j]$; these are the only invariant
premises required by \Cref{thm:closure-soundness} here.
Since $C_j,\Lambda_j.(L_j,\nextop M_g^{j+1})$ is a closure, by \Cref{thm:closure-soundness},
$\sigma,\partition{J},\alpha_j,i+j \models_\tau C_j,\Lambda_j.(L_j,\nextop M_g^{j+1})$.
Since $\sigma,\partition{J},\alpha_j,i+j \models_\tau C_j,\Lambda_j.L_j$
and $\sigma,\partition{J},\alpha_{j+1},i+j+1 \models_\tau \ftrace{\pi^{j+1}}$,
we obtain
$\sigma,\partition{J},\alpha_j,i+j \models_\tau C_j,\Lambda_j. (L_j,\nextop\ftrace{\pi^{j+1}})$.
Hence $\sigma,\partition{J},\alpha_j,i+j \models_\tau \ftrace{\pi^j}$.
Since $M_g^j \sqsubseteq C_j,\Lambda_j.(L_j,\nextop M_g^{j+1})$,
\Cref{lem:tnf_subsumption} gives
\[
\sigma,\partition{J},\alpha_j,i+j \models_\tau M_g^j.
\qedhere
\]
\end{proofrules}
\end{proof}

\restatetheorem{lem:generator-path-existence}

\begin{proof}
We show the following.
Given a signal $\sigma$ with time horizon $\tau$,
a state-form judgement $M_g=C,\Lambda.L$,
a fully stable partition $\partition{J}$ of $\domain{0}{\tau}$ for $\varphi$,
a clock valuation $\alpha$, and a position $i$,
if
$\sigma,\partition{J},\alpha,i\models_\tau M_g$
and
$\predInv_{\intvOf{J}{i}}(L)$
is satisfied by $\alpha[\Lambda]$,
then there exist a closure $\Phi$ of $M_g$ and an accepting path $\pi$
of $\tabG{\varphi}$ starting from the state
$s=\langle\Phi,M_g\rangle$
such that
$\sigma,\partition{J},\alpha,i\models_\tau\ftrace{\pi}$.

We prove the claim by backward induction on $i$.
For the induction only, extend the claim to the terminal position
$i=|\partition{J}|$ for satisfied state-form judgements containing no
interval predicates; no interval invariant is required there.

\begin{proofrules}[topsep=3pt,itemsep=4pt]
\item
(Base case)
Suppose $i=|\partition{J}|$.
Assume $\sigma,\partition{J},\alpha,i\models_\tau M_g$, where
$M_g=C,\Lambda.L$ contains no interval predicates.
There is no predicate to expand, so $M_g$ is its own closure.
Choose $\Phi=M_g=C_c,\Lambda_c.L_c$.
Only clock constraints, type judgements, and possibly $\ast$ are
evaluated here, as permitted by the terminal-position semantics.
Then $s=\langle\Phi,M_g\rangle$ is a final state.
Let $\pi=(C_c,\Lambda_c)\,s$.
Then $\pi$ is an accepting path starting from $s$.
By definition, $\ftrace{\pi}=C_c,\Lambda_c.L_c$.
Since $\sigma,\partition{J},\alpha,i\models_\tau\Phi$,
we conclude $\sigma,\partition{J},\alpha,i\models_\tau\ftrace{\pi}$.

\item
(Inductive case)
Suppose the claim holds for $i+1$.
We show the claim for $i$.
Assume $\sigma,\partition{J},\alpha,i\models_\tau M_g$ and
$\predInv_{\intvOf{J}{i}}(L)$ is satisfied by $\alpha[\Lambda]$, where $M_g=C,\Lambda.L$.
Choose the closure constructed in the proof of
\Cref{thm:closure-completeness}, namely a closure
$\Phi$ of $M_g$ such that
$\sigma,\partition{J},\alpha,i\models_\tau\Phi$
and its current and next predicate invariants are satisfied.
By the subformula restriction in the tableau construction, the STL formulas
underlying the interval predicates of $M_g$ are subformulas of $\varphi$.
Thus the fully stable partition for $\varphi$ also supplies the stability
needed for closure completeness.
Expansion introduces only predicates whose underlying STL formulas are
subformulas of those in $M_g$, so the next part of $\Phi$, when present,
is again a state-form judgement in the same class.
If $\Phi$ is in state form, then $s=\langle\Phi,M_g\rangle$ is final,
and the one-state path with the incoming clock constraints and resets
of $\Phi$ proves the claim, just as in the base case.
Otherwise, write $\Phi=C_c,\Lambda_c.(L_c,\nextop M_g')$.
Let $M_g'=C',\Lambda'.L'$.
Let $\alpha'=\alpha[\Lambda_c]+|\intvOf{J}{i}|$.
Then $\sigma,\partition{J},\alpha',i+1 \models_\tau M_g'$.
The invariant-preserving construction in the proof of
\Cref{thm:closure-completeness} gives the current invariant of $\Phi$ under
$\alpha[\Lambda_c]$ and its successor invariant under $\alpha'[\Lambda']$.
If $i+1<|\partition{J}|$, then
$\predInv_{\intvOf{J}{i+1}}(L')$ is satisfied by $\alpha'[\Lambda']$.
If $i+1=|\partition{J}|$, satisfaction of the next part permits only
terminal clock and type judgements and $\ast$, not interval predicates,
so the terminal case applies without an invariant premise.
No invariant assertion about arbitrary closures is used.
By the induction hypothesis applied to $M_g'$ at position $i+1$,
there exist a closure $\Phi'$ of $M_g'$ and an accepting path $\pi'$
starting from $s'=\langle\Phi',M_g'\rangle$ such that
$\sigma,\partition{J},\alpha',i+1 \models_\tau \ftrace{\pi'}$.
Let $s=\langle\Phi,M_g\rangle$ and define
$\pi=(C_c,\Lambda_c)\,s\,\pi'$.
By definition of tableau transitions, $\pi$ is an accepting path starting from $s$.
Since $\ftrace{\pi} = C_c,\Lambda_c.(L_c,\nextop\ftrace{\pi'})$,
together with $\sigma,\partition{J},\alpha,i \models_\tau C_c,\Lambda_c.L_c$
and $\sigma,\partition{J},\alpha',i+1 \models_\tau \ftrace{\pi'}$,
we conclude $\sigma,\partition{J},\alpha,i \models_\tau \ftrace{\pi}$.
\end{proofrules}

For a path constructed from the initial generator $\ipstl{\varphi}$,
the full trace, and hence the trace, has no free clocks.
Indeed, each rule branch uses only clocks inherited from its source
predicate or clocks bound by its own resets.
The compatible bound-clock renaming and reset merging preserve these
bindings. Choosing a $\sqsubseteq$-minimal judgement in the closure
construction cannot expose a formerly bound
clock that still occurs, by the no-capture conditions of subsumption.
Thus closure construction and path concatenation introduce no new
free clocks. Since $\ipstl{\varphi}$ has none, every clock occurrence
in the constructed full trace remains bound by an enclosing clock reset.
By \Cref{lem:irrelevant-reset}, the satisfaction of this fixed path's
trace is therefore independent of the initial clock valuation.
\qedhere
\end{proof}

\subsection{Finiteness of the Tableau}
\label{sec:app-tab-finiteness}

\begin{restatable}{lemma}{boundedalter}
\label{lem:bounded-alternation-stl}
Let $\varphi = \bothtemp_I \psi$ with $\bothtemp \in \{\always{}, \eventually{}\}$,
where $I \subseteq \mathbb{R}_{\ge 0}$ is a non-singular closed interval.
If $\sup I<\infty$, the number of satisfaction changes of
$\varphi$ on a signal over any interval of length $\sup I$ is
$O(\lceil \frac{\sup I}{\sup I-\inf I} \rceil)$.
If $\sup I=\infty$, $\varphi$ changes its satisfaction at most
once over the signal domain.
\end{restatable}

\begin{proof}
We first prove the claim for $\eventually{I}\psi$.
Let $I=[a,b]$. Suppose $b<\infty$. Fix an interval $K\subseteq\domain{0}{\tau}$ of
length at most $b$, and let $T_1,\dots,T_m$ be any $m$ distinct maximal subintervals of $K$
on which $\eventually{I}\psi$ is satisfied. For each $i$, choose
$t_i\in T_i$ and a witness $w_i\in t_i+I$ such that
$\sigma,w_i\models_\tau\psi$. Then every $s\in(w_i-I)\cap\domain{0}{\tau}$ also satisfies
$\eventually{I}\psi$, since $w_i\in s+I$. Hence $(w_i-I)\cap\domain{0}{\tau}$ is a
satisfied interval containing $t_i$, and $w_i-I$ has length $b-a$.

At most two witness intervals $w_i-I$ are not contained in $K$:
otherwise two would contain points outside $K$ on the same side and
overlap inside $K$, contradicting the distinctness of the maximal
true intervals $T_i$.
Suppose
$m>\lceil b/(b-a)\rceil+2$. After excluding those at most two
intervals, more than $\lceil b/(b-a)\rceil$ intervals of length
$b-a$ remain inside $K$, so two must overlap. Hence
$(w_i-I)\cap(w_j-I)\neq\emptyset$ for some $i\neq j$. Since
$t_i\in w_i-I$ and $t_j\in w_j-I$, this implies
$T_i\cap T_j\neq\emptyset$, contradicting the maximality of the $T_i$'s.
Since this bound holds for every finite selection, the total number
of maximal satisfied subintervals of $K$ is at most
$\lceil b/(b-a)\rceil+2$. Since each maximal satisfied
interval contributes at most two satisfaction changes, the
number of satisfaction changes of $\eventually{I}\psi$ inside $K$ is
also $O(\lceil b/(b-a)\rceil)$.

Now suppose $b=\infty$, so $I=[a,\infty)$. If
$\sigma,t\models_\tau\eventually{I}\psi$, then there exists
$w\ge t+a$ such that $\sigma,w\models_\tau\psi$. Hence for every
$s\le t$ in the signal domain, the same $w$ witnesses
$\sigma,s\models_\tau\eventually{I}\psi$. Therefore
$\eventually{I}\psi$ can change its satisfaction at most once.

Finally,
$\always{I}\psi\equiv\neg\eventually{I}\neg\psi$, and negation
preserves the number of satisfaction changes. Hence the same bounds
hold for $\always{I}\psi$.
\end{proof}

\paragraph{Construction of the simplified tableau.}
We combine interval predicates using \Cref{lem:intv-merge} when
$d_1,c_2$ are reset to $0$ and its side conditions hold, and rewrite
judgements using \Cref{prop:judg-equiv}. Before combining predicates
for $\bothtemp_I\psi$, the judgement $\ipstl{\bothtemp_I\psi}$ may
be added if absent.
For a state-form judgement $M$, let $\mathcal C(M)$ contain its original closures
and all completed transition-form judgements obtained from $M$ using the
transformations above. Unused resets are removed.

\begin{definition}[Simplified tableau]\label{def:intermediate-tableau}
Given $\tabG{\varphi}=(X,X_c,S,T,T_0,S_F,\mathcal L)$, for each
$s=\langle\Phi,M_g\rangle\in S$, let
$\tilde s=\langle\tilde\Phi,M_g\rangle$ range over all choices of
$\tilde\Phi\in\mathcal C(M_g)$.
The \emph{simplified tableau} of $\tabG{\varphi}$ is
$\simptabG{\varphi}=(X,X_c,\tilde S,\tilde T,\tilde T_0,
\tilde S_F,\tilde{\mathcal L})$, where:
\begin{itemize}
\item $X$ is the set of real-valued variables occurring in $\varphi$;
\item $X_c$ is a set of clock variables;
\item $\tilde S=\{\tilde s\mid s\in S\}$, including every choice
of the operation above;
\item
$(\langle\tilde\Phi,M_g\rangle,(C',\Lambda'),
\langle\tilde\Phi',M_g'\rangle)\in\tilde T$
iff $\tilde\Phi=C,\Lambda.(L,\nextop M)$,
$\tilde\Phi'=C',\Lambda'.(L',\nextop M')$, and
$\tilde\Phi'\in\mathcal C(M)$;
\item
$((C_0,\Lambda_0),\langle\tilde\Phi_0,M_g\rangle)\in\tilde T_0$
iff $\tilde\Phi_0\in\mathcal C(\ipstl{\varphi})$ and
$\tilde\Phi_0=C_0,\Lambda_0.(L_0,\nextop M_g')$;
\item $\langle\tilde\Phi,M_g\rangle\in\tilde S_F$
iff $\tilde\Phi$ is in state form;
\item
$\tilde{\mathcal L}(\langle\tilde\Phi,M_g\rangle)
=\{\,l\mid\tilde\Phi=C,\Lambda.(L,\nextop M),\;l\in L,\;
l\text{ is }\ast,\ \lty{\upLc},\text{ or an atomic predicate}\,\}$.
\end{itemize}
\end{definition}

The full trace is defined as in the original tableau.

Each state retains its generator as a second component. The local
argument uses the actual incoming requirement of its first component;
the stored generator does not constrain an edge.

\begin{lemma}\label{lem:intermediate-sound}
Let $\pi$ be an accepting run of $\simptabG{\varphi}$.
If $\sigma,\partition{J}, \alpha, 0 \models_\tau \trace{\pi}$, then
$\sigma,\partition{J}, \alpha, 0 \models_\tau \ftrace{\pi}$
and $\sigma,0\models_\tau\varphi$.
\end{lemma}
\begin{proof}
Write $\pi=(C_0,\Lambda_0)s_0\cdots(C_n,\Lambda_n)s_n$, with
$s_i=\langle\tilde\Phi_i,M_g^i\rangle$ and
$\tilde\Phi_i=C_i,\Lambda_i.(L_i,\nextop M_{i+1})$,
omitting the next part at $i=n$. Put
\[
M_0=\ipstl{\varphi},\qquad A_i=\tilde{\mathcal L}(s_i),\qquad
\alpha_0=\alpha,\quad
\alpha_{i+1}=\alpha_i[\Lambda_i]+|\intvOf{J}{i}|.
\]
By \Cref{def:intermediate-tableau} and the trace hypothesis,
\begin{equation}\label{eq:intermediate-atomic}
\tilde\Phi_i\in\mathcal C(M_i),\qquad
\sigma,\partition{J},\alpha_i,i\models_\tau C_i,\Lambda_i.A_i.
\end{equation}

The proof of \Cref{cor:closure-invariant} extends to these
constructions. Indeed, combination preserves the left clock and
marker and enlarges the interval represented by the scope, preserving
conditions \invmark{1} and \invmark{3}; an added predicate of the form $\ipstl{\psi}$ has invariant
$\true$. Rewriting and unused resets are handled by
\Cref{prop:judg-equiv,lem:reset-agreement,lem:irrelevant-reset}.
The required guards and type judgements follow backwards from
\eqref{eq:intermediate-atomic} and the next input, using
\Cref{lem:type-subsumption,lem:tnf_subsumption} for expansions;
combination preserves the surrounding conditions.
Starting with the trivial invariant of $M_0$, forward induction
therefore establishes conditions \invmark{1} and \invmark{3}
at every construction step, in particular for $L_i$ at
$\alpha_i[\Lambda_i]$.

We prove backwards on $i$ that
\begin{equation}\label{eq:intermediate-backward}
\sigma,\partition{J},\alpha_i,i\models_\tau
\ftrace{\pi^i},M_i.
\end{equation}
By \eqref{eq:intermediate-atomic} and the induction hypothesis,
\[
\sigma,\partition{J},\alpha_i,i\models_\tau
C_i,\Lambda_i.(A_i,\nextop M_{i+1}),
\]
with the next part omitted in the base case $i=n$.
Fix a finite completed construction witnessing
$\tilde\Phi_i\in\mathcal C(M_i)$ and induct on this construction.
Completion expands the current non-atomic predicates, leaving the
next judgement unexpanded. Its atomic and next parts are satisfied
by the preceding display; its guards, resets, and type judgements
are retained under the rewritings of \Cref{prop:judg-equiv}.
For an expansion $p\to\Psi_1\mid\cdots\mid\Psi_k$, apply the
induction hypothesis to the completed selected branch $\Psi_j$
and its surrounding context. This recovers $\Psi_j$ without
assuming the retained source predicate $p$. The established
invariants and \Cref{thm:expansion-rule-soundness} then give $p$.
This is the local recovery argument of
\Cref{thm:closure-soundness}; current branch predicates have smaller
complexity by \Cref{lem:complexity-reduced}.
If branch predicates were subsequently combined, first recover
the combined predicate by induction, then recover its inputs by
\Cref{lem:intv-merge} under its side conditions. If a predicate of the form
$\ipstl{\psi}$ was added, discard that conjunct when recovering the
preceding input. Rewriting and unused resets preserve satisfaction
by \Cref{prop:judg-equiv,lem:irrelevant-reset}.
Thus the induction recovers both the retained current predicates
and the input, without requiring $\tilde\Phi_i$ itself to be an
original closure:

\[
\sigma,\partition{J},\alpha_i,i\models_\tau
C_i,\Lambda_i.L_i
\quad\text{and}\quad
\sigma,\partition{J},\alpha_i,i\models_\tau M_i.
\]
Together with the full-trace assertion at $i+1$, this proves
\eqref{eq:intermediate-backward}; at $i=n$ it holds directly.
Taking $i=0$ gives the full trace and
$\ipstl{\varphi}$, hence $\sigma,0\models_\tau\varphi$.
\end{proof}

\restatetheorem{lem:simp-trace-equiv}

\begin{proof}
Fix a signal $\sigma$ with finite variability for $\varphi$,
a fully stable partition $\partition{J}$ for $\varphi$,
and a clock valuation $\alpha$.
Precisely, we prove that there exists an accepting run $\pi$ of
$\tabG{\varphi}$ with
$\sigma,\partition{J},\alpha,0\models_\tau\ftrace{\pi}$
if and only if there exists an accepting run $\tilde\pi$ of
$\simptabG{\varphi}$ with
$\sigma,\partition{J},\alpha,0\models_\tau\ftrace{\tilde\pi}$.

$(\Rightarrow)$ Let $\pi$ be an accepting run of $\tabG{\varphi}$
with $\sigma,\partition{J},\alpha,0\models_\tau\ftrace{\pi}$.
Every original closure of $M$ belongs to $\mathcal C(M)$.
Hence, by \Cref{def:intermediate-tableau}, the same states and
transitions form an accepting run $\tilde\pi$ of
$\simptabG{\varphi}$ with $\ftrace{\tilde\pi}=\ftrace{\pi}$.

$(\Leftarrow)$ Let $\tilde\pi$ be an accepting run of
$\simptabG{\varphi}$ with
$\sigma,\partition{J},\alpha,0\models_\tau\ftrace{\tilde\pi}$.
Since $\trace{\tilde\pi}\sqsubseteq\ftrace{\tilde\pi}$,
\Cref{lem:tnf_subsumption,lem:intermediate-sound} give
$\sigma,0\models_\tau\varphi$.
By \Cref{thm:tableau-complete}, there is an accepting run $\pi$
of $\tabG{\varphi}$ with
$\sigma,\partition{J},\alpha,0\models_\tau\trace{\pi}$;
\Cref{lem:trace-equivalence} yields
$\sigma,\partition{J},\alpha,0\models_\tau\ftrace{\pi}$.
\end{proof}

\restatetheorem{lem:bounded-accepting-run}

\begin{proof}[Proof sketch]
Fix a signal $\sigma$ with finite variability for $\varphi$, a fully
stable partition $\partition{J}$, and a clock valuation $\alpha$
at which the given run's full trace is satisfied.
By \Cref{lem:intermediate-sound}, $\sigma,0\models_\tau\varphi$.
We follow the accepting-path construction of
\Cref{lem:generator-path-existence}, making the following choices.

Process each formula before its proper subformulas, sharing identical
judgements $\ipstl{\psi}$ by \Cref{prop:judg-equiv}.
While a timed subformula $\theta=\bothtemp_I\psi$ remains true,
combine its incoming predicate
$\intvpredT{c}{d}{\upLc}{\dnLc}{\theta}$ with $\ipstl{\theta}$ before
expansion, adding the latter if absent. The preceding branch for
$\intvpredupT{c}{\upLc}{\theta}$
supplies the zero right reset and complementary endpoint marker
required by \Cref{lem:intv-merge}; its side conditions also hold for
singleton partition intervals. The combination preserves the full
predicate invariant, so \Cref{prop:pred_invariant} supplies a satisfied
expansion branch. Each branch for either clocked form has at most one
successor predicate of the form $\intvpredT{c}{d}{\upLc}{\dnLc}{\theta}$,
so at most one such predicate tracks each interval
over which $\theta$ remains true.

These choices yield completed outputs with literally matching next
inputs. Each satisfied input has an original closure by
\Cref{lem:generator-path-existence}, so the resulting pairs are states
of $\simptabG{\varphi}$ by \Cref{def:intermediate-tableau}.
The same accepting-path argument then gives an accepting run whose
full trace is satisfied on the fixed $\sigma,\partition{J},\alpha$.

To count predicates of the form $\intvpredT{c}{d}{\upLc}{\dnLc}{\theta}$,
suppose one still tracks an interval on which $\theta$ was true and
which ended at $e$. Its clock $d$ was reset at $e$ and is not reset
by subsequent expansions of such predicates.
At current time $u$, its value is therefore $u-e$.
For $\sup I<\infty$, the invariants in
\Cref{tab:predicate-invariants} give $u-e\le\sup I$.
Thus \Cref{lem:bounded-alternation-stl} bounds the number of such
predicates by $O(r(I))$, with at most one more for the interval on
which $\theta$ is currently true. For $\sup I=\infty$, the truth
changes at most once, giving the same bound with $r(I)=1$.

The current part retains at most one predicate per incoming predicate
$\intvpredT{c}{d}{\upLc}{\dnLc}{\theta}$,
one $\intvpredupT{c}{\upLc}{\theta}$ per timed subformula, and one
judgement $\ipstl{\psi}$ per subformula. Summing gives $O(|\varphi|r(\varphi))$
interval predicates per state. The stored input and the next judgement
satisfy the same bound.
\end{proof}

\restatetheorem{thm:finite-tableau}

\begin{proof}
Let $\tableau{\varphi}{}$ be the finite subgraph described above.
By \Cref{lem:bounded-accepting-run}, satisfied accepting runs can
be chosen with $O(r(I))$ predicates per timed subformula
$\bothtemp_I\psi$. Assign a fixed collection of $O(r(I))$ clock
triples to each such subformula. In each triple, use the first clock $c$
in $\intvpredupT{c}{\upLc}{\bothtemp_I\psi}$ and pair it with either
remaining clock $d$ in $\intvpredT{c}{d}{\upLc}{\dnLc}{\bothtemp_I\psi}$.
Names are retained while occurring in the next judgement, including
its constraints and resets, and reused afterwards. The bounded number
of incoming predicates and expansions ensures that sufficiently many
triples and a fresh right clock are always available.
Each triple permits constantly many predicates, constraints, and resets,
using only zero and the finite endpoints of $I$.
Together with the judgements $\ipstl{\psi}$ for subformulas $\psi$
and the type and terminal judgements,
this gives $O(|\varphi|\cdot r(\varphi))$ possible entries in the
current part, next part, and stored generator.
Thus the tableau uses $O(|\varphi|\cdot r(\varphi))$ clocks and has
at most $2^{O(|\varphi|\cdot r(\varphi))}$ states.
The assignments preserve transitions and satisfaction; all introduced
clocks are reset, so the selected runs work for every initial valuation.
Their actual incoming judgements can serve as stored generators by
\Cref{lem:generator-path-existence}.
It remains to prove (1) and (2).

\begin{enumerate}[label=(\arabic*)]
\item
Assume that $\sigma$ is recognized by $\tableau{\varphi}{}$
with respect to $\partition{J}$.
Its accepting run is also a run of $\simptabG{\varphi}$ with the
same satisfied trace. Hence \Cref{lem:intermediate-sound} gives
$\sigma,0\models_\tau\varphi$.

\item
Assume that $\sigma$ has finite variability for $\varphi$ and
$\sigma,0\models_\tau\varphi$, and fix any fully stable partition
$\partition{J}$.
By \Cref{thm:tableau-complete,lem:trace-equivalence,lem:simp-trace-equiv},
$\simptabG{\varphi}$ has an accepting run with a satisfied full
trace on $\sigma,\partition{J}$.
By \Cref{lem:bounded-accepting-run} and the assignment above,
such a run $\tilde\pi$ exists within $\tableau{\varphi}{}$ for
which $\sigma,\partition{J},\alpha,0\models_\tau\trace{\tilde\pi}$
for every initial valuation $\alpha$.
Thus $\sigma$ is recognized by $\tableau{\varphi}{}$ with respect
to $\partition{J}$.
\qedhere
\end{enumerate}
\end{proof}

\clearpage
\section{Constructing Hybrid Automata for STL}
\label{sec:app-automata}

Induced partitions relate automaton executions to tableau runs in the translation of \Cref{sec:transform}. We use this correspondence to prove the correctness results required for model checking.

\subsection{Executions and Induced Partitions}
\label{sec:app-translate-ha}

\paragraph{$n$-bounded extension.}
\label{app:n-bounded-extension}
In $\hat H^{(n)}$, initial discrete transitions are represented by $\init$,
and each sequence of at most $n$ consecutive zero-time jumps by one added
discrete transition. System executions can thus be taken to alternate
continuous and discrete transitions, starting with a continuous transition.

\paragraph{Goal reachability.}
A goal witness is an execution of $\hatableau{\varphi}{\tau}$ ending in
a mode in $S_F$ at some $b\le\tau$. Its projection onto $X$ agrees with
$\sigma$ on $[0,b)$ and also at $b$ for a final zero-time visit before
$\tau$. A terminal jump at $\tau$ follows the last positive flow
and has target $q'\in S_F$ with $\mathcal L(q')\subseteq\{\ast,\lty{\strue}\}$.
For the last flow of duration $r$ ending at $\tau$, we check $\flow$ on
$[0,r]$ and $\inv$ on $[0,r)$. The execution can end at the endpoint
valuation or use it for the guard and reset of a terminal jump.
At the target of such a jump, we check the invariant with the clock bound
$c_g\le\tau$ in place of $c_g<\tau$.

\begin{definition}[Boundary forms and induced partition]
\label{def:translation-induced-partition}
The boundary forms of a partition are $\ldots)\{t\}(\ldots$ at a
singleton and $\ldots)[\ldots$ at every other internal boundary $t$.
A partition begins with either a singleton at $0$ or a left-closed
non-singular interval; its last interval is right-open.

Retain every mode visit before $\tau$ through the chosen cut.
At an internal boundary $t$, one jump separates two positive visits;
two jumps at $t$ give a single zero-time visit. At most two jumps are
allowed at such a boundary, and at most one at $0$.
Assign a zero-time visit at $t$, including a final one before $\tau$,
the singleton $\{t\}$. There is at most one such visit at each time.
Assign a positive visit from $a$ to $b$ the interval $[a,b)$, or
$(a,b)$ if it follows a singleton.

For an execution observing $\sigma$ to induce a partition, every tableau
left-end requirement must agree with the assignment. At each singleton
$\{t\}$, the zero-time visit must have $X$-valuation $\sigma(t)$,
as must the start of the following continuous transition when present.
This is a condition on how the execution observes the signal, not a
reset of $X$. No marker variable is added to the automaton.

If the witness ends at $b<\tau$, append $[b,\tau)$ to the partition
after a positive flow and $(b,\tau)$ after a zero-time visit.
For a terminal jump, take the witness through that jump. Its tableau
edge is the last edge of the run, with its target at the terminal
position after the last interval.
A trajectory induces exactly the partitions obtained in this way
from its goal witnesses and cuts.
\end{definition}

For \Cref{lem:well-behaved-suff} and model checking, the system execution
observes the signal throughout $\domain{0}{\tau}$. In \Cref{sec:red_thm},
a goal-reaching execution of
$\composeha{\varphi}{\tau}=H\parallel\hatableau{\varphi}{\tau}$
combines a prefix of such a system execution with an execution of the
translated automaton that induces a partition of the same signal.
The translated component uses the terminal convention above.
The last flow of $H$ satisfies its flow and invariant
conditions through the endpoint, and $H$ stutters there during the
terminal jump. This jump can keep the shared valuation because its
conditions concern only clocks and its target has no atomic predicates.
We assume that the system's continuous transitions are closed under
restriction. For Boolean STL model checking, the tableau input is a
Boolean-equivalent normalization of $\neg\varphi$ with only non-strict
atomic predicates defined by continuous functions. Finite variability for
$\varphi$ carries over to this input. In the robust no-goal case, finite
variability is with respect to the formula used to construct
$\tableau{\neg(\varphi^{+\epsilon})}{}$.

For the contrapositive of \Cref{sec:red_thm}(2), refine a fully stable
partition at the system execution's jump times. Then replace each
right-closed non-singular interval ending at $t$ by its right-open part
and $\{t\}$.
This preserves full stability. Use the accepting run given by
\Cref{thm:finite-tableau} on this refinement in the construction of
\Cref{lemma:translate-correct}(1). On $X$, take the restrictions of the
system flows, including their endpoint valuations. Continuity ensures
that the non-strict atomic predicates hold at both endpoints.
In the synchronized prefix, each system jump coincides with the first
translated jump at its boundary. At the other translated jumps, $H$ stutters.

\subsection{Proofs of Correctness}

\restatetheorem{lem:well-behaved-suff}

\begin{proof}
Let $\sigma$ be a system trajectory witnessed by an execution covering
$\domain{0}{\tau}$.
We show that $\sigma$ is well-behaved with respect to $\varphi$.
This execution gives times $0=t_1<\cdots<t_{n+1}=\tau$
and evolutions $\gamma_i:[0,r_i]\to\mathbb{R}^X$, where
$r_i=t_{i+1}-t_i$, such that $\sigma(t_i+u)=\gamma_i(u)$ for
$u\in[0,r_i)$. At $t_1=0$ this equality holds because the first continuous
transition starts at the initial state, as in the
\hyperref[app:n-bounded-extension]{paragraph on the $n$-bounded extension}.
The flow conditions make each $\gamma_i$ continuous.

Let $t\in\domain{0}{\tau}$.
Choose $i$ with $t\in[t_i,t_{i+1})$. For all sufficiently small
$h>0$, both $\sigma(t)$ and $\sigma(t+h)$ are given by $\gamma_i$.
Hence, by continuity of $\gamma_i$,
$\lim_{h\downarrow0}\sigma(t+h)=\sigma(t)$.
Thus $\sigma$ is right-continuous.
For every function $f$ appearing in $\varphi$, since $f$ is continuous,
$f\circ\sigma_f$ is right-continuous.
Therefore, $\sigma$ is well-behaved with respect to $\varphi$.
\qedhere
\end{proof}

\restatetheorem{lemma:translate-correct}

\begin{proof}
Let $\tableau{\varphi}{} = (X, X_c, S, T, T_0, S_F, \mathcal{L})$
and $\hatableau{\varphi}{\tau} = (Q, G, E, X^\sharp, \init, \inv, \flow, \jump)$.
\begin{enumerate}
  \item 
Assume that $\sigma$ is recognized by $\tableau{\varphi}{}$
with respect to a partition $\partition{J}$ of $\domain{0}{\tau}$.
Fix an accepting run
$\pi=(C_0,\Lambda_0)s_0\cdots(C_n,\Lambda_n)s_n$
such that, for every clock valuation $\alpha$,
$\sigma,\partition{J},\alpha,0\models_\tau\trace{\pi}$.
Since $Q=S$ and $E=T$, the same modes and edges occur in
$\hatableau{\varphi}{\tau}$. Let $q_i=s_i$ and $A_i=\mathcal L(q_i)$.
We use the given partition and accepting path, without requiring the
partition to have the boundary forms of an induced partition.
Each $\nextop$ advances the partition position once, so $n\le|\partition{J}|$.
Put $m=\min\{n,|\partition{J}|-1\}$.
The visits $q_0,\ldots,q_m$ use the endpoints of $\partition{J}$ as jump
times. If $n=|\partition{J}|$, the last edge is taken at $\tau$.

Fix an arbitrary clock valuation
$\alpha:X_c\to\realset_{\ge0}$ and define
$\alpha_0=\alpha$ and
$\alpha_{i+1}=\alpha_i[\Lambda_i]+|\intvOf{J}{i}|$ for $i<n$.
Thus $\alpha_i$ is the clock valuation immediately before entering
$q_i$. Since
$\sigma,\partition{J},\alpha_i,i\models_\tau C_i,\Lambda_i.A_i$,
we have $\sigma,\partition{J},\alpha_i,i\models_\tau C_i$ and
$\sigma,\partition{J},\alpha_i[\Lambda_i],i\models_\tau A_i$.

For each interval position $i$, let $t_i=\inf\intvOf{J}{i}$.
The terminal position has no corresponding interval.
For $0\le i\le m$, define the entry valuation $v_i:X^\sharp\to\realset$ by
$v_i(x)=\sigma(t_i)(x)$ for $x\in X$,
$v_i(c)=\alpha_i[\Lambda_i](c)$ for $c\in X_c$,
and $v_i(c_g)=t_i$. No value of $\sigma(\tau)$ is used.

We construct the execution
\[
\langle q_0,v_0\rangle
 \xrightarrow{r_0;\gamma_0}\langle q_0,v_0'\rangle
 \xrightarrow{\boldsymbol{\cdot}}\cdots
 \xrightarrow{\boldsymbol{\cdot}}\langle q_m,v_m\rangle
 \xrightarrow{r_m;\gamma_m}\langle q_m,v_m'\rangle,
\]
where $r_i=|\intvOf{J}{i}|$ for interval visits.
Zero-duration flow arrows are omitted, so the execution may start or
end with a discrete transition.

We first show that $\init$ holds.
Let $v:X^\sharp\to\realset$ be defined by
$v(x)=\sigma(0)(x)$ for $x\in X$,
$v(c)=\alpha_0(c)$ for $c\in X_c$,
and an arbitrary value on $c_g$.
Since $v\restriction X_c=\alpha_0$, we also have
$\sigma,\partition{J},v\restriction X_c,0\models_\tau C_0$.
Moreover, since $t_0=0$,
$v_0=v[\Lambda_0][\cset{\creset{c_g}{0}}]$.
Thus $\init(q_0,v_0)$.
Now fix $0\le k\le m$.
We verify $\inv$, $\flow$, and the jumps between interval visits.

\begin{proofrules}[topsep=3pt,itemsep=3pt]
\item $(\inv)$
For every $\ipstl{f(\vec x)\ge0}\in A_k$,
$\sigma,\partition{J},\alpha_k[\Lambda_k],k\models_\tau A_k$
implies $f(\sigma(t))\ge0$ for all $t\in\intvOf{J}{k}$.
If $\intvOf{J}{k}$ is left-open, well-behavedness implies that the same
predicates also hold at the left endpoint. Hence every valuation induced
by $\sigma$ satisfies these predicates with the left endpoint included.
The global clock satisfies $c_g<\tau$ at every time before $\tau$,
including entry times and all flow endpoints before $\tau$.
The last flow ending at $\tau$ uses the invariant convention of
\Cref{sec:app-translate-ha}.

\item $(\flow)$
If $\intvOf{J}{k}$ is singleton, no continuous evolution is needed and
we put $v_k'=v_k$.
Otherwise, use the duration $r_k$ chosen above.
If $t_k+r_k\in\intvOf{J}{k}$, take $w=\sigma(t_k+r_k)$.
Otherwise, take $w=\sigma(t_k)$. In both cases $w$ satisfies the atomic
predicates in $A_k$.
The flow observes the signal on $[t_k,t_k+r_k)$.
Define
$\gamma_k:[0,r_k]\to\realset^{X^\sharp}$ by
$\gamma_k(r)(x)=\sigma(t_k+r)(x)$ for $x\in X$ and $r<r_k$,
$\gamma_k(r_k)(x)=w(x)$ for $x\in X$,
$\gamma_k(r)(c)=v_k(c)+r$ for $c\in X_c$,
$\gamma_k(r)(c_g)=t_k+r$.
The flow condition and the atomic invariants of $q_k$ hold throughout
$[0,r_k]$. Put $v_k'=\gamma_k(r_k)$.

\item $(\jump)$
If $k<m$, the clock part of $v_k'$ is $\alpha_{k+1}$ and
$v_k'(c_g)=t_{k+1}$, so
$\sigma,\partition{J},\alpha_{k+1},k+1\models_\tau C_{k+1}$.
The clock part satisfies
\[
 v_{k+1}\restriction(X_c\cup\{c_g\})
 =v_k'[\Lambda_{k+1}]\restriction(X_c\cup\{c_g\}).
\]
Trace satisfaction excludes $\ast$ from $A_{k+1}$, so the tableau edge
is enabled at $t_{k+1}<\tau$.
If the source is a singleton, $t_k=t_{k+1}$ and both $v_k'$ and $v_{k+1}$
observe $\sigma(t_k)$ on $X$. These jumps do not test the type labels.
\end{proofrules}
The flows and zero-time visits observe $\sigma$ on the executed prefix.
If $n<|\partition{J}|$, the execution ends in $q_n\in S_F$, observing
$\sigma(t_n)$ at entry when the final visit is a singleton.

Otherwise, $n=|\partition{J}|$ and the final tableau state $q_n$
is at the terminal position.
The last interval is nonempty and right-open, hence has positive duration.
With its end valuation $v_{n-1}'=\gamma_{n-1}(r_{n-1})$, we have
$v_{n-1}'(c_g)=\tau$ and
$v_{n-1}'\restriction X_c=\alpha_n$.
The accepting path supplies
$(q_{n-1},(C_n,\Lambda_n),q_n)\in T$ and $q_n\in S_F$.
Trace satisfaction supplies $\alpha_n\models C_n$ and
$A_n\subseteq\{\ast,\lty{\strue}\}$, the only labels true at the
terminal position. Put $w=v_{n-1}'[\Lambda_n]$.
The clock guard and reset hold, and $w(c_g)=\tau$ satisfies the horizon
guard of the jump and the closed clock bound at the target.
With the terminal condition on $q_n$, the last flow and the jump along
$(q_{n-1},(C_n,\Lambda_n),q_n)$ yield a terminal step to
$\langle q_n,w\rangle$ at $\tau$.

\item
Fix an induced partition $\partition{J}$ and choose a goal-reaching
execution of $\hatableau{\varphi}{\tau}$ observing $\sigma$
and a cut that induce it.
We show that $\sigma$ is recognized by $\tableau{\varphi}{}$
with respect to this $\partition{J}$: there is an accepting run $\pi$
such that $\sigma,\partition{J},\alpha,0\models_\tau\trace{\pi}$
for every clock valuation $\alpha$.

\begin{proofrules}[topsep=3pt,itemsep=3pt]
  \item (existence of $\pi$) 
Retain the visits $q_0,\ldots,q_n$ through the goal, treating a terminal
jump at $\tau$ separately.
For each such visit let $(C_i,\Lambda_i)$ be its entering label,
$t_i$ its entry time, and $r_i$ its residence time.
Then $t_{i+1}=t_i+r_i$.
The global clock records elapsed time: it starts at zero, advances at
rate one, and is unchanged by jumps.
Every jump between the retained visits occurs before $\tau$.
The corresponding modes and edges are in
$S$ and $T$, and $\init$ gives $((C_0,\Lambda_0),q_0)\in T_0$.
Thus the observed path
$\pi_{\mathrm{obs}}=(C_0,\Lambda_0)q_0\cdots(C_n,\Lambda_n)q_n$
is a tableau run.
If the last retained visit is in $S_F$ and ends the witness,
put $\pi=\pi_{\mathrm{obs}}$.
For a terminal jump $(q_n,(C',\Lambda'),q')$, let $v_n'$ be the
end valuation of the last flow. Its guard gives
$v_n'\restriction X_c\models C'$, and the terminal convention gives
$q'\in S_F$ and $\mathcal L(q')\subseteq\{\ast,\lty{\strue}\}$.
Use this jump as the last tableau edge, putting
$\pi=\pi_{\mathrm{obs}}(C',\Lambda')q'$. In either case $\pi$ is accepting.
  \item (properties of the induced partition $\partition{J}$)
Each positive visit gives a right-open interval, left-open precisely
after a zero-time visit. At such a boundary the singleton owns the
point; at any other boundary the interval on the right owns it.
The first piece includes $0$. With the suffix when the cut is early,
the pieces are nonempty, disjoint, and cover $[0,\tau)$.
The witness is finite, so this is a finite induced partition.
For a terminal jump, its target is at the terminal position.
  \item  (satisfaction of $\trace{\pi}$)
Fix an arbitrary clock valuation $\alpha$ and put $\alpha_0=\alpha$.
For $i<n$, put
$\alpha_{i+1}=\alpha_i[\Lambda_i]+|\intvOf{J}{i}|
=\alpha_i[\Lambda_i]+r_i$.
For a terminal jump also put
$\alpha_{n+1}=\alpha_n[\Lambda_n]+r_n$.
By construction of the tableau, $\trace{\pi}$ has no free clocks.
Hence each clock read by $C_i$ was reset strictly before position $i$
(the guard precedes $\Lambda_i$). At its last such reset, the clock
received the same value in the execution and in the trace; since then
the same residence times have elapsed. Thus $\alpha_i$ agrees with
the execution on every clock read by $C_i$. The entering jump therefore
gives $\sigma,\partition{J},\alpha_i,i\models_\tau C_i$; at $i=0$
there is no clock read and the guard holds by $\init$.

Since every label $A_i=\mathcal L(q_i)$ consists only of atomic
predicates of the form $\ipstl{f(\vec x)\ge0}$, the proposition
$\ast$, and interval-type propositions $\lty{\cdot}$, it suffices to
verify these cases.

If $\intvOf{J}{i}=\{t_i\}$, the entry valuation $v_i$ of $q_i$ has
$v_i\restriction X=\sigma(t_i)$ by \Cref{def:translation-induced-partition}.
Thus $\inv(q_i,v_i)$ gives $f(\sigma(t_i))\ge0$ for every
$\ipstl{f(\vec x)\ge0}\in A_i$.
On a positive visit, atomic truth follows from the flow invariant on its
observed interval; at time $0$ it follows from the initial state's
invariant. No positive visit owns its right endpoint.

The induced partition agrees with every tableau left-end requirement,
so every type label holds.
The rules introduce $\ast$ only under $\nextop$.
The global-clock guard excludes $\ast$ from later entries strictly before
$\tau$.
The last flow may end at $\tau$, but its mode was entered earlier
and its atomic predicates are evaluated on its actual interval.
Combining these facts, we conclude
$\sigma,\partition{J},\alpha_i[\Lambda_i],i\models_\tau A_i$.
Together with $\sigma,\partition{J},\alpha_i,i\models_\tau C_i$, this yields
$\sigma,\partition{J},\alpha_i, i
\models_\tau C_i,\Lambda_i.A_i$
for every $i\le n$. Thus,
the trace of $\pi_{\mathrm{obs}}$ is satisfied.
For a terminal jump, clock agreement also gives
$\alpha_{n+1}\models C'$ from
$v_n'\restriction X_c\models C'$.
The labels of $q'$ hold at $|\partition{J}|$ and do not depend on
$\Lambda'$; they contain only $\ast$ and the closed marker.
Appending this terminal frame therefore gives
$\sigma,\partition{J},\alpha_0,0\models_\tau\trace{\pi}$.
Without a terminal jump this already follows from $\pi=\pi_{\mathrm{obs}}$.
Since $\alpha_0=\alpha$ was arbitrary and $\pi$ is accepting,
$\sigma$ is recognized by $\tableau{\varphi}{}$
with respect to $\partition{J}$.
Since $\partition{J}$ was arbitrary, this holds for every partition
induced by the trajectory.
\qedhere
\end{proofrules}
\end{enumerate}
\end{proof}

\restatetheorem{prop:zero-jump}

\begin{proof}
The continuous dynamics and invariants are unchanged.
A trajectory of $\hat H^{(n)}$ unfolds into one of $H$ by restoring the
initial discrete transitions represented by $\init$ and replacing each
added jump with its corresponding sequence of zero-time jumps.
Conversely, the state at the start of the first continuous transition
of an execution of $H$ is initial in $\hat H^{(n)}$, and each subsequent
sequence of discrete transitions has an added jump by the choice of $n$.
Both preserve the observed signal.
\end{proof}

\clearpage
\section{Supplementary Figures}

The figures below illustrate how clock-based constructions may require discrete
transitions at timing bounds even when no relevant subformula changes truth value.

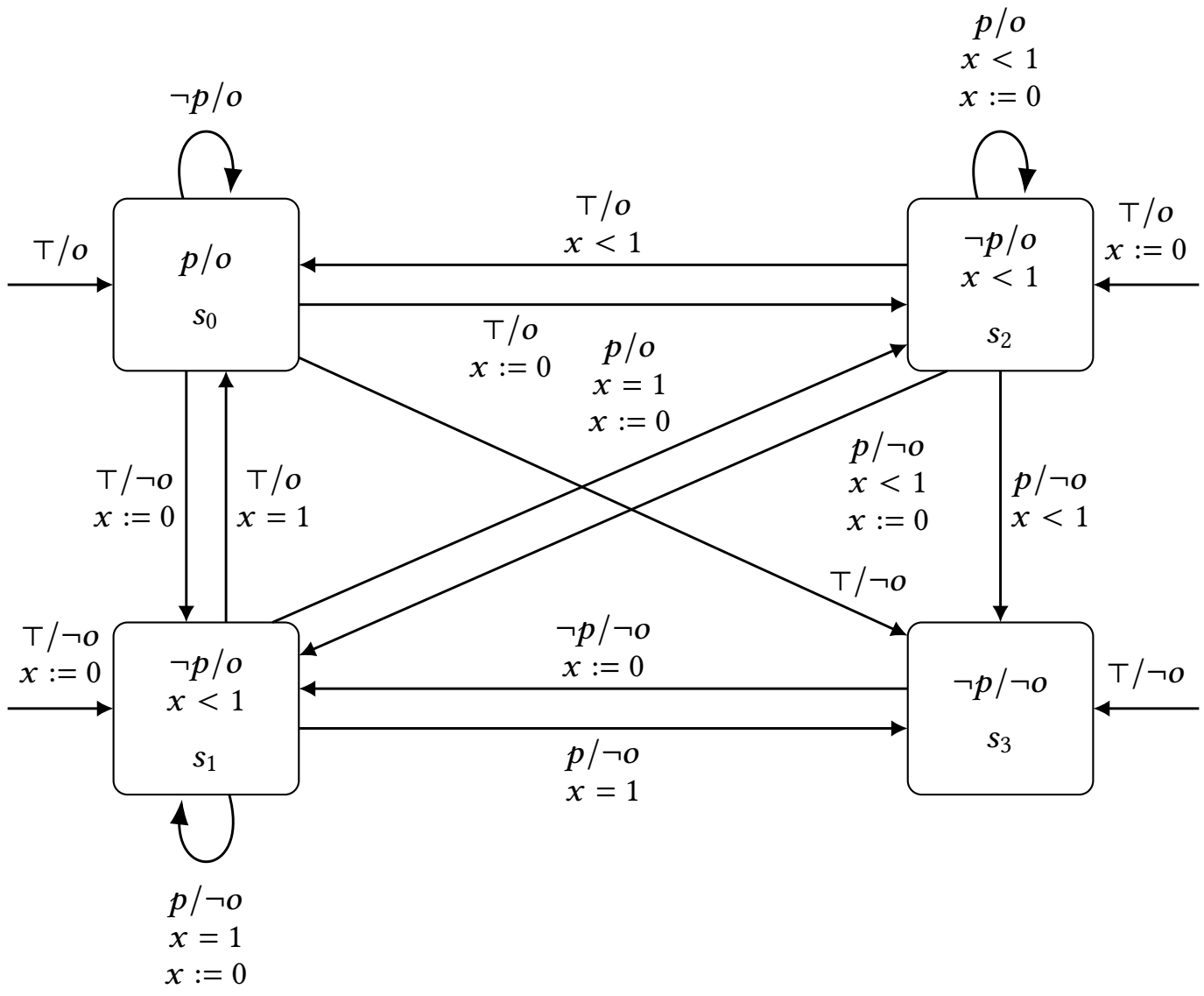
\begin{figure}[htbp]
\centering
\begin{tikzpicture}[%
  >=Latex,
  font=\scriptsize,
  x=2.5cm,y=2cm,
  box/.style={
    draw,
    rounded corners=3pt,
    minimum width=14mm,
    minimum height=13mm,
    align=center,
    inner xsep=2pt,
    inner ysep=2pt,
    outer sep=0pt
  },
  edge/.style={line width=0.55pt,-{Latex[length=1.25mm,width=1.25mm]}},
  lab/.style={
    font=\scriptsize,
    inner sep=2pt,
    text opacity=1,
    align=center
  },
  every node/.style={text depth=0.1ex},
]
\node[box] (s0) at (0,1.6)
{$p/o$\\[4pt]$s_0$};

\node[box] (s1) at (0,0)
{$\neg p/o$\\$x<1$\\[4pt]$s_1$};

\node[box] (s2) at (2.4,1.6)
{$\neg p/o$\\$x<1$\\[4pt]$s_2$};

\node[box] (s3) at (2.4,0)
{$\neg p/\neg o$\\[4pt]$s_3$};

\draw[edge] (-0.6,1.6) --
  node[lab,above, yshift=1mm] {$\top/o$} (s0);

\draw[edge] (-0.6,0) --
  node[lab,above, yshift=1mm] {$\top/\neg o$\\$x:=0$} (s1);

\draw[edge] (3,1.6) --
  node[lab,above, yshift=1mm] {$\top/o$\\$x:=0$} (s2);

\draw[edge] (3,0) --
  node[lab,above, yshift=1mm] {$\top/\neg o$} (s3);

\draw[edge] ([xshift=-1.5mm]s0.south) --
  node[lab,left] {$\top/\neg o$\\$x:=0$}
  ([xshift=-1.5mm]s1.north);

\draw[edge] ([xshift=1.5mm]s1.north) --
  node[lab,right] {$\top/o$\\$x=1$}
  ([xshift=1.5mm]s0.south);

\draw[edge] ([yshift=1.5mm]s2.west) --
  node[lab,above] {$\top/o$\\$x<1$}
  ([yshift=1.5mm]s0.east);

\draw[edge] ([yshift=-1.5mm]s0.east) --
  node[lab,below, xshift=-7mm] {$\top/o$\\$x:=0$}
  ([yshift=-1.5mm]s2.west);

\draw[edge] (s2.south) --
  node[lab,right] {$p/\neg o$\\$x<1$}
  (s3.north);

\draw[edge] ([yshift=1.5mm]s3.west) --
  node[lab,above] {$\neg p/\neg o$\\$x:=0$}
  ([yshift=1.5mm]s1.east);

\draw[edge] ([yshift=-1.5mm]s1.east) --
  node[lab,below] {$p/\neg o$\\$x=1$}
  ([yshift=-1.5mm]s3.west);

\draw[edge] ([yshift=1mm]s0.south east) -- node[lab,above, xshift=2cm,yshift=-0.8cm] {$\top/\neg o$} ([yshift=-1mm]s3.north west);

\draw[edge] ([xshift=5mm]s1.north) --
  node[lab,above, xshift=3mm, yshift=3mm] {$p/o$\\$x=1$\\$x:=0$}
  ([yshift=2mm]s2.south west);

\draw[edge] ([xshift=3mm]s2.south west) --
  node[lab,below, xshift=2cm,yshift=0.7cm] {$p/\neg o$\\$x<1$\\$x:=0$}
  ([yshift=4mm]s1.east);

\draw[edge, distance=7mm]
(s0) edge[loop above]
node[lab, yshift=1mm] {$\neg p/o$} ();

\draw[edge, distance=7mm]
(s1) edge[loop below]
node[lab, yshift=-1mm]
{$p/\neg o$\\$x=1$\\$x:=0$} ();

\draw[edge, distance=7mm]
(s2) edge[loop above]
node[lab, yshift=1mm]
{$p/o$\\$x<1$\\$x:=0$} ();
\end{tikzpicture}%
\caption{Temporal tester for $\eventually{(0,1)}$.}
\label{fig:temporal-tester}
\Description{A temporal tester with four states relates input p to output o.
State invariants and jump guards compare clock x with 1; arrows and
loops back to the same state may reset x to 0.}
\end{figure}

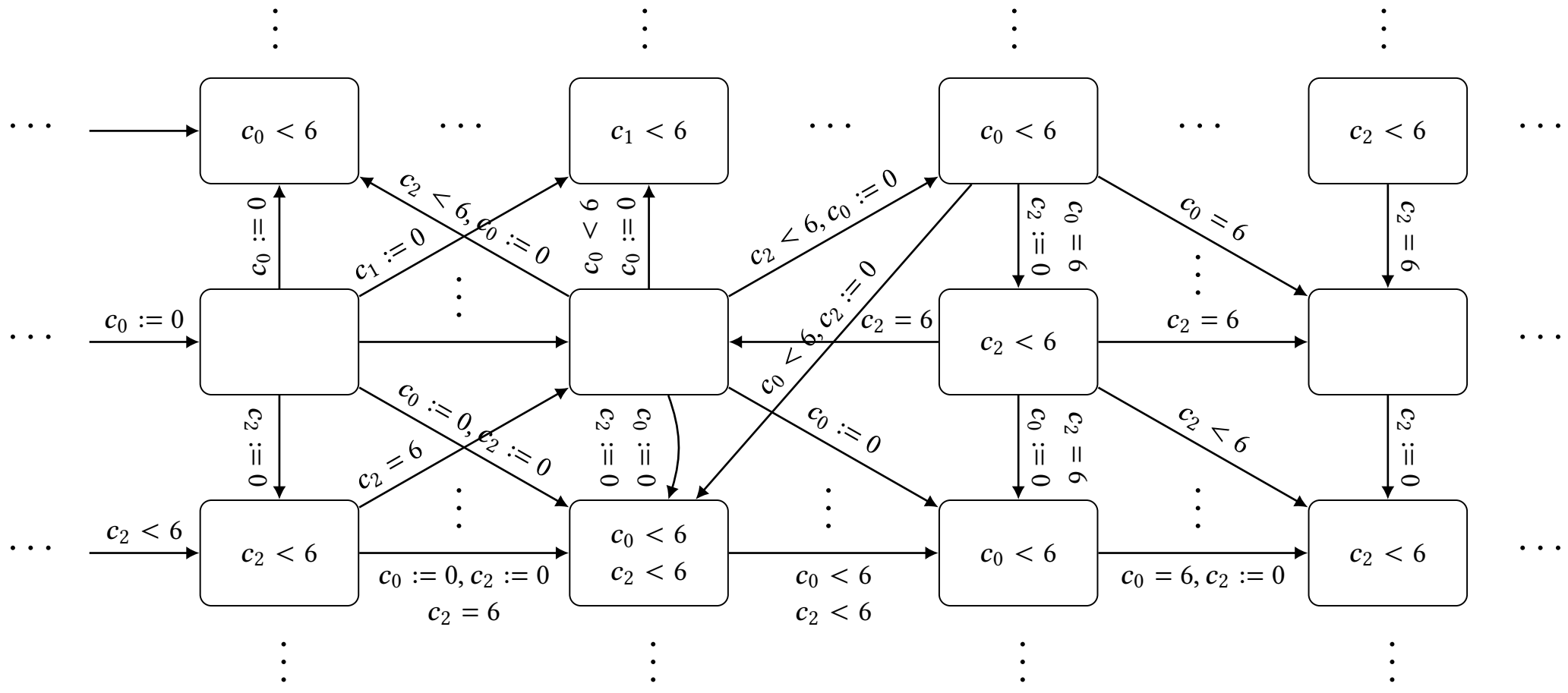
\begin{figure}[htbp]
\centering
\resizebox{\linewidth}{!}{%
\begin{tikzpicture}[
  >=Latex,
  font=\footnotesize,
    x=3.5cm, y=2cm,
  box/.style={
    draw,
    rounded corners=3pt,
    minimum width=15mm,
    minimum height=10mm,
    align=center,
    inner xsep=0pt,
    inner ysep=0pt
  },
  accept/.style={
    circle,
    draw,
    double,
    double distance=1.2pt,
    minimum size=10mm,
    inner sep=0pt,
  },
  edge/.style={line width=0.55pt,-{Latex[length=1.25mm,width=1.25mm]}},
  lab/.style={
    font=\footnotesize,
    inner sep=3pt,
    text opacity=1,
    align=center
  },
  every node/.style={text depth=0.1ex},
]

\node[box] (02) at (0, 2) {$c_0 < 6$};
\node[box] (01) at (0, 1) {};
\node[box] (00) at (0, 0) {$c_2 < 6$};

\node[box] (12) at (1, 2) {$c_1 < 6$};
\node[box] (11) at (1, 1) {};
\node[box] (10) at (1, 0) {$c_0 < 6$\\$c_2 < 6$};

\node[box] (22) at (2, 2) {$c_0 < 6$};
\node[box] (21)at (2, 1) {$c_2 < 6$};
\node[box] (20) at (2, 0) {$c_0 < 6$};

\node[box] (32) at (3, 2) {$c_2 < 6$};
\node[box] (31) at (3, 1) {};
\node[box] (30) at (3, 0) {$c_2 < 6$};

\node[left=1.8cm of 02] (dl2) at (0, 2) {};
\node[left=1.8cm of 02] (dl1) at (0, 1) {};
\node[left=1.8cm of 02] (dl0) at (0, 0) {};

\node[rotate=90, above=of 02, yshift=-0.2cm] at (0, 2) {\large $\cdots$};
\node[rotate=90, above=of 02, yshift=-0.2cm] at (1, 2) {\large $\cdots$};
\node[rotate=90, above=of 02, yshift=-0.2cm] at (2, 2) {\large $\cdots$};
\node[rotate=90, above=of 02, yshift=-0.2cm] at (3, 2) {\large $\cdots$};
\node at ($(02)!0.5!(12)$) {\large $\cdots$};
\node at ($(12)!0.5!(22)$) {\large $\cdots$};
\node at ($(22)!0.5!(32)$) {\large $\cdots$};

\node[rotate=90,xshift=13pt] at ($(01)!0.5!(11)$) {\large $\cdots$};
\node[rotate=90,xshift=19pt] at ($(21)!0.5!(31)$) {\large $\cdots$};
\node[rotate=90,xshift=13pt] at ($(00)!0.5!(10)$) {\large $\cdots$};
\node[rotate=90,xshift=13pt] at ($(10)!0.5!(20)$) {\large $\cdots$};
\node[rotate=90,xshift=13pt] at ($(20)!0.5!(30)$) {\large $\cdots$};

\node[left=-0.1cm of dl2] {\large $\cdots$};
\node[left=-0.1cm of dl1] {\large $\cdots$};
\node[left=-0.1cm of dl0] {\large $\cdots$};

\node[right=of 32, xshift=0.1cm] at (3, 2) {\large $\cdots$};
\node[right=of 32, xshift=0.1cm] at (3, 1) {\large $\cdots$};
\node[right=of 32, xshift=0.1cm] at (3, 0) {\large $\cdots$};

\node[rotate=90, below=of 00, yshift=0.1cm] at (0, 0) {\large $\cdots$};
\node[rotate=90, below=of 00, yshift=0.1cm] at (1, 0) {\large $\cdots$};
\node[rotate=90, below=of 00, yshift=0.1cm] at (2, 0) {\large $\cdots$};
\node[rotate=90, below=of 00, yshift=0.1cm] at (3, 0) {\large $\cdots$};

\draw[edge] (dl2) --
  node[lab, midway, above, sloped] {} (02);

\draw[edge] (dl1) --
  node[lab, midway, above, sloped] {$c_0 := 0$} (01);

\draw[edge] (dl0) --
  node[lab, midway, above, sloped] {$c_2 < 6$} (00);

\draw[edge] (01) --
  node[lab, midway, above, sloped] {$c_0 := 0$} (02);

\draw[edge] (01) --
  node[lab, midway, sloped, below] {$c_2 := 0$} (00);

\draw[edge] (01) --
  node[lab, midway, above, sloped,xshift=-0.7cm] {$c_1 := 0$} (12);

\draw[edge] (01) --
  node[lab, midway, above, sloped] {} (11);

\draw[edge] (01) --
  node[lab, midway, above, sloped] {$c_0 := 0,c_2 := 0$} (10);

\draw[edge] (00) --
  node[lab, midway, above, sloped, xshift=-0.7cm] {$c_2 = 6$} (11);

\draw[edge] (00) --
  node[lab, midway, below, sloped] {$c_0:=0, c_2:=0$\\$c_2 = 6$} (10);

\draw[edge] (11) --
  node[lab, midway, above, sloped] {$c_2 < 6,c_0 := 0$} (02);

\draw[edge] (11) --
  node[lab, midway, above, sloped] {$c_0 < 6$\\$c_0 := 0$} (12);

\draw[edge] (11) to[bend left=20]
  node[lab, midway, below, sloped, yshift=-0.1cm] {$c_0:=0$\\$c_2:=0$} (10);

\draw[edge] (11) --
  node[lab, midway, above, sloped] {$c_2 < 6,c_0:=0$} (22);

\draw[edge] (11) --
  node[lab, midway, above, sloped] {$c_0:=0$} (20);

\draw[edge] (10) --
  node[lab, midway, below, sloped] {$c_0 < 6$\\$c_2 < 6$} (20);

\draw[edge] (22) --
  node[lab, midway, above, sloped] {$c_0 < 6,c_2:=0$} (10);

\draw[edge] (22) --
  node[lab, midway, above, sloped] {$c_0 = 6$\\$c_2:=0$} (21);

\draw[edge] (22) --
  node[lab, midway, above, sloped] {$c_0 = 6$} (31);

\draw[edge] (21) --
  node[lab, midway, above, sloped, xshift=0.6cm] {$c_2 = 6$} (11);

\draw[edge] (21) --
  node[lab, midway, above, sloped] {$c_2 = 6$\\$c_0:=0$} (20);

\draw[edge] (21) --
  node[lab, midway, above, sloped] {$c_2 = 6$} (31);

\draw[edge] (21) --
  node[lab, midway, above, sloped] {$c_2 < 6$} (30);

\draw[edge] (20) --
  node[lab, midway, below, sloped] {$c_0 = 6,c_2 := 0$} (30);

\draw[edge] (31) --
  node[lab, midway, above, sloped] {$c_2 := 0$} (30);

\draw[edge] (32) --
  node[lab, midway, above, sloped] {$c_2 = 6$} (31);
\end{tikzpicture}%
}
\caption{Timed automaton for
$\varphi = (\always{(6,12)} u \ge 4) \release{} (t \le 5)$.}
\label{fig:timed-am}
\Description{A fragment of a timed automaton uses three clocks.
Its invariants keep selected clock values below 6, and some jumps reset
clocks or test equality to 6. Jumps at timing bounds need not coincide
with changes in atomic propositions.}
\end{figure}

\end{document}